\documentclass[12pt]{article}
\pdfoutput=1

\usepackage{draft,hyperref,tikz,cancel,subfig,float,multirow,caption}

\usepackage{graphicx,etoolbox}
\robustify{\includegraphics}

\usepackage[nosort]{cite}

\usetikzlibrary{snakes}
\usetikzlibrary{shapes.misc}

\newsavebox{\ns}
\newsavebox{\dbrane}
\newsavebox{\dbshort}
\renewcommand{\arraystretch}{1.2}
\def\be{\begin{equation}}
\def\ee{\end{equation}}
\def\bea{\begin{eqnarray}}
\def\eea{\end{eqnarray}}

\newcommand{\nn}{\nonumber}

\newcommand\diff{\mathrm{d}}

\newcommand\Ie{\textit{i.e.}}

\newcommand{\ii}{\mathrm{i}}

\newcommand{\Real}{\mathrm{Re}\, }

\newlength{\sswidth}

\newcommand\cI{\mathcal{I}}

\usepackage{color}
\usepackage{latexsym}
\usepackage{epsfig,amssymb,euscript, mathrsfs,cite}
\usepackage{arydshln} 
\usepackage{amsmath}
\definecolor{MyDarkBlue}{rgb}{0.15,0.15,0.45}

\begin{document}

\begin{titlepage}

\preprint{CALT-TH 2017-030 \\ PUPT-2539}

\begin{center}

\hfill \\
\hfill \\

\title{Spheres, Charges, Instantons, and Bootstrap:
\\
A Five-Dimensional Odyssey
}

\author{Chi-Ming Chang,$^a$ Martin Fluder,$^b$ Ying-Hsuan Lin,$^b$ Yifan Wang$^c$
}

\address{${}^a$Center for Quantum Mathematics and Physics (QMAP) \\
University of California, Davis, CA 95616, USA}
\address{${}^b$Walter Burke Institute for Theoretical Physics \\ California Institute of Technology,
Pasadena, CA 91125, USA}
\address{${}^c$Joseph Henry Laboratories\\ Princeton University, Princeton, NJ 08544, USA}

\email{wychang@ucdavis.edu, fluder@caltech.edu, yhlin@caltech.edu, yifanw@princeton.edu}

\end{center}

\abstract{We combine supersymmetric localization and the conformal bootstrap to study five-dimensional superconformal field theories. To begin, we classify the admissible counter-terms and derive a general relation between the five-sphere partition function and the conformal and flavor central charges. Along the way, we discover a new superconformal anomaly in five dimensions. We then propose a precise triple factorization formula for the five-sphere partition function, that incorporates instantons and is consistent with flavor symmetry enhancement.  We numerically evaluate the central charges for the rank-one Seiberg and Morrison-Seiberg theories, and find strong evidence for their saturation of bootstrap bounds, thereby determining the spectra of long multiplets in these theories. Lastly, our results provide new evidence for the $F$-theorem and possibly a $C$-theorem in five-dimensional superconformal theories.}

\vfill

\end{titlepage}

\eject

\tableofcontents

\section{Introduction}

Interacting conformal field theories in dimensions greater than four are notoriously difficult to study. Unlike the lower-dimensional cases, where there exist constructions based on renormalization group flows connecting weakly coupled theories in the ultraviolet to nontrivial fixed points in the infrared (such as the Ising model in $4-\epsilon$ dimensions~\cite{Wilson:1971dc} and the Caswell-Banks-Zaks fixed point in four dimensions~\cite{Caswell:1974gg,Banks:1981nn}), analogous considerations in five dimensions and beyond usually have weakly coupled descriptions in the infrared instead of in the ultraviolet, and are much more difficult to control.\footnote{Some evidence for the existence of nontrivial {\it infrared} fixed points in five dimensions was given in~\cite{Fei:2014yja}, using the $\epsilon$-expansion for the critical ${\rm O}(N)$ model in $d=6-\epsilon$ dimensions. The fixed point was shown to be unitary to all orders in ${1\over N}$, but it remained unclear whether unitarity holds non-perturbatively. In fact, numerical studies by the exact renormalization group equations suggested that the fixed point is metastable~\cite{Mati:2014xma,Mati:2016wjn}.
}

A powerful way to gain control over the renormalization group flows is to introduce supersymmetry and restrict to supersymmetric flows. Given $\cN=1$ supersymmetry in five dimensions, any weakly coupled quantum field theory is described by (gauged) vector multiplets coupled to (matter) hypermultiplets. Since the five-dimensional Yang-Mills coupling $g^2_{\text{\tiny YM}}$ has negative mass dimension, the theory is free in the infrared but strongly coupled in the ultraviolet. When special conditions on the matter content in relation to the gauge group are satisfied, nontrivial ultraviolet fixed points are argued to exist, by combining effective field theory arguments with string and M-theory constructions \cite{Seiberg:1996bd,Morrison:1996xf,Intriligator:1997pq}.\footnote{These conditions are often referred to as the Intriligator-Morrison-Seiberg conditions.  Their refinements were further given in~\cite{Jefferson:2017ahm}.
}
A class of examples comes from ${\rm USp}(2N)$ gauge theory coupled to $N_{\bf f} \leq 7$ hypermultiplets in the fundamental representation, and a single hypermultiplet in the antisymmetric representation.  Such theories turn out to have enhanced $E_{N_{\bf f}+1}$ flavor symmetry at their ultraviolet fixed points, and are referred to as the Seiberg exceptional superconformal field theories~\cite{Seiberg:1996bd,Morrison:1996xf,Intriligator:1997pq}.
 
The past two decades have seen developments in both extending the construction of~\cite{Seiberg:1996bd,Morrison:1996xf,Intriligator:1997pq} to much larger classes of five-dimensional $\cN=1$ superconformal field theories, and understanding the fixed-point physics.  A plethora of theories have been constructed using five-brane webs in type IIB string theory, and using isolated canonical three-fold singularities in M-theory; in cases, string dualities provide different infrared effective descriptions for the same ultraviolet fixed point~\cite{Witten:1996qb,Douglas:1996xp,Katz:1996fh,Aharony:1997ju,Aharony:1997bh,Leung:1997tw,Xie:2017pfl}.  Thanks to advances in the topological vertex formalism~\cite{Iqbal:2002we,Aganagic:2003db,Awata:2005fa,Iqbal:2007ii,Taki:2007dh,Awata:2008ed,Awata:2011ce} and supersymmetric localization~\cite{Nekrasov:2002qd,Nekrasov:2003rj,Pestun:2007rz,Kapustin:2009kz}, many BPS quantities can be computed systematically from either the infrared gauge theory or the five-brane web diagram, providing various nontrivial consistency checks of field theory dualities, symmetry enhancement, and the holographic correspondence for such theories~\cite{Kallen:2012cs,Hosomichi:2012ek,Kallen:2012va,Kim:2012ava,Kim:2012gu,Terashima:2012ra,Jafferis:2012iv,Imamura:2012xg,Iqbal:2012xm,Lockhart:2012vp,Imamura:2012bm,Kim:2012qf,Assel:2012nf,Bergman:2013ala,Bao:2013pwa,Hayashi:2013qwa,Taki:2014pba,Alday:2014rxa,Alday:2014bta,Hwang:2014uwa,Mitev:2014jza,Hayashi:2015xla,Bergman:2015dpa,Alday:2015lta,Alday:2015jsa,Gaiotto:2015una,Zafrir:2015ftn,Chang:2016iji,DHoker:2016ysh,DHoker:2017mds,Hayashi:2017jze}.  One notable quantity is the superconformal index that counts the ${1\over 8}$-BPS operators at the fixed point~\cite{Kim:2012gu,Iqbal:2012xm,Bergman:2013ala,Hwang:2014uwa}.  A subset of these BPS operators preserving extra supercharges furnish the Higgs branch chiral ring and are further studied in~\cite{Lambert:2014jna,Tachikawa:2015mha, Zafrir:2015uaa, Yonekura:2015ksa,Cremonesi:2015lsa,Bergman:2016avc}.  Another key quantity is the supersymmetric five-sphere partition function~\cite{Kallen:2012cs,Hosomichi:2012ek,Kallen:2012va,Kim:2012ava,Jafferis:2012iv,Imamura:2012xg,Lockhart:2012vp,Imamura:2012bm,Kim:2012qf}, which is conjectured to decrease monotonically under renormalization group flows, and plays a similar role to the trace anomalies in even spacetime dimensions and the sphere partition function in three dimensions \cite{Cardy:1988cwa,Capper:1974ic,Zamolodchikov:1986gt,Komargodski:2011vj,Myers:2010tj,Jafferis:2010un,Casini:2011kv,Jafferis:2011zi,Klebanov:2011gs,Elvang:2012st,Giombi:2014xxa,Cordova:2015vwa,Cordova:2015fha,Pufu:2016zxm}.

A new angle to study (super)conformal field theories is the conformal bootstrap of local correlation functions.  
 This method exploits unitarity, (super)conformal symmetry, and crossing symmetry to constrain the local operator spectrum and OPE coefficients, and has been successfully applied to various theories in dimensions from two to four~\cite{Rattazzi:2008pe,Rychkov:2009ij,Poland:2010wg,Poland:2011ey,ElShowk:2012ht,ElShowk:2012hu,Beem:2013qxa,Kos:2013tga,El-Showk:2014dwa,Chester:2014fya,Kos:2014bka,Caracciolo:2014cxa,Chester:2014mea,Beem:2014zpa,Simmons-Duffin:2015qma,Kos:2015mba,Chester:2015qca,Iliesiu:2015qra,Chang:2015qfa,Lemos:2015awa,Kim:2015oca,Lin:2015wcg,Kos:2016ysd,Chang:2016ftb,Collier:2016cls,Lin:2016gcl,Lemos:2016xke,Kravchuk:2016qvl,Li:2017ddj,Collier:2017shs,Li:2017agi,Cuomo:2017wme,Karateev:2017jgd,Dymarsky:2017yzx,Kravchuk:2017dzd}, to non-supersymmetric five-dimensional theories such as the ${ \rm O}(N)$ models~\cite{Nakayama:2014yia,Bae:2014hia,Chester:2014gqa,Li:2016wdp}, and to six-dimensional superconformal field theories~\cite{Beem:2015aoa,Chang:2017xmr}.  It is then natural to ask what can be learned about five-dimensional superconformal field theories, if we combine the power of supersymmetric localization with the bootstrap machinery.

Up until now, all known localization results in five dimensions have made no contact with correlation functions of local operators in the fixed-point theory.  One aim of the current paper is to develop a systematic procedure for extracting such correlation functions from the supersymmetric partition function, which can be computed by localization.  A reasonable starting point is to consider the two-point functions of the stress tensor and the conserved currents.  Because conformal symmetry and the conservation laws fix their tensor structures uniquely, and supersymmetry generates the two-point functions of all other operators in the associated superconformal multiplets, the only independent physical quantities are the overall coefficients, namely, the conformal central charge $C_T$ and the flavor central charge $C_J$.\footnote{Of course, the two-point functions by themselves are {\it a priori} ambiguous due to the freedom of rescaling the operators. We choose the canonical normalization for the stress tensor. For a non-Abelian flavor symmetry, we normalize the conserved currents as in Appendix~\ref{sec:NorCTCJ}; for an Abelian one, we demand that the minimal charge is one.
}
We shall study conformal field theories on the five-sphere, which is conformally flat and naturally regulates infrared divergences. Since the stress tensor is coupled to the spacetime metric, it is no surprise that under small perturbations of the five-sphere metric $ g_{\m\n}$, the variations in the free energy\footnote{In \cite{Klebanov:2011gs}, a general $F$-theorem in odd dimensions was conjectured for $\tilde F_{{\rm S}^d} \equiv (-1)^{d-1\over 2} \log |\cZ_{{\rm S}^d}|$, where $\tilde F_{{\rm S}^d}$ should be positive for unitary conformal field theories and decrease under renormalization group flows. Here we define $F_{{\rm S}^5}$ with the opposite sign of $\tilde F_{{\rm S}^5}$.
} 
\ie
F_{{\rm S}^5}\equiv -\log \cZ_{{\rm S}^5}
\fe
will depend on $C_T$ \cite{Closset:2012ru,Bobev:2017asb}. Similarly, the dependence of $F_{{\rm S}^5}$ on a background vector field $W_\m$ coupled to the conserved current will capture $C_J$~\cite{Closset:2012vg}.\footnote{See also \cite{Giombi:2015haa,Diab:2016spb,Giombi:2016fct} where $F_{{\rm S}^d}$,  $C_J$ and $C_T$ were computed directly for certain classes of non-supersymmetric weakly-coupled fixed points that exist for a range of spacetime dimensions including $d=5$.}  In this procedure, one should also carefully take into account the potential ambiguities in $F_{{\rm S}^5}$ due to local counter-terms in terms of the  background fields.

Although the connection between the five-sphere free energy and the central charges exists in bosonic conformal field theories as well, the relation is not practically useful unless the free energy {\it at the fixed point} can be reliably computed in a weakly-coupled (effective) description, such as in the infrared gauge theory.  Supersymmetry thus comes as a crucial ingredient.  Most of the known five-dimensional superconformal field theories admit gauge theory descriptions after mass deformations and renormalization group flows. The particular mass parameter $m_{\rm I}$ that triggers the desired flow is identified (up to a numerical constant) with ${1/g_{\text{\tiny YM}}^2}$~\cite{Seiberg:1996bd}.
This same mass parameter is also associated with the ${\rm U}(1)_{\rm I}$ instanton symmetry with the conserved current
\ie
J_{{\rm U}(1)_{\rm I}} = {\ii\over 8\pi^2} * {\rm Tr}(F \wedge F),
\fe
and sets the mass scale for the instanton particles in the gauge theory.  It has been conjectured that the supersymmetric five-sphere partition function of the superconformal field theory deformed by the instanton particle mass $m_{\rm I}$ is exactly captured by the infrared gauge theory\cite{Kim:2012ava,Jafferis:2012iv,Kim:2012qf}.\footnote{{\it A priori}, one may worry about potential higher derivative deformations of the gauge theory Lagrangian that may modify the localization result. 
	Because of the preserved supersymmetry, the partition function is protected from the $Q$-exact deformations of the localization Lagrangian. On general grounds, $Q$-closed deformations of the gauge theory Lagrangian should come from integrating out BPS particles (\emph{e.g.} W-bosons and point-like instantons). Since such 
	BPS states are all realized in the gauge theory and incorporated in the localization computation, it is plausible that the ``bare'' Lagrangian suffices for our purpose here. We emphasize that this is merely a heuristic argument.
	\label{fn:protect}
}
In particular, the undeformed free energy $F_{{\rm S}^5}$ for the fixed-point theory is obtained by sending $m_{\rm I}\to 0$, which is the strong coupling limit in the gauge theory.
Supersymmetry also puts strong constraints on the admissible counter-terms and their contributions to the deformed free energy.
 
We now summarize the key results and discoveries of this paper.

\paragraph{Extracting $C_T$ and $C_J$ from the five-sphere free energy (Section \ref{sec:PFtoCC})}
 
We can consider supersymmetric deformations of the five-sphere partition function by turning on (real) squashing parameters $\omega_i=1+a_i$ for the metric ($a_i=0$ gives the round-sphere metric), and mass parameters $M^a$ for the flavor symmetries. These can be systematically studied by coupling the superconformal field theory to off-shell background supergravity~\cite{Festuccia:2011ws,Hosomichi:2012ek}. It is straightforward to compute the dependence of the free energy $F_{{\rm S}^5}$ on these small supersymmetric deformations using conformal perturbation theory. We obtain the following expression for the quadratic terms in the squashing parameters $a_i$,
 \ie
 \left.F_{{\rm S}^5}\right|_{a_i^2}=-{\pi^2 C_T \over 1920} \left( \textstyle \sum_{i=1}^3 a_i^2-\sum_{i<j} a_i a_j \right), 
\label{CTi}
\fe
and similarly for the quadratic terms in the mass parameters $M^a$,
\ie
 \left. F_{{\rm S}^5}\right|_{M^2}={3\pi^2r^{2} C_J \over 256 } \D_{ab} M^a M^b ,
\label{CJi}
\fe
where $r$ is the radius of the sphere. These formulae allow us to determine the central charges $C_T$ and $C_J$ from the deformed five-sphere partition function, which is computable by localization in the infrared gauge theory.

\paragraph{Supersymmetric counter-terms (Section \ref{sec:S5PF})}

A subtlety in the computation of the five-sphere free energy is that there may be potential ambiguities due to finite local counter-terms, given by supersymmetric completions of various (mixed) Chern-Simons terms. For example, there are the counter-terms
\ie
W \wedge \diff W\wedge \diff W, \quad
W\wedge {\rm Tr} ({{R}} \wedge {{R}}) ,
\fe
for $W$ the ${\rm U}(1)$ background gauge field and ${{R}}$ the Riemannian curvature two-form. Such contributions capture our ignorance of contact-terms in the short distance limit of correlation functions. Their coefficients are quantized by demanding invariance under large background gauge transformations. Although the Chern-Simons terms are perfectly conformal, it turns out that in some cases their supersymmetric completions break conformal invariance. This means that if the relevant non-conformal contact-terms have fractional coefficients, we can choose to restore conformal invariance at coincident points at the expense of introducing a \emph{non-quantized} supersymmetric Chern-Simons counter-term, thereby breaking large background gauge symmetry. In other words, we \emph{cannot} simultaneously fulfill conformal invariance and background gauge invariance with supersymmetric regulators. This tension between conformal symmetry, background gauge symmetry, and supersymmetry is analogous to what happens in three dimensions~\cite{Closset:2012vg,Closset:2012vp}, and signals the existence of a new superconformal anomaly in $\cN=1$ theories in five dimensions. It turns out that these counter-terms do not affect the relations \eqref{CTi} and \eqref{CJi}, but are crucial in understanding flavor symmetry enhancement on the five-sphere.

\paragraph{Instanton contributions to the five-sphere partition function (Section \ref{Sec:En})}

Having understood how to extract the central charges from localizing the gauge theory path integral, we apply this procedure to the ``simplest" class of five-dimensional superconformal field theories, namely the rank-one Seiberg $E_{N_{\bf f}+1}$ and Morrison-Seiberg $\widetilde E_1$ theories. These theories are described in the infrared by a ${\rm USp}(2)$ super Yang-Mills theory coupled to $N_{\bf f}$ fundamental hypermultiplets. The infrared flavor symmetry ${\rm SO}(2N_{\bf f}) \times {\rm U}(1)_{\rm I}$ is enhanced to $E_{N_{\bf f}+1}$ at the ultraviolet fixed point.\footnote{Such ${\rm USp}(2)$ gauge theories come with a discrete theta parameter $\theta\in\{0,\pi\}$, prescribing how the two topological sectors characterized by $\pi_4({\rm USp}(2)) \cong \bZ_2$ are summed together in the path integral. For $N_{\bf f} > 0$, the massless infrared Lagrangian has symmetry ${\rm O}(2 N_{\bf f}) \cong \mZ_2 \rtimes {\rm SO}(2N_{\bf f})$ acting on the hypermultiplets, and the $\mZ_2$ normal subgroup which flips the sign of a single mass is equivalent to exchanging $\theta=0$ with $\theta=\pi$ \cite{Douglas:1996xp,Bergman:2013ala,Tachikawa:2015mha}.  Therefore, for $N_{\bf f}>0$, the two theories at $\theta=0$ and $\theta=\pi$ are equivalent. This $\bZ_2$ symmetry of the Lagrangian is not a symmetry of the gauge theory.
} 

As usual in localization, the gauge theory path integral reduces to an integral over BPS configurations weighted by their classical action and one-loop determinants. The novelty in five dimensions is that the BPS locus also involves the so-called \emph{contact instantons} for the gauge fields.\footnote{Upon circle compactification, the five-dimensional contact instantons reduce to the usual Yang-Mills instantons in four dimensions.
}
It was conjectured that for generic squashing, the contributions from contact instantons on the five-sphere are captured by gluing together three copies of the Nekrasov partition function $\cZ^{1,2,3}_{{\rm S}^1\times \mR^4}$~\cite{Lockhart:2012vp,Kim:2012qf}. In other words, the conjecture states that\footnote{Notice that each $\cZ_{{\rm S}^1\times \mR^4}$ contains perturbative and instanton contributions.
}
\ie
\cZ_{{\rm S}^5}=\int [\diff\lambda] e^{-{\cal F}^\vee_{\rm eff}(\lambda)}\cZ^1_{{\rm S}^1\times \mR^4}(\lambda) \cZ^2_{{\rm S}^1\times \mR^4}(\lambda)
 \cZ^3_{{\rm S}^1\times \mR^4}(\lambda),
 \label{s5fac}
\fe
where ${\cal F}^\vee_{\rm eff}(\lambda)$ is a cubic polynomial in $\lambda$, and is equal to the one-loop effective prepotential when every Coulomb branch parameter $\lambda$ is larger than all the masses of the hypermultiplets. We shall address some key issues   with \eqref{s5fac} and write down a precise formula, where the chemical potentials in the Nekrasov partition function $\cZ_{{\rm S}^1\times \mR^4}$ are substituted by the deformation parameters (squashing and masses).  Although it is well understood that the flavor chemical potentials for the hypermultiplets involve imaginary shifts,
 the shift for the U(1)$_{\rm I}$ instanton number chemical potential has not been determined in the literature.  We fix this ambiguity by demanding that the five-sphere partition function exhibits enhanced flavor symmetry $E_{N_{\bf f}+1}$ in the ultraviolet.  Our formula also renders key physical quantities real at arbitrary instanton order.

A generic squashed five-sphere can be viewed as a circle-fibration over a compact base manifold. The three ${\rm S}^1\times \mR^4$ patches are located at the three fixed points of the $ {\rm U}(1)\times {\rm U}(1)$ isometry of the base. In the weak coupling limit $m_{\rm I}\to \infty$, each $\cZ_{{\rm S}^1\times \mR^4}$ is a series in $e^{-m_{\rm I}}$, whose coefficients are computed by the Witten index of a certain ``generalized" ADHM quantum mechanics. However, we are interested in the strong coupling limit $m_{\rm I}\to 0$ in order to probe the superconformal fixed point. Ideally, one would hope to re-sum the entire series in $e^{-m_{\rm I}}$ and then re-expand around $m_{\rm I}=0$, akin to what has been done for the five-dimensional $\cN=2$ super Yang-Mills theory~\cite{Kim:2012ava,Lockhart:2012vp,Kim:2012qf}.  However, the re-summation for $\cN=2$ super Yang-Mills relies on the known modularity properties of the instanton series, that follow from a relation to the six-dimensional ${\cal N} = (2,0)$ ``parent'' theories.  Since such properties are not known for the ${\cal N} = 1$ gauge theories, we proceed by numerically evaluating the instanton contributions up to fourth order in the $e^{-m_{\rm I}}$ expansion, and setting $m_{\rm I} \to 0$.  Miraculously, we shall observe that the free energy $F_{{\rm S}^5}$ appears to converge (with squashing and mass deformations turned on), at least up to cubic order in the deformation parameters.  We thus reliably compute the undeformed free energy and the central charges $C_T$ and $C_J$.

\paragraph{Symmetry enhancement in the five-sphere partition function (Section \ref{Sec:En})}

Based on the fiber-base duality of the M-theory or type-IIB five-brane setup, it was observed in~\cite{Mitev:2014jza} that the Nekrasov partition function $\cZ_{\mR^4\times{\rm S}^1}$ for the Seiberg theories exhibit not just the manifest ${\rm SO}(2N_{\bf f}) \times {\rm U}(1)_{\rm I}$ flavor symmetry but the full enhanced $E_{N_{\bf f}+1}$.  This directly implies that the superconformal index
\ie
\cZ_{{\rm S}^1 \times {\rm S}^4}=\int [\diff \phi] \, \left|\cZ_{\mR^4\times{\rm S}^1}(\phi) \right|^2,
\fe
is also invariant under the Weyl group action of the enhanced flavor symmetry on the flavor fugacities~\cite{Kim:2012gu,Hwang:2014uwa}. Here $\phi$  collectively denotes the gauge field holonomies. The situation is less clear for the five-sphere free energy $F_{{\rm S}^5}$, due to 
the appearance of ${\cal F}^\vee_{\rm eff}(\lambda)$ in \eqref{s5fac}. However, we shall see that after incorporating appropriate local counter-terms to the gauge theory action, $F_{{\rm S}^5}$ is indeed invariant under the enhanced Weyl group action on the mass parameters. Our arguments for flavor symmetry enhancement in $F_{{\rm S}^5}$ formally requires the inclusion of instanton contributions to \emph{all} orders in $e^{-m_{\rm I}}$. Nevertheless, at low orders in $e^{-m_{\rm I}}$, we already find solid numerical evidence for the enhancement.  In particular, we compare the values of $C_J$ for the $E_{N_{\bf f}+1}$ flavor symmetry computed on one hand by ${\rm SO}(2N_{\bf f})$ mass deformations, and on the other by a ${\rm U}(1)_{\rm I}$ mass deformation, and find beautiful agreement.  These checks provide extra confidence for the convergence of the instanton expansion at strong coupling.

\paragraph{Superconformal bootstrap (Section \ref{Sec:Bootstrap})}

After reliably computing the values of the central charges $C_T$ and $C_J$ in the Seiberg theories, we make a connection to the conformal bootstrap of the four-point functions of moment map operators, the superconformal primaries that generate the flavor current multiplets. The semi-definite programming technique generates constraints on the intermediate spectrum and the OPE coefficients, some of which involving the stress tensor and the flavor current multiplets depend on the central charges $C_T$ and $C_J$. More specifically, demanding that the flavor symmetry be SU(2), $E_6$, $E_7$, and $E_8$, we obtain lower bounds on the central charges $C_T$ and $C_J$.  The bootstrap method is most powerful when a conformal field theory saturates the bound, in which case, one can use the extremal functional method to determine the non-BPS spectrum and their OPE coefficients \cite{ElShowk:2012hu,El-Showk:2014dwa}.  We find strong evidence for the saturation by the rank-one Seiberg theories, and thereby make predictions for the dimension of the lowest non-BPS operator appearing in the intermediate channel.  One can also in principle solve for the OPE data in these theories, going much beyond the data that supersymmetric localization alone can provide.  We demonstrate the power of this method in this paper, and invite for further explorations.

\section{Supersymmetric five-sphere partition function }
\label{sec:S5PF}

Any five-dimensional conformal field theory can be put on a five-sphere in a canonical manner by the stereographic mapping from $\mR^5$. 
The five-sphere free energy $F_{{\rm S}^5}$ should then capture physical information about the conformal field theory.  However, such quantities generally suffer from divergences and require regularization. The sphere being compact eliminates potential infrared divergences. However, at short distances the theory is indistinguishable from that on flat space, which leads to ultraviolet divergences of the form~\cite{Gerchkovitz:2014gta} 
\ie
F_{{\rm S}^5}\sim (\Lambda r)^5 +(\Lambda r)^3+(\Lambda r)+{\rm finite}.
\label{FS5div}
\fe
The explicit $r$-dependence breaks conformal invariance, but these divergent pieces can always be removed by introducing local diffeomorphism-invariant counter-terms of the schematic form
\ie
S_{\rm ct}=\Lambda^5 \int_{{\rm S}^5} \diff^5 x \sqrt{g}  +\Lambda^3 \int_{{\rm S}^5} \diff^5 x \sqrt{g} R +\Lambda \int_{{\rm S}^5} \diff^5 x \sqrt{g} R^2,
\label{S5ct}
\fe
where $R$ denotes the background curvature. Such counter-terms are obviously non-conformal. However, they compensate the non-conformal pieces in \eqref{FS5div}, giving rise to a regularization scheme, which actually \emph{preserves} conformal invariance on the sphere.  

This is to be contrasted with the even-dimensional case, in which the five-sphere free energies of conformal field theories are of the form
\ie
F_{{\rm S}^{2n}}\sim    (\Lambda r)^{2n}+(\Lambda r)^{2n-2} +\dots + a \log (\Lambda r)+{\rm finite},
\fe
where the logarithmic divergence \emph{cannot} be cancelled by a local diffeomorphism-invariant counter-term. In that case, the coefficient $a$ becomes a physical observable of the conformal field theory, the well-known trace anomaly. The presence of the $\log$-term also means that the finite term in $F_{{\rm S}^{2n}}$ is ambiguous. In contrast, the finite piece of $F_{{\rm S}^5}$ is unambiguous and an intrinsic observable of the conformal field theory, since there is no diffeomorphism-invariant counter-term dependent on the background curvature that could shift its value. In other words, with any ultraviolet diffeomorphism-invariant regularization scheme, we obtain the same answer for the \emph{finite} part of $F_{{\rm S}^5}$. From now on, we write $F_{{\rm S}^5}$ to mean the finite part after subtraction by the counter-terms in \eqref{S5ct}. In unitarity theories, $F_{{\rm S}^5}$ is a real number.

We can deform from the conformal five-sphere background either by (a) putting the conformal field theory on a general Riemannian manifold $(\cM,g_{\m\n})$, 
or (b) if the conformal field theory has global symmetry, by coupling the global symmetry currents $J_\m$ to background gauge fields $W_\m$. Upon such deformations, the free energy $F_{\cM}(g,W)$ typically suffers from ambiguities due to \emph{contact-terms} among the conserved currents and the stress tensor of the form
\ie
J_\m(p_1) J_\n(p_2) J_\rho (-p_1-p_2) &\ni {\ii\over 24\pi^2} \kappa_{JJJ} \epsilon_{\m\n\rho\sigma\lambda} p_1^\sigma p_2^\lambda,
\\
T_{\m\n}(p_1) T_{\rho\sigma}(p_2) J_\lambda(-p_1-p_2)  &\ni {\ii\over 384\pi^2}\kappa_{TTJ} \epsilon_{\lambda\A\B(\underline\m(\rho} p_1^\A p_2^\B
\left[p_1\cdot p_2 \D_{
\underline{\n})\sigma)}-p_{1\sigma)} p_{2\underline\n)}\right],
\label{nonSUSYct}
\fe
written in  momentum space for convenience. The coefficients $\kappa_{JJJ}$ and $\kappa_{TTJ}$ are real due to unitarity.\footnote{The factors of $\ii$ come from Wick-rotating the unitary Lorentzian theory.
}

Such ambiguities can be classified by the possible local counter-terms, which involve the background metric and gauge fields, subject to diffeomorphism  and gauge invariance.\footnote{We focus on finite (marginal) counter-terms as opposed to divergent (dimensionful) counter-terms.
}
For example, at cubic order in the background fields, we can have (mixed) five-dimensional Chern-Simons terms\footnote{Note that there is no five-dimensional pure gravitational Chern-Simons term.
}
\ie
{\ii\over 24\pi^2} \kappa_{JJJ} W\wedge F \wedge F,
\quad 
{\ii\over 192\pi^2} \kappa_{TTJ}W\wedge {\rm Tr} ({{R}}\wedge {{R}}),
\fe
which are conformal. On generic backgrounds, these counter-terms give nonzero contributions to $F_{\cM}(g,W)$, which is generally complex. 

In supersymmetric theories, the contact-term ambiguities are further constrained by insisting on a supersymmetric regularization scheme. Such a scheme can be systematically formulated by coupling the superconformal field theory to off-shell background $\cN=1$ supergravity, and its flavor symmetries to background $\cN=1$ vector multiplets.
We should only consider counter-terms in the supergravity that are manifestly supersymmetric. Due to the presence of additional fields in the background multiplets, there are a number of \emph{non-conformal} supersymmetric counter-terms.\footnote{Stated differently, the correlation functions of currents themselves have conformal contact-terms as in \eqref{nonSUSYct}, but supersymmetry implies that the correlation functions of the other operators in the current multiplets must have non-conformal contact-terms, if the former are present.
}
Depending on which symmetries of the theory we want to preserve at short distances (on nontrivial backgrounds), we have different choices for the coefficients of the counter-terms. As we shall see, the tension between supersymmetry, conformal invariance and invariance under large background gauge transformations leads to a new superconformal anomaly for five-dimensional superconformal field theories.\footnote{This is analogous to the superconformal anomaly of three-dimensional $\cN=2$ superconformal field theories studied in~\cite{Closset:2012vg,Closset:2012vp}. In that case, the anomalies originate from contact-terms in the two-point functions of current multiplets, whereas here they appear in the three-point functions.
}

\subsection{Off-shell supergravity and conformal field theories on five-spheres}
\label{sec:5dsugra}

Before going into the details of the supersymmetric free energy of five-dimensional superconformal field theories, let us first clarify how supersymmetry is preserved on a round sphere, and discuss the relevant off-shell supergravity theories.

When we place a five-dimensional superconformal field theory on the sphere via the stereographic map, the full superconformal symmetry $F(4)$ is {\rm formally} preserved. Hence, the background coupling is naturally described by conformal supergravity. In particular, the  superconformal stress tensor multiplet couples to the standard Weyl multiplet
\ie
g_{\m\n},~~ D,~~V_\m^{ij},~~v_{\m\n},~~\psi_\m^i,~~\chi^i,
\label{Wsm}
\fe
which contains the metric $g_{\m\n}$, the dilaton $D$,  an $\mf{su}(2)_{\rm R}$ gauge field $V_\m^{ij}$, a two-form field $v_{\m\n}$, the gravitino $\psi_\m^i$, and the dilatino $\chi^i$. Similarly, the superconformal current multiplets couple to five-dimensional vector multiplets
\ie
W^a_\m,~~M^a,~~\Omega^{ai}_\A,~~ Y^{aij},
\label{Vm}
\fe
where $W_\m$ is the background gauge field for the flavor symmetry, $M$ denotes the scalar mass parameter, $\Omega^i_\A$ is the gaugino, and $Y^{ij}$ is a triplet of auxiliary scalars. 

However, due to ultraviolet divergences, the partition function and the correlation functions need to be regulated.  On a round sphere, the maximal subalgebra that can be preserved by the regulators is ${\mf{su}}(4|1)$.  The eight supercharges are parametrized by symplectic-Majorana Killing spinors $\ve^i$ satisfying
\ie
\nabla_\m \ve^i= \C_\m t^i{}_j \ve^j,
\label{S5ks}
\fe
where $t= {\ii\over 2r}\sigma_3$, and $r$ is the radius of the sphere.  Therefore, to describe the coupling between the (regulated) superconformal field theory and the (deformed) five-sphere background, we should consider Poincar\'e supergravity, whose gauge algebra is ${\mf{su}}(4|1)$, instead of conformal supergravity which gauges the entire $F(4)$. 

Off-shell Poincar\'e supergravity can be obtained by introducing two gauge-fixing compensators to the standard Weyl multiplet in conformal supergravity~\cite{Bergshoeff:2004kh,Ozkan:2013nwa}. One of them needs to be a vector multiplet
\ie
\hat W_\m,~~\hat M,~~\hat \Omega_\A^i,~~ \hat Y^{ij},
\label{Cvm}
\fe
and the other can either be a hypermultiplet, a linear multiplet, or a non-linear multiplet. We choose the other compensator to be a linear multiplet
\ie
\hat L_{ij},~~\hat \varphi_\A^{i},~~\hat E^\m,~~\hat N,
\label{Clm}
\fe
where $\hat N$ and $\hat L_{ij}$ are scalars, $\hat E^\m$ is a divergenceless vector, and $\hat \varphi_\A^{i}$ is their fermionic partner. As we shall see, this ensures that upon gauge-fixing, the resulting Poincar\'e supergravity admits the desired supersymmetric five-sphere background. 

We gauge-fix the dilatation and superconformal transformations by setting\footnote{The standard Weyl multiplet also contains a gauge field $b_\m$ for the dilatation symmetry. Here we set $b_\m=0$ in order to fix the special conformal gauge symmetry.
} 
\ie
\hat M=\m ,~~ \hat \Omega^i_\A=0,~~
\label{gfDS}
\fe
and break the ${\rm SU}(2)_{\rm R}$ symmetry down to ${\rm U}(1)_{\rm R}$ by imposing
\ie
\hat L_{ij}=L t_{ij}, \quad \text{with}\quad t_{ij}= {\ii\over 2r}(\sigma_3)_{ij}.
\label{gfU}
\fe
Here $\m$ and $L$ are arbitrary constants of mass dimension one and two, respectively. To summarize, the off-shell Poincar\'e supergravity multiplet contains the component fields~\cite{Ozkan:2013nwa}
\ie
&g_{\m\n},~~ D,~~V_\m^{ij},~~v_{\m\n},~~\hat W_\m,~~\hat Y^{ij},~~\hat E^\m,~~\hat N,
~~\psi_{\m\A}^i,~~\chi^i_\A,~~\hat\varphi^i_\A.
\label{Psm}
\fe

A supersymmetric five-sphere background can be obtained by setting all fermionic fields in the Weyl multiplet and the compensator multiplets to zero, as well as by requiring the vanishing of their supersymmetric variations
\ie
\hspace{-.05in} \D \hat\varphi^i &=-\sD \hat  L^{ij} \ve_j+{1\over 2} \C^\m \ve^i \hat  E_\m +{1\over 2}\ve^i \hat  N 
+
2\C^{\m\n} v_{\m\n} \ve_j \hat  L^{ij}-6\hat   L^{ij} \eta_j , 
\\
\hspace{-.05in} \D \hat\Omega^i &=-{1\over 4} \C^{\m\n} F_{\m\n}(\hat W) \ve^i -{1\over 2}\sD \hat M \ve^i+\hat Y^i{}_j \ve^j-\hat  M\eta^i , 
\\
\hspace{-.05in} \D\psi_\m^i &=D_\m \ve^i +{1\over 2} v^{\n\rho} \C_{\m \n\rho} \ve^i-\C_\m \eta^i ,
\\
\hspace{-.05in} \D\chi^i &=\ve^i D-2\C^\rho \C^{\m\n}\ve^i   D_\m v_{\n \rho}+\C^{\m\n} F_{\m\n}{}^i{}_j(V) \ve^j-2\C^\m \ve^i \epsilon_{\m\n \rho\sigma\lambda}v^{\n \rho}v^{\sigma\lambda}+4\C^{\m\n} v_{\m\n} \eta^i.
\label{PSfv}
\fe
Here $D_\m$ denotes the covariant derivative
$
D_\m \equiv \nabla_\m  -V_\m .
$
 Furthermore, $F(\hat W)$ and $F(V)$ are the field strengths of $\hat W_{\mu}$ and $V_{\m}^{ij}$ respectively.
In particular, given the Killing spinor equation \eqref{S5ks}, all variations in \eqref{PSfv} vanish if we set
\ie
&g_{\m\n}=g_{\m\n}^{{\rm S}^5},\quad \eta^i=t^i{}_j \ve^j,
\quad \hat Y_{ij}=\m t_{ij},\quad \hat N={6\over r^2}L,
\label{S5psbkgd} 
\fe
and all other background fields to zero.

\subsection{Counter-terms and a new superconformal anomaly}
\label{sec:ctt}

The counter-term ambiguities in the supersymmetric five-sphere free energy are characterized by local diffeormorphism- and supersymmetry-invariant functions of the Poincar\'e supergravity multiplet \eqref{Psm}, and also the vector multiplets \eqref{Vm} if the superconformal field theory has global symmetries. Equivalently, we can write these counter-terms in terms of the Weyl multiplet \eqref{Wsm} and the two compensator multiplets, while keeping in mind the gauge-fixing conditions \eqref{gfDS} and \eqref{gfU}. For ease of notation, we take the latter approach. 

Let us first examine the dimensionful bosonic counter-terms in \eqref{S5ct}.  There is no supersymmetric completion of the leading $\Lambda^5$ divergence, which simply reflects the universal feature of vanishing cosmological constant for supersymmetric field theories. The $\Lambda^3$ divergent term can be completed by the supersymmetric Einstein-Hilbert action~\cite{Zucker:1999ej,Kugo:2000af,Kugo:2000hn,Bergshoeff:2004kh}. Finally, the $\Lambda$ divergent term is completed by the supersymmetrized ${{R}}^2$ interactions, which come with three parameters associated to the three independent structures: Ricci squared $R_{\m\n}R^{\m\n}$, Ricci-scalar squared $R^2$, and Weyl tensor squared $C_{\m\n\rho\sigma}C^{\m\n\rho\sigma}$~\cite{Ozkan:2013nwa,Bergshoeff:2011xn,Ozkan:2013uk,Butter:2014xxa}.\footnote{We adopt the convention $R_{\m\n\rho\sigma}g^{\m\rho}=-R_{\n\rho}$, such that the Riemannian curvature decomposes as
	\ie
	R_{\m\n\rho\sigma}=C_{\m\n\rho\sigma}
	-{2\over 3} R_{\m[\rho} g_{\sigma]\n} 
	+{2\over 3} R_{\n[\rho} g_{\sigma]\m} 
	+{1\over 6} g_{\m[\rho} g_{\sigma]\n} R.
	\fe
}
These divergent counter-terms allow us to regularize the five-sphere free energy while preserving the massive subalgebra $\mf{su}(4|1)$.  Due to the absence of marginal counter-terms purely in terms of the Poincar\'e supergravity multiplet \eqref{Psm}, the regularized supersymmetric five-sphere free energy is real and free of (finite) ambiguities.\footnote{We emphasize here that this reality property is true for unitary five-dimensional $\cN=1$ field theories that are not necessarily conformal.  In contrast, in the three-dimensional $\cN=2$ case, the free energy has a purely imaginary ambiguity~\cite{Closset:2012vg,Closset:2012vp}.
}

This freedom from ambiguities is no longer the case once we couple the conserved currents in the superconformal field theory to background vector multiplets $W_f$. The regularized free energy $F_{{\rm S}^5}$ can be shifted by marginal counter-terms that are all supersymmetric completions of various (mixed) Chern-Simons terms. Below we shall only keep track of the couplings between the scalars $M_f$ in the vector multiplet and the Riemannian curvature ${{R}}$.\footnote{We do not prove here, but conjecture that these furnish a complete basis of marginal counter-terms preserving $\cN=1$ supersymmetry. This conjecture is natural based on dimensional analysis and the available Chern-Simons terms.
}
By unitarity, the overall coefficients $\kappa_{TTJ}^{1,2,3}$ and $\kappa_{JJJ}$ are real numbers.

\paragraph{Flavor-${\rm \bf R^2}$ counter-term}

The first counter-term we find is given by
\ie
{\ii\over 24\pi^2} \kappa_{TTJ}^1 \int_{{\rm S}^5} \diff^5 x  \left( {9\over 64}\sqrt{g}
M_f R^2 -   W_f \wedge F(\bar V)\wedge F(\bar V)+\dots
\right),
\label{ctfr}
\fe
where $\bar V\equiv  2 r t_{ij} V^{ij}$. This counter-term is solely written in terms of the Weyl multiplet \eqref{Wsm}, the compensator linear multiplet \eqref{Clm}, and the vector multiplet $W_f$~\cite{Ozkan:2013nwa}. The $M_f R^2$ term breaks conformal invariance on the sphere and evaluates to a linear term in $M_f$ on the supersymmetric five-sphere background.

\paragraph{Flavor-${\rm \bf Ric^2}$ counter-term}
The second counter-term given by\footnote{Recall from \eqref{Cvm} and \eqref{Clm} that the hat denotes compensator fields.
}
\ie
\hspace{-.15in}
{\ii\over 24\pi^2} \kappa_{TTJ}^2\int_{{\rm S}^5}  \diff^5 x  \left( {15\over 8}\sqrt{g}
M_f \left(R_{\m\n}R^{\m\n} -{1\over 8}R^2\right)  
- W_f \wedge F(\hat W )\wedge F(\hat W)+\dots
\right),
\label{ctfric}
\fe
is another non-conformal counter-term built from the compensator vector multiplet in \eqref{Cvm}~\cite{Butter:2014xxa}.  On the round sphere, \eqref{ctfr} and \eqref{ctfric} give the same contribution.

\paragraph{Flavor-${\rm \bf Weyl^2}$ counter-term}

The third counter-term is given by
\ie
{\ii\over 12\pi^2} \kappa_{TTJ}^3\int_{{\rm S}^5}\diff^5 x & \left( 
{1\over 8}\sqrt{g}
M_f  C_{\m\n\rho\sigma}C^{\m\n\rho\sigma} - {1\over 16} W_f \wedge {\rm Tr} \left({{R}}\wedge {{R}}\right) \right.
\\
& 
\left.\hspace{.25in} - {1\over 12} W_f \wedge {\rm Tr} \big(F(V)\wedge F(V)\big)+\dots
\right),
\label{ctfweyl}
\fe
which is a conformal counter-term constructed in~\cite{Hanaki:2006pj}. It does not involve any compensator multiplets and has manifest conformal invariance.  This term vanishes on the supersymmetric round five-sphere background \label{S5psbkgd} since it is conformally flat.

\paragraph{${\rm \bf Flavor^3}$ counter-term}

Finally, the last counter-term reads \cite{Fujita:2001kv}
\ie
{\ii\over 24\pi^2} \kappa_{JJJ}^{abc} \int_{{\rm S}^5}\diff^5 x  \left(
{1\over 4}\sqrt{g} M_f^a M_f^b M_f^c R
-W_f^a \wedge  F(W_f^b)\wedge F(W_f^c)+\dots
\right),
\label{ctfff}
\fe
which is possible in the presence of a set of background vector multiplets $W_f^a$. It does not involve any of the compensator multiplets. The curvature coupling breaks conformal invariance and evaluates to a cubic term in $M_f^a$ on the supersymmetric five-sphere background.

The non-conformal counter-terms \eqref{ctfr}, \eqref{ctfric}, and \eqref{ctfff} have crucial consequences. Since they are supersymmetric completions of Chern-Simons terms, their coefficients $\kappa_{TTJ}^1$, $\kappa_{TTJ}^2$, and $\kappa_{JJJ}^{abc}$ must be properly quantized to respect large background gauge invariance.  Therefore, while the integral parts of the non-conformal contact-terms between the current and stress tensor multiplet can be cancelled by counter-terms, the fractional part is unambiguous and physical.  There is a tension between large background gauge invariance and conformal symmetry, analogous to what happens in three-dimensional superconformal field theories~\cite{Closset:2012vg,Closset:2012vp}.  This tension leads to a new superconformal anomaly in five-dimensional superconformal field theories.\footnote{We assume that the partition function is defined using an ultraviolet regulator that preserves background diffeomorphism invariance, background gauge invariance, and supersymmetry.
}

Phrased differently, if we start with a given five-dimensional superconformal field theory regularized by a particular scheme (\emph{e.g.} specified by some renormalization group flow), the contact-term in the three-point function between the stress tensor $T_{\m\n}$ and the scalar $N$ in the current multiplet that couples to $M_f$ takes the form\footnote{The factor of $\ii$ is due to $N$ being a pseudo-scalar in the Lorentzian superconformal field theory.
} 
\ie
& T_{\m\n}(p_1) T_{\rho\sigma}(p_2) N(-p_1-p_2) 
\\
&
\hspace{0in} \ni
{5\ii \over 64\pi^2}
\kappa^2_{TTJ}\left[
\left(p_{1\rho}p_{2(\m}-\D_{\rho(\m}(p_1\cdot p_2)\right) \left(p_{2\n)}p_{1\sigma}-\D_{\n)\sigma}(p_1\cdot p_2)\right)
\right]
\\
&
\hspace{.25in} + 
{3\ii \over 512\pi^2}
(3\kappa^1_{TTJ}-5\kappa^1_{TTJ})\left[
\left(p_{1\m}p_{1\n}-\D_{\m\n}(p_1\cdot p_1)\right) \left(p_{2\rho}p_{2\sigma}-\D_{\rho\sigma}(p_2\cdot p_2)\right)\right].
\fe
This contact-term breaks conformal invariance at short distances, which leads to a non-vanishing one-point function of the scalar $N$ on the five-sphere.\footnote{The operator $N$ has scaling dimension four, thus on the five-sphere it can mix with the identity operator by
	\ie
	N \to N+ c_1 R^2 \, \mathbf{1}.
	\fe
	This is another way to interpret the fact that the one-point function $\la N \ra_{{\rm S}^5}$ is non-vanishing, which is similar to what happens to exactly marginal operators in four-dimensional $\cN=2$ superconformal field theories~\cite{Gerchkovitz:2016gxx}.
}
If $\kappa^{1,2}_{TTJ}$ is non-integral, we have the option of removing this non-conformal contact-term by adding a counter-term of the form \eqref{ctfr} with the non-integral coefficient thereby breaking the background large gauge invariance explicitly.\footnote{The more precise definition of the five-dimensional Chern-Simons terms involve the extension to an auxiliary six-manifold. The fact that we are using a fractional Chern-Simons counter-term means that the partition function now depends on the data of the bulk six-manifold.}
Either way, the fractional part of $\kappa_{TTJ}$ is a measure of the superconformal anomaly and a physical quantity of the theory. The same can be said about the contact
term in the four point function
\ie
& T_{\m\n}(-p_1-p_2-p_3)   N(p_1) N(p_2)  N(p_3)  
\\
&\hspace{.5in} \ni 
{\ii\over 192\pi^2}
\kappa_{JJJ}\left[
(p_1+p_2+p_3)_\m (p_1+p_2+p_3)_\n -\D_{\m\n}(p_1+p_2+p_3)^2
\right].
\fe
As is usual with anomalies, the change of these contact-term coefficients under renormalization group flows will be protected and their fractional parts should also be preserved under dualities. We shall discuss these anomalies in general five-dimensional $\cN=1$ field theories in more details in a separate paper~\cite{Chang1}.

The fact that the above marginal counter-terms always involve either one or three (non-compensator) vector multiplets implies that $F_{{\rm S}^5}$ is \emph{unambiguous} when there is no coupling between the current multiplets and background gauge fields. With background vector multiplets, the ambiguities of $F_{{\rm S}^5}(W_f)$ are constrained to be $\cO(W_f)$ and $\cO(W_f^3)$. Most importantly, the $\cO(W_f^2)$ piece, which -- as we shall observe in Section~\ref{sec:CJderiv} -- carries information about the flavor central charge $C_J$, is free from ambiguities. The linear and cubic ambiguities in $F_{{\rm S}^5}(W_f)$ will also play an important role in Section~\ref{Sec:Enhancement} in identifying the symmetry enhancement in five-sphere localization.

\section{From the partition function to central charges}
\label{sec:PFtoCC}

In a conformal field theory in $d$ spacetime dimensions, the stress tensor two-point function is constrained by conformal symmetry to take the form
\ie\label{eqn:TT}
\la T_{\m\n}(x)T_{\rho\sigma}(0)\ra
= {C_T \over V^2_{\widehat {\rm S}^{d-1}}} {{\cal I}_{\m\n,\sigma\rho}(x)\over x^{2d}},
\fe
where $V_{\widehat {\rm S}^{d-1}} = {2\pi^{d/2} \over \Gamma(d/2)}$ is the volume of the unit $(d-1)$-sphere.  Similarly, if we consider a conformal field theory that has a flavor symmetry $G$ generated by the flavor currents $J_\m^a$. Then the two-point functions of the flavor currents are constrained by conformal symmetry and conservation laws to take the form
\ie\label{eqn:JJ}
&\la J^a_\mu(x)J^b_\nu(0)\ra = {C_J \over V^2_{\widehat {\rm S}^{d-1}}} {\D^{ab} I_{\mu\nu}(x)\over x^{8}}.
\fe
The normalization of the stress tensor $T_{\m\n}$ and the flavor currents $J_\m^a$ are specified in Appendix~\ref{sec:Convention}, and the explicit forms of the conformally covariant structures ${\cal I}_{\m\n,\sigma\rho}(x)$ and $I_{\mu\nu}(x)$ are given in \eqref{eqn:structureI}.  In the following, we denote by $C_J^G$ the flavor central charge of the flavor group $G$.

This section introduces precise relations between the central charges $C_T$ and $C_J$ and supersymmetric partition functions on five-spheres for five-dimensional superconformal field theories.  In particular, $C_T$ can be probed by adding a nontrivial squashing to the five-sphere metric, and $C_J$ can be probed by turning on a mass-deformation for flavor symmetries.  We first present the explicit formulae relating $C_T$ and $C_J$ to the respective deformations of the five-sphere partition functions, and test them for the (simple) case of a free hypermultiplet. In Sections~\ref{sec:CTderiv},~\ref{sec:CJderiv} and~\ref{sec:5dGaugeMass}, we prove these relations by explicitly coupling the theory to background supergravity. An important ingredient in the proof is the supersymmetric background for a five-sphere with generic squashing, which is discussed in Appendix~\ref{App:sqbg}.

Supersymmetric field theories can be put on nontrivial geometries by coupling to off-shell background supergravity~\cite{Festuccia:2011ws}.  When appropriate background superfields are turned on and enough supersymmetry is preserved, the partition function on such backgrounds is generally well-behaved along renormalization group flows.  Preserving supersymmetry is crucial because generic interacting superconformal field theories in five dimensions are strongly coupled, and computations are only viably performed in the infrared Lagrangian theory.  Supersymmetry allows one to extrapolate these infrared computations to the ultraviolet fixed points (see Section~\ref{Sec:En} for explicit examples).  In such curved backgrounds, the stress tensor is coupled to the background metric. Thus, the dependence of the partition function under infinitesimal deformations of the geometry away from the round sphere is captured by correlation functions of the stress tensor.  Conversely, the deformed partition function encodes the stress tensor correlators, in particular the stress tensor two-point function, and therefore determines the conformal central charge $C_T$ at the superconformal fixed points. The coupling of the stress tensor multiplet to the background supergravity multiplet will be introduced in Section~\ref{sec:CTderiv}. It depends on the squashing parameters $\omega_i = 1 + a_i, ~ i = 1, 2, 3$, which appear in the squashed sphere metric \eqref{5dmetric}.  According to the above discussion, we expect $C_T$ to be proportional to the quadratic order coefficients in the free energy $F=-\log \cZ$ under infinitesimal deformations from the round sphere.  The precise relation we derive is
\ie
\label{CTRelation}
\left.F\right|_{a_i^2}=-{\pi^2 C_T \over 1920} \left(\textstyle\sum_{i=1}^3 a_i^2-\sum_{i<j} a_i a_j\right).
\fe
From here on out, we shall set the five-sphere radius $r=1$.

For theories with flavor symmetry $G$, one can perform a different type of deformation, by coupling the flavor current multiplet to a background vector multiplet. The detailed coupling depends crucially on the mass parameters $M^a$ ($a=1,\ldots, {\rm dim}(G)$) and will be introduced in Sections~\ref{sec:CJderiv} and~\ref{sec:5dGaugeMass}. The precise relation between the flavor central charge $C^{G}_J$ for $G$ and the mass-deformed round-sphere partition function is given by
\ie
\label{CJRelation1}
\left. F\right|_{M^2}={3\pi^2  C^{G}_J\over 256} \D_{ab} M^a M^b.
\fe
Suppose the superconformal field theory flows to a gauge theory in the infrared with an (infrared) flavor group $K\subseteq G$ acting on the hypermultiplets. Then the mass term of the hypermultiplet in the action on the round ${\rm S}^5$ ($\omega_i=1$) is given by
\ie\label{eqn:hyperMassAction}
S_{\cal M}=\int \diff^5 x \sqrt{g} \Big(-\epsilon^{ij}\bar q_i \cM^2 q_j+2\ii  t^{ij}  \bar q_i  \cM q_j-2  \bar \psi \cM \psi\Big),
\fe
where ${\cal M} \in {\mathfrak k}\equiv Lie(K)$ is the mass matrix. The relation between the flavor central charge $C_J^{K}$ of the flavor group $K$ and the mass-deformed free energy is
\ie\label{CJRelation2}
\, \left. F\right |_{\cM^2}={3\pi^2 C_J^{K}\over 256   }{\rm Tr}(\cM^2),
\fe
where ${\rm Tr}(\cdot)$ is the Killing form defined in \eqref{eqn:Killing}. If the embedding $\iota:{\mathfrak k} \hookrightarrow {\mathfrak g}\equiv Lie(G)$ is known, the flavor central charge $C_J^{G}$ of the ultraviolet flavor group $G$ is related by the embedding index (reviewed in Appendix~\ref{Sec:Embedding})
\ie
C_J^{K} = I_{\mf {k} \hookrightarrow \mf{g} } C_J^{G}.
\fe

\paragraph{Free hypermultiplet: a check}

In $d$ dimensions, the values of $C_T$ for a scalar $\phi$ and a Dirac spinor $\psi$ are~\cite{Osborn:1993cr}
\ie
C_T^\phi = {d \over d-1}, 
\quad C_T^\psi = 2^{\lfloor {d\over2} \rfloor-1} d.
\fe
A free hypermultiplet in five dimensions consists of four scalars and a single Dirac spinor, and hence has central charge
\ie
C_T = 15.
\fe
For the flavor central charge $C_J$ for the ${\rm SU}(2)$ flavor symmetry of a free hypermultiplet, a calculation analogous to~\cite{Osborn:1993cr} gives
\ie
C_J^{{\rm SU}(2)} = {8 \over 3}.
\fe

The partition function of a mass-deformed free hypermultiplet on a squashed sphere is given by \cite{Imamura:2012xg,Imamura:2012bm}
\ie
{\cal Z} = {1 \over S_3( {\omega_1+\omega_2+\omega_3\over2} + \ii m | \omega_1, \omega_2, \omega_3 )},
\fe
where $\omega_1, \omega_2, \omega_3$ are the squashing parameters of the squashed five-sphere metric \eqref{5dmetric}, and the mass matrix in \eqref{eqn:hyperMassAction} is chosen to be $
{\cal M} = \ii m \sigma_3$.  The definition and relevant properties of the triple sine function $S_3$ are reviewed in Appendix~\ref{App:triplesine}.  Using the integral representation \eqref{SNint} for $S_3$, we find that the free energy $F \equiv -\log \cZ$ is given by the expression
\ie
F = - {\pi \ii \over 6} B_{3,3} \left( \tfrac{|\omega|}{2} + \ii m \mid \omega_1, \omega_2, \omega_3 \right) - \int_{\bR + \ii 0^+} {\diff\ell \over \ell} {e^{|\omega| \ell \over 2} e^{\ii m \ell} \over \prod_{i=1}^3 (e^{\omega_i \ell} - 1) }.
\fe
In the second term, the series expansions of the integrand in small $a_i \equiv \omega_i - 1$ and in $m$ commute with the integral.  To quadratic order in these parameters we thus find
\ie
F = F_0 - {\pi^2 \over 128} \left( \textstyle \sum_{i=1}^3 a_i^2 - \prod_{i \leq j} a_i a_j \right) - {\pi^2 \over 16} m^2+\cO(m^4),
\fe
which indeed gives the correct $C_T$ and $C_J$ applying the formulas \eqref{CTRelation} and \eqref{CJRelation2}.

\subsection{Linearized coupling for the stress tensor multiplet and $C_T$}
\label{sec:CTderiv}

The primary operator content of the five-dimensional $\cN=1$ stress tensor multiplet ${\cal B}[3;0,0;0]$ (Appendix~\ref{App:SuperRep} explains the notation) can be summarized as
\ie
\begin{array}{c@{\ \ \to\ \ }c@{\ \ \to\ \ }c@{\ \ \to\ \ }c@{\ \ \to\ \ }c  }
[0,0]_3^{(0)} 
& [0,1]_{7/2}^{(1)}
& [1,0]_4^{(2)} \oplus [0,2]_4^{(0)}
& [1,1]_{9/2}^{(1)}
& [2,0]_5^{(0)} \\
\Phi
& \Psi^i_\A
&  J^{ij}_\m \oplus B_{\m\n}
& S^i_{\m \A}
& T_{\m\n}
\end{array}
\fe
where each entry $[d_1,d_2]_\Delta^{(2J_{\rm R})}$ labels the representation of the corresponding bosonic conformal primary under the bosonic subgroup of $F(4)$~\cite{Buican:2016hpb,Cordova:2016emh}. The $d_1$, $d_2$ and $2J_{\rm R}$ are the Dynkin labels of the $\mf{so}(5)\times \mf{su}(2)_{\rm R}$. The Poincar\'e supercharges 
\ie
Q\in [0,1]^{(1)}_{1/2} 
\fe
acting on the superconformal primary $\Phi$ generate the entire stress tensor multiplet.

The normalization of the individual operators are fixed by requiring
\ie
& J_\m^{ij}=Q^{(i} \C_\m Q^{j)} \Phi,
\quad
B_{\m\n}=\epsilon_{ij}Q^i \C_{\m\n} Q^j \Phi,
\\
& T_{\m\n}=\epsilon_{ik}\epsilon_{j\ell}Q^{(i} \C_\m Q^{j)}Q^{(k} \C_\n Q^{\ell)} \Phi + \text{(conformal descendants)}, 
\label{stn}
\fe
where $Q$ satisfies the five-dimensional supersymmetry algebra (see Appendix~\ref{App:5dSUSY}), and $T_{\m\n}$ is canonically normalized as in Appendix~\ref{sec:NorCTCJ}.

Any five-dimensional superconformal field theory can be put on a curved manifold by coupling the stress tensor multiplet to the background $\cN=1$ standard Weyl multiplet, which contains the dilaton $D$, the metric $g_{\m\n}$, an $\mf{su}(2)_{\rm R}$ gauge field $V_\m^{ij}$, a 2-form  field $v_{\m\n}$, the gravitino $\psi_\m^i$, and the dilatino $\chi^i$. To preserve supersymmetry in the ``rigid limit" \cite{Festuccia:2011ws}, we require the background fields to have trivial variations under a subset of the $\cN=1$ supercharges. This requirement boils down to the conditions $\psi_\m^i=\chi^i=0$ as well as
\ie
\D\psi_\m^i &= D_\m \ve^i +{1\over 2} v^{\n\rho}\C_{\m \n\rho} \ve^i-\C_\m \eta^i=0,
\\
\D\chi^i &= \ve^i D-2\C^\rho \C^{\m\n}\ve^i D_\m v_{\n\rho}+\C^{\m\n} F_{\m\n}{}^i{}_j(V) \ve^j
-2\C^\m \ve^i \epsilon_{\m\n\rho\sigma\tau}v^{\n\rho}v^{\sigma\tau}+4\C^{\m\n} v_{\m\n} \eta^i=0.
\label{gvdv}
\fe
The preserved supercharges are parametrized by $\xi^i$ and $\eta^i$ subject to the above conditions. The round five-sphere background is given by \eqref{S5psbkgd}, but we do not need to worry about the compensator multiplets here.\footnote{Note that the compensator multiplets \eqref{Cvm} and \eqref{Clm} do not couple to physical operators in the superconformal field theory, and therefore do not enter the linearized coupling. The only way they could show up is in the potential counter-term ambiguities, which is absent here as explained in Section~\ref{sec:5dsugra}.
}  
In Appendix~\ref{App:sqbg}, we derive the general squashed supersymmetric five-sphere background, which is parametrized by three real parameters $\omega_{1,2,3}$. Here we shall show that the leading order effect of infinitesimal squashing on $F_{{\rm S}^5}$ is precisely captured by the conformal central charge $C_T$.

To linear order in perturbation theory around the round ${\rm S}^5$ background, we have the coupling\footnote{The coefficients in the linearized coupling are fixed by \eqref{stn} and \eqref{gvdv}. One can also extract the coefficients from the coupling between standard Weyl multiplet and hypermultiplets in \cite{Fujita:2001kv,Hanaki:2006pj,Alday:2015lta}.}
\ie
\D S=\int \diff^5 x  \sqrt{g}
\bigg(
-h^{\m\n}T_{\m\n}
+2
V_\m^{ij} J_{ij}^\m
-
iv_{\m\n} B^{\m\n}
+
{1\over 8}D \Phi
\bigg)+\cO(h_{\m\n}^2)
\label{linearweylstress}
\fe
We expand the squashed five-sphere partition function for the conformal field theory in the squashing parameters $a_i=\omega_i-1$ around $a_i = 0$. The first order terms in this expansion vanish as a consequence of the vanishing one-point functions of the stress tensor multiplet operators. At second order, the contributions are captured by the integrated two-point functions of $\Phi$, $B_{\m\n}$ and $J_\m^{ij}$. Note that a general feature of the supersymmetric squashed five-sphere background is that
\ie
h_{\m\n}\sim \cO(a_i^2),\quad v_{\m\n}\sim \cO(a_i),\quad V_{\m}^{ij}\sim \cO(a_i),\quad D\sim \cO(a_i).
\fe
Hence, we do not need to consider the two-point functions of $T_{\m\n}$ to second order in $a_i$.  Thus, from \eqref{linearweylstress}, it follows immediately that
\ie
\left.F\right|_{a_i^2}
&=-{1\over 2}\int \diff^5 x \sqrt{g}\int \diff^5 y \sqrt{g} \bigg(
4 V^\m_{ij} (x)V^\n_{k\ell} (y)\la J_\m^{ij}(x) J_\n^{k\ell}(y)\ra_{{\rm S}^5}
\\
&
-v^{\m\n}(x) v^{\rho\sigma}(y)\la B_{\m\n}(x) B_{\rho\sigma}(y)\ra_{{\rm S}^5}
+{1\over 64} D(x)D(y)\la \Phi(x) \Phi(y)\ra_{{\rm S}^5}
\bigg)
\label{squashedF}
\fe
to second order in the squashing parameters.

Now the squashed five-sphere background in Appendix~\ref{App:sqbg} expanded to linear order in $a_i$ gives
\ie
V^{ij}&= {\ii\over 2}(\sigma_3)^{ij} \sum_{i=1}^3 a_i y_i^2 \diff\phi_i + {\cal O}(a_i^2),
\\
v&=-{\ii\over 8}\diff((\sigma_3)_{ij}V^{ij}) + {\cal O}(a_i^2)=-{1\over 2} \sum_{i=1}^3 a_i y_i \diff y_i\wedge  \diff\phi_i + {\cal O}(a_i^2),
\\
D&=-4 \sum_{i=1}^3 a_i + {\cal O}(a_i^2).
\fe
To evaluate the integrated two-point functions (contracted with spacetime tensors) \eqref{squashedF}, it is convenient to transform to the stereographic coordinates (see Appendix~\ref{app:stereo}), in which case we have
\ie
V^{ij}&= {\ii(\sigma_3)^{ij}\over (1+x^2)^2}
\Big[
2a_1 \left(x_1 \diff x_2-x_2 \diff x_1\right)-  2a_2 \left(x_3 \diff x_4-x_4 \diff x_3\right) \\
&\qquad \qquad \qquad +a_3\left((x^2-2x_5^2-1)\diff x_5-2x_5 x_\m \diff x_\m\right)
\Big].
\fe 
The flat-space two-point functions of $\Phi$, $J^{ij}_\m$, and $B_{\m\n}$ are given by
\ie
\la \Phi(x) \Phi(y) \ra &= 
{C_T\over 480 \pi^4}{1\over |x-y|^6},
\\
\la J_\m^{ij}(x) J_\n^{k\ell}(y)\ra &= 
{C_T\over 640 \pi^4}\epsilon^{(i k} \epsilon^{j)\ell}{\D_{\m\n}-2{(x-y)_\m (x-y)_\n\over |x-y|^2} \over |x-y|^8}, 
\\
\la B_{\m\n}(x) B_{\rho\sigma}(y)\ra &= -{3C_T\over 1280 \pi^4}{{\rm Tr}[ \C_{\m\n}({\slashed x}-{\slashed y})\C_{\rho\sigma}({\slashed x}-{\slashed y})] \over |x-y|^{10}},
\fe
which can be derived from \eqref{stn} and the superconformal Ward identities. Using stereographic projection, we obtain the two-point functions on the five-sphere,
\ie
\la \Phi(x) \Phi(y) \ra_{{\rm S}^5}&={C_T\over 480 \pi^4}{1\over s(x,y)^6},
\\
\la J_\m^{ij}(x) J_\n^{k\ell}(y)\ra_{{\rm S}^5}
&={C_T\over 640 \pi^4}\epsilon^{(i k} \epsilon^{j)\ell}{\D_{ab}-2{(x-y)_a (x-y)_b\over |x-y|^2} \over s(x,y)^8} e^a_\m(x) e^b_\n(y),
\\
\la B_{\m\n}(x) B_{\rho\sigma}(y)\ra_{{\rm S}^5}
&=
-{3C_T\over 1280 \pi^4}{{\rm Tr}[ \C_{ab}({\slashed x}-{\slashed y})\C_{c d}({\slashed x}-{\slashed y})] \over |x-y|^2s(x,y)^{8}} e^a_\m(x) e^b_\n(x)e^c_\rho(y)e^d_\sigma(y),
\label{stress2pf}
\fe
where we introduced a frame
\ie
e^a_\m=\D^a_\m {2\over  1+x^2}, \quad  g_{\m\n}=e^a_\m e^b_\n \D_{ab},
\fe
and the ${\rm SO}(6)$ invariant distance,
\ie
s(x,y)={2|x-y|\over \sqrt{1+x^2}\sqrt{1+y^2}}.
\label{sxy}
\fe

To simplify the integrated two-point functions, we first take advantage of the ${\rm SO}(6)$ invariance of the measure, to rotate $y$ to $0$. This allows us to extract a factor of the unit five-sphere volume
\ie
V_{\widehat{\rm S}^5}=\pi^3 .
\fe
The remaining integral over $\int \diff^5 x \sqrt{g}$ has power law divergences at small $x$ and needs to be regularized. Here we use the dimensional regularization (analytic continuation using the Gamma function)
\ie
\int_0^\infty {\diff x \over (1+|x|^2)^a |x|^b}={{\CC({1-b\over 2})\CC({b+2a-1\over 2}})\over 2\CC(a)} .
\label{gammar}
\fe
The choice of regularization does not affect the result, as follows from the counter-term analysis in Section~\ref{sec:ctt}.

Let us start with the first term in \eqref{squashedF}, which has contributions from $\la JJ\ra$. It can be evaluated to
\ie
& 4\int \diff^5 x \sqrt{g} \,    V^\m_{ij} (x)V^\n_{k\ell} (0)\la J_\m^{ij}(x) J_\n^{k\ell}(0)\ra_{{\rm S}^5} =  {C_T\over 640\pi} a_3^2.
\fe
Next we consider the contribution from $\la BB\ra$, which leads to
\ie
& -\int \diff^5 x \sqrt{g}  \,
v^{\m\n}(x) v^{\rho\sigma}(y)\la B_{\m\n}(x) B_{\rho\sigma}(0)\ra_{{\rm S}^5} = {C_T\over 640 \pi} (a_1^2+a_2^2).
\fe
Finally, the remaining $\la\Phi\Phi\ra$ term integrates to
\ie
&  {1\over 64}\int \diff^5 x \sqrt{g}  \,
D(x)D(0)\la \Phi(x) \Phi(0)\ra_{{\rm S}^5} = -{C_T \over 1920 \pi}(a_1+a_2+a_3)^2 .
\fe
Putting everything together, we end up with the previously advertised equation
\ie
\boxed{
 \,	\left.F\right|_{a_i^2}=-{\pi^2 C_T \over 1920} \left(\textstyle\sum_{i=1}^3 a_i^2-\sum_{i<j} a_i a_j\right) . \,
}
\fe

\subsection{Linearized coupling for the current multiplet and $C_J$}
\label{sec:CJderiv}

Now, let us consider a five-dimensional $\cN=1$ superconformal field theory with flavor symmetry $G$. The primary operator content of a flavor current multiplet $\cD[3; 0, 0; 2]$ (Appendix~\ref{App:SuperRep} explains the notation) is
\ie
\label{FlavorContent}
\begin{array}{c@{\ \ \  \to \ \ \  }c@{\ \ \ \to \ \ \ }c }
[0,0]_3^{(2)} 
& [0,1]_{7/2}^{(1)}
& [1,0]_4^{(0)} \oplus [0,0]_4^{(0)} \\
L^a_{ij}
& \varphi^{ai}_\A
&  J^a_\m \oplus N^a
\end{array}
\fe
where $a$ is the adjoint label.  More explicitly, the primary operators $N^a$ and $J^a_\m$ are related to the superconformal primary $L^a_{ij}$ by 
\ie
N^a 
= {\ii\over 12}C^{\A\B} 
Q_\A^j Q_\B^k L^a_{jk}
,\quad 
J^a_\m  ={\ii\over 12}(\C_\m C)^{\A\B}  Q_\A^j Q_\B^k L^a_{jk}.
\label{Qdflavor}
\fe

We can introduce mass deformations for the superconformal field theory whilst still preserving supersymmetry, by coupling to background off-shell supergravity and vector multiplets. The values of the background fields will be constrained by requiring that their off-shell supersymmetry variations vanish. Here we are interested in mass-deformed five-dimensional superconformal field theory on the five-sphere, in which case the background supergravity fields take values as in \eqref{S5psbkgd}.  The coupling to a background vector field with components
\ie
W^a_\m,~~M^a,~~\Omega^{ai},~~ Y^{aij},
\fe
takes the form~\cite{Fujita:2001kv}
\ie
\D S=\int \diff^5 x \sqrt{g}
\left( 2Y^{aij}  L_{ij}^a - W^a_\m  J^{a\m}+ M^a N^a+\dots
\right),
\label{c2bv}
\fe
where the terms involving fermions are suppressed. The supersymmetry variation of the background gaugino is
\ie
\D\Omega^i=-{1\over 4}\C^{\m\n} F_{\m\n}(W) \ve^i -{1\over 2} \C^\m D_\m M \ve^i +Y^i{}_j \ve^j+M t^{i}{}_j  \ve^j,
\label{gauginovar}
\fe
where $\ve^j$ is a symplectic-Majorana spinor satisfying the Killing spinor equation \eqref{S5ks} on the five-sphere. It is easy to see that
\ie
W_\m=\Omega^i=0, \quad Y^{ij}=-M t^{ij},
\fe
with $M$ a constant element of $\mf g=Lie(G)$, gives the desired supersymmetry-preserving mass deformation
\ie\label{eqn:SUSYMassDeformation}
\D S=\int \diff^5 x \sqrt{g}
M^a \left( N^a-2 t^{ij} L_{ij}^a 
\right).
\fe

As explained in Section~\ref{sec:ctt}, while the $\cO(M)$ and $\cO(M^3)$ terms in the mass-deformed five-sphere free energy are ambiguous, the $\cO(M^2)$ dependence is not, and is in particular determined by
\ie
& \hspace{-.15in} \left.F\right|_{M^2}
= -{1\over 2}M^a M^b \int \diff^5 x \sqrt{g}\int \diff^5 y \sqrt{g}\left(4 t^{ij} t^{k\ell} \la L_{ij}^a(x) L_{k\ell}^b(y) \ra_{{\rm S}^5}+\la N^a(x)N^b(y)
\ra_{{\rm S}^5}\right) .
\fe
The two-point functions of $N^a$ and $L^a_{ij}$ on flat space are related to the two-point function of $J_\m$ by \eqref{Qdflavor} and the superconformal Ward identities. The two-point functions on the five-sphere are then given by stereographically projecting the flat space two-point functions, which results in
\ie
\la L^a_{ij}(x) L^b_{k\ell}(y) \ra_{{\rm S}^5}={3C^{G}_J\over 1024\pi^4}{ \D^{ab} \epsilon_{i(\underline{k}}\epsilon_{j\underline{\ell})}\over s(x,y)^6},\quad\la N^a(x)N^a(y)\ra_{{\rm S}^5}= -{27C^{G}_J\over 256\pi^4}{ \delta^{ab}\over s(x,y)^8},
\fe
where $s(x,y)$ is the geodesic distance \eqref{sxy} on the sphere. Performing the double integral by dimensional regularization using \eqref{gammar}, we end up with
\ie
\boxed{ \, \left.F\right|_{M^2}={3\pi^2   C^{G}_J\over 256} \D_{ab} M^a M^b .
	\, }\label{FCJ}
\fe

\subsection{Relation to the gauge theory partition function}
\label{sec:5dGaugeMass}

As explained in the introduction, part of the ultraviolet global symmetries are preserved in the infrared gauge theory phase. They are realized either by the flavor symmetries of the hypermultiplets or by the topological ${\rm U}(1)_{\rm I}$ instanton symmetry. To compute $C_J$ from \eqref{FCJ} by localization for either of these subgroups (see Section~\ref{Sec:En}), we need to correctly normalize the mass matrix, or equivalently normalize the conserved current multiplet (now no longer superconformal) in the infrared gauge theory.  

We fix the normalization by demanding that the current $J_\m^a$ is canonically normalized as in \eqref{eqn:QJS} and \eqref{eqn:QfJ} . This prescription is unambiguous (irrespective of the gauge coupling) thanks to the Ward identity of the flavor symmetry. The normalization of the other bosonic conformal primaries in the current multiplet are fixed by supersymmetry \eqref{Qdflavor}.  In particular, the OPE between the moment map operators, at the ultraviolet fixed point, takes the form
\ie
\hspace{-.15in} L^a_{ij}(x) L^b_{k\ell}(y) &= {3 C_{J} \over 1024\pi^4} { \D^{ab} \epsilon_{i(\underline{k}}\epsilon_{j\underline{\ell})}\over |x-y|^6}
+ {2 \ii \over \sqrt{3}} f^{abc}
\left(\epsilon_{i(k} L^c_{\ell )j}(y)+\epsilon_{j(k}L^c_{\ell )i}(y)\right)
{1\over |x-y|^3}
+\dots ,
\label{LLope}
\fe
with the prescribed normalization. Below we shall identify the normalized infrared current multiplets associated to both the hypermultiplet flavor symmetry and the ${\rm U}(1)_{\rm I}$ instanton symmetry.

Suppose we gauge a subgroup $H \subset {\rm USp}(2N)$.  The commutant (centralizer) then gives the flavor symmetry of the hypermultiplets. 
For simplicity, let us focus on one simple factor $K$ of the commutant subgroup, and consider the charged hypermultiplets that transform in the representation $(R_H,R_K)$ of $H \times K$. The action for $N$ massive (gauged) hypermultiplets on the five-sphere is \cite{Hosomichi:2012ek}
\ie
S&=\int \diff^5 x \sqrt{g} 
\bigg(
\epsilon^{ij}D_\m \bar q_i D^\m q_j-\epsilon^{ij}\bar q_i \sigma^2 q_j+{15\over 4}\epsilon^{ij}\bar q_i q_j
-2\ii\bar\psi \sD \psi-2\bar\psi \sigma \psi-4\epsilon^{ij}\bar\psi \lambda_i q_j
\\
&
\hspace{1.5in} -\ii \bar q_i D^{ij} q_j -\epsilon^{ij}\bar q_i \cM^2 q_j+2\ii  t^{ij}  \bar q_i  \cM q_j-2  \bar \psi \cM \psi
\bigg),
\label{hyperaction}
\fe
where we labeled fields in the dynamical vector multiplets as
\ie
A_\m,~~\sigma,~~\lambda_i,~~D^{ij}.
\fe
We have suppressed the $R_H$ and $R_K$ indices $\dot A=1,\cdots ,{\rm dim}(R_H)$ and $ A=1,\cdots ,{\rm dim}(R_K)$, and defined
\ie
\bar\psi_{\dot B B} \equiv \psi^{\dot A A}{\cal I}_{\dot A\dot B}\widetilde {\cal I}_{ A B},
\quad
\bar q_{\dot BB} \equiv q^{\dot AA}{\cal I}_{\dot A\dot B}\widetilde {\cal I}_{ A B},
\fe
with the invariant tensors ${\cal I}_{\dot A\dot B}$ and $\widetilde {\cal I}_{A B}$ given by restrictions of the invariant tensor of ${\rm USp}(2N)$.  To be explicit, the ${\cal I}_{\dot A\dot B}$ and $\widetilde {\cal I}_{ A B}$ are normalized such that
\ie
&\cI^{\dot A \dot B}\equiv \cI_{\dot A \dot B}^*,\quad 
\cI_{\dot A \dot B} \cI^{\dot B \dot C}=-\D_{\dot A}^{ \dot C},
\\
&\widetilde\cI^{\dot A \dot B}\equiv \widetilde\cI_{\dot A \dot B}^*,\quad 
\widetilde\cI_{\dot A \dot B} \widetilde\cI^{\dot B \dot C}=-\D_{\dot A}^{ \dot C}.
\fe
The mass matrix $\cM$ is Hermitian, takes values in $\mf k=Lie(K)$, and couples in \eqref{hyperaction} to the moment map operators as
\ie
L^a_{ij} (T^a)^A{}_B \propto \ii  q_i^{\dot A A}q_j^{\dot B}{}_B{\cI}_{\dot A\dot B},
\label{Lntbd}
\fe
where $T^a$ denote the Hermitian generators of $\mf k$ with normalization given in \eqref{eqn:GTnor}.

The normalization in \eqref{Lntbd} needs to be specified before we use \eqref{FCJ} to compute $C_J$ from localization. Since the normalization is independent of the gauge coupling, we can work in the weak coupling limit, and use the two-point functions for free hypermultiplets,\footnote{The normalization here is fixed by \eqref{hyperaction}.}
\ie
\la   q^A_i(x) q^B_j(y) \ra={1\over 2}C_\phi {\Omega^{AB}\epsilon_{ij}\over |x-y|^3},\quad
\la  \psi^A (x) \psi^B (y) \ra={\ii\over 4}C_\psi\Omega^{AB} {C\C_\m (x^\m-y^\m) \over |x-y|^5},\quad
\label{fh2p}
\fe
with \ie
C_\phi={1\over 3}C_\psi={1\over 8\pi^2}.
\fe
Therefore, the normalized current is given by
\ie
J_\m=  2   (\ii q^{i\dot A A} \pa_\m q_i^{\dot B B} +\psi^{\dot A A} \C_\m \psi^{\dot B B}){\cI}_{\dot A\dot B} T^a_{AB},
\fe
and the normalized moment map operators are fixed by supersymmetry to be
\ie
L^a_{ij}=  \ii  q_i^{\dot A A}q_j^{\dot B B}{\cI}_{\dot A\dot B} T^a_{AB}.
\fe
Hence, $M^a$ in \eqref{c2bv} is identified with
\ie
M^a= {\rm Tr}(\cM T^a).
\fe
From \eqref{FCJ}, we have 
\ie
\left. F\right |_{\cM^2}={3\pi^2 C_J^K\over 256   }{\rm Tr}(\cM^2),
\fe
or equivalently,
\ie
\boxed{
	\, \left. F\right |_{\cM^2}={3\pi^2 C_J^K \over 512  C_2(R_{K}) }{\rm tr}_{R_{K}}(\cM^2) ,\,
}
\fe
where $\cM$ is the mass matrix in \eqref{hyperaction}, and $C_2(R_{K})$ is the Dynkin index associated with $R_K$.

At the ultraviolet fixed points of five-dimensional theories, the global symmetry is typically enhanced to a larger group $G \supset K$. If $G$ is simple (or a simple factor), then we can obtain $C_J^G$ from $C_J^K$ through the embedding index
\ie
C_J^{K} = I_{\mf {k} \hookrightarrow \mf{g} } C_J^G,
\fe
and similarly when $K$ is replaced by the ${\rm U}(1)_{\rm I}$ instanton symmetry. In Appendix~\ref{Sec:Embedding}, we provide some details about the embedding indices appearing in the Seiberg exceptional superconformal field theories.

\section{Seiberg and Morrison-Seiberg exceptional theories}
\label{Sec:En}

A special class of five-dimensional superconformal field theories are the theories with exceptional $E_{n}$ ($n = 1,\dotsc,8$) flavor symmetry proposed by Seiberg~\cite{Seiberg:1996bd}, and $\widetilde E_1 = {\rm U}(1)$ by Morrison and Seiberg~\cite{Morrison:1996xf}.  For $n<6$, the  flavor groups are $E_5={\rm {\rm SO}(10)}$, $E_4={\rm {\rm SU}(5)}$, $E_3={\rm {\rm SU}(3)}\times {\rm {\rm SU}(2)}$, $E_2={\rm {\rm SU}(2)}\times {\rm U}(1)$, and $E_1={\rm {\rm SU}(2)}$. For each flavor group, there is a family of theories of labeled by their ranks, and the rank-$N$ theory has an $N$-dimensional Coulomb branch, ${\mathbb R}^N/W_{{\rm USp}(2N)}$, where $W_{{\rm USp}(2N)}$ is the Weyl group of ${{\rm USp}(2N)}$. 
 The family of Seiberg $E_8$ theories arise in the low energy limit of $N$ D4 branes probing a nine-dimensional $E_8$ singularity, which is constructed out of seven D8-branes on top of an O8-orientifold plane at infinite string coupling (Table~\ref{table:BraneDirections}).

\begin{table}[h]
\centering
\begin{tabular}{|c|cccccccccc|} \hline
&0&1&2&3&4&5&6&7&8&9\\
\hline\hline
D8/O8 &$\times$&$\times$&$\times$&$\times$&$\times$&$\times$&$\times$&$\times$&$\times$&  \\
\hline
D4&$\times$&$\times$&$\times$&$\times$&$\times$&&&&&\\
\hline
\end{tabular}
\caption{The D4-D8/O8 brane system in type I' string theory.}
\label{table:BraneDirections}
\end{table}

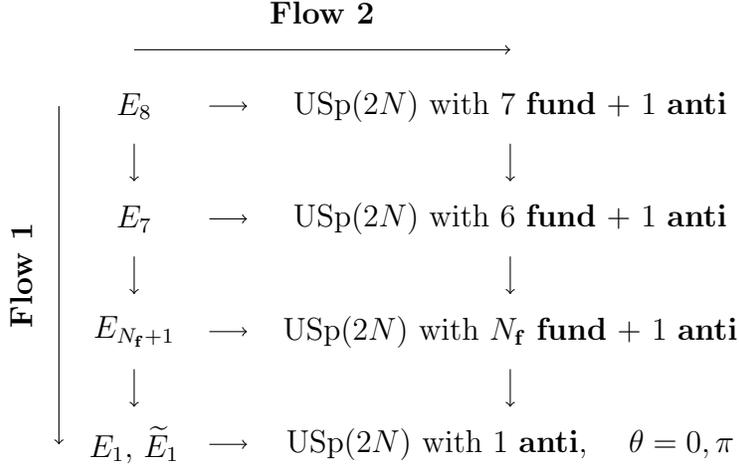
\begin{figure}[h]
\centering
\begin{tikzpicture}

\draw [->] (0,.75)--(5,.75);
\node at (2.5, 1.25) {{\bf Flow 2}};

\draw [->] (-1,0)--(-1,-4.5);
\node [rotate=90] at (-1.5, -2.25) {{\bf Flow 1}};

\node at (0, 0) {$E_8$};
\draw [->] (0,-.5)--(0,-1);
\node at (0, -1.5) {$E_7$};
\draw [->] (0,-2)--(0,-2.5);
\node at (0, -3) {$E_{N_{\bf f}+1}$};
\draw [->] (0,-3.5)--(0,-4);
\node at (0, -4.5) {$E_1$, $\widetilde E_1$};

\node at (5, 0) {${\rm USp}(2N)$ with 7 {\bf fund} + 1 {\bf anti}};
\draw [->] (5,-.5)--(5,-1);
\node at (5, -1.5) {${\rm USp}(2N)$ with 6 {\bf fund} + 1 {\bf anti}};
\draw [->] (5,-2)--(5,-2.5);
\node at (5, -3) {${\rm USp}(2N)$ with $N_{\bf f}$ {\bf fund} + 1 {\bf anti}};
\draw [->] (5,-3.5)--(5,-4);
\node at (5, -4.5) {${\rm USp}(2N)$ with 1 {\bf anti}, \quad $\theta = 0, \pi$};

\draw [->] (1,0)--(1.5,0);
\draw [->] (1,-1.5)--(1.5,-1.5);
\draw [->] (1,-3)--(1.5,-3);
\draw [->] (1,-4.5)--(1.5,-4.5);

\end{tikzpicture}
\caption{Renormalization group flows across various theories, starting from the Seiberg $E_8$ theories.  On the right, {\bf fund} and {\bf anti} denote hypermultiplets transforming in the fundamental and antisymmetric representations of ${\rm USp}(2N)$.  Note that for rank-one, there is no antisymmetric representation of ${\rm USp}(2)$.}
\label{Fig:RG}
\end{figure}

The Seiberg $E_8$ theories can be deformed by the dimension-four R-symmetry-singlet scalar primary $N^a$ in the $E_8$ flavor current multiplet (see \eqref{FlavorContent}), and flow to low energy theories with smaller flavor symmetry.  The flows fall into two categories, as depicted in Figure~\ref{Fig:RG}:
\begin{itemize}
\item {\bf Flow 1}  If the deformation by $N^a$ breaks the $E_8$ flavor group to $E_n$ ($n < 8$), then the infrared fixed point is the Seiberg $E_n$ theory of the same rank. There exist similar flows from $E_n$ to $E_m$ (and $\widetilde E_1$) for $n > m$.
\item {\bf Flow 2}  If the deformation by $N^a$ breaks the $E_{N_{\bf f}+1}$ flavor group to ${\rm SO}(2N_{\bf f}) \times {\rm U}(1)_{\rm I}$ ($N_{\bf f}=0,1,\ldots,7$), then the low-energy theory is ${\cal N}=1$ {\rm USp}($2N$) Yang-Mills coupled to $N_{\bf f}$ fundamental hypermultiplets together with a single antisymmetric hypermultiplet (the latter is decoupled for rank-one).\footnote{The antisymmetric hypermultiplet transforms under an additional mesonic ${\rm SU(2)}_{\rm m}$ flavor symmetry.  However, since our focus in this paper is on the rank-one theories, we shall omit this additional ${\rm SU(2)}_{\rm m}$ from most of our discussions (in particular, we omit its corresponding chemical potential from the partition function), and refer the reader to~\cite{Chang:2017mxc}.  }
\end{itemize}

Along {\bf Flow 2}, the operator $N^a$ flows to the Yang-Mills Lagrangian density in the infrared.  In the language of the infrared, the inverse Yang-Mills coupling squared $1/g_{\text{\tiny YM}}^2$ is the mass parameter that parameterizes the renormalization group flow.  The infrared gauge theories contain instanton particles of mass $m_{\rm I}=4\pi^2/g_{\text{\tiny YM}}^2$, which are charged under the instanton symmetry ${\rm U}(1)_{\rm I}$.   Furthermore, along {\bf Flow 2}, it was assumed in~\cite{Kim:2012ava,Jafferis:2012iv,Kim:2012qf} that with respect to a particular supercharge $Q$, the Yang-Mills term is $Q$-closed, while all other irrelevant operators are $Q$-exact.\footnote{Similar arguments were made for five-dimensional maximal ${\cal N}=2$ super Yang-Mills\cite{Kim:2012ava,Kim:2012qf}. In fact, the deformations of maximally supersymmetric gauge theories that preserve 16 supersymmetries have been classified in~\cite{Movshev:2003ib,Movshev:2004aw,Movshev:2005ei,Movshev:2009ba,Chang:2014kma}. Based on this classification, there is only one irrelevant deformation that could potentially not be $Q$-exact -- the supersymmetric completion of ${\rm Tr}(F_{\m\n}F_{\n\rho}F_{\rho\sigma}F_{\sigma\m})$. Thus the claim of~\cite{Kim:2012ava,Kim:2012qf} is essentially the statement that this particular term does not arise in the compactification of the six dimensional ${\cal N} = (2,0)$ theory to five dimensions.
}
Another possibility is as we explained in footnote~\ref{fn:protect}. 
Consequently, the partition function is insensitive to those irrelevant couplings. The supersymmetric partition function can be computed by standard localization techniques, in which a $Q$-exact term is added to the action that formally does not change the partition function~\cite{Nekrasov:2002qd,Nekrasov:2003rj,Pestun:2007rz,Kapustin:2009kz,Kallen:2012cs,Hosomichi:2012ek}. When the coefficient of the $Q$-exact term is tuned to be large, the path integral localizes to the fixed-point loci of the action of the supercharge $Q$. The result from localization formally takes the form
\ie\label{eqn:instExpan}
{\cal Z}
= {\cal Z}^{\rm pert}  + \sum_{n} e^{- S_n}
{\cal Z}^k,
\fe
where ${\cal Z}^{\rm pert}$ is the one-loop partition function at the perturbative fixed point, $n$ labels the nontrivial fixed points, $S_n \propto m_{\rm I}$ is the classical action and ${\cal Z}^n$ the one-loop determinant of the $Q$-exact localizing term, both at the $n$th fixed point.

The goal of this section is to explicitly compute the supersymmetric five-sphere partition function for the Seiberg exceptional theories. We borrow the matrix model expression for the localized path integral~\cite{Kallen:2012va,Kim:2012ava,Imamura:2012bm} and a recipe for computing the instanton contributions\cite{Lockhart:2012vp,Kim:2012qf,Hwang:2014uwa} from the literature.  The resulting free energy
\ie
F(\omega_i,m_f,m_{\rm I}) = - \log {\cal Z}(\omega_i,m_f,m_{\rm I}),
\fe
is a function of the squashing parameters $\omega_i=1+a_i$ ($i=1,2,3$), the instanton particle mass $m_{\rm I}$, and the masses $m_f$ ($f=1,\ldots, N_{\bf f}$) of the fundamental hypermultiplets. The $m_f$ appear in the mass matrix in the hypermultiplet action \eqref{hyperaction} as\footnote{Although the physical mass matrix $\cM$ in~\eqref{eqn:hyperMassAction} is Hermitian, when performing localization, $\cM$ has be to analytically continued to an anti-Hermitian matrix for the convergence of the localized path integral~\cite{Hosomichi:2012ek,Kallen:2012va}.
}
\ie\label{eqn:fundMassMatrix}
{\cal M}_{\bf f} =\ii  \sigma_2 \otimes \begin{pmatrix}m_{1} & & \\ & \ddots & \\ & & m_{N_{\bf f}}\end{pmatrix},
\fe
which is a Cartan generator of ${\rm SO}(2N_{\bf f})$ in the vector representation.

In Section~\ref{Sec:Weyl}, we first discuss the dependence of the partition function on the central charges, and certain properties that follow from the Weyl group of the exceptional flavor symmetry.  Since only the ${\rm SO}(2N_{\bf f}) \times {\rm U}(1)_{\rm I}$ subgroup is manifest in the gauge theory, these properties provide highly nontrivial checks on the precise triple factorization formula, which is presented in Section~\ref{Sec:Five}, and explicitly (numerically) evaluated in Section~\ref{Sec:InstantonNumerics}.

\subsection{Central charges and constraints from Weyl group}
\label{Sec:Weyl}

We apply the relations derived in Section~\ref{sec:PFtoCC} between the central charges $C_T$, $C_J$ and the five-sphere free energy to the specific case of the Seiberg $E_{N_{\bf f} +1}$ theories, with the mass matrix given in  \eqref{eqn:fundMassMatrix}.

\paragraph{$\bf \textit{N}_{\bf f} = 0, 3, 4, \dotsc, 7$}  These $E_{N_\textbf{f}+1}$ groups are simple, so there is only one independent flavor central charge $C_J^{E_{N_{\bf f} + 1}}$ for each $N_{\bf f}$.  However, in the gauge theory description, only the ${\rm SO}(2N_{\bf f}) \times {\rm U}(1)_{\rm I}$ subgroup is manifest. So the five-sphere free energy to quadratic order in the deformation parameters depends on two flavor central charges, $C_J^{{\rm SO}(2N_{\bf f})}$ and $C_J^{{\rm U}(1)_{\rm I}}$, as well as the conformal central charge $C_T$.\footnote{In Appendix~\ref{App:mI}, we fix the normalization of the ${\rm U}(1)_{\rm I}$ instanton current multiplet by requiring that the minimal charge of the ${\rm U}(1)_{\rm I}$ symmetry is one. This gives $
M=\ii m_{\rm I} \equiv {4\pi^2 \ii / g_{\rm YM}^2}$ in \eqref{c2bv}.
}
By \eqref{CTRelation}, \eqref{CJRelation1} and \eqref{CJRelation2}, the precise dependence is
\ie\label{eqn:2ndFree}
F|_{a_i^2,m_f^2,m_{\rm I}^2} &= -{\pi^2\over 1920} C_T\left(  \sum_{i=1}^3 a_i^2-\sum_{i<j} a_i a_j \right)
 -{3\pi^2 \over 256} C_J^{{\rm SO}(2N_{\bf f})} {\sum_{f=1}^{N_{\bf f}}m_f^2}-{3\pi^2 \over 256} C_J^{{\rm U}(1)_{\rm I}} m_{\rm I}^2.
\fe
The $C_J^{{\rm SO}(2N_{\bf f})}$ and $C_J^{{\rm U}(1)_{\rm I}}$ are related to the flavor central charge $C_J^{E_{N_{\bf f}+1}}$ of the enhanced flavor symmetry $E_{N_{\bf f}+1}$ by
\ie\label{eqn:enhCJ}
C_J^{{E_{N_{\bf f}+1}}} &= {C_J^{{\rm SO}(2N_{\bf f})} \over I_{\mathfrak{so}(2N_{\bf f}) \hookrightarrow \mathfrak{e}_{N_{\bf f}+1}}}={C_J^{{\rm U}(1)_{\rm I}} \over I_{\mathfrak{u}(1)_{\rm I} \hookrightarrow \mathfrak{e}_{N_{\bf f}+1}}},
\fe
where the embedding indices are computed in Appendix~\ref{Sec:Embedding} to be
\ie
I_{\mathfrak{so}(2N_{\bf f}) \hookrightarrow \mathfrak{e}_{N_{\bf f}+1}}=1, \quad I_{\mathfrak{u}(1)_{\rm I} \hookrightarrow \mathfrak{e}_{N_{\bf f}+1}}={4\over8-N_{\bf f}}.
\fe

\paragraph{$\bf\textit{N}_{\bf f}=1$}  The flavor symmetry is enhanced from ${\rm {\rm SO}(2)}\times {\rm U}(1)_{\rm I}$ to $E_2={\rm {\rm SU}(2)}\times{\rm U}(1)$. By the embedding map given in~\cite{Kim:2012gu,Hwang:2014uwa}, the {\rm SU}(2) mass matrix is $({1\over 4}m_1-{1\over 2}m_{\rm I})i\sigma^3$, and the U(1) mass parameter is $M={7\over 4}m_1+{1\over 2}m_{\rm I}$.  From \eqref{CJRelation1} and \eqref{CJRelation2}, the free energy at quadratic order in the mass parameters is
\ie
\label{Nf=1Free}
F^{N_{\bf f}=1}|_{m_f^2,m_{\rm I}^2}&=- {3\pi^2 \over 128} C_J^{\rm {\rm SU}(2)} \left({1\over 4}m_1-{1\over 2}m_{\rm I}\right)^2-{3\pi^2 \over 256} C_J^{{\rm U}(1)} \left({7\over 4}m_1+{1\over 2}m_{\rm I}\right)^2.
\fe

\paragraph{$\bf \textit{N}_{\bf f}=2$}  The flavor symmetry in the infrared is ${\rm {\rm SO}(4)}\cong {\rm {\rm SU}(2)}\times {\rm {\rm SU}(2)}'$, with mass matrix ${1\over 2}(m_1-m_2)i\sigma_3\oplus{1\over 2}(m_1+m_2)i\sigma_3$.  The ${\rm {\rm SU}(2)}'$ part is enhanced to ${\rm SU}(3)$ at the ultraviolet fixed point, with embedding index $I_{\mathfrak{su}(2)' \hookrightarrow \mathfrak{su}(3)}=1$.  By \eqref{CJRelation1} and \eqref{CJRelation2}, the free energy at quadratic order in the mass parameters is
\ie
\label{Nf=2Free}
\hspace{-.15in} F^{N_{\bf f}=2}|_{m_f^2,m_{\rm I}^2} &= -{3\pi^2 \over 512} \left[C_J^{{\rm {\rm SU}(2)}}(m_1-m_2)^2+C_J^{{\rm {\rm SU}(2)}'}(m_1+m_2)^2\right]-{3\pi^2\over 256} C_J^{{\rm U}(1)_{\rm I}} m_{\rm I}^2.
\fe

The Weyl group of the enhanced flavor group $E_{N_{\bf f}+1}$ acts on the mass parameters as linear transformations on $m_f$ and $m_{\rm I}$. Up to possible counter-terms at linear and cubic orders in the mass parameters (see Section~\ref{sec:ctt}), the free energy $F(\omega_i,m_f,m_{\rm I})$ should be invariant under the enhanced Weyl group. For example, there is a second degree Weyl group invariant polynomial
\ie\label{eqn:WIPD2}
\sum_{f=1}^{N_{\bf f}} m_f^2+{4\over 8-N_{\bf f}} m_{\rm I}^2,
\fe
which is unique for $N_{\bf f}=0,3,4,\dotsc,7$, and the second equality in \eqref{eqn:enhCJ} is a consequence of the enhanced Weyl group.  The Weyl group actions on the mass parameters $m_f$ and $m_{\rm I}$ are the same as the Weyl group actions on the chemical potentials ${\mathfrak m}_f$ and ${\mathfrak m}_{\rm I}$, which are specified in \eqref{eqn:WelyReflection}.

The above properties are only expected to be true for the {\it exact} five-sphere partition function at the ultraviolet fixed point.  When computed using the gauge theory description, all instanton saddles must be taken into account, and the instanton particle mass $m_{\rm I}$ should be taken to zero in order to hit the ultraviolet fixed point.  A finitely truncated free energy would not exhibit enhanced Weyl group invariance; in particular, the flavor central charges $C_J^{{\rm SO}(2N_{\bf f})}$ and $C_J^{{\rm U}(1)_{\rm I}}$ would not satisfy the relations \eqref{eqn:enhCJ}.  Since the instanton expansion \eqref{eqn:instExpan} becomes uncontrolled at $m_{\rm I} = 0$, {\it a priori}, one could not hope that the partition function computed by a finite-order truncation of the series gives a good approximation to the exact partition function.  However, by explicit numerical computation up to four instantons in Section~\ref{Sec:InstantonNumerics}, we see {\it miraculously} that the expansion \eqref{eqn:instExpan} seems to converge even at $m_{\rm I} = 0$, and the coefficients of the terms violating enhanced Weyl group diminish at higher and higher orders!  In particular, the violation of the relation \eqref{eqn:enhCJ} between $C_J^{{\rm SO}(2N_{\bf f})}$ and $C_J^{{\rm U}(1)_{\rm I}}$ diminishes.

\subsection{Five-sphere partition function}
\label{Sec:Five}
\subsubsection{Heuristic argument for triple factorization}
It was conjectured in~\cite{Kim:2012ava,Lockhart:2012vp,Kim:2012qf} (also see~\cite{Qiu:2016dyj,Kim:2016usy} for reviews) that the five-sphere partition function of five-dimensional ${\cal N}=1$ gauge theories can be computed by a Coulomb branch integral that ``glues" three copies of the Nekrasov partition function.  The heuristic argument of this triple factorization goes as follows. In the localization computation of the partition function on the squashed five-sphere background \eqref{5dmetric}, one adds a large $Q$-exact term to the action.  At the fixed-point loci of $Q$, the hypermultiplets have vanishing vacuum expectation values, while the gauge fields in the vector multiplets satisfy the ``contact instanton equation"~\cite{Kallen:2012cs,Hosomichi:2012ek},
\ie\label{eqn:CIE}
\xi^\m \widehat F_{\m\n}=0,\quad\widehat F_{\m\n}={1\over 2\tilde\beta\tilde\kappa}\epsilon_{\m\n\rho\sigma\delta}\widehat F^{\rho\sigma}\xi^\delta,
\fe
and the scalar fields in the vector multiplets satisfy
\ie\label{eqn:ScalarEq}
D_\m(\tilde\kappa^{-1}\phi)=0,\quad D_{ij}={\ii\over 2}(\sigma^3)_{ij}\tilde\kappa\phi.
\fe
Here, $\widehat F \equiv F-\ii\tilde\kappa^{-1}\phi\,\diff {\cal Y}$; the Reeb vector $\xi^\m$, scalars $\tilde\kappa$ and $\tilde\beta$, and one-form ${\cal Y}$ are defined in Appendix~\ref{sec:SquashedBackground}. A class of solutions to the scalar field equations \eqref{eqn:ScalarEq} is
\ie
\phi = \tilde\kappa \lambda,
\fe
with $\lambda$ being a constant that takes values in the Cartan subalgebra of the gauge group, and $[\widehat F_{\m\n},\lambda]=0$, by the first equation in \eqref{eqn:ScalarEq}. The five-sphere partition function is computed by an integral over the Coulomb branch parameters $\lambda$, of the schematic form
\ie
{\cal Z}_{{\rm S}^5}=\int \diff\lambda\,\sum_n e^{-S_{{\rm cl},n}}{\cal Z}_n,
\fe
where $S_{{\rm cl},n}$ is the classical action of the $n$-th solution to the contact instanton equation, and ${\cal Z}_n$ is the one-loop determinant in that background.

The most general solution to the contact instanton equation is unknown, but there is a class of solutions given by embedding the flat space instanton solutions as follows. We can view the squashed five-sphere metric as a (singular) ${\rm T}^3$ fibration over the base ${\rm S}^2$, which we parameterize with constrained coordinates $y_i$ satisfying $y_1^2+y_2^2+y_3^2=1$. The ${\rm T}^3$ fiber degenerates to S$^1$ at three points on the S$^2$,
\ie\label{eqn:degePoints}
 (y_1,y_2,y_3)=(1,0,0),\quad (0,1,0),\quad{\rm and}\quad (0,0,1).
 \fe
At these three points, the Reeb vector $\xi^\m$ generates closed orbits along the S$^1$, and the contact instanton equation \eqref{eqn:CIE} reduces to the instanton equation on ${\mathbb R}^4$. One can embed the ${\mathbb R}^4$ small instanton solution into S$^5$, localized at one of the three points and constant along the direction of the Reeb vector. When the Coulomb branch parameters $\lambda$ take generic values in the Cartan subalgebra, the field strength of the small instanton must also be in the Cartan subalgebra, such that it commutes with $\lambda$. The instanton solutions at different degenerate points can be superposed since the field strengths are all in the Cartan subalgebra. Consider a solution with instanton numbers $k_1$, $k_2$, and $k_3$ at the three degenerate points.  The classical action was computed in~\cite{Kim:2012qf} to be
\ie\label{eqn:onShellAction}
S_{\rm cl}=\frac{\pi m_{\rm I}}{\omega_1\omega_2\omega_3}{\rm Tr}(\lambda^2)+{2\pi  m_{\rm I}\over\omega_1}k_1+{2\pi m_{\rm I}\over\omega_2}k_2 +{2\pi m_{\rm I}\over\omega_3}k_3,
\fe
where $m_{\rm I}=4\pi^2/g_{\text{\tiny YM}}^2$ is the mass of the instanton particle.  At each degenerate point, the geometry is locally ${\rm S}^1 \times \mathbb{R}^4$, and it was conjectured that the one-loop determinants factorize into three copies of the twisted partition function on ${\rm S}^1 \times \mathbb{R}^4$ associated to the three degenerate points. The ${\rm S}^1 \times \mathbb{R}^4$ partition function suffers from infrared divergences, which can be regularized by the twisted boundary condition $(z_1,z_2,t)\sim(z_1 e^{-\epsilon_1},z_2 e^{-\epsilon_2},t+1)$, where $t$ is the coordinate of the S$^1$ with unit radius, $z_1$ and $z_2$ are the complex coordinates on the two two-planes $\bR^{12}, \bR^{34} \in \bR^4$. The twisted ${\rm S}^1 \times \mathbb{R}^4$ partition function is nothing but the Nekrasov partition function ${\cal Z}_{{\rm S}^1 \times \mathbb{R}^4}$~\cite{Nekrasov:2002qd,Nekrasov:2003rj}.

\subsubsection{Triple factorization formula for rank-one}
\label{sec:TriFac}

We focus on the rank-one Seiberg exceptional theories. The five-sphere partition function is computed by the Nekrasov partition function via the formula
\ie\label{eqn:S5partition}
\hspace{-.15in} \mathcal{Z}_{{\rm S}^5} &
 = \int_{\cal C} {\diff\lambda \over 4\pi} \, e^{-{\cal F}^\vee_{\rm eff}} \bigg[ {\cal Z}_{{\rm S}^1 \times \mathbb{R}^4}\left(\frac{ 2\pi \ii \omega_2}{\omega_1}, \frac{2\pi \ii \omega_3}{\omega_1},\frac{2\pi\lambda}{\omega_1},\frac{2\pi m_f}{\omega_1}-\pi \ii,\frac{2\pi m_{\rm I}}{\omega_1}-{ N_{\bf f}\over 2}\pi \ii\right)
\\
& \hspace{1.5in}  \times \text{(2 cyclic perms on $\omega_i$)} \bigg],
\fe
where the exponent $S_0$ is
\ie
{\cal F}^\vee_{\rm eff}&= \frac{(8-N_{\bf f})\pi  \lambda ^3 }{3 \omega _1 \omega _2 \omega _3}+\frac{ 2\pi m_{\rm I}\lambda^2}{\omega_1\omega_2\omega_3}
- \frac{  \left[\sum_{f=1}^{N_{\bf f}}m^2_f +{N_{\bf f}+4 \over 12} \sum_{i=1}^3 \omega_i^2 + \sum_{i<j} \omega_i \omega_j \right]\pi  \lambda}{\omega _1 \omega _2 \omega
   _3}.
   \label{s5effpp}
\fe
The contour $\cal C$ in the undeformed case lies slightly above the real axis, $-\infty + \ii\epsilon$ to $\infty + \ii\epsilon$ for small $\epsilon > 0$, and in the general case is such that when continuously deforming from the undeformed case, no pole crosses the contour.\footnote{We do not have a first-principle derivation of this contour prescription, but it seems to be the only prescription that gives sensible results.  For instance, naively choosing the contour to be on the real axis gives a free energy that diverges in the undeformed limit (when the instantons are included).
}
In the region $\lambda \geq m_f$ for all $f$, what we call ${\cal F}^\vee_{\rm eff}$ has the interpretation as the one-loop effective prepotential on the squashed five-sphere.\footnote{We thank Hee-Cheol Kim for correspondence on this point.
}
In ${\cal F}^\vee_{\rm eff}$, the quadratic term in $\lambda$ comes from the on-shell action \eqref{eqn:onShellAction}, and the linear and cubic terms in $\lambda$ are from the one-loop determinants of the vector and hypermultiplets.\footnote{Without loss of generality, let us assume $m_{N_{\bf f}} \leq m_{N_{\bf f}-1} \leq \cdots \leq m_{1}$. We conjecture that the one-loop effective prepotential on the squashed five-sphere is
\ie\label{s5effpp-c}
{\cal F}_{\rm eff} &= \frac{ 2\pi m_{\rm I}\lambda^2}{\omega_1\omega_2\omega_3}+{\pi \ii\over 6}\bigg[B_{3,3}(2\ii\lambda\mid \vec{\omega})-B_{3,3}(-2\ii\lambda\mid \vec{\omega})-\sum_{f=1}^{N_{\bf f}} B_{3,3}(\ii\lambda+  \ii m_f + \tfrac{\omega_1+\omega_2+\omega_3}{2}\mid \vec{\omega})
\\
& \quad +\sum_{f=n+1}^{N_{\bf f}}B_{3,3}(-\ii\lambda+  \ii m_f + \tfrac{\omega_1+\omega_2+\omega_3}{2}\mid \vec{\omega})-\sum_{f=1}^{n}B_{3,3}(-\ii\lambda+  \ii m_f +\tfrac{\omega_1+\omega_2+\omega_3}{2}\mid \vec{\omega})\bigg],
\fe
for $m_{n+1} \leq \lambda \leq m_{n}$.  This expression reduces to \eqref{s5effpp} when $\lambda \geq m_f$ for all $f$.  The functions $B_{3,3}(z\mid \vec{\omega})$ come from the triple sine functions in \eqref{eqn:PertPartition} by applying \eqref{eqn:S3inShiftedFactorial1} for ${\rm Im}(z)\ge0$ and \eqref{eqn:S3inShiftedFactorial2} for ${\rm Im}(z)<0$. In the flat space limit $\omega_i\to 0$, the five-sphere one-loop effective prepotential behaves as
$\cF_{\rm eff}\propto \cF_{\rm eff}^{\mR^4} / (\omega_1\omega_2 \omega_3) $ (with the identification of $m_{\rm I}$ with ${1\over 2} m_0$ in \cite{Intriligator:1997pq}). Thus, it is plausible that $\cF_{\rm eff}$ 
is the generalization of $ \cF_{\rm eff}^{\mR^4}$ to the squashed sphere. In particular, the pieces of  \eqref{s5effpp-c} inhomogeneous in $(\lambda,m_{\rm I},m_{ f})$  should be produced by the one-loop effective coupling between the background supergravity multiplet and the ${\rm U}(1)$ vector multiplet (on the Coulomb branch).
}
The arguments of the Nekrasov partition function ${\cal Z}_{{\rm S}^1 \times \mathbb{R}^4}(\epsilon_1, \epsilon_2, \A, {\mathfrak m}_f, {\mathfrak m}_{\rm I})$ are the following: $\A$ is the Coulomb branch parameter, ${\mathfrak m}_f$ with $f=1,\dotsc,N_{\bf f}$ are the chemical potentials for the {\rm SO}($2N_{\bf f}$) flavor symmetry, and ${\mathfrak m}_{\rm I}$ is the chemical potential for the ${\rm U}(1)_{\rm I}$ instanton symmetry. 

The Nekrasov partition function ${\cal Z}_{{\rm S}^1 \times \mathbb{R}^4}(\epsilon_1, \epsilon_2, \A, {\mathfrak m}_f, {\mathfrak m}_{\rm I})$ receives perturbative and instanton contributions,
\ie\label{eqn:Nekrasov}
{\cal Z}_{{\rm S}^1 \times \mathbb{R}^4}(\epsilon_1, \epsilon_2, \A, {\mathfrak m}_f, {\mathfrak m}_{\rm I}) = {\cal Z}^{\rm pert}_{{\rm S}^1 \times \mathbb{R}^4}(\epsilon_1, \epsilon_2, \A, {\mathfrak m}_f) \, {\cal Z}^{\rm inst}_{{\rm S}^1 \times \mathbb{R}^4}(\epsilon_1, \epsilon_2, \A, {\mathfrak m}_f, {\mathfrak m}_{\rm I}).
\fe
The perturbative part was computed in~\cite{Kim:2012gu} using the Atiyah-Singer equivariant index theorem, and the result is summarized in Appendix~\ref{App:Factorization}.  The instanton partition function organizes into a sum of contributions from different instanton numbers,
\ie
&{\cal Z}^{\rm inst}_{{\rm S}^1 \times \mathbb{R}^4}(\epsilon_1,\epsilon_2,\A,{\mathfrak m}_f,{\mathfrak m}_{\rm I})=1+\sum_{k=1}^\infty \exp\left({- k {\mathfrak m}_{\rm I}}\right) {\cal Z}^{{\rm inst},k}_{{\rm S}^1 \times \mathbb{R}^4}(\epsilon_1,\epsilon_2,\A,{\mathfrak m}_f),
\fe
which was obtained in~\cite{Kim:2012gu,Hwang:2014uwa} by computing the Witten indices of the ADHM quantum mechanics arising in the D-brane configuration of Table~\ref{table:BraneDirections}, with the D0-brane worldline along the zeroth direction.  In Appendix~\ref{sec:ADHM}, we review the ADHM quantum mechanics and the computation of the Witten indices. 

The substitution rules for the chemical potentials $\epsilon_1$, $\epsilon_2$, $\A$, and ${\mathfrak m}_f$ in terms of the Coulomb branch parameter $\lambda$, the squashing parameters $\omega_i$, and the mass parameters $m_f$ and $m_{\rm I}$ can be determined  by studying the $m_{\rm I} \to \infty$ limit of this formula, and demanding that it reduces to the perturbative formula\footnote{The factor of $S_3'(0\mid \vec{\omega})$ coming from the contributions of the zero modes in the vector multiplet is of critical importance, but is often ignored in the literature. In particular, only when this factor is included will the five-sphere free energy take the form \eqref{CTRelation}. Furthermore, the $S_3'(0\mid \vec{\omega})$ factor is essential for the gauge theory partition function \eqref{eqn:PertPartition} to be invariant under a simultaneous rescaling $\omega_i\to c \, \omega_i$ with $c \in \mR$ when the mass parameters $m_{\rm I}$ and $m_f$ are turned off; this property is consistent with (but not implied by) \eqref{CTRelation} and continues to hold when instanton contributions are included.}
\ie\label{eqn:PertPartition}
 \mathcal{Z}^{\rm pert}_{{\rm S}^5} &= \int_{-\infty}^\infty { \diff\lambda \over 4\pi} \exp \left(- {2\pi m_{\rm I}\lambda^2\over \omega_1\omega_2\omega_3} \right){ S_3'(0\mid \vec{\omega})S_3(\pm 2\ii \lambda\mid \vec{\omega})
\over
 \prod_f S_3( \pm \ii \lambda +  i m_f +{\omega_1+\omega_2+\omega_3\over 2}\mid \vec{\omega} )},
\fe
which was obtained by a direct one-loop computation in~\cite{Kallen:2012va,Kim:2012ava,Imamura:2012bm}.  The triple sine function $S_3 \left( z \mid \vec{\omega} \right)$ is defined and discussed in some detail in Appendix~\ref{App:triplesine}, and $S^{\prime}_3 \left(z \mid \vec{\omega} \right)$ is its $z$-derivative.  The comparison between the $m_{\rm I} \to \infty$ limit of \eqref{eqn:S5partition} and \eqref{eqn:PertPartition} is reviewed in Appendix~\ref{App:Factorization}.  Given the on-shell action \eqref{eqn:onShellAction}, we expect that ${\mathfrak m}_{\rm I}$ should be substituted by ${2\pi m_{\rm I}\over \omega_i}$.  However, we conjecture that there is an additional imaginary shift $-{N_{\bf f}\over 2}\pi \ii$ in the substitution rule for the U(1)$_{\rm I}$ chemical potential ${\mathfrak m}_{\rm I}$.

\subsubsection{Instanton particle mass shift and flavor symmetry enhancement}
\label{Sec:Enhancement}

We presently give arguments for the $-{N_{\bf f}\over 2}\pi \ii$ shift in the substitution rule for ${\mathfrak m}_{\rm I}$ in the triple factorization formula \eqref{eqn:S5partition} for the five-sphere partition function of rank-one Seiberg exceptional theories. As shown in~\cite{Mitev:2014jza}, the Nekrasov partition function exhibits manifest enhanced flavor symmetry when expanded in the shifted Coulomb branch parameter
\ie
\widetilde\A=\A +{2\over 8-N_{\bf f}}{\mathfrak m}_{\rm I}.
\fe
We define
\ie
\label{Ztilde}
\widetilde{\cal Z}_{{\rm S}^1 \times \mathbb{R}^4}\left(\epsilon_1,\epsilon_2,\widetilde\A ,{\mathfrak m}_f,{\mathfrak m}_{\rm I}\right)={\cal Z}_{{\rm S}^1 \times \mathbb{R}^4}\left(\epsilon_1,\epsilon_2,\widetilde\A -{2\over 8-N_{\bf f}}{\mathfrak m}_{\rm I},{\mathfrak m}_f,{\mathfrak m}_{\rm I}\right).
\fe
At finite order in the $\widetilde w$-expansion (or $w$-expansion), where $\widetilde w=e^{-\widetilde \A}$ ($w=e^{-\A}$), the Nekrasov partition function only receives contributions up to finite instanton number. The coefficients of the $\widetilde w$-expansion can be organized into characters of the enhanced flavor group, 
\ie\label{eqn:EnCharacter}
\chi_{{R}}(\vec{\mathfrak m})=\sum_{\vec\rho\in {{R}}}e^{\vec\rho\cdot\vec{\mathfrak m}},
\fe
where $\vec{\mathfrak m}=({\mathfrak m}_1,\ldots,{\mathfrak m}_{N_{\bf f}}, {2(-1)^{N_{\bf f}}\over \sqrt{8-N_{\bf f}}} {\mathfrak m}_{\rm I})$, and ${{R}}$ is a representation of the enhanced flavor group, here to be interpreted as the set of weights of $R$ in the weight lattice $\Lambda^{{\mathfrak e}_{N_{\bf f}+1}}_{\rm weight}$.  Our choice of basis is specified in Appendix \ref{sec:Basis}.  The character is invariant under the shift
\ie
\vec{\mathfrak m}\to \vec{\mathfrak m}+2\pi \ii\vec \A ,
\fe
for any root vector $\vec\A\in\Lambda^{{\mathfrak e}_{N_{\bf f}+1}}_{\rm root}$. In other words, the imaginary part of $\vec{\mathfrak m}$ takes values in the space $\bR^{N_{\bf f}+1}/(2\pi \Lambda^{{\mathfrak e}_{N_{\bf f}+1}}_{\rm root})$. A Weyl reflection $w_{\vec\A}$ associated to a root vector $\vec\A$ acts on a weight vector $\vec\rho$ and the chemical potential $\vec{\mathfrak m}$ as\footnote{For $N_{\bf f}=1$, the Weyl group is only generated by the Weyl reflection associated to the simple root on the second line of the $2\times2$ matrix in \eqref{eqn:SimpleRoots}.
}
\ie\label{eqn:WelyReflection}
w_{\vec\A}(\vec\rho)&=\vec\rho-2{\vec\A\cdot\vec\rho\over |\vec\A|^2}\vec\A,
\quad
w_{\vec\A}(\vec{\mathfrak m})&=\vec{\mathfrak m}-2{\vec\A\cdot\vec{\mathfrak m}\over |\vec\A|^2}\vec\A.
\fe
The character is invariant under the Weyl reflections
\ie
\chi_{{{R}}}(w_{\vec\A}(\vec{\mathfrak m}))= \chi_{{{R}}}(\vec{\mathfrak m}).
\fe 
We define the shifted character $\chi_{{{R}}}(\vec{\mathfrak m}+2\pi \ii\vec\rho)$ with $\vec\rho\in\Lambda^{{\mathfrak e}_{N_{\bf f}+1}}_{\rm weight}$, which is also invariant under the Weyl reflections on $\vec{\mathfrak m}$,
\ie\label{eqn:ShiftedCharInv}
\chi_{{{R}}}( w_{\vec\A}(\vec{\mathfrak m})+2\pi \ii\vec\rho)
=\chi_{{{R}}}(\vec{\mathfrak m}+2\pi \ii\vec\rho).
\fe

Rewriting the triple factorization formula \eqref{eqn:S5partition} in terms of $\widetilde{\cal Z}_{{\rm S}^1 \times \mathbb{R}^4}$, we obtain
\ie\label{eqn:PartitionRewrited}
\mathcal{Z}_{{\rm S}^5} 
& = \int_{\cal C} { \diff\widetilde \lambda \over 4\pi} \,\exp\Big(-{\cal F}^\vee_{\rm eff}|_{\lambda=\widetilde \lambda-{2\over 8-N_{\bf f}}m_{\rm I}}\Big)
\\
& \quad\times\Big[\widetilde{\cal Z}_{{\rm S}^1 \times \mathbb{R}^4}\Big(\frac{2\pi \ii\omega_2}{\omega_1},\frac{2\pi \ii\omega_3}{\omega_1},\frac{2\pi\widetilde\lambda}{\omega_1}-{N_{\bf f}\over8-N_{\bf f}}\pi \ii,{2\pi m_f\over \omega_1}-\pi \ii,{2\pi m_{\rm I}\over \omega_1}-{N_{\bf f}\over2}\pi \ii\Big)
\\
&\hspace{.5in} \times \text{(2 cyclic perms on $\omega_i$)}\Big].
\fe
The substitution rule for the vector $\vec{\mathfrak m}=({\mathfrak m}_1,\ldots,{\mathfrak m}_{N_{\bf f}}, {2(-1)^{N_{\bf f}}\over \sqrt{8-N_{\bf f}}} {\mathfrak m}_{\rm I})$ can be written as
\ie
\vec{\mathfrak m} = {2\pi \vec m\over \omega_i}+ 2\pi \ii\vec \rho,
\fe
where $\vec m = (m_1,\ldots, m_{N_{\bf f}}, {2(-1)^{N_{\bf f}}\over \sqrt{8-N_{\bf f}}}  m_{\rm I})$, and $\vec\rho =-({1\over 2},\ldots,{1\over 2}, {(-1)^{N_{\bf f}}N_{\bf f}\over 2\sqrt{8-N_{\bf f}}} )$ is a weight vector. By \eqref{eqn:ShiftedCharInv}, we find that the expression in the square bracket in \eqref{eqn:PartitionRewrited} is invariant under the enhanced Weyl group actions \eqref{eqn:WelyReflection}. Below we find that ${\cal F}^\vee_{\rm eff}$ is invariant under the enhanced Weyl group after taking into appropriate counter-terms, thus the full partition function respects the enhanced symmetry. This gives strong evidence for the  $-{N_{\bf f}\over 2}\pi \ii$ chemical potential shift in $\widetilde{\cal Z}_{{\rm S}^1 \times \mathbb{R}^4}$.

We now show that ${\cal F}^\vee_{\rm eff}$ is indeed invariant under the enhanced Weyl group, after an appropriate subtraction of counter-terms.  In terms of the shifted Coulomb branch parameter $\widetilde \lambda$, we have
\ie\label{eqn:1-loopPrepotentialShifted}
&{\cal F}^\vee_{\rm eff}\Big|_{\lambda=\widetilde \lambda-{2\over 8-N_{\bf f}}m_{\rm I}}
 = \frac{(8-N_{\bf f})\pi  \widetilde \lambda^3 }{3 \omega _1 \omega _2 \omega _3}
\\
& \quad -\frac{  \left[ 12 \sum_{f=1}^{N_{\bf f}}m^2_f +{48\over 8-N_{\bf f}}m_{\rm I}^2+(N_{\bf f}+4) \sum_{i=1}^3 \omega_i^2 + 12 \sum_{i<j} \omega_i \omega_j \right] \pi  \widetilde \lambda}{12 \omega _1 \omega _2 \omega_3}
\\
&\quad+\frac{  \left[ 12 \sum_{f=1}^{N_{\bf f}}m^2_f+{32\over 8-N_{\bf f}}m_{\rm I}^2 +(N_{\bf f}+4) \sum_{i=1}^3 \omega_i^2 +12\sum_{i<j} \omega_i \omega_j \right] \pi  m_{\rm I}}{6(8-N_{\bf f}) \omega _1 \omega _2 \omega
   _3}.
\fe
The cubic term in $\widetilde\lambda$ is obviously Weyl-invariant.  After recognizing the Weyl-invariant polynomial \eqref{eqn:WIPD2}, we see that the linear term in $\widetilde\lambda$ is also Weyl-invariant.  The $\widetilde\lambda$-independent terms are not Weyl-invariant, but they can be canceled by the admissible counter-terms classified in Section~\ref{sec:ctt}.\footnote{We emphasize here that these Weyl non-invariant terms should not be confused with the potential superconformal anomalies of the fixed point superconformal field theory with $E_{N_{\bf f}+1}$ flavor symmetry. Instead, they imply that the regularization scheme used in the gauge theory localization computation only preserves the ${\rm SO}(2N_{\bf f}) \times {\rm U}(1)_{\rm I}$ subgroup. In other words, the $E_{N_{\bf f}+1}$ preserving scheme can be implemented in the localization computation by including the corresponding counter-terms from the beginning. Note that this is not an issue for the superconformal index because these counter-terms all vanish on the supersymmetric ${\rm S}^1\times {\rm S}^4$ background.}

We end this section by pointing out an interesting observation on the Nekrasov partition function ${\cal Z}_{{\rm S}^1 \times \mathbb{R}^4}$.  We conjecture that the states contributing to the Nekrasov partition function for the Seiberg exceptional theories satisfy\footnote{This relation between the gauge and the $\bZ_{8-N_{\bf f}}$ center charges was also observed in the ray operator indices \cite{Chang:2016iji}.
}
\ie
\label{Observation}
q \equiv q_{ R}\mod (8-N_{\bf f}),
\fe
where $q$ is the charge under the USp(2) gauge group, $R$ is the representation of the $E_{N_{\bf f}+1}$ flavor group and $q_R$ is the charge under its center.

For $N_{\bf f}>1$, the simply connected Lie group $E_{N_{\bf f}+1}$ with Lie algebra $\mathfrak{e}_{N_{\bf f}+1}$ has center
\ie\label{eqn:center}
\Lambda^{{\mathfrak e}_{N_{\bf f}+1}}_{\rm weight}/\Lambda^{{\mathfrak e}_{N_{\bf f}+1}}_{\rm root}\cong\bZ_{8-N_{\bf f}}.
\fe
For $E_2$, the quotient \eqref{eqn:center} is identified with the $\bZ_7$ subgroup of the center ${\rm U(1)}\subset E_2$.
The generators of $\bZ_{8-N_{\bf f}}$ are represented by the following weight vectors $\vec\rho_e$,
\ie
&N_{\bf f}=1:\,\vec\rho_e=\left(1,-{1\over \sqrt{7}}\right),&& N_{\bf f}=2:\,\vec\rho_e=\left(1,0,{1\over \sqrt{6}}\right),
\\
&N_{\bf f}=3:\,\vec\rho_e=\left(1,0,0,-{1\over \sqrt{5}}\right),&& N_{\bf f}=4:\,\vec\rho_e=\left(1,0,0,0,{1\over 2}\right),
\\
&N_{\bf f}=5:\,\vec\rho_e=\left(1,0,0,0,0,-{1\over \sqrt{3}}\right),&& N_{\bf f}=6:\,\vec\rho_e=\left(1,0,0,0,0,0,{1\over \sqrt{2}}\right).
\fe
Any weight vector $\vec\rho\in\Lambda_{\rm weight}^{{\mathfrak e}_{N_{\bf f}+1}}$ can be written as $\vec\rho=n\vec\rho_e+\vec\A$, for some $n\in\bZ_{8-N_{\bf f}}$ and $\vec \A\in\Lambda^{{\mathfrak e}_{N_{\bf f}+1}}_{\rm root}$. If $R$ is an irreducible representation, then the characters obey
\ie
\chi_{{ R}}(\vec{\mathfrak m}+2\pi i\vec\rho) = \xi_R \chi_{{ R}}(\vec{\mathfrak m}),
\fe
where $\xi_R$ is an $(8-N_{\bf f})$-th root of unity.
The $\bZ_{8-N_{\bf f}}$ center charge $q_{R}$ of a representation $R$ is defined by
\ie
\chi_{{R}}(\vec{\mathfrak m}+2\pi \ii\vec \rho_e) = \exp\left({{2\pi \ii q_{{R}}\over 8-N_{\bf f}}}\right) \chi_{{ R}}(\vec{\mathfrak m}).
\fe
Finally, the property \eqref{Observation} follows from a conjectural identity for the $\widetilde{\cal Z}_{{\rm S}^1 \times \mathbb{R}^4}$ defined in \eqref{Ztilde},
\ie
\widetilde{\cal Z}_{{\rm S}^1 \times \mathbb{R}^4}\left(\epsilon_1,\epsilon_2,\widetilde\A+{2\pi \ii \over 8-N_{\bf f}},{\mathfrak m}_f,{\mathfrak m}_{\rm I}\right)\big|_{\vec{\mathfrak m}\to \vec{\mathfrak m} +2\pi \ii\vec\rho_e}=\widetilde {\cal Z}_{{\rm S}^1 \times \mathbb{R}^4}\left(\epsilon_1,\epsilon_2,\widetilde\A,{\mathfrak m}_f,{\mathfrak m}_{\rm I}\right),
\fe
which in terms of the ordinary Nekrasov partition function ${\cal Z}_{{\rm S}^1 \times \mathbb{R}^4}$ is equivalent to\footnote{We checked this identity up to instanton number five.
}
\ie
{\cal Z}_{{\rm S}^1 \times \mathbb{R}^4}(\epsilon_1,\epsilon_2,\A,{\mathfrak m}_f,{\mathfrak m}_{\rm I})\big|_{\vec{\mathfrak m}\to \vec{\mathfrak m} +2\pi \ii \vec\rho_e}= {\cal Z}_{{\rm S}^1 \times \mathbb{R}^4}\left(\epsilon_1,\epsilon_2,\A,{\mathfrak m}_f,{\mathfrak m}_{\rm I}\right).
\fe

\subsection{Numerical results}
\label{Sec:InstantonNumerics}

We proceed with the direct numerical evaluation of the triple factorization formula \eqref{eqn:S5partition}, and compute the undeformed free energies $F_0$ as well as the central charges $C_T$ and flavor central charges $C_J$ using \eqref{eqn:2ndFree}, \eqref{Nf=1Free}, and \eqref{Nf=2Free} for the rank-one Seiberg exceptional and Morrison-Seiberg $\widetilde E_1$ theories.  The results up to four instantons are summarized in Tables~\ref{Tab:InstantonF},~\ref{Tab:InstantonCT}, and~\ref{Tab:InstantonCJ}.  We close this section with the following remarks.
\begin{itemize}
\item  The instanton contributions are small compared to the perturbative result, especially for larger $N_{\bf f}$.
\item  For $E_n$, $n = 3, 4, \dotsc, 8$, the flavor central charge $C_J^{E_n}$ can be obtained by the dependence of the free energy on either the masses of the fundamental hypermultiplets, or on the mass of the instanton particle.  The two methods may give differing results at finite instanton number, but should ultimately agree to be consistent with the enhanced flavor group.  Table~\ref{Tab:InstantonCJDiff} tracks the differences of the two up to four instantons.  We see that the difference indeed diminishes with higher instanton numbers, especially in the $E_5$ case, providing a strong check of the triple factorization formula \eqref{eqn:S5partition}.  This also suggests that the instanton expansion \eqref{eqn:instExpan} may be convergent even at the ultraviolet fixed point, where $m_{\rm I} = 0$.
\item In general, the coefficients of the terms which are not invariant under the Weyl group can be nonzero at finite instanton number. We computed the expansion of the free energy in both $m_{\rm I}$ and $m_f$ up to cubic order, and after adding counter-terms to cancel the last line of \eqref{eqn:1-loopPrepotentialShifted}, observed that such coefficients all appear to diminish at higher and higher instanton numbers.  This check is another piece of evidence for the convergence of the instanton series at $m_{\rm I} = 0$.
\item  The values of the undeformed free energy given in Table~\ref{Tab:InstantonF} show that the Seiberg exceptional theories connected by renormalization group flows ({\bf Flow 1} in Figure~\ref{Fig:RG}) have larger $-F_0$ and $C_T$ in the ultraviolet than in the infrared.  This observation hints towards a five-dimensional version of the $F$- or $C$-theorem \cite{Cardy:1988cwa,Capper:1974ic,Zamolodchikov:1986gt,Komargodski:2011vj,Myers:2010tj,Jafferis:2010un,Casini:2011kv,Jafferis:2011zi,Klebanov:2011gs,Elvang:2012st,Cordova:2015vwa,Cordova:2015fha}.\footnote{In three-dimensional $\cN=2$ superconformal field theories, there are counter-examples to a $C$-theorem for $C_T$ \cite{Nishioka:2013gza}. In five-dimensional non-supersymmetric conformal field theories, there is also a counter-example described in \cite{Fei:2014yja}.  However, it is plausible that with eight supercharges, a $C$-theorem could hold for five-dimensional $\cN=1$ superconformal field theories. We hope to investigate this possibility in the near future.
}

\end{itemize}

\begin{table}[H]
\centering
\begin{tabular}{|c||c|c|c|c|c||c|}
\hline
$G$ & $-F_0^{\rm pert}$ & $-\Delta F_0^{\text{1-inst}}$ & $-\Delta F_0^{\text{2-inst}}$ & $-\Delta F_0^{\text{3-inst}}$ & $-\Delta F_0^{\text{4-inst}}$ & $-F_0^{\text{4-inst}}$
\\\hline\hline
 $ \widetilde{E}_1 $ & $ 5.0967 $ & $ -1.1\times10^{-2} $ & $ 1.6\times10^{-4} $ & $ 3.2\times10^{-6} $ & $ 3.1\times10^{-7} $ & $ 5.0855 $ \\\hline
 $ {E}_1 $ & $ 5.0967 $ & $ 1.5\times10^{-3} $ & $ -2.9\times10^{-4} $ & $ -7.7\times10^{-5} $ & $ -3.2\times10^{-5} $ & $ 5.0978 $ \\\hline
 $ {E}_2 $ & $ 6.1401 $ & $ 7.9\times10^{-3} $ & $ -7.9\times10^{-5} $ & $ -4.5\times10^{-5} $ & $ -1.6\times10^{-5} $ & $ 6.1478 $ \\\hline
 $ {E}_3 $ & $ 7.3949 $ & $ 9.7\times10^{-3} $ & $ 4.5\times10^{-5} $ & $ -1.7\times10^{-5} $ & $ -4.5\times10^{-6} $ & $ 7.4046 $ \\\hline
 $ {E}_4 $ & $ 8.9590 $ & $ 1.1\times10^{-2} $ & $ 2.4\times10^{-4} $ & $ 4.9\times10^{-7} $ & $ 3.8\times10^{-7} $ & $ 8.9706 $ \\\hline
 $ {E}_5 $ & $ 11.007 $ & $ 1.2\times10^{-2} $ & $ 3.3\times10^{-4} $ & $ -4.8\times10^{-6} $ & $ 6.0\times10^{-9} $ & $ 11.019 $ \\\hline
 $ {E}_6 $ & $ 13.898 $ & $ 9.0\times10^{-3} $ & $ 7.6\times10^{-4} $ & $ 3.8\times10^{-5} $ & $ -2.1\times10^{-6} $ & $ 13.907 $ \\\hline
 $ {E}_7 $ & $ 18.538 $ & $ 3.7\times10^{-3} $ & $ 1.4\times10^{-3} $ & $ 3.8\times10^{-5} $ & $ 8.4\times10^{-6} $ & $ 18.544 $ \\\hline
 $ {E}_8 $ & $ 28.473 $ & $ 1.9\times10^{-4} $ & $ 2.9\times10^{-4} $ & $ 1.9\times10^{-4} $ & $ 4.3\times10^{-5} $ & $ 28.474 $ \\
\hline
\end{tabular}
\caption{The contributions to the values of the undeformed free energy $-F_0$ at each instanton number in the rank-one Seiberg exceptional theories and the Morrison-Seiberg $\widetilde E_1$ theory.}
\label{Tab:InstantonF}
\end{table}

\begin{table}[H]
\centering
\begin{tabular}{|c||c|c|c|c|c||c|}
\hline
$G$ & $C_T^{\rm pert}$ & $\Delta C_T^{\text{1-inst}}$ & $\Delta C_T^{\text{2-inst}}$ & $\Delta C_T^{\text{3-inst}}$ & $\Delta C_T^{\text{4-inst}}$ & $C_T^{\text{4-inst}}$
\\\hline\hline
 $ \widetilde{E}_1 $ & $ 333.39 $ & $ 6.6\times10^{-1} $ & $ 3.7\times10^{-1} $ & $ -6.1\times10^{-2} $ & $ -1.3\times10^{-2} $ & $ 334.35 $ \\\hline
 $ {E}_1 $ & $ 333.39 $ & $ -4.5 $ & $ -1.3 $ & $ -8.4\times10^{-1} $ & $ -6.6\times10^{-1} $ & $ 326.04 $ \\\hline
 $ {E}_2 $ & $ 422.94 $ & $ -2.4 $ & $ -3.5\times10^{-1} $ & $ -3.4\times10^{-1} $ & $ -2.9\times10^{-1} $ & $ 419.58 $ \\\hline
 $ {E}_3 $ & $ 529.78 $ & $ -1.2 $ & $ 3.8\times10^{-2} $ & $ -1.1\times10^{-1} $ & $ -8.2\times10^{-2} $ & $ 528.39 $ \\\hline
 $ {E}_4 $ & $ 662.00 $ & $ 5.9\times10^{-1} $ & $ 2.3\times10^{-1} $ & $ -2.7\times10^{-2} $ & $ -6.8\times10^{-3} $ & $ 662.78 $ \\\hline
 $ {E}_5 $ & $ 834.00 $ & $ 2.2 $ & $ -5.5\times10^{-2} $ & $ 2.4\times10^{-3} $ & $ 2.1\times10^{-3} $ & $ 836.17 $ \\\hline
 $ {E}_6 $ & $ 1075.1 $ & $ 2.6 $ & $ -2.7\times10^{-2} $ & $ 6.4\times10^{-2} $ & $ -4.3\times10^{-3} $ & $ 1077.8 $ \\\hline
 $ {E}_7 $ & $ 1459.5 $ & $ 1.5 $ & $ 5.3\times10^{-1} $ & $ -2.2\times10^{-2} $ & $ 1.2\times10^{-2} $ & $ 1461.6 $ \\\hline
 $ {E}_8 $ & $ 2274.4 $ & $ 1.4\times10^{-1} $ & $ 1.8\times10^{-1} $ & $ 1.4\times10^{-1} $ & $ 3.0\times10^{-2} $ & $ 2274.9 $ \\
\hline
\end{tabular}
\caption{The contributions to the values of the conformal central charge $C_T$ at each instanton number in the rank-one Seiberg exceptional theories and the Morrison-Seiberg $\widetilde E_1$ theory.}
\label{Tab:InstantonCT}
\end{table}

\newpage
\vspace*{\fill}

\begin{table}[H]
\centering
\caption*{\bf Fundamental hypermultiplet masses}
\begin{tabular}{|c||c|c|c|c|c||c|}
\hline
$G$ & $C_J^{\rm pert}$ & $\Delta C_J^{\text{1-inst}}$ & $\Delta C_J^{\text{2-inst}}$ & $\Delta C_J^{\text{3-inst}}$ & $\Delta C_J^{\text{4-inst}}$ & $C_J^{\text{4-inst}}$
\\\hline\hline
 ${\rm {\rm SU}(3)} \subset E_3$ & $ 23.700 $ & $ -8.3\times10^{-2} $ & $ -3.5\times10^{-2} $ & $ -1.7\times10^{-2} $ & $ -1.1\times10^{-2} $ & $ 23.554 $
 \\
  ${\rm {\rm SU}(2)} \subset E_3$ & $ 23.700 $ & $ 1.4\times10^{-1} $ & $ -1.4\times10^{-2} $ & $ -1.1\times10^{-2} $ & $ -6.1\times10^{-3} $ & $ 23.818 $
\\\hline
 $ {E}_4 $ & $ 26.413 $ & $ 1.9\times10^{-1} $ & $ -1.4\times10^{-3} $ & $ -2.5\times10^{-3} $ & $ -7.1\times10^{-4} $ & $ 26.594 $ \\\hline
 $ {E}_5 $ & $ 30.131 $ & $ 2.1\times10^{-1} $ & $ 6.5\times10^{-3} $ & $ -6.9\times10^{-5} $ & $ 1.4\times10^{-7} $ & $ 30.345 $ \\\hline
 $ {E}_6 $ & $ 35.587 $ & $ 1.7\times10^{-1} $ & $ 8.2\times10^{-4} $ & $ 5.9\times10^{-4} $ & $ 3.1\times10^{-6} $ & $ 35.760 $ \\\hline
 $ {E}_7 $ & $ 44.657 $ & $ 7.3\times10^{-2} $ & $ 1.6\times10^{-2} $ & $ 8.4\times10^{-4} $ & $ 1.8\times10^{-4} $ & $ 44.747 $ \\\hline
 $ {E}_8 $ & $ 64.752 $ & $ 3.9\times10^{-3} $ & $ 5.9\times10^{-3} $ & $ 3.9\times10^{-3} $ & $ 3.3\times10^{-4} $ & $ 64.766 $ \\
\hline
\end{tabular}
\\
\vspace{.15in}
\caption*{\bf Instanton particle mass}
\begin{tabular}{|c||c|c|c|c|c||c|}
\hline
$G$ & $C_J^{\rm pert}$ & $\Delta C_J^{\text{1-inst}}$ & $\Delta C_J^{\text{2-inst}}$ & $\Delta C_J^{\text{3-inst}}$ & $\Delta C_J^{\text{4-inst}}$ & $C_J^{\text{4-inst}}$
\\\hline\hline
 $ \widetilde{E}_1 $ & $ 18.409 $ & $ -8.5\times10^{-1} $ & $ 7.3\times10^{-2} $ & $ 5.8\times10^{-3} $ & $ 9.5\times10^{-4} $ & $ 17.636 $ \\\hline
 $ {E}_1 $ & $ 18.409 $ & $ -2.7\times10^{-1} $ & $ -2.3\times10^{-1} $ & $ -1.6\times10^{-1} $ & $ -1.3\times10^{-1} $ & $ 17.605 $ 
\\\hline
 ${\rm {\rm SU}(3)} \subset E_3$ & $ 23.120 $ & $ 3.8\times10^{-1} $ & $ 1.4\times10^{-3} $ & $ -1.7\times10^{-2} $ & $ -1.0\times10^{-2} $ & $ 23.477 $ \\\hline
 $ {E}_4 $ & $ 26.190 $ & $ 3.6\times10^{-1} $ & $ 3.3\times10^{-2} $ & $ 1.1\times10^{-3} $ & $ 7.2\times10^{-4} $ & $ 26.589 $ \\\hline
 $ {E}_5 $ & $ 30.128 $ & $ 2.1\times10^{-1} $ & $ 6.5\times10^{-3} $ & $ -6.9\times10^{-5} $ & $ 1.4\times10^{-7} $ & $ 30.345 $ \\\hline
 $ {E}_6 $ & $ 35.664 $ & $ 2.3\times10^{-2} $ & $ 6.7\times10^{-2} $ & $ 7.0\times10^{-3} $ & $ -6.1\times10^{-4} $ & $ 35.760 $ \\\hline
 $ {E}_7 $ & $ 44.707 $ & $ -2.6\times10^{-2} $ & $ 6.4\times10^{-2} $ & $ 1.2\times10^{-4} $ & $ 1.3\times10^{-3} $ & $ 44.747 $ \\\hline
 $ {E}_8 $ & $ 64.756 $ & $ 3.9\times10^{-3} $ & $ -3.0\times10^{-3} $ & $ 3.9\times10^{-3} $ & $ 4.8\times10^{-3} $ & $ 64.766 $ \\
\hline
\end{tabular}
\\
\vspace{.15in}
\caption*{\bf Both}
\begin{tabular}{|c||c|c|c|c|c||c|}
\hline
$G$ & $C_J^{\rm pert}$ & $\Delta C_J^{\text{1-inst}}$ & $\Delta C_J^{\text{2-inst}}$ & $\Delta C_J^{\text{3-inst}}$ & $\Delta C_J^{\text{4-inst}}$ & $C_J^{\text{4-inst}}$
\\\hline\hline
 $ {\rm {\rm SU}(2)} \subset {E}_2 $ & $ 20.406 $ & $ 3.6\times10^{-1} $ & $ -7.3\times10^{-2} $ & $ -8.1\times10^{-2} $ & $ -6.1\times10^{-2} $ & $ 20.547 $
 \\
 $ {\rm U}(1) \subset {E}_2 $ & $ 6.2326 $ & $ 2.4\times10^{-2} $ & $ -1.9\times10^{-3} $ & $ -1.1\times10^{-3} $ & $ -5.4\times10^{-4} $ & $ 6.2531 $
 \\\hline
\end{tabular}
\\
\vspace{.15in}
\caption{The contributions to the values of the flavor central charge $C_J^G$ at each instanton number in the rank-one Seiberg exceptional theories and the Morrison-Seiberg $\widetilde E_1$ theory, computed using the fundamental hypermultiplet masses and the instanton particle mass.}
\label{Tab:InstantonCJ}
\end{table}

\vspace*{\fill}
\newpage

\begin{table}[H]
\centering
\begin{tabular}{|c||c|c|c|c|c|}
\hline
$G$ & Diff $C_J^{\rm pert}$ & Diff $C_J^{\text{1-inst}}$ & Diff $C_J^{\text{2-inst}}$ & Diff $C_J^{\text{3-inst}}$ & Diff $C_J^{\text{4-inst}}$
\\\hline\hline
 $ {\rm {\rm SU}(3)} \subset {E}_3 $ & $ 5.8\times10^{-1} $ & $ 1.2\times10^{-1} $ & $ 7.9\times10^{-2} $ & $ 7.8\times10^{-2} $ & $ 7.7\times10^{-2} $ \\
 $ {E}_4 $ & $ 2.2\times10^{-1} $ & $ 4.5\times10^{-2} $ & $ 1.1\times10^{-2} $ & $ 7.0\times10^{-3} $ & $ 5.5\times10^{-3} $ \\
 $ {E}_5 $ & $ 3.2\times10^{-3} $ & $ 5.9\times10^{-6} $ & $ 1.8\times10^{-9} $ & $ 2.8\times10^{-14} $ & $ o(10^{-15}) $ \\
 $ {E}_6 $ & $ -7.7\times10^{-2} $ & $ 7.2\times10^{-2} $ & $ 5.9\times10^{-3} $ & $ -5.9\times10^{-4} $ & $ 1.6\times10^{-5} $ \\
 $ {E}_7 $ & $ -5.1\times10^{-2} $ & $ 4.9\times10^{-2} $ & $ 4.9\times10^{-4} $ & $ 1.2\times10^{-3} $ & $ 1.1\times10^{-4} $ \\
 $ {E}_8 $ & $ -4.5\times10^{-3} $ & $ -4.5\times10^{-3} $ & $ 4.4\times10^{-3} $ & $ 4.4\times10^{-3} $ & $ 2.1\times10^{-5} $ \\
\hline
\end{tabular}
\caption{The differences between the values of the flavor central charge $C_J^G$ computed using the fundamental hypermultiplet masses and using the instanton particle mass, at each instanton number in the rank-one Seiberg exceptional theories.}
\label{Tab:InstantonCJDiff}
\end{table}

\section{Superconformal bootstrap}
\label{Sec:Bootstrap}

In previous sections, the values of the conformal central charge $C_T$ and the flavor central charge $C_J$ have been computed for the Seiberg exceptional theories, based on computations of the squashed sphere partition functions.  In this section, we study these theories by the superconformal bootstrap, exploiting the superconformal and flavor symmetries in these theories.  The values of $C_T$ and $C_J$ are related to certain OPE coefficients involving the BPS scalars residing in the flavor current multiplets, and allow us to pinpoint the Seiberg exceptional theories in the space of unitary solutions to bootstrap.  We then present the results of the numerical bootstrap.  As a check of our numerics, we first consider theories with {\rm SU}(2) flavor symmetry that may have higher spin conserved currents, and compare the bootstrap bounds with a single free hypermultiplet.  Then we go on to consider interacting theories with flavor groups $E_1 \cong {\rm SU}(2)$, $E_6$, $E_7$, and $E_8$, and compare with the Seiberg exceptional theories.

We stress here that our goal is not to compute $C_T$ or $C_J$ in certain theories by bootstrap, but to demonstrate that the bootstrap (the extremal functional method) in principle systematically solves certain strongly interacting theories such as the Seiberg exceptional theories, by providing evidence that their central charges saturate bootstrap bounds.

\subsection{Superconformal bootstrap with flavor symmetry}

Flavor symmetries of a five-dimensional superconformal field theory are realized by the conserved currents in the ${\cal D}[2]$ superconformal multiplets~\cite{Buican:2016hpb,Cordova:2016emh}. The superconformal primary of the ${\cal D}[2]$ multiplet is the moment map operator $L^a_{ij}$, which transforms in the adjoint representation of the flavor group. Its top component $L^a_{11}$ is ${1\over 2}$-BPS. These moment map operators furnish the so-called Higgs branch chiral ring, and their expectation values, subject to the ring relations, parametrize the Higgs branch vacuum moduli space $\cM_H$ of the superconformal field theory.

The superconformal bootstrap of the four-point functions of the moment map operators can be treated in a uniform fashion across spacetime dimensions three, five, and six~\cite{Chang:2017xmr}.\footnote{The four-point functions of other operators in the ${\cal D}[2]$ flavor current multiplet do not contain extra information, since they are related to the four-point function of the moment map operators by the superconformal Ward identity.
}
This section reviews the setup.  For simplicity, we assume for now that the moment map operators reside in a single flavor current multiplet.  After contracting the ${\rm SU}(2)_{\rm R}$ indices on each operator $L_{ij}(x)$ with auxiliary variables $Y^i$ to form $L(x,Y) \equiv L_{ij}(x) Y^{i} Y^{j}$, the four-point function takes the form
\ie
& \langle L(x_1, Y_1) L(x_2, Y_2) L(x_3, Y_3) L(x_4, Y_4) \rangle = \left( (Y_1 \cdot Y_2) (Y_3 \cdot Y_4) \over x_{12}^{2\epsilon} x_{34}^{2\epsilon} \right)^2 G(u, v; w),
\\
& G(u, v; w) = G_0(u, v) + G_1(u, v) w^{-1} + G_2(u, v) w^{-2},
\fe
where $\epsilon$ is related to the number of spacetime dimensions by $\epsilon = {d-2\over2}$, and the cross ratios $u, v, w$ are defined as\footnote{Due to the identity $(Y_1 \cdot Y_2)(Y_3 \cdot Y_4) - (Y_1 \cdot Y_3)(Y_2 \cdot Y_4) + (Y_1 \cdot Y_4)(Y_2 \cdot Y_3)=0$, there is only a single cross ratio $w$ formed out of $Y_1$, $Y_2$, $Y_3$, and $Y_4$.
}
\ie
& u = {x_{12}^2 x_{34}^2 \over x_{13}^2 x_{24}^2}, \quad v = {x_{14}^2 x_{23}^2 \over x_{13}^2 x_{24}^2}, \quad w = {(Y_1 \cdot Y_2)(Y_3 \cdot Y_4) \over (Y_1 \cdot Y_4)(Y_2 \cdot Y_3)},
\\
& x_{12}^2 = (x_1 - x_2)^2, \quad Y_1 \cdot Y_2 = Y_1^A Y_2^B \epsilon_{BA}.
\fe

By superconformal symmetry and the fact that $L_{11}$ is $1\over 2$-BPS, the OPE coefficients in $L \times L$ for all operators residing in each superconformal multiplet are linearly related to a single structure constant; otherwise, without the $1\over2$-BPS condition, there would be multiple superconformal blocks and multiple independent structure constants.  This implies that the four-point function $G$ can be expanded in superconformal blocks,
\ie
G(u, v; w) = \sum_{\cal X} \lambda_{\cal X}^2 {\cal A}^{\cal X}(u, v; w),
\fe
where ${\cal X}$ labels superconformal multiplets allowed by the selection rules for $L \times L$.  We refer the reader to~\cite{Chang:2017xmr,Bobev:2017jhk} for the expressions for the blocks.  Like the four-point function, the blocks are second order polynomials in $w^{-1}$,
\ie
{\cal A}^{\cal X}(u, v; w) = \widetilde{A}^{\cal X}_0(u, v) + \widetilde{A}^{\cal X}_1(u, v) w^{-1} + \widetilde{A}^{\cal X}_2(u, v) w^{-2}.
\fe
In unitary theories, the expansion coefficients $\lambda_{\cal X}^2$ are non-negative.

In a consistent conformal field theory, the operator product expansions must be associative, which entails the crossing symmetry constraints,
\ie
G\left( {u \over v}, {1 \over v}, - {w \over w+1} \right) = G(u, v; w) = \left( u^\epsilon \over v^\epsilon w \right)^2 G(v, u; w^{-1}).
\fe
The first equality is solved by imposing certain selection rules on the intermediate primary operators, so we only need to consider the second equality.  By the use of superconformal Ward identities, this second equation can be reduced to
\ie
u^{-2\epsilon} G_2(u, v) = v^{-2\epsilon} G_0(v, u).
\fe
Expanding this equation in superconformal blocks, we have
\ie
& 0 = \sum_{{\cal X} \in {\cal I} \cup \{ {\cal D}[0] \}} \lambda_{\cal X}^2 {\cal K}^{\cal X}(u, v), \quad {\cal K}^{\cal X}(u, v) \equiv v^{2\epsilon} \widetilde{A}^{\cal X}_2(u, v) - u^{2\epsilon} \widetilde{A}^{\cal X}_0(v, u),
\\
& \lambda_{{\cal D}[0]}^2 = 1, \quad \lambda_{\cal X}^2 \geq 0 \text{ for } {\cal X} \in {\cal I},
\fe
where ${\cal I}$ is the putative spectrum of non-identity superconformal multiplets.

Let us relax the assumption that the scalars reside in a single flavor current multiplet, and label them by an adjoint index $a$.  The four-point function now has extra indices
\ie
& \langle L_a(x_1, Y_1) L_b(x_2, Y_2) L_c(x_3, Y_3) L_d(x_4, Y_4) \rangle = {(Y_1 \cdot Y_2)^2 (Y_3 \cdot Y_4)^2 \over x_{12}^{4\epsilon} x_{34}^{4\epsilon}} G_{abcd}(u, v; w),
\fe
and its superconformal block decomposition takes the form
\ie
& G_{abcd}(u, v; w) = \sum_{{{R}}_I \in {\bf adj} \otimes {\bf adj}} P_I^{abcd} G_I(u, v; w),
\\
& G_I(u, v; w) = \sum_{\cal X} \lambda_{{\cal X}, I}^2 {\cal A}^{\cal X}(u, v; w),
\fe
where $P_I^{abcd}$ is the projection matrix that projects onto the contributions from intermediate operators transforming in the representation ${{R}}_I$.  Under crossing, the contributions from different ${{R}}_I$ mix together, and thus the crossing equation with flavors becomes
\ie
F_I^{\,J} G_J(u, v; w) = {u^{2\epsilon} \over v^{2\epsilon} w^2} G_I(v, u; w),
\fe
where the crossing matrix $F_I^{\,J}$ is defined as
\ie
F_I^{\,J} = {1 \over \text{dim}({{R}}_I)} P_I^{dabc} P_J^{abcd}.
\fe
The crossing matrices can be computed by the methods of~\cite{cvitanovic2008group}, and the results for the flavors groups of interest are listed in Table~\ref{Tab:Crossing}.

Putting everything together, the full system of bootstrap equations are
\ie
\label{FlavorBootstrapEquations}
& 0 = \sum_{({\cal X},J) \in {\cal I} \cup \{ ({\cal D}[0],{\bf 1}) \}} \lambda_{{\cal X},J}^2 ({\cal K}^{\cal X})_I^{\,J}(u, v),
\\
& ({\cal K}^{\cal X})_I^{\,J}(u, v) \equiv F_I^{\,J} v^{2\epsilon} \widetilde{A}^{\cal X}_2(u, v) - \D_I^{\,J} u^{2\epsilon} \widetilde{A}^{\cal X}_0(v, u),
\\
& \lambda_{{\cal D}[0],I}^2 = \delta_{I}^{\,0}, \quad \lambda_{{\cal X},I}^2 \geq 0 \text{ for } ({\cal X},I) \in {\cal I}.
\fe
The putative spectrum ${\cal I}$ contains a subset of
\ie
{\cal D}[2], \quad {\cal D}[4], \quad {\cal B}[0]_{\ell}, \quad {\cal B}[2]_{\ell}, \quad {\cal L}[0]_{\Delta,\ell},
\fe
subject to the following selection rules:
\begin{enumerate}
\item  Symmetric representations in ${\bf adj} \times {\bf adj}$ appear with $\ell + J_{\rm R}$ even, and antisymmetric ones with $\ell + J_{\rm R}$ odd.
\item  ${\cal D}[0]$ must be in the trivial representation.
\item  ${\cal D}[2]$ must be in the adjoint representation of the flavor group.
\item  In interacting theories with a unique stress tensor, ${\cal B}[0]_0$ must be in the trivial representation, and ${\cal B}[0]_{\ell>0}$ cannot appear.
\end{enumerate}

The bootstrap makes contact with the previous sections by a relation between the OPE coefficients for the stress tensor and flavor current multiplets and the central charges $C_T$ and $C_J$. The formulae are~\cite{Chang:2017xmr}
\ie
\label{lambdaToCTCJ}
\lambda_{{\cal B}[0]_0,{\bf 1}}^2 = {4(2\epsilon+2)(2\epsilon+3) \over 2\epsilon+1} {1 \over C_T}, \quad \lambda_{{\cal D}[2],{\bf adj}}^2 = {4(2\epsilon+1) h^\vee \over 2\epsilon} { 1\over C_J}.
\fe

\begin{table}[h]
\centering
\begin{tabular}{|c|c|c|c|c|}
\hline
$G_F$ & $h^\vee$ & {\bf adj} $\otimes_S$ {\bf adj} & {\bf adj} $\otimes_A$ {\bf adj} & $F$
\\\hline\hline
{\rm SU}(2) & 2 & {\bf 1} $\oplus$ {\bf 5} & {\bf 3} & $\left(
\begin{array}{ccccc}
 {1 \over 3} & {5 \over 3} & 1 \\
 {1 \over 3} & {1 \over 6} & -{1 \over 2} \\
 {1 \over 3} & -{5 \over 6} & {1 \over 2} \\
\end{array}
\right)$
\\
$E_6$ & 12 & {\bf 1} $\oplus$ {\bf 650} $\oplus$ {\bf 2430} & {\bf 78} $\oplus$ {\bf 2925} & $\left(
\begin{array}{ccccc}
 {1 \over 78} & {25 \over 3} & {405 \over 13} & 1 & {75 \over 2} \\
 {1 \over 78} & -{7 \over 24} & {81 \over 104} & {1 \over 4} & -{3 \over 4} \\
 {1 \over 78} & {5 \over 24} & {29 \over 104} & -{1 \over 12} & -{5 \over 12} \\
 {1 \over 78} & {25 \over 12} & -{135 \over 52} & {1 \over 2} & 0 \\
 {1 \over 78} & -{1 \over 6} & -{9 \over 26} & 0 & {1 \over 2} \\
\end{array}
\right)$
\\
$E_7$ & 18 & {\bf 1} $\oplus$ {\bf 1539} $\oplus$ {\bf 7371} & {\bf 133} $\oplus$ {\bf 8645} & $\left(
\begin{array}{ccccc}
 {1 \over 133} & {81 \over 7} & {1053 \over 19} & 1 & 65 \\
 {1 \over 133} & -{23 \over 70} & {78 \over 95} & {2 \over 9} & -{13 \over 18} \\
 {1 \over 133} & {6 \over 35} & {61 \over 190} & -{1 \over 18} & -{4 \over 9} \\
 {1 \over 133} & {18 \over 7} & -{117 \over 38} & {1 \over 2} & 0 \\
 {1 \over 133} & -{9 \over 70} & -{36 \over 95} & 0 & {1 \over 2} \\
\end{array}
\right)$
\\
$E_8$ & 30 & {\bf 1} $\oplus$ {\bf 3875} $\oplus$ {\bf 27000} & {\bf 248} $\oplus$ {\bf 30380} & $\left(
\begin{array}{ccccc}
 {1 \over 248} & {125 \over 8} & {3375 \over 31} & 1 & {245 \over 2} \\
 {1 \over 248} & -{3 \over 8} & {27 \over 31} & {1 \over 5} & -{7 \over 10} \\
 {1 \over 248} & {1 \over 8} & {23 \over 62} & -{1 \over 30} & -{7 \over 15} \\
 {1 \over 248} & {25 \over 8} & -{225 \over 62} & {1 \over 2} & 0 \\
 {1 \over 248} & -{5 \over 56} & -{90 \over 217} & 0 & {1 \over 2} 
\\
\end{array}
\right)$
\\\hline
\end{tabular}
\caption{
The crossing matrices for {\bf adj} $\otimes$ {\bf adj} of {\rm SU}(2), $E_6$, $E_7$, and $E_8$ flavor groups.  The basis for the crossing matrices are in the order as shown in the {\bf adj} $\otimes_S$ {\bf adj} and {\bf adj} $\otimes_A$ {\bf adj} columns.
}
\label{Tab:Crossing}
\end{table}

\subsection{Solving theories by the extremal functional method}

The linear functional method exploits the non-negativity of the coefficients in the (super)conformal block expansion in unitary theories, and puts nontrivial bounds on the operator dimensions and OPE coefficients.  This section reviews this method. We then explain how the extremal functional method {\it solves} the theories that saturate bootstrap bounds.

The idea is to consider a vector valued linear functional $\A^I$ on functions of $u, v$, and act on the bootstrap equations \eqref{FlavorBootstrapEquations} to obtain
\ie
0 = \sum_{({\cal X},J) \in {\cal I} \cup \{ ({\cal D}[0],{\bf 1}) \}} \sum_I \lambda_{{\cal X},J}^2 \A^I[({\cal K}^{\cal X})_I^{\,J}].
\fe
In the following, we keep the sum over the $I$ index implicit, and write $({\cal K}^{\cal X})^{J}$ as ${\cal K}^{{\cal X},J}$.  Each linear functional that satisfies
\ie
\label{FunctionalConstraints}
\A[{\cal K}^{{\cal D}[0],{\bf 1}}] = -1, \quad \A[{\cal K}^{{\cal X},J}] \geq 0 \text{ for } ({\cal X}, J) \in {\cal I}
\fe
implies a bound on the OPE coefficients
\ie\label{eqn:OPEbound}
\lambda_{{\cal X},J}^2 = {\lambda_{{\cal X},J}^2 \over \sum_{({\cal X}',J') \in {\cal I}} \lambda_{{\cal X}',J'}^2 \A[{\cal K}^{{\cal X'},J'}]} \leq {1 \over \A[{\cal K}^{{\cal X},J}]}.
\fe
By maximizing $\A[{\cal K}^{{\cal X},J}]$ within the space of linear functionals satisfying \eqref{FunctionalConstraints}, we obtain the most stringent upper bound on $\lambda_{{\cal X},J}^2$. The functional that maximizes $\A[{\cal K}^{{\cal X},J}]$ is called the {\it extremal functional}~\cite{ElShowk:2012hu}, which we denote by $\A_{{\cal X},J}$. If there exists a four-point function that saturates the bound \eqref{eqn:OPEbound}, then the OPE coefficients satisfy
\ie
0 = \sum_{({\cal X}',J') \in {\cal I}\setminus \{({\cal X},J)\}} \lambda_{{\cal X}',J'}^2 \A_{{\cal X},J}[{\cal K}^{{\cal X'},J'}],
\fe
which, given \eqref{FunctionalConstraints}, means that the multiplets $({\cal X}',J')$ other than $({\cal X},J)$ contributing to this four-point function have vanishing $\A_{{\cal X},J}[{\cal K}^{{\cal X'},J'}]$. Such a four-point function is called an {\it extremal four-point function}~\cite{ElShowk:2012hu,El-Showk:2014dwa}.

In practice, the above extremization procedure can only be performed within a finite-dimensional subspace of linear functionals.  The following is a convenient basis.  Define $z$ and $\bar z$ by
\ie
u = z \bar z, \quad v = (1-z) (1-\bar z),
\fe
so that crossing $u \leftrightarrow v$ is equivalent to $(z, \bar z) \leftrightarrow (1-z, 1-\bar z)$, and consider the space of linear functionals at derivative order $\Lambda$:
\ie
\A = \sum_{m,n=0}^\Lambda \A_{m,n} \partial_z^m \partial_{\bar z}^n |_{z = \bar z = {1 \over 2}}, \quad \A_{m,n} \in \bR.
\fe
The optimal bounds at higher and higher derivative orders become tighter and tighter, and the most stringent bound is obtained by extrapolating $\Lambda$ to infinity. 
Other practicalities with the bootstrap numerics have been discussed in~\cite{Chang:2017xmr}, to which the reader is referred.

The most interesting bounds to consider are perhaps on the OPE coefficients for the stress tensor and flavor current multiplets, since they are related to the central charges $C_T$ and $C_J$, which have been computed in previous sections.

\subsection{Numerical bounds}

This section presents the results of the numerical application of the linear functional method to bootstrap superconformal field theories with flavor groups ${\rm SU}(2)$, $E_6$, $E_7$, and $E_8$. In all cases considered, the bounds on central charges are extrapolated to infinite derivative order $\Lambda \to \infty$ using two ansatzes, quadratic
\ie
{\rm min} \, C_{T/J} = a + {b \over \Lambda} + {c \over \Lambda^2}, \quad b < 0, \quad \Lambda \geq 24, 28, 32
\label{QuadAnsatz}
\fe
and linear
\ie
{\rm min} \, C_{T/J} = a + {b \over \Lambda}, \quad \Lambda \geq 36.
\label{LinearAnsatz}
\fe
If we assume that a theory saturates a bound, then the gap in the spectrum of long multiplets is determined by the first zero of the extremal functional acted on the contribution of the spin-zero long multiplet to the crossing equation, $\A_{{\cal D}[2],{\bf adj}}[{\cal K}^{{\cal L}[0]_{\Delta,0}}]$.  The resulting gaps are also extrapolated to infinite derivative order $\Lambda \to \infty$ using two ansatzes, exponential
\ie
\Delta_{\rm gap}^{{\cal L}[0]} = a + b \exp{c \over \Lambda}, \quad \Lambda \geq 24,
\label{GapAnsatz}
\fe
and linear
\ie
\Delta_{\rm gap}^{{\cal L}[0]} = a + {b \over \Lambda}, \quad \Lambda \geq 36.
\label{GapAnsatzLinear}
\fe
The variation among extrapolations with different ansatzes serves as an estimate for the extrapolation error.

\subsubsection{Free hypermultiplet}
\label{Sec:BootstrapFreeHyper}

\begin{figure}[h]
\centering
\subfloat{
\includegraphics[width=.45\textwidth]{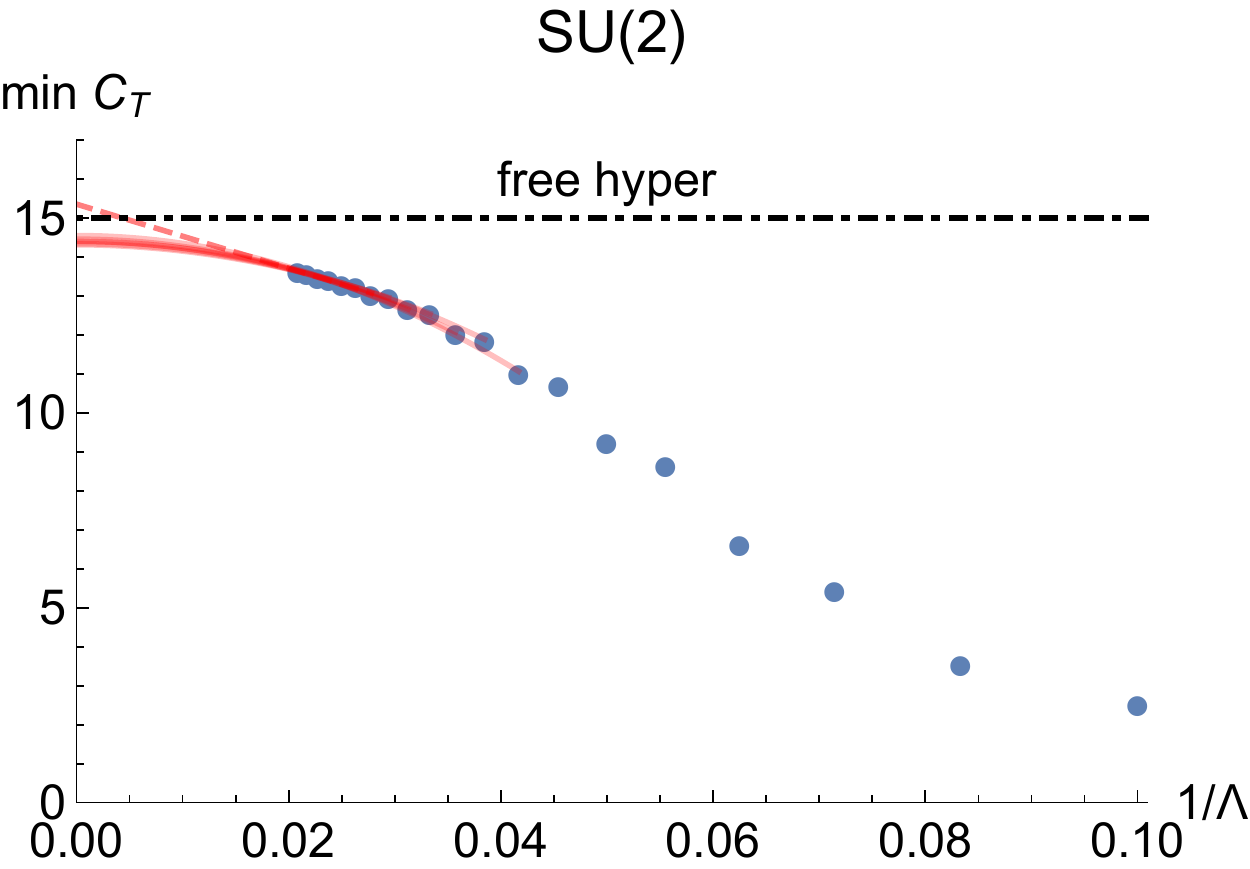}
}
\quad
\subfloat{
\includegraphics[width=.45\textwidth]{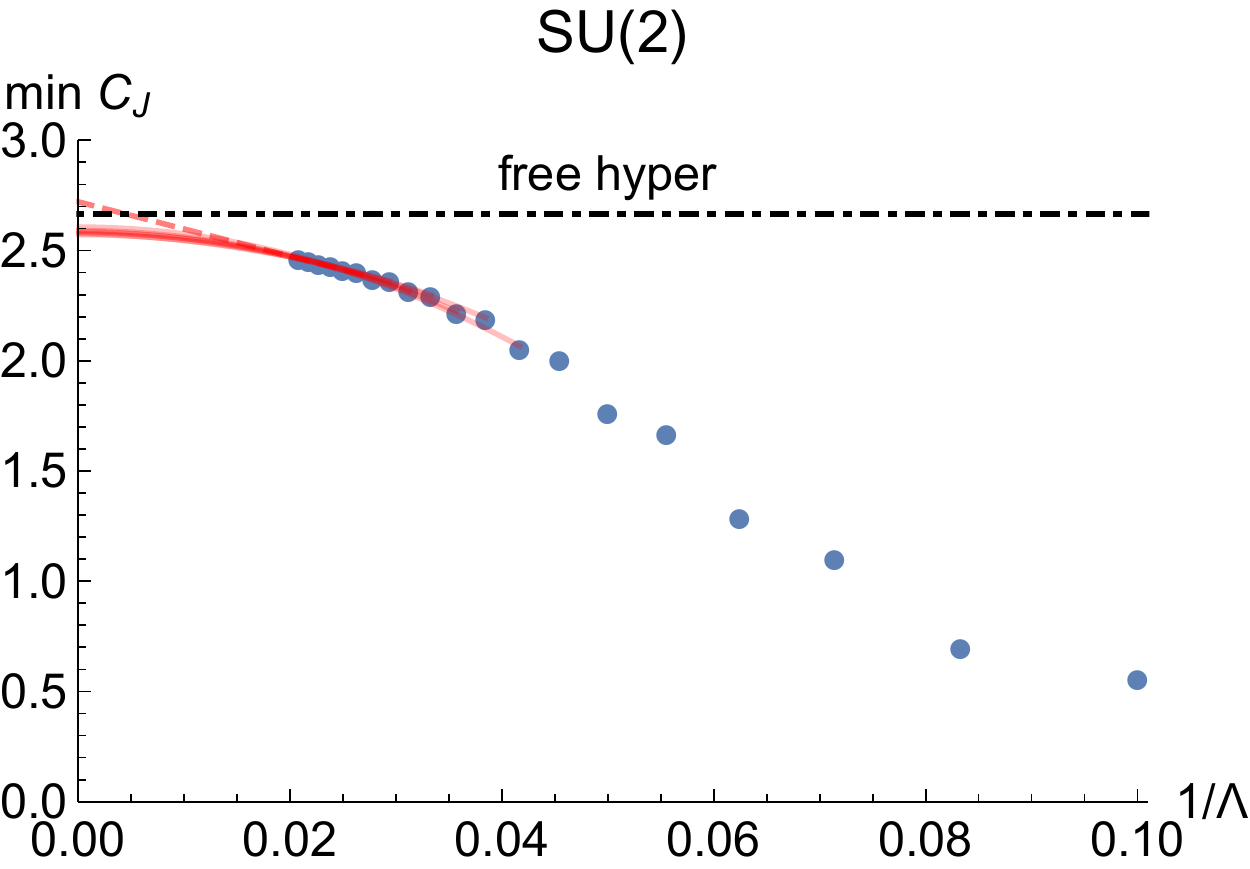}
}
\caption{Lower bounds on $C_T$ and $C_J$ for general (free or interacting) theories with {\rm SU}(2) flavor group and twist gap $\Delta - s \geq 6$, at various derivative orders $\Lambda$ and extrapolated to infinite order using the quadratic ansatz \eqref{QuadAnsatz} (solid) and the linear ansatz \eqref{LinearAnsatz} (dashed).  Also shown are the values $C_T = 15$ and $C_J = {8 \over 3}$ for a free hypermultiplet (dotdashed).}
\label{Fig:CTandCJSU2HS}
\end{figure}

As a first step, we make assumptions that should single out the free hypermultiplet as a solution to the crossing equations -- ${\rm SU}(2)$ flavor symmetry, the existence of higher spin conserved currents (residing in ${\cal B}[0]_{\ell>0}$), and twist gap $\Delta - s \geq 6$ in the spectrum of long multiplets ({\it cf.} $\Delta - s \geq 4$ is the unitarity bound).  With these assumptions, Figure~\ref{Fig:CTandCJSU2HS} shows the upper bounds on $C_T$ and $C_J$ at derivative orders $\Lambda = 4, 6, \dotsc, 48$, and extrapolations to infinite derivative order.  A free hypermultiplet has $C_T = 15$ and $C_J = {8 \over 3}$, saturating both the infinite-derivative-order bounds on $C_T$ and $C_J$ to within extrapolation errors,
\ie
\lim_{\Lambda \to \infty} {\rm min}\,C_T = 14.9(5), \quad \lim_{\Lambda \to \infty} {\rm min}\,C_J = 2.65(7).
\fe

\begin{figure}[H]
\centering
\subfloat{
\includegraphics[width=.45\textwidth]{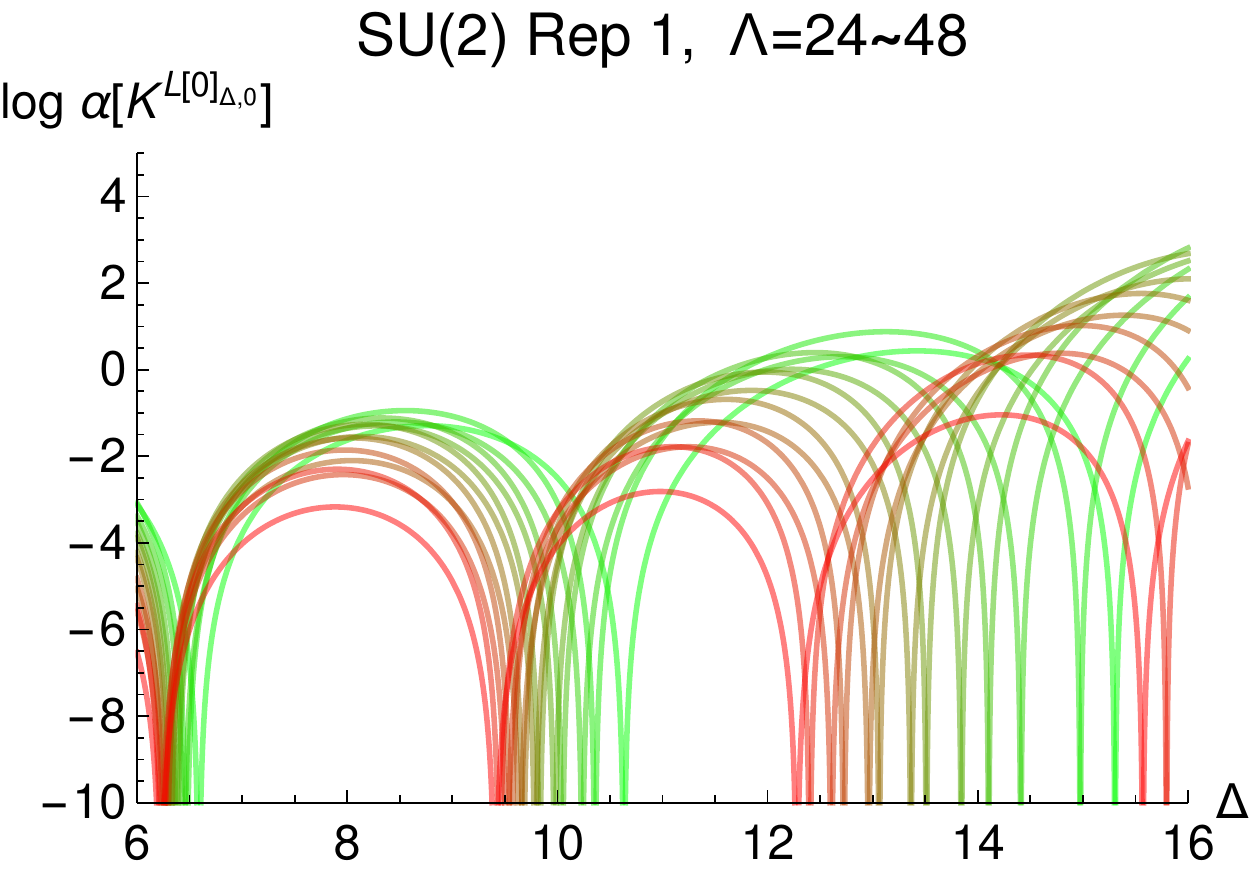}
}
\quad
\subfloat{
\includegraphics[width=.45\textwidth]{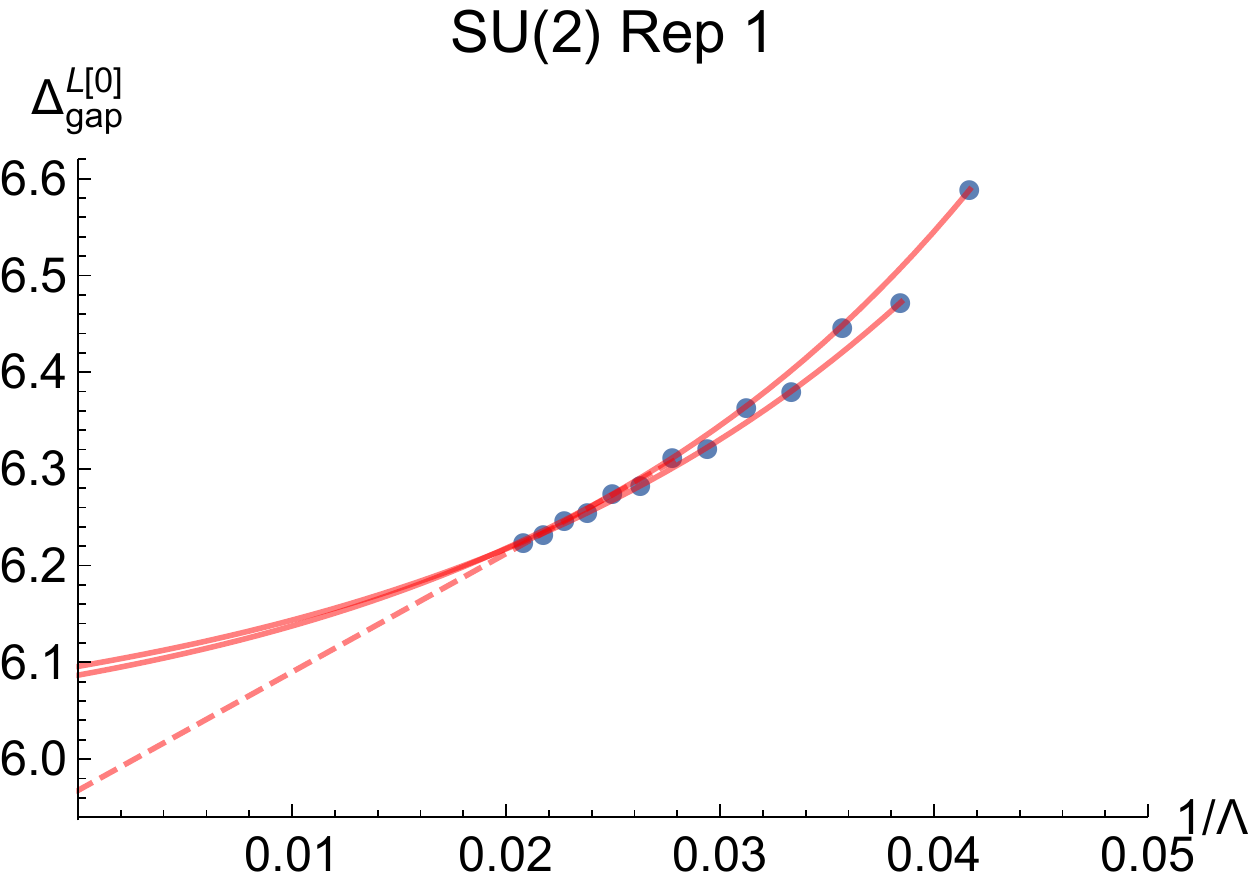}
}
\\
\subfloat{
\includegraphics[width=.45\textwidth]{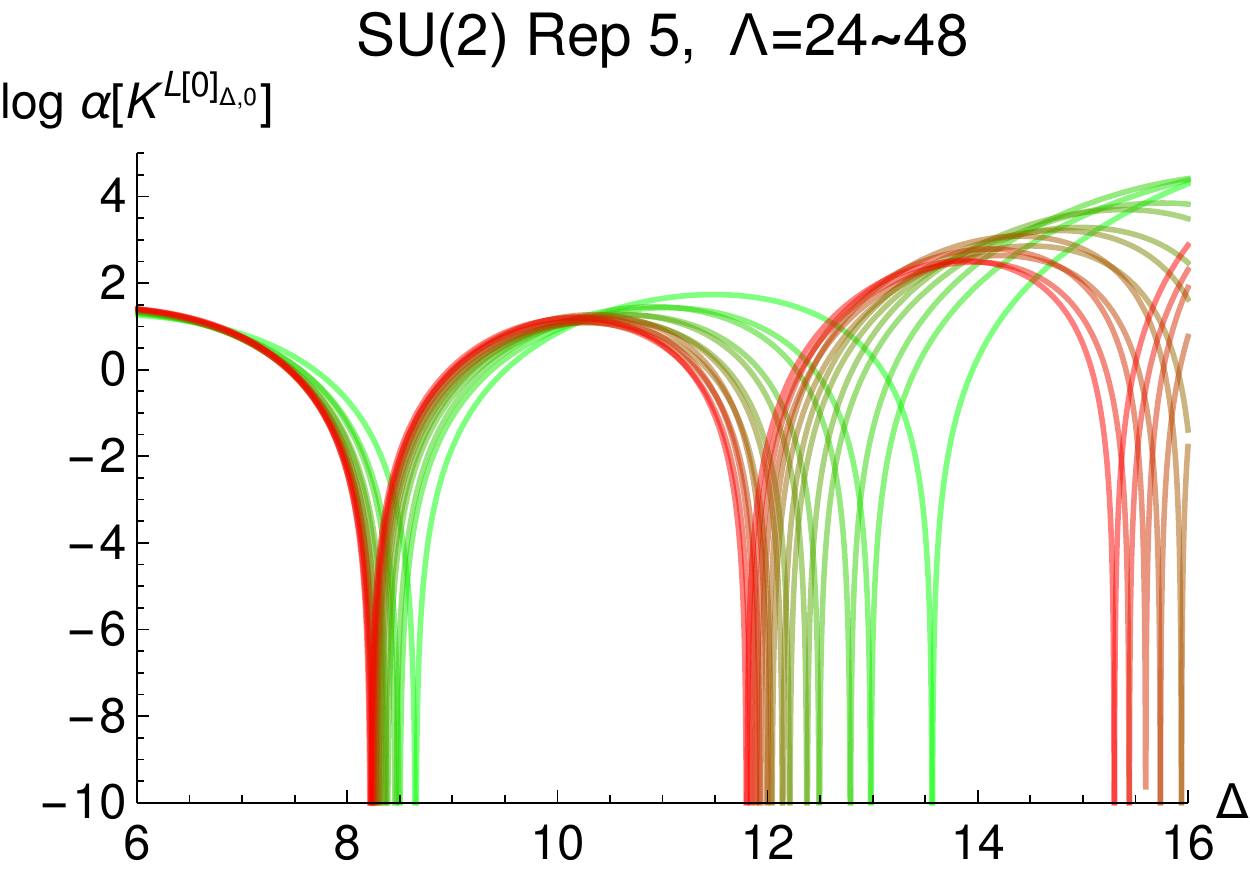}
}
\quad
\subfloat{
\includegraphics[width=.45\textwidth]{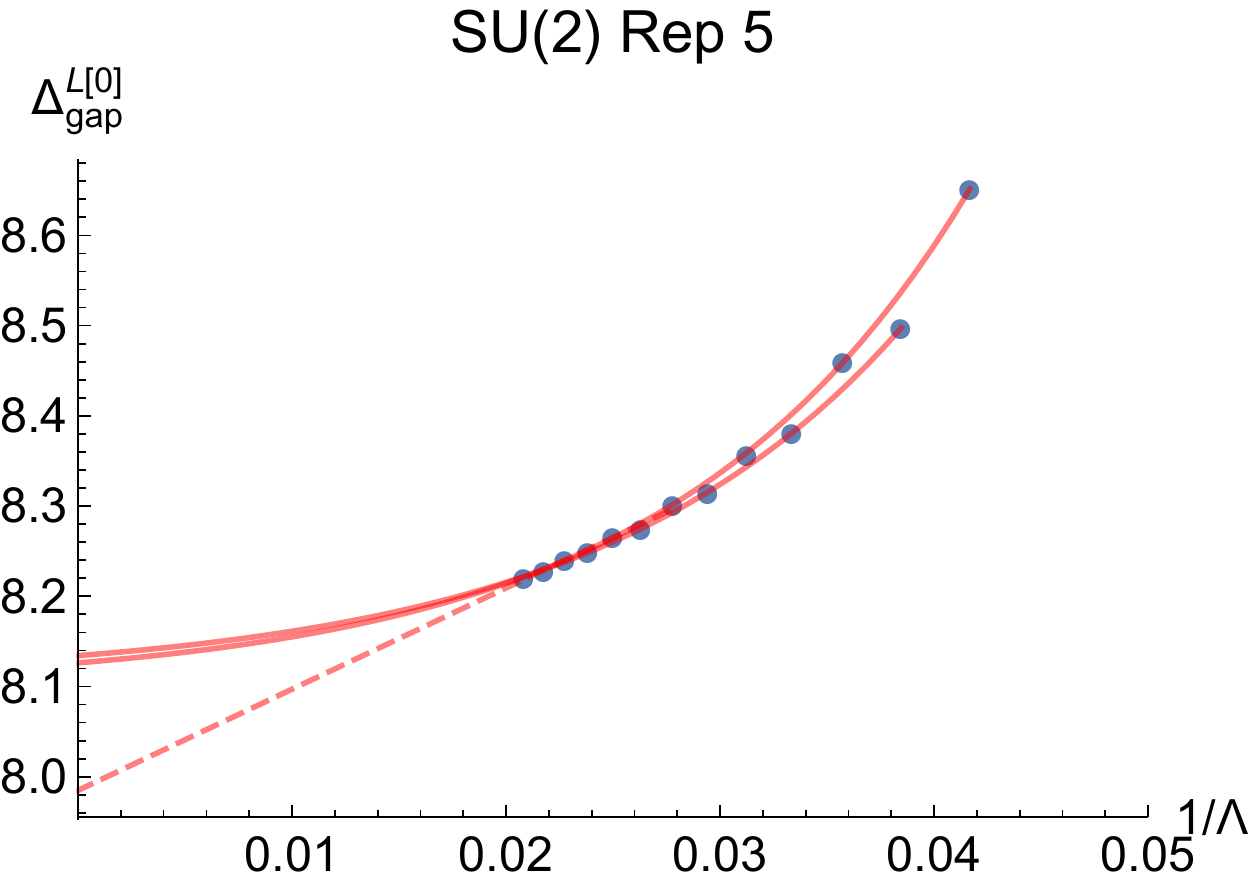}
}
\caption{{\bf Left:} The extremal functional optimizing the lower bound on $C_J$ for general (free or interacting) theories with {\rm SU}(2) flavor group and twist gap $\Delta - s \geq 6$, acted on the contribution of the spin-zero long multiplet to the crossing equation, $\A_{{\cal D}[2],{\bf 3}}[{\cal K}^{{\cal L}[0]_{\Delta,0}}]$, in the {\bf 1} and {\bf 5} of ${\rm SU}(2)$, plotted in logarithmic scale.  Increasing derivative orders $\Lambda = 24, 26, \dotsc, 48$ are shown from green to red.  {\bf Right:} The gaps at different $\Lambda$, and extrapolations to $\Lambda \to \infty$ using the quadratic ansatz \eqref{GapAnsatz} for $\Lambda \in 4\bZ$ and $\Lambda \in 4\bZ+2$, separately (solid), and using the linear ansatz \eqref{GapAnsatzLinear} (dashed).}
\label{Fig:HyperFunctional}
\end{figure}

To further check that the extremal theory minimizing $C_J$ is indeed a free hypermultiplet, we employ the extremal functional method to determine the spectrum appearing in the ${\cal D}[2] \times {\cal D}[2]$ OPE in the theory with minimal $C_J$, and compare with the known spectrum of a free hypermultiplet.  Figure~\ref{Fig:HyperFunctional} shows the gaps (lowest scaling dimension) at various derivative orders, and extrapolations to infinite derivative order.  We find the gaps in the $\bf 1$ and $\bf 5$ channels to be
\ie
\Delta_{\rm gap}^{{\cal L}[0], \bf 1} = 6.03(6), \quad \Delta_{\rm gap}^{{\cal L}[0], \bf 5} = 8.06(7),
\fe
respectively, consistent with the actual gaps 6 and 8.

\vspace*{\fill}

\vspace*{\fill}

\subsubsection{$\text{SU(2)}$ flavor symmetry}
\label{Sec:E1}

\begin{figure}[H]
\centering
\subfloat{
\includegraphics[width=.8\textwidth]{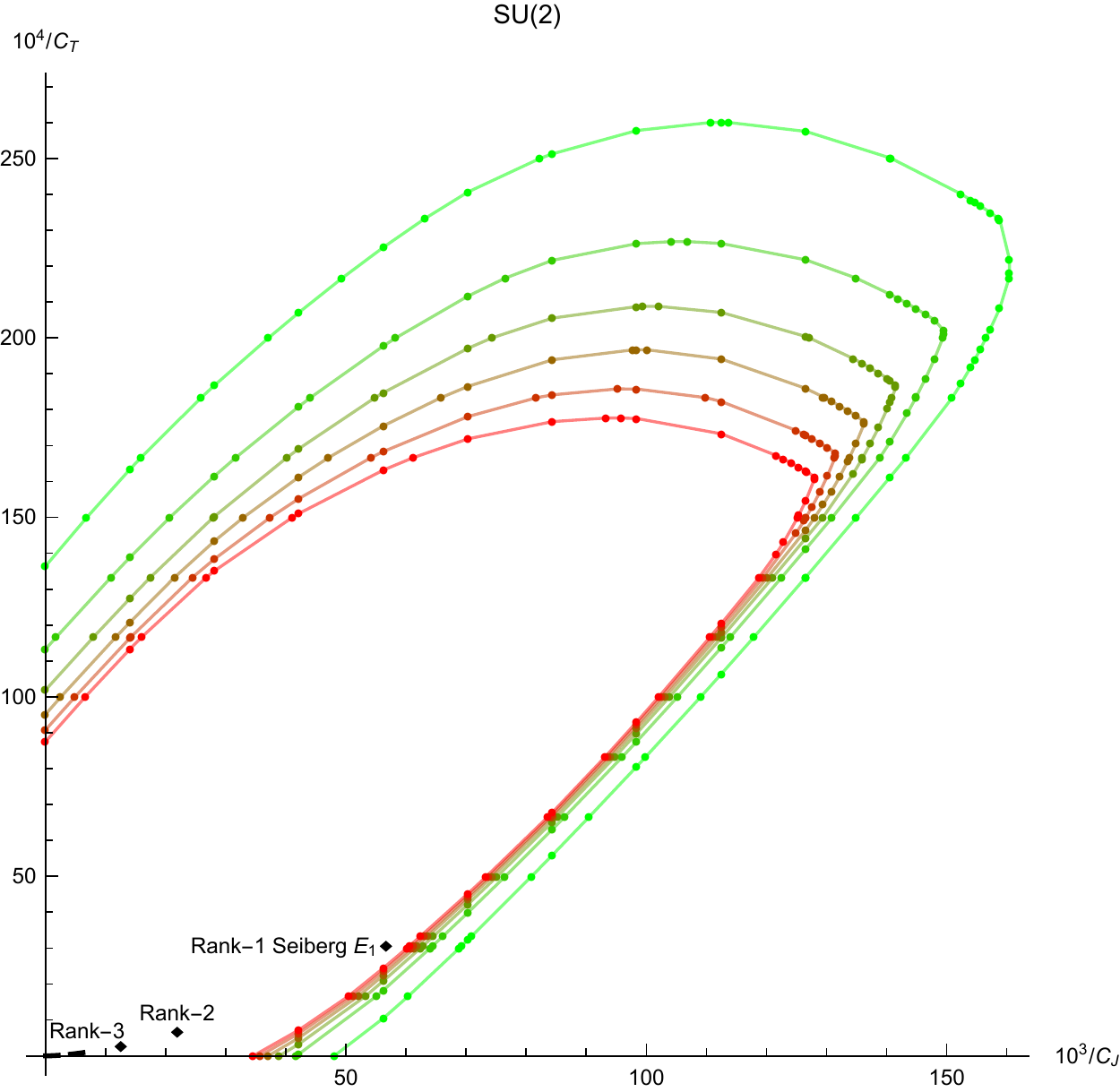}
}
\caption{Allowed region in the $C_T-C_J$ plane for interacting theories with {\rm SU}(2) flavor group, at derivative orders $\Lambda = 20, 24, \dotsc, 40$, shown from green to red.  Also shown are the four-instanton values in the rank-one Seiberg $E_1$ theory, the perturbative values in the rank-two and three, and the values according to the large-rank formula (dashed line).}
\label{Fig:CTCJSU2}
\end{figure}

Next, we bootstrap interacting unitary superconformal field theories with ${\rm SU}(2)$ flavor symmetry.  Interacting means that the ${\cal B}[0]_{\ell>0}$ multiplets containing higher spin conserved currents are absent.  Figure~\ref{Fig:CTCJSU2} maps out the allowed region in the $C_T-C_J$ plane, where there appears to be a sharp corner with the minimal allowed $C_J$.  Also shown are the four-instanton values of $(C_J, C_T)$ in the rank-one Seiberg $E_1$ theory, and the perturbative values in the higher-rank theories borrowed from upcoming work~\cite{Chang:2017mxc}.  Figure~\ref{Fig:CTandCJEn} shows the lower bounds on $C_T$ and $C_J$ at various derivative orders, and extrapolations to infinite derivative order.  Assuming that the four-instanton values of $C_T$ and $C_J$ are good approximations to the exact values, the extrapolations suggest that neither the bound on $C_T$ nor that on $C_J$ is saturated by the rank-one Seiberg $E_1$ theory,
\ie
\lim_{\Lambda \to \infty} {\rm min}\,C_T = 87(6). < 326., \quad \lim_{\Lambda \to \infty} {\rm min}\,C_J = 10.1(3) < 17.6.
\fe
In particular, they suggest that the sharp corner does not approach the rank-one Seiberg $E_1$ theory at infinite derivative order.  We are not aware of a candidate theory that sits at the sharp corner.

\begin{figure}[H]
\centering
\subfloat{
\includegraphics[width=.45\textwidth]{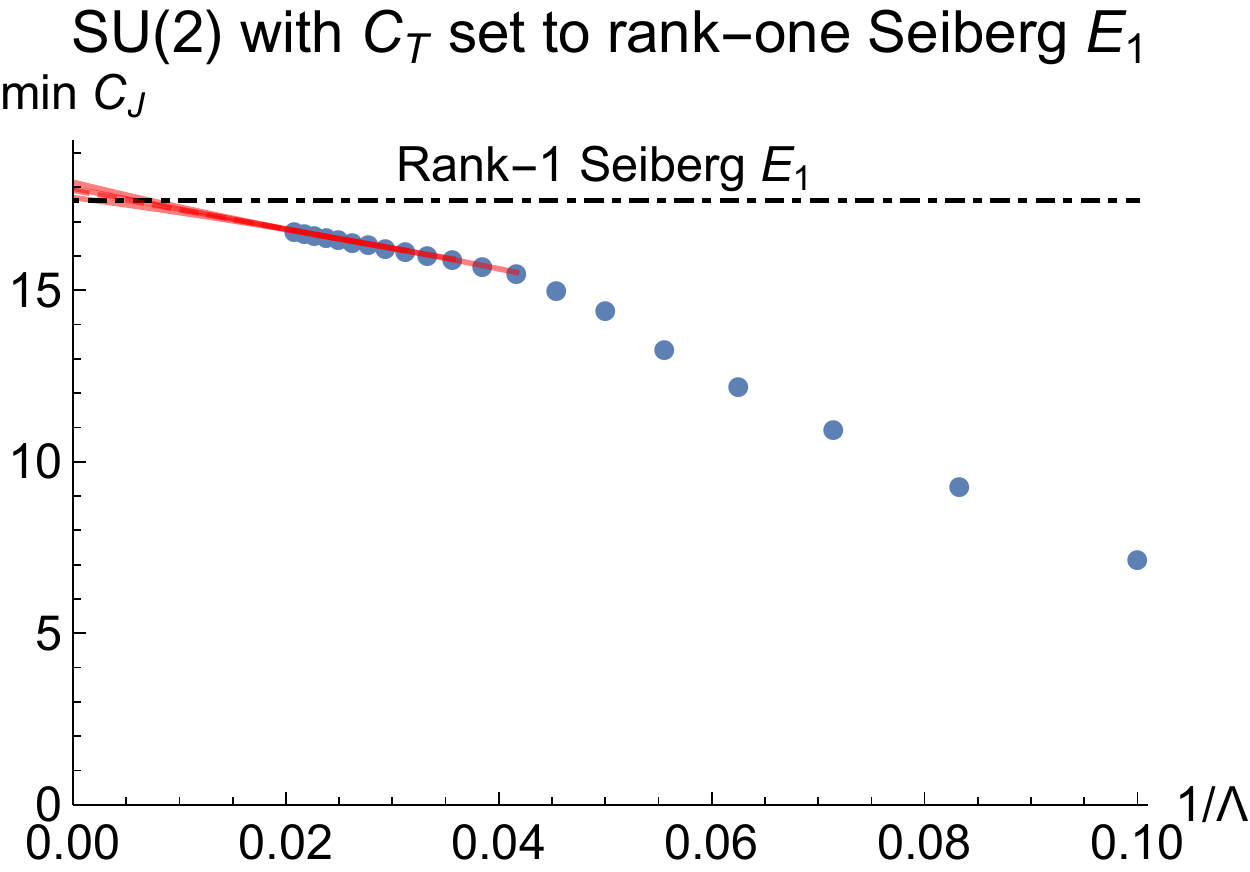}
}
\caption{Lower bounds on $C_J$ with $C_T$ set to the value in the rank-one Seiberg $E_1$ theory, for interacting theories with {\rm SU}(2) flavor group, at various derivative orders $\Lambda$ and extrapolated to infinite order using the quadratic ansatz \eqref{QuadAnsatz} (solid) and the linear ansatz \eqref{LinearAnsatz} (dashed).  Also shown is the value of $C_J$ in the rank-one Seiberg $E_1$ theory (dotdashed).
}
\label{Fig:MinSU2}
\end{figure}

\begin{figure}[H]
\centering
\subfloat{
\includegraphics[width=.45\textwidth]{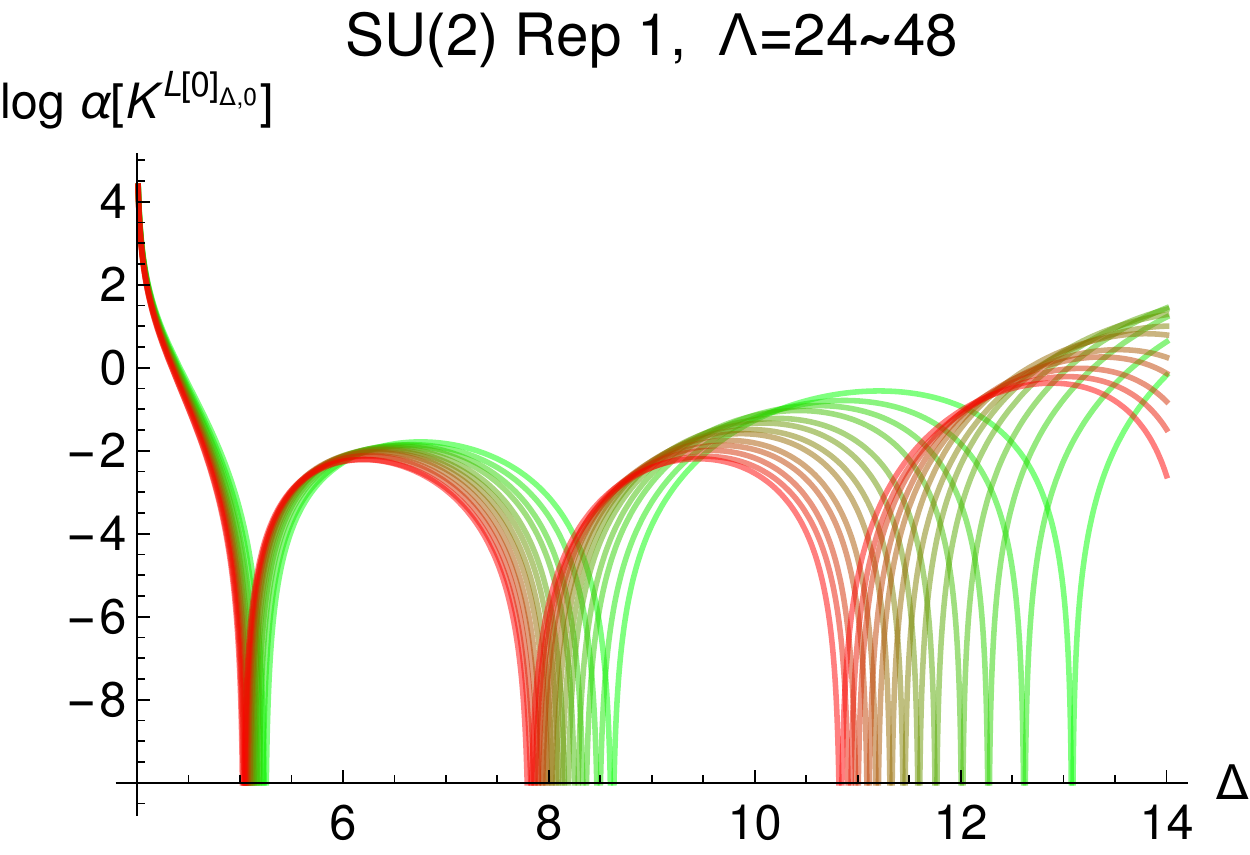}
}
\quad
\subfloat{
\includegraphics[width=.45\textwidth]{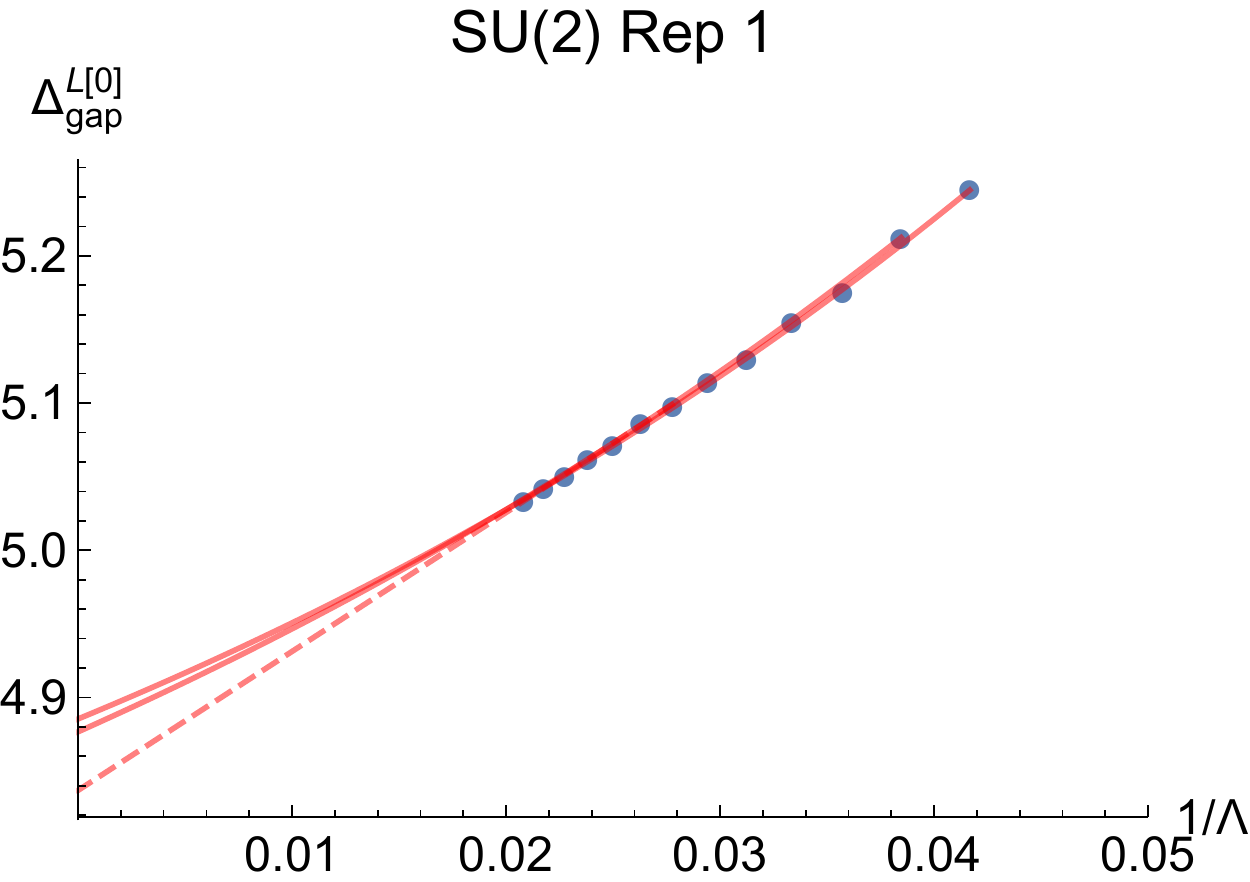}
}
\\
\subfloat{
\includegraphics[width=.45\textwidth]{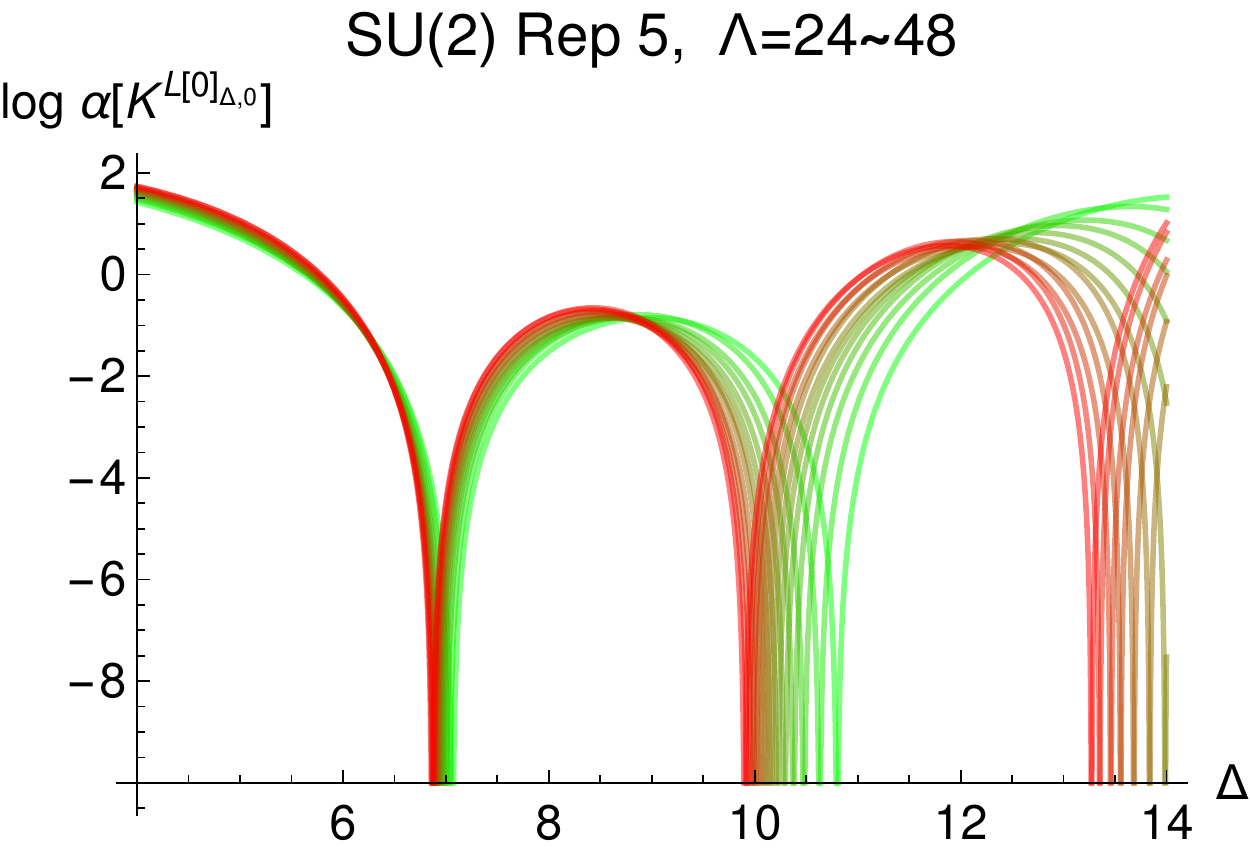}
}
\quad
\subfloat{
\includegraphics[width=.45\textwidth]{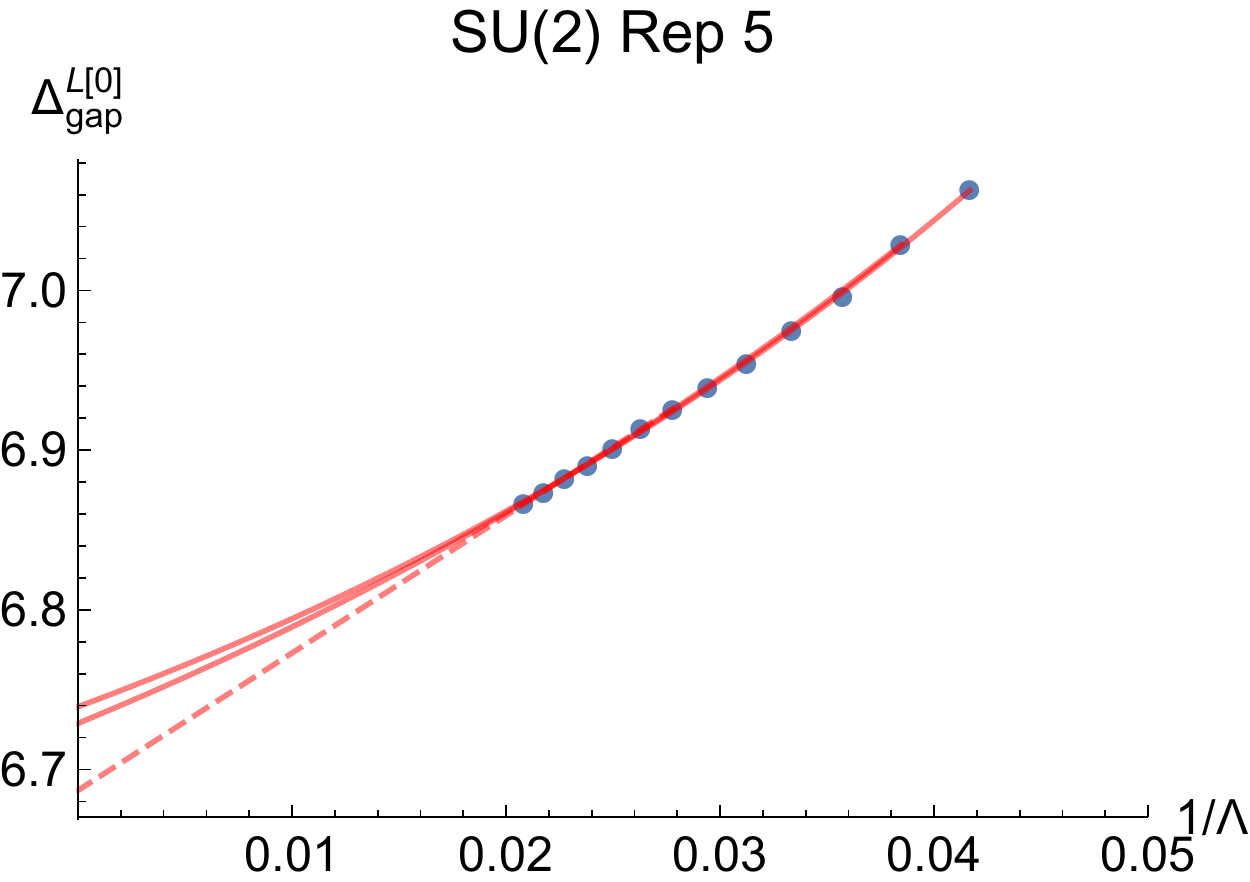}
}
\caption{{\bf Left:} The extremal functional optimizing the lower bound on $C_J$ when $C_T$ is set to the four-instanton value in the rank-one Seiberg $E_1$ theory, acted on the contribution of the spin-zero long multiplet to the crossing equation, $\A_{{\cal D}[2],{\bf 3}}[{\cal K}^{{\cal L}[0]_{\Delta,0}}]$, in the {\bf 1} and {\bf 5} of ${\rm {\rm SU}(2)}$, plotted in logarithmic scale.  Increasing derivative orders $\Lambda = 24, 26, \dotsc, 48$ are shown from green to red.  {\bf Right:} The gaps at different $\Lambda$, and extrapolations to $\Lambda \to \infty$ using the exponential ansatz \eqref{GapAnsatz} for $\Lambda \in 4\bZ$ and $\Lambda \in 4\bZ+2$, separately (solid), and using the linear ansatz \eqref{GapAnsatz} (dashed).}
\label{Fig:E1Functional}
\end{figure}

Nonetheless, the rank-one Seiberg $E_1$ theory does appear to sit close to the boundary of the allowed region.  To examine this further, Figure~\ref{Fig:MinSU2} shows the lower bounds on $C_J$ when $C_T$ is set to the four-instanton value of $C_T$ in the rank-one Seiberg $E_1$ theory, and extrapolated to infinite derivative order, giving
\ie
\lim_{\Lambda \to \infty} {\rm min}\,C_J = 17.9(1).
\fe
Assuming that the four-instanton values of $C_T$ and $C_J$ are good approximations, the extrapolation supports the hypothesis that the rank-one Seiberg $E_1$ theory sits at the boundary of the allowed region.  If this hypothesis is true, then we can employ the extremal functional method to determine the spectrum of long multiplets appearing the ${\cal D}[2] \times {\cal D}[2]$ OPE in the rank-one Seiberg $E_1$ theory.  Figure~\ref{Fig:E1Functional} shows the gaps (the lowest scaling dimension) at various derivative orders, and extrapolations to $\Lambda \to \infty$ using the quadratic ansatz \eqref{GapAnsatz} and linear ansatz \eqref{GapAnsatzLinear}.  We find the gaps in the $\bf 1$ and $\bf 5$ channels to be
\ie
\Delta_{\rm gap}^{{\cal L}[0],\bf 1} = 4.86(2), \quad \Delta_{\rm gap}^{{\cal L}[0],\bf 5} = 6.71(2).
\fe

\newpage

\vspace*{\fill}

\vspace*{\fill}

\newpage

\vspace*{\fill}

\begin{figure}[H]
\centering
\subfloat{
\includegraphics[width=.45\textwidth]{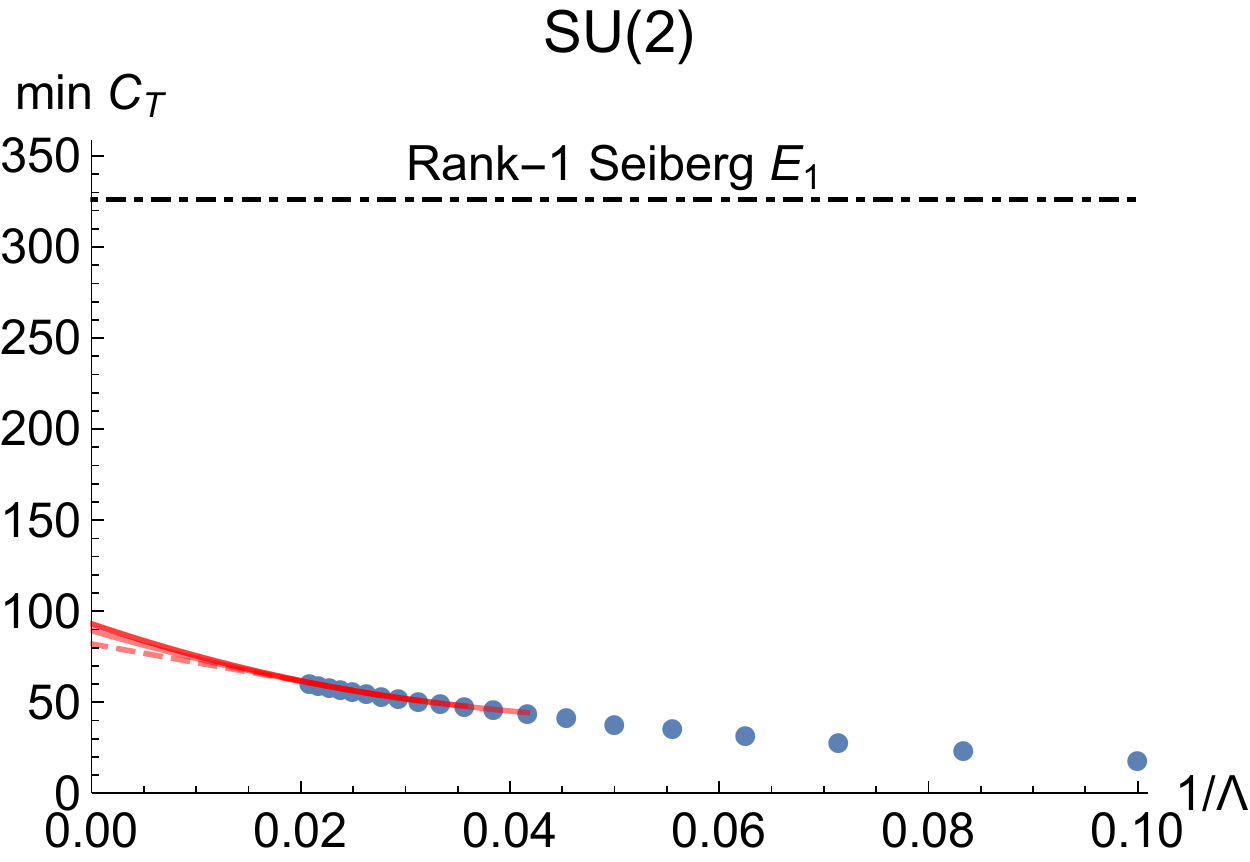}
}
\quad
\subfloat{
\includegraphics[width=.45\textwidth]{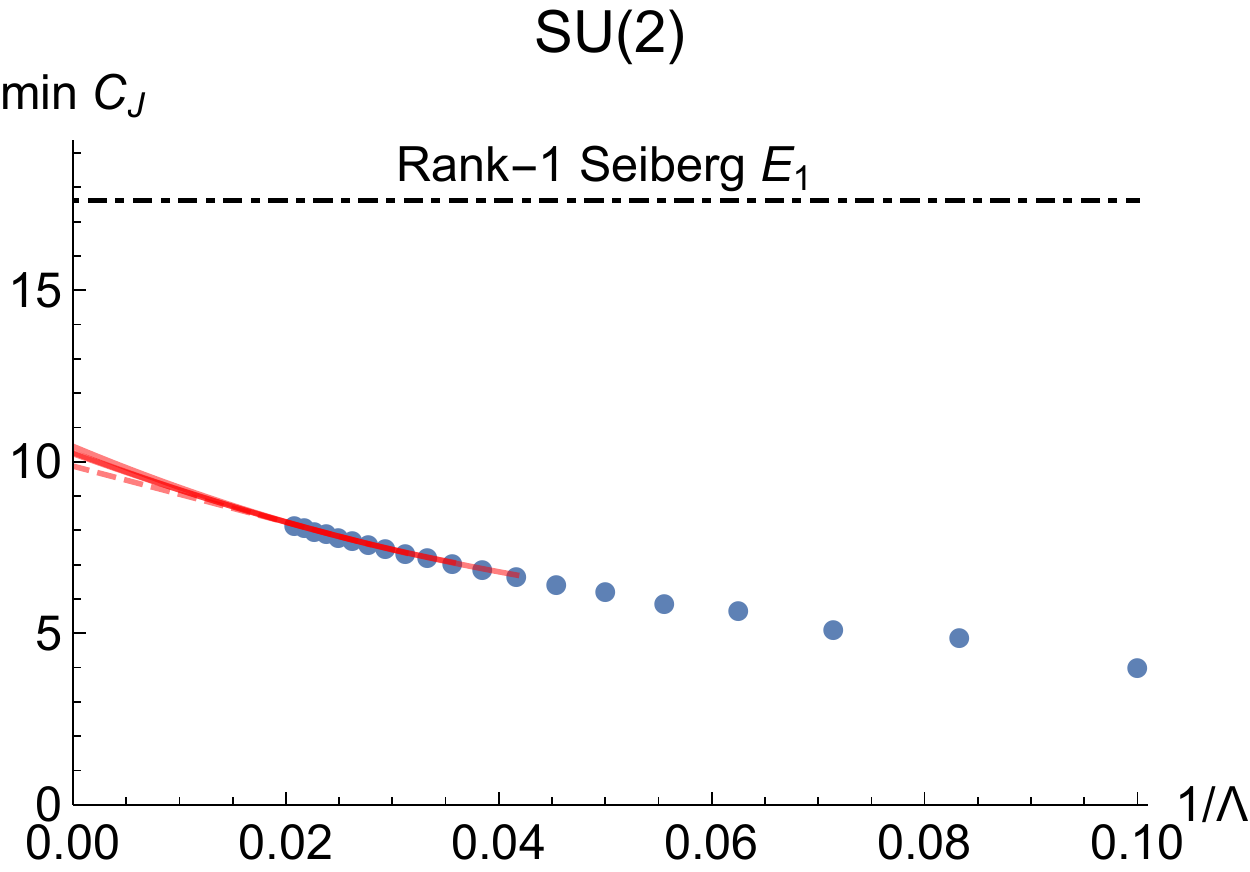}
}
\caption{Lower bounds on $C_T$ and $C_J$ for interacting theories with {\rm SU}(2) flavor group, at various derivative orders $\Lambda$ and extrapolated to infinite order using the quadratic ansatz \eqref{QuadAnsatz} (solid) and the linear ansatz \eqref{LinearAnsatz} (dashed).  Also shown are the values of $C_T$ and $C_J$ in the rank-one Seiberg $E_1$ theory (dotdashed).}
\label{Fig:CTandCJSU2}
\end{figure}

\vspace*{\fill}

\vspace*{\fill}

\newpage

\vspace*{\fill}

\vspace*{\fill}

\newpage

\subsubsection{$E_6$, $E_7$, and $E_8$ flavor symmetry}

\begin{figure}[H]
\centering
\subfloat{
\includegraphics[width=.45\textwidth]{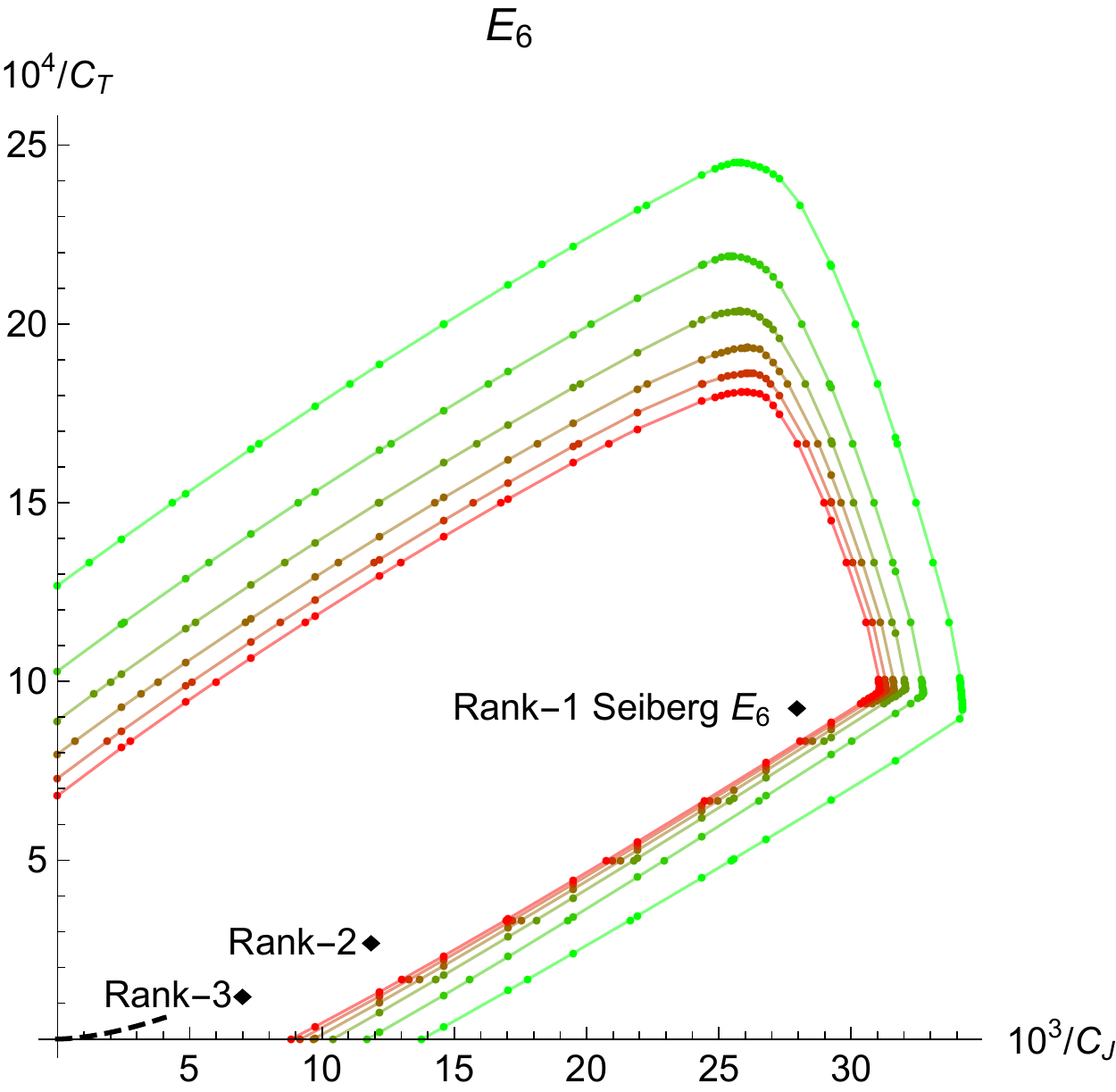}
}
\\
\subfloat{
\includegraphics[width=.45\textwidth]{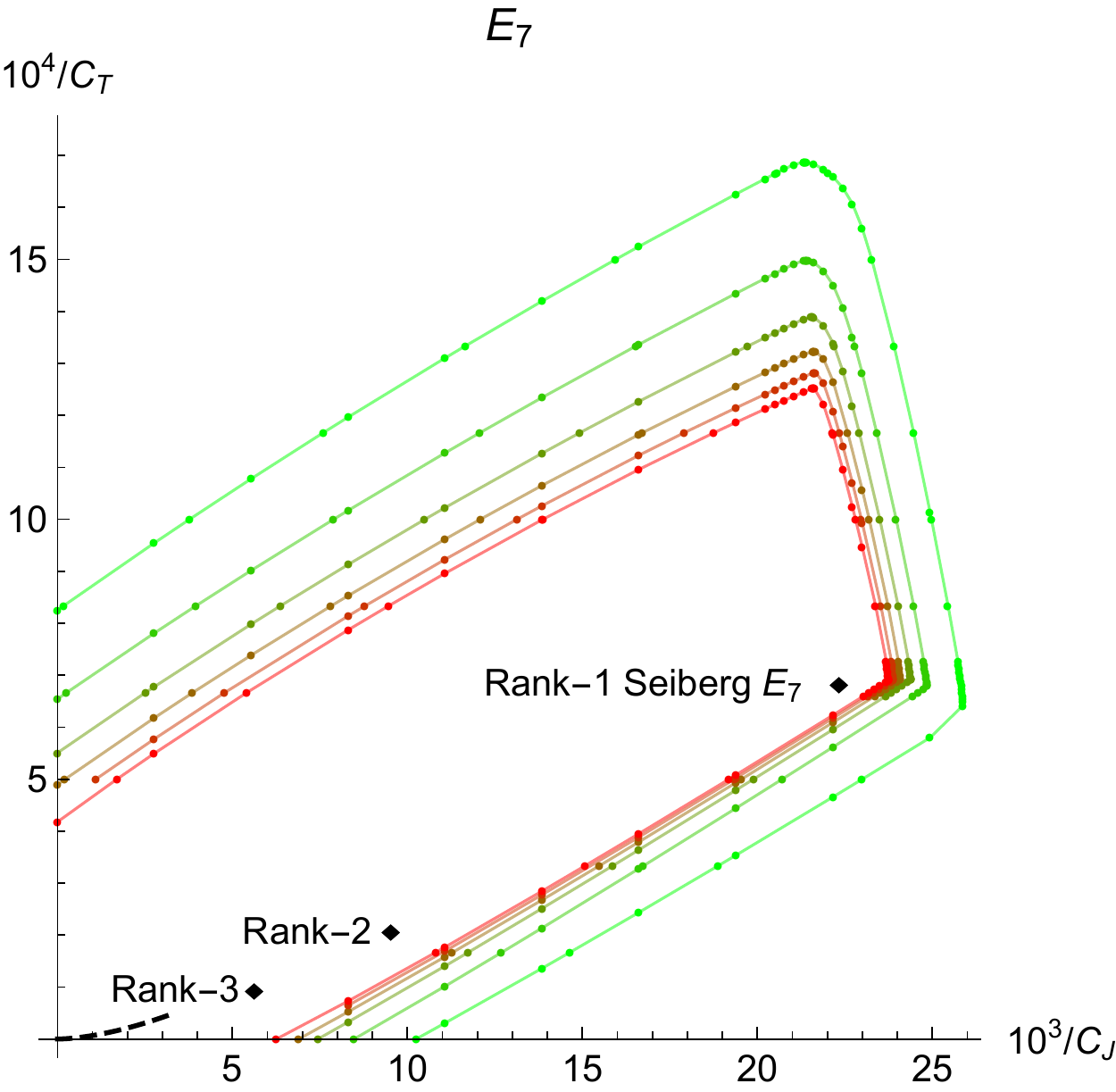}
}
\quad
\subfloat{
\includegraphics[width=.45\textwidth]{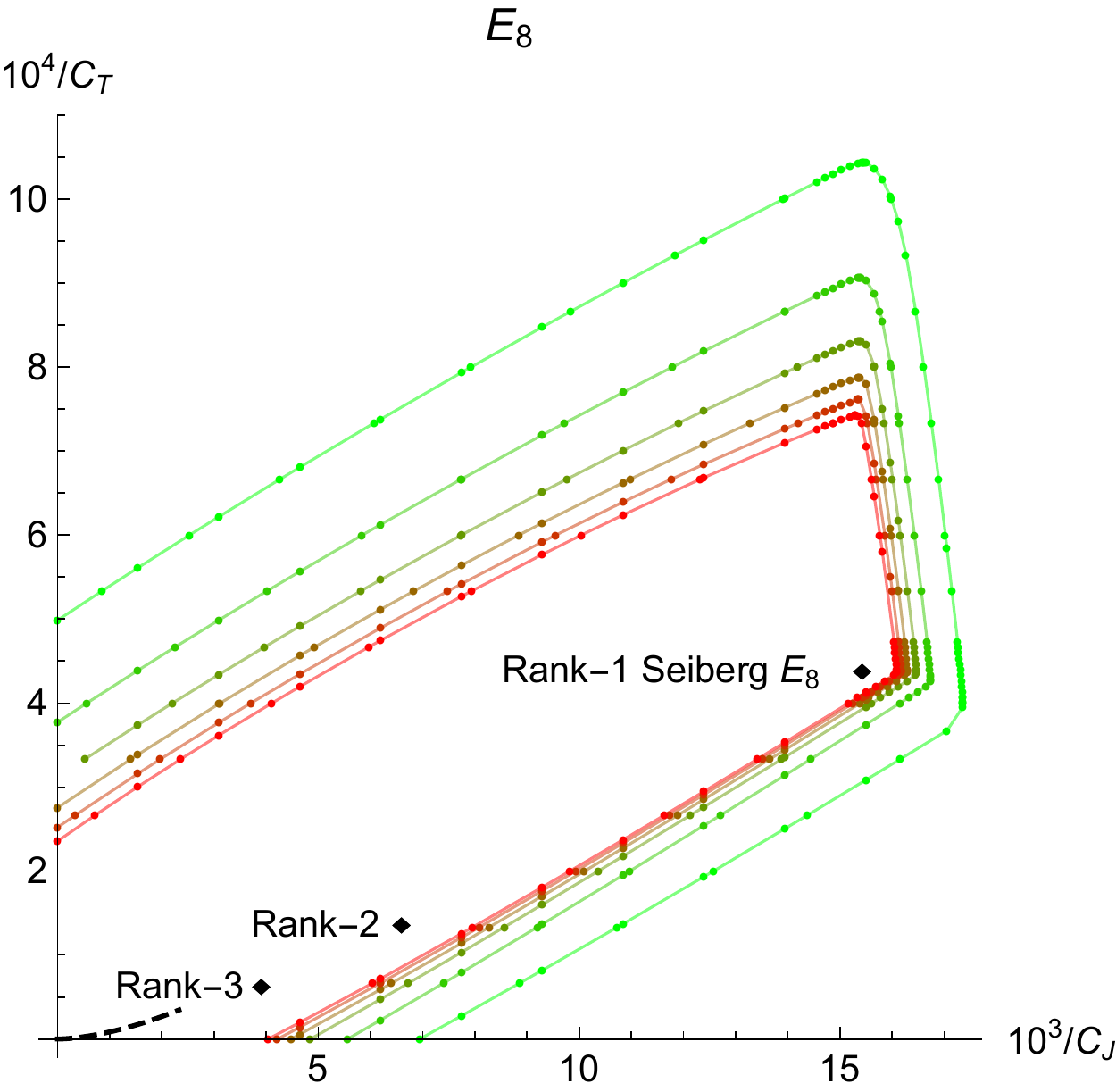}
}
\caption{Allowed regions in the $C_T-C_J$ plane for interacting theories with $E_1$, $E_6$, $E_7$, and $E_8$ flavor symmetry, at derivative orders $\Lambda = 20, 24, \dotsc, 40$, shown from green to red.  Also shown are the four-instanton values in the rank-one Seiberg theories, the perturbative values in the rank-two and three, and the values according to the large-rank formula (dashed line).}
\label{Fig:CTCJEn}
\end{figure}

\begin{figure}[H]
\centering
\subfloat{
\includegraphics[width=.45\textwidth]{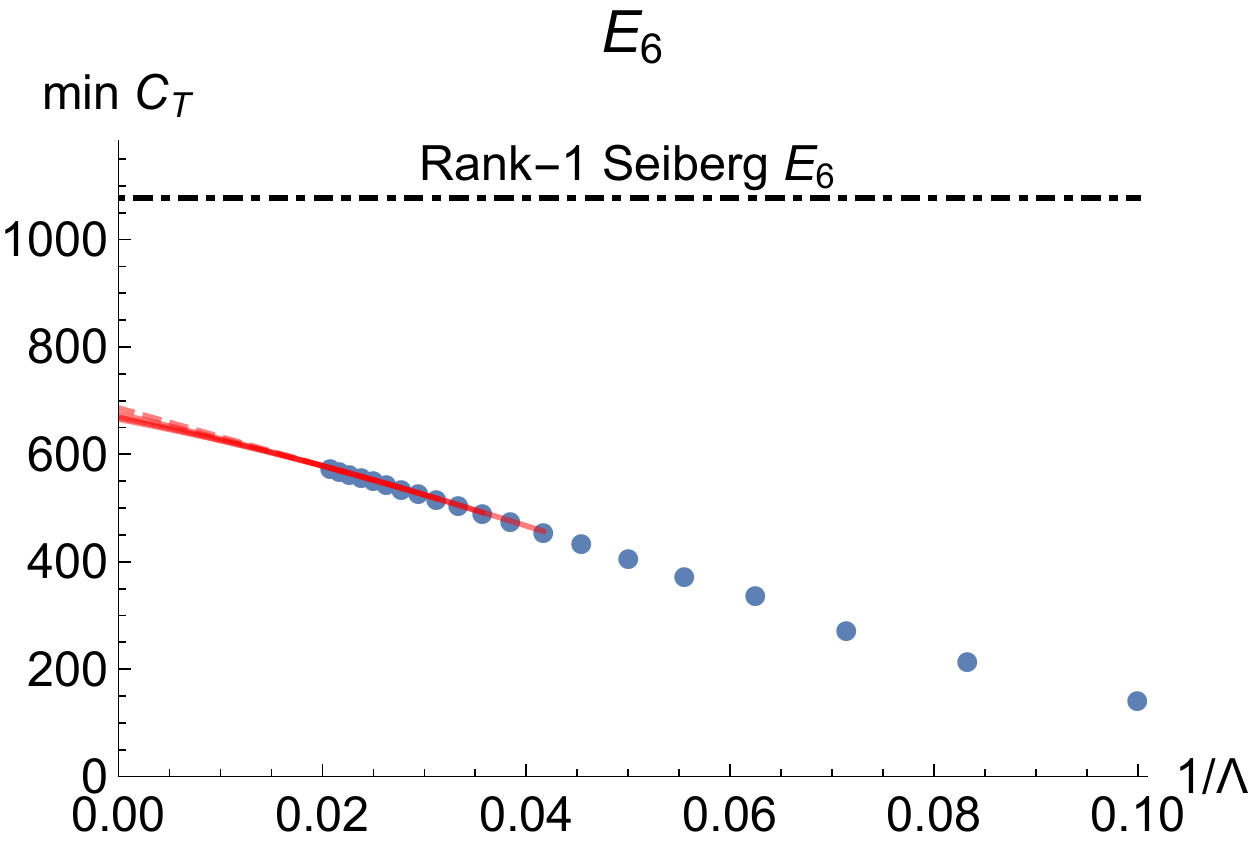}
}
\quad
\subfloat{
\includegraphics[width=.45\textwidth]{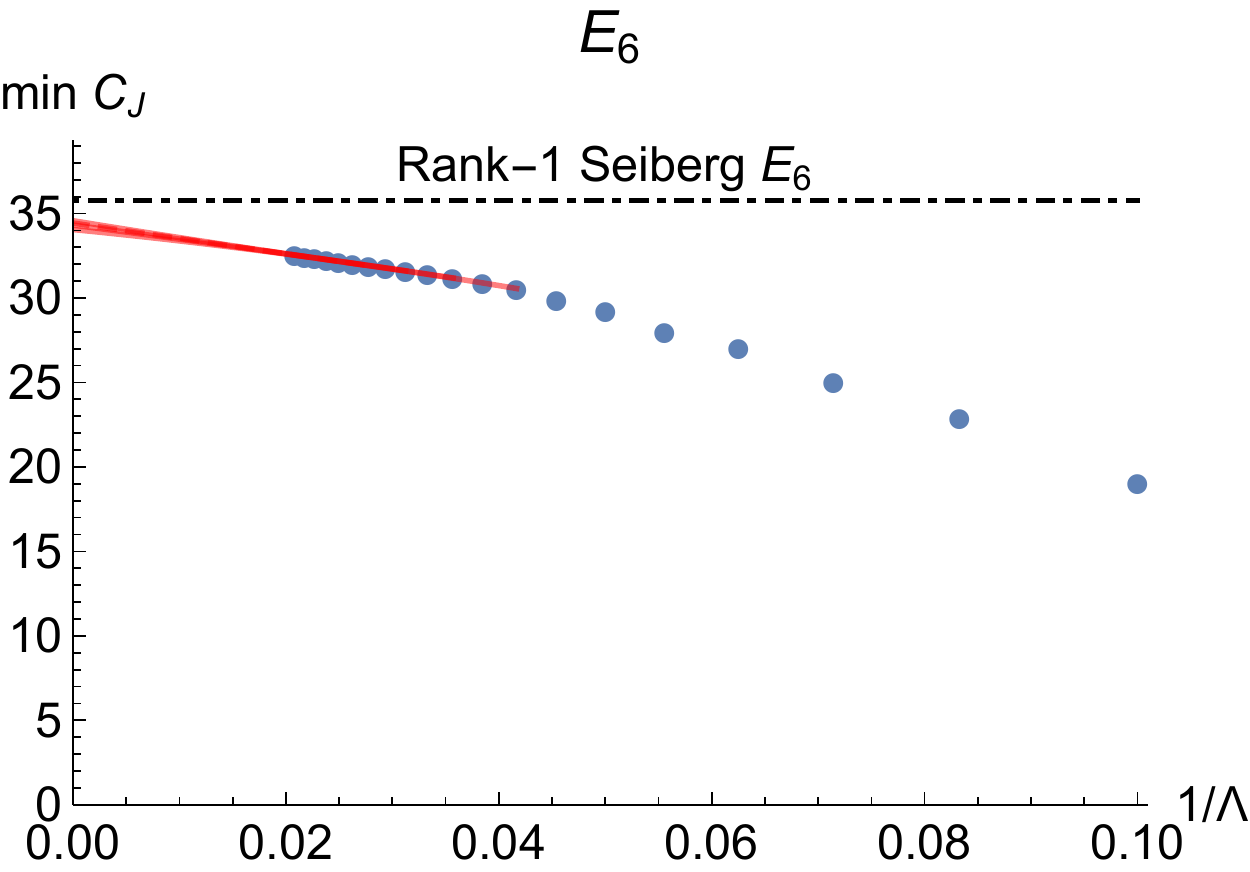}
}
\\
\subfloat{
\includegraphics[width=.45\textwidth]{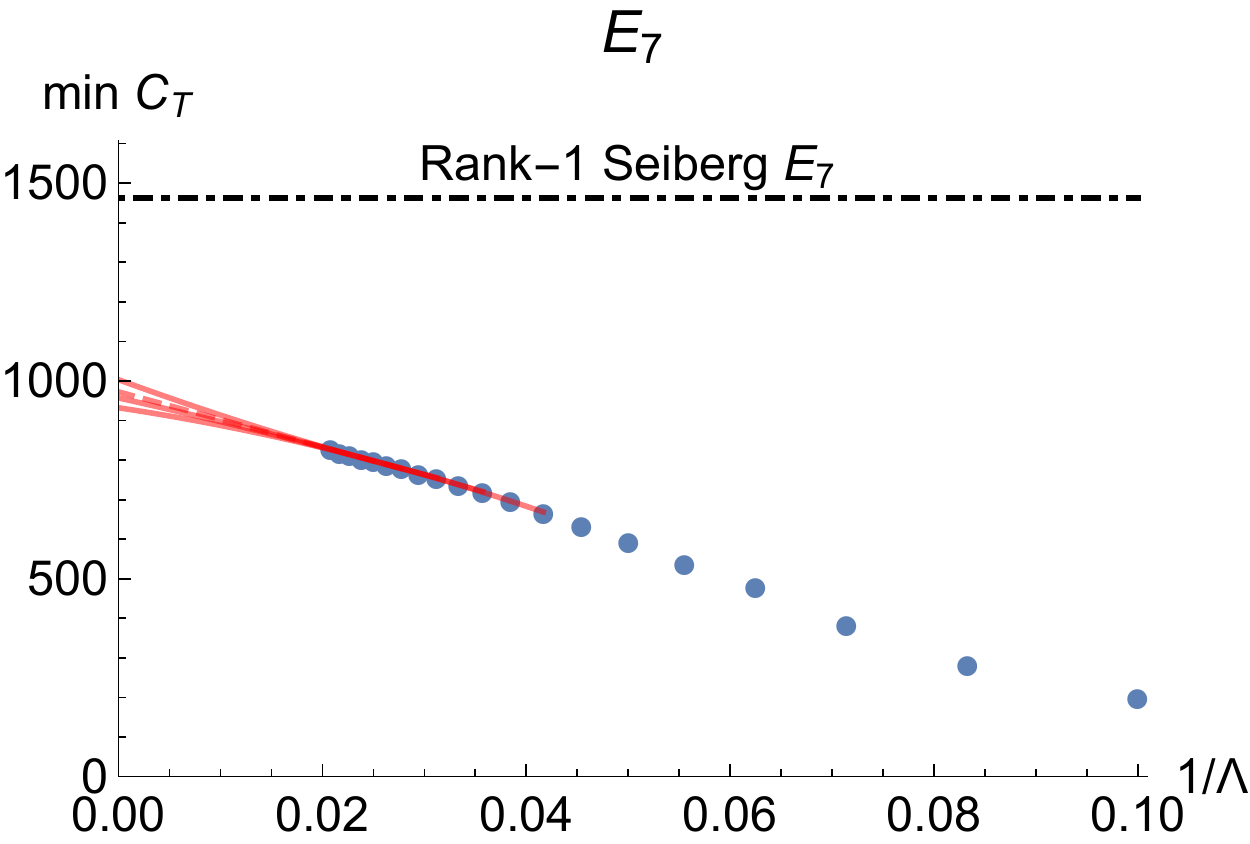}
}
\quad
\subfloat{
\includegraphics[width=.45\textwidth]{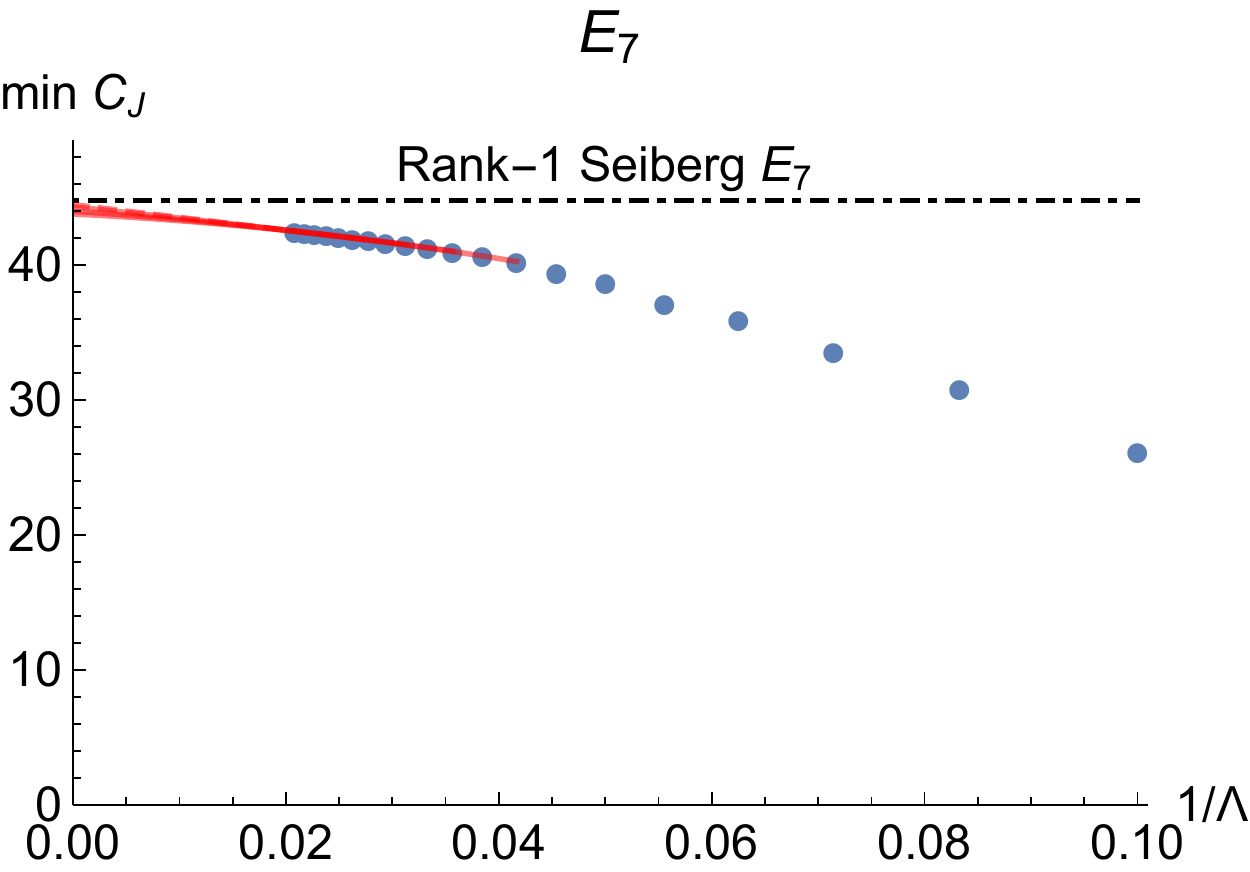}
}
\\
\subfloat{
\includegraphics[width=.45\textwidth]{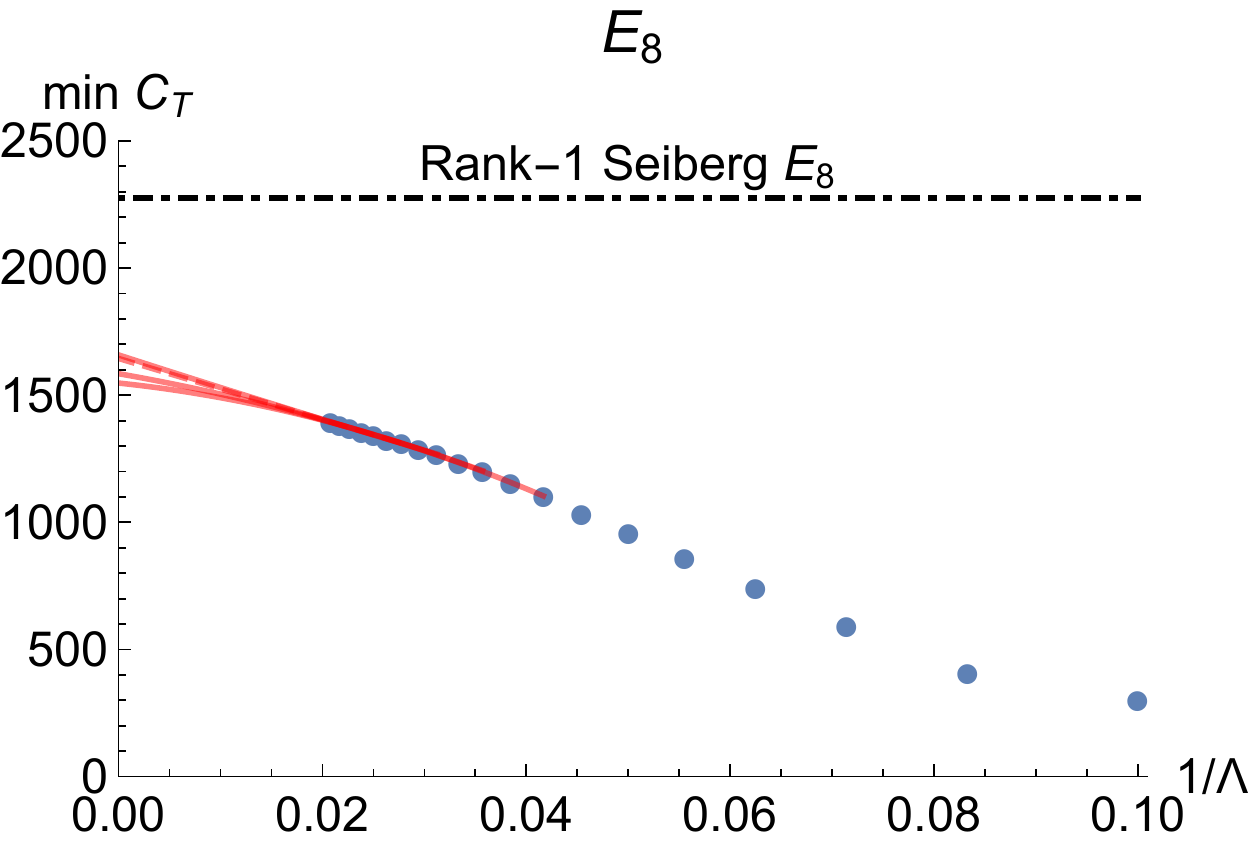}
}
\quad
\subfloat{
\includegraphics[width=.45\textwidth]{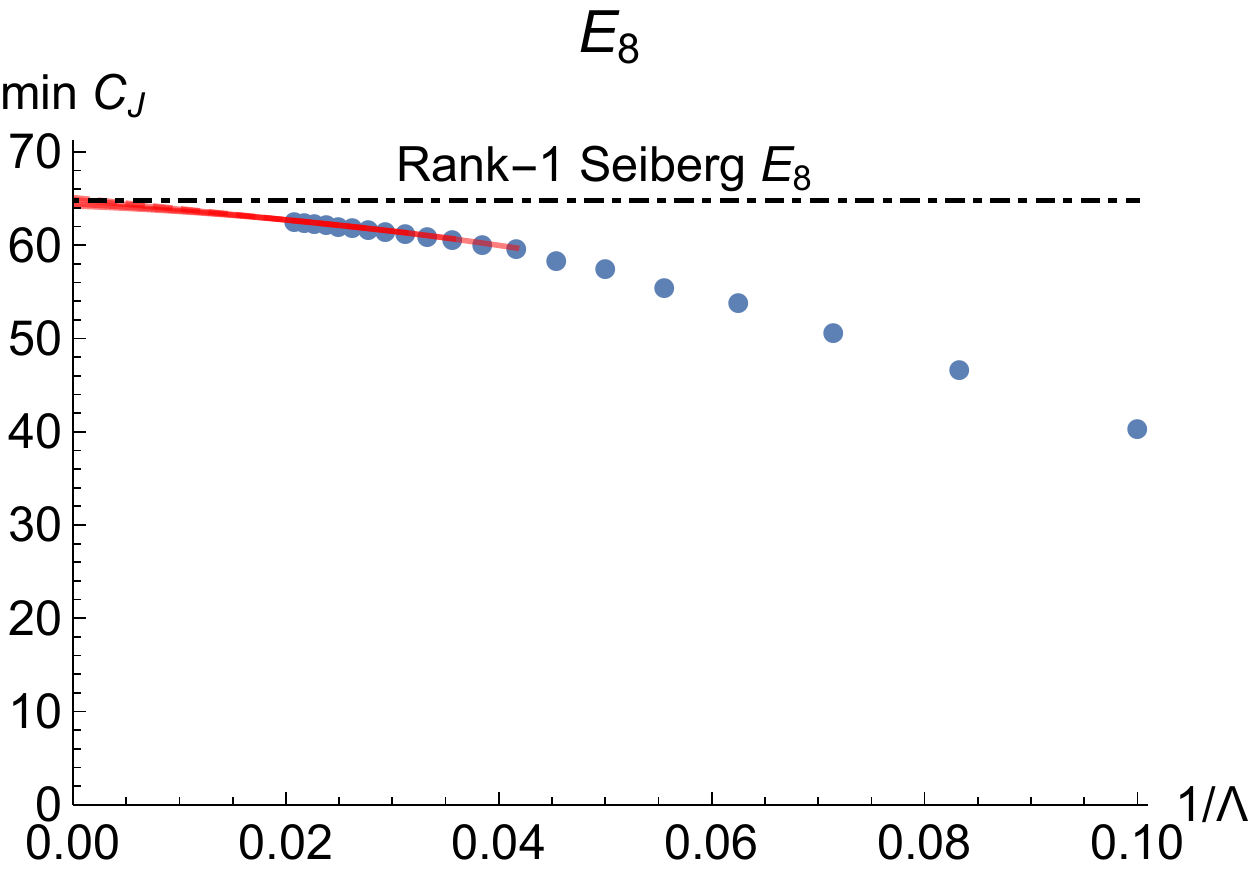}
}
\caption{Lower bounds on $C_T$ (left) and $C_J$ (right) for interacting theories with $E_6$, $E_7$, and $E_8$ flavor groups, at various derivative orders $\Lambda$ and extrapolated to infinite order using the quadratic ansatz \eqref{QuadAnsatz} (solid) and the linear ansatz \eqref{LinearAnsatz} (dashed).  Also shown are the four-instanton values of $C_T$ and $C_J$ in the rank-one Seiberg exceptional theories (dotdashed).}
\label{Fig:CTandCJEn}
\end{figure}

\begin{figure}[H]
\centering
\subfloat{
\includegraphics[width=.45\textwidth]{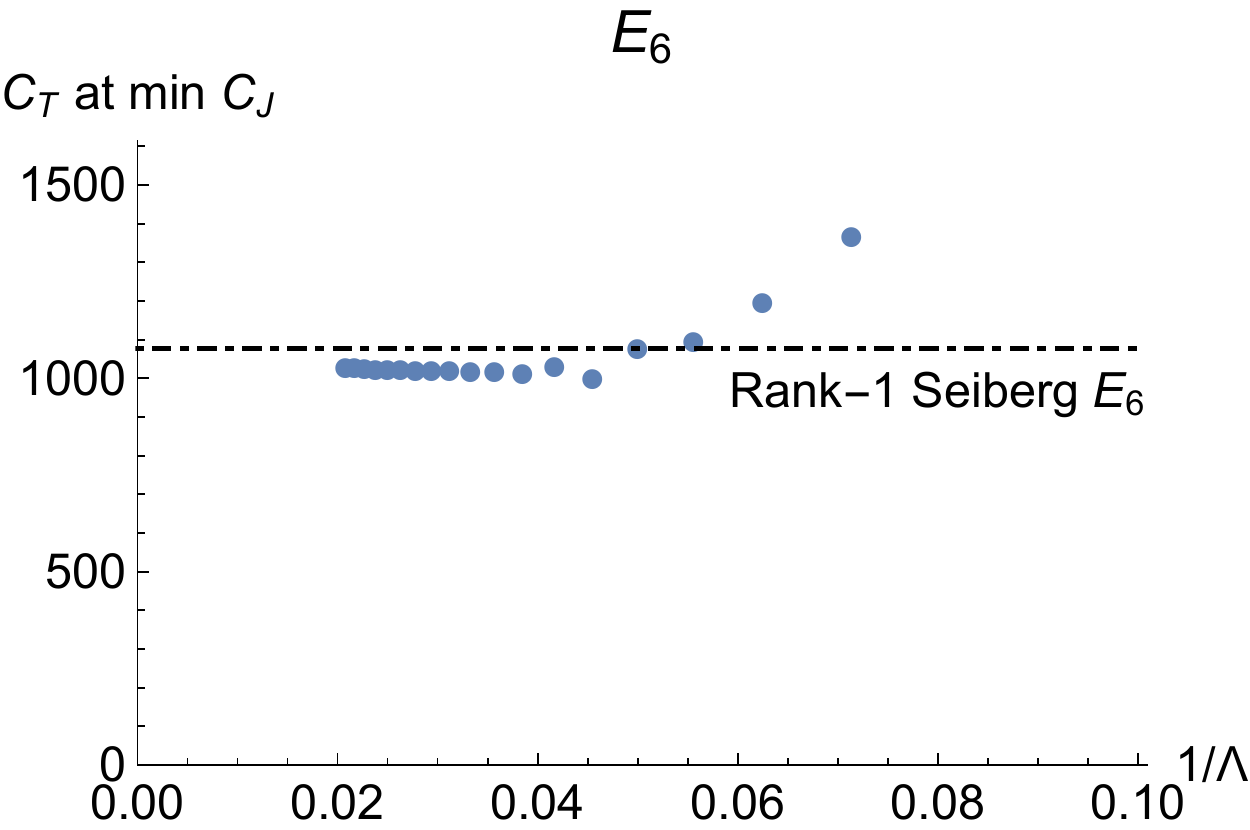}
}
\quad
\subfloat{
\includegraphics[width=.45\textwidth]{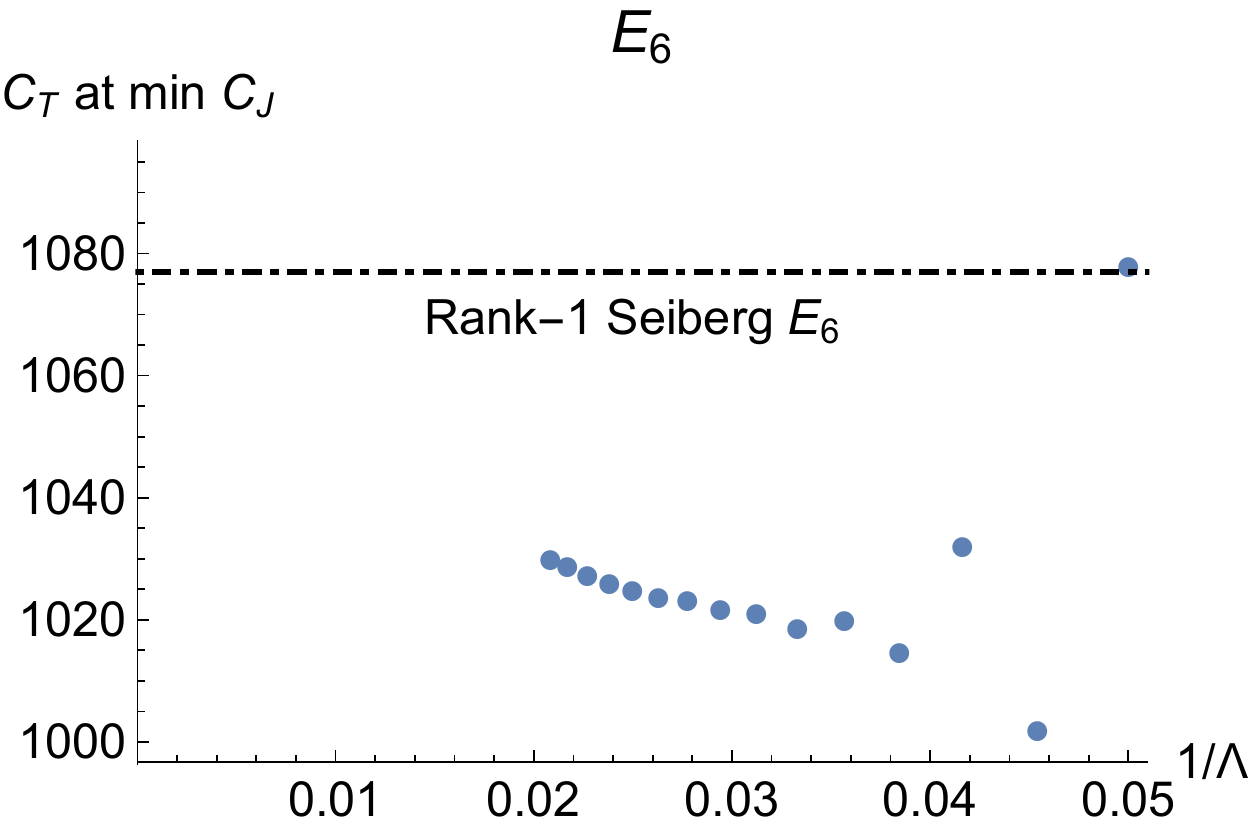}
}
\\
\subfloat{
\includegraphics[width=.45\textwidth]{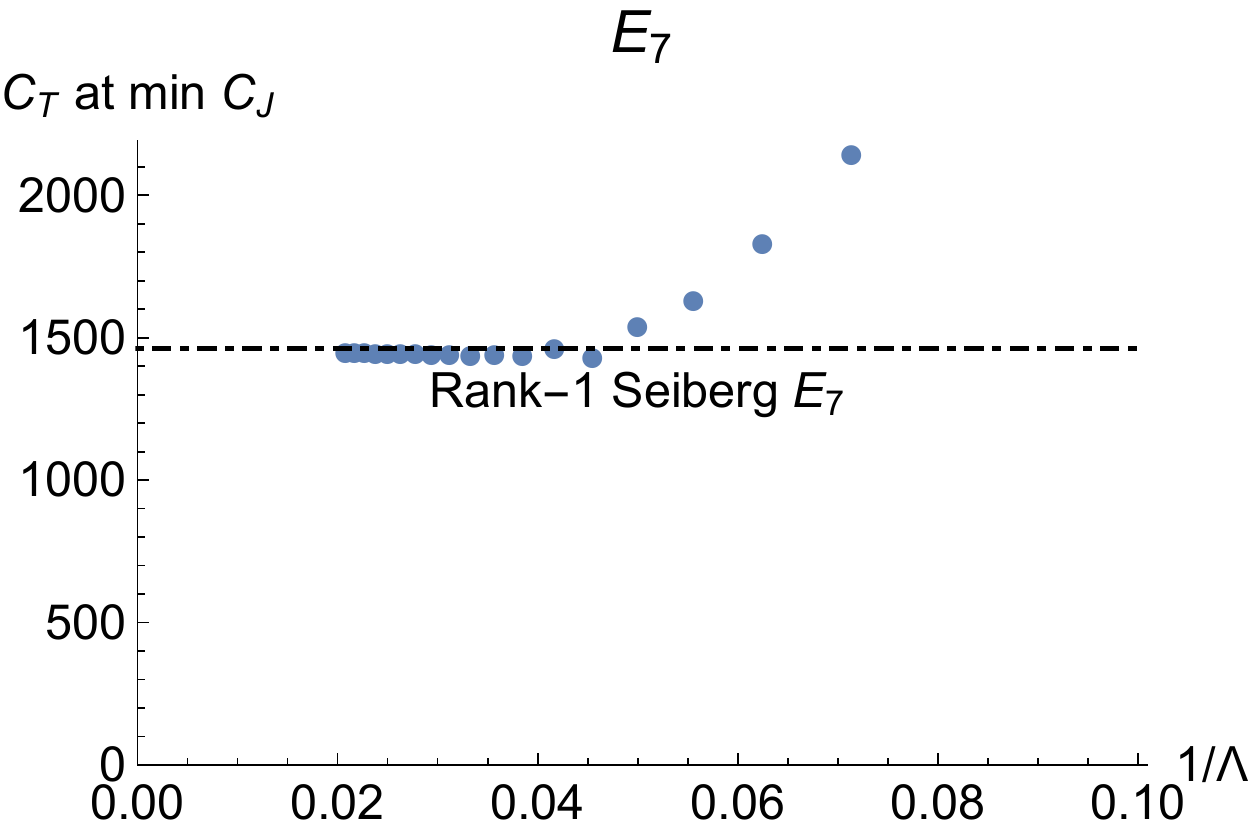}
}
\quad
\subfloat{
\includegraphics[width=.45\textwidth]{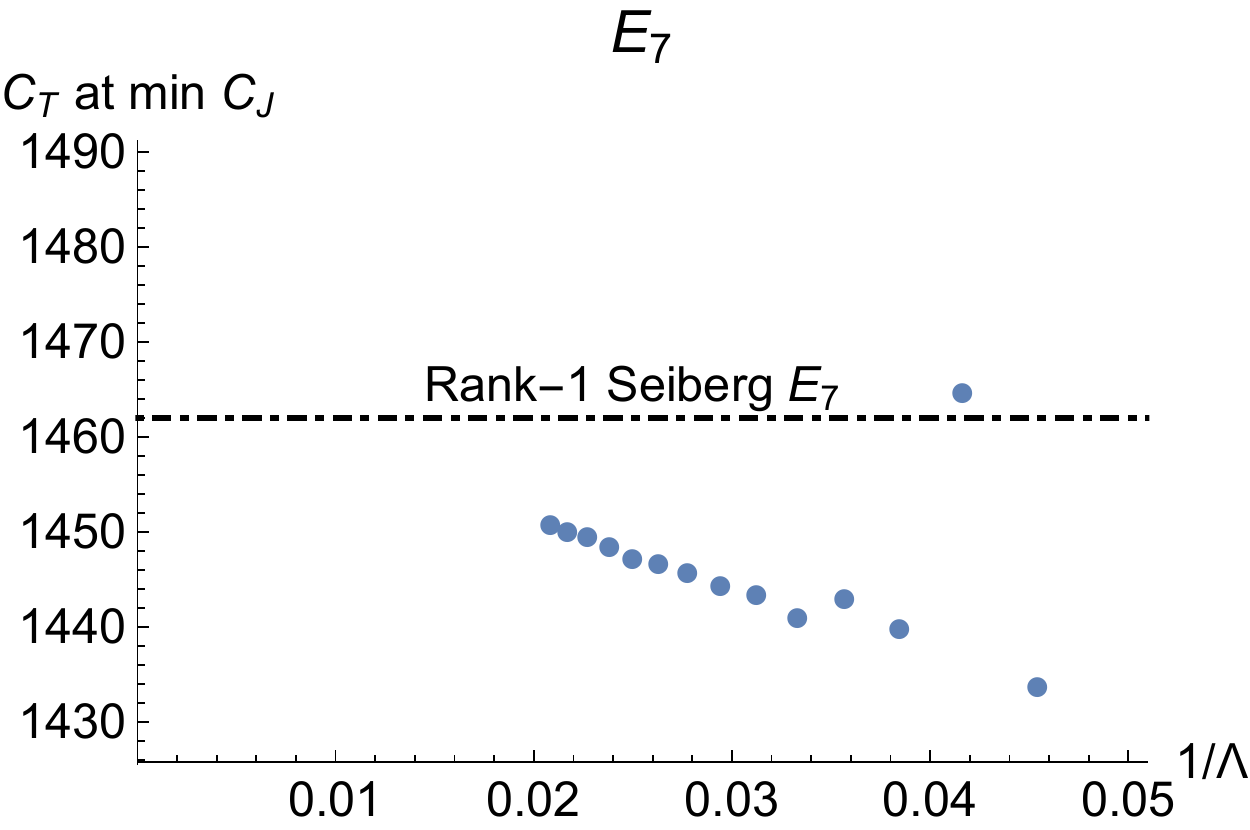}
}
\\
\subfloat{
\includegraphics[width=.45\textwidth]{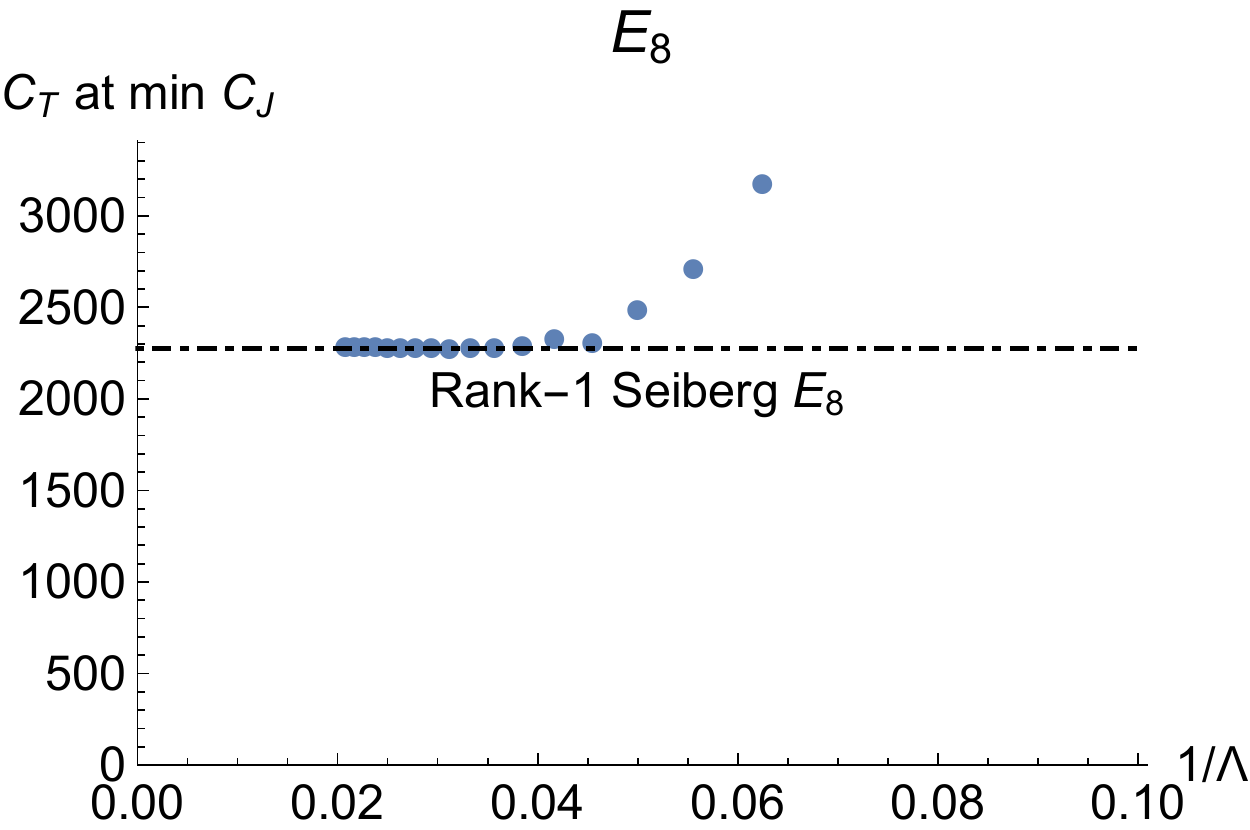}
}
\quad
\subfloat{
\includegraphics[width=.45\textwidth]{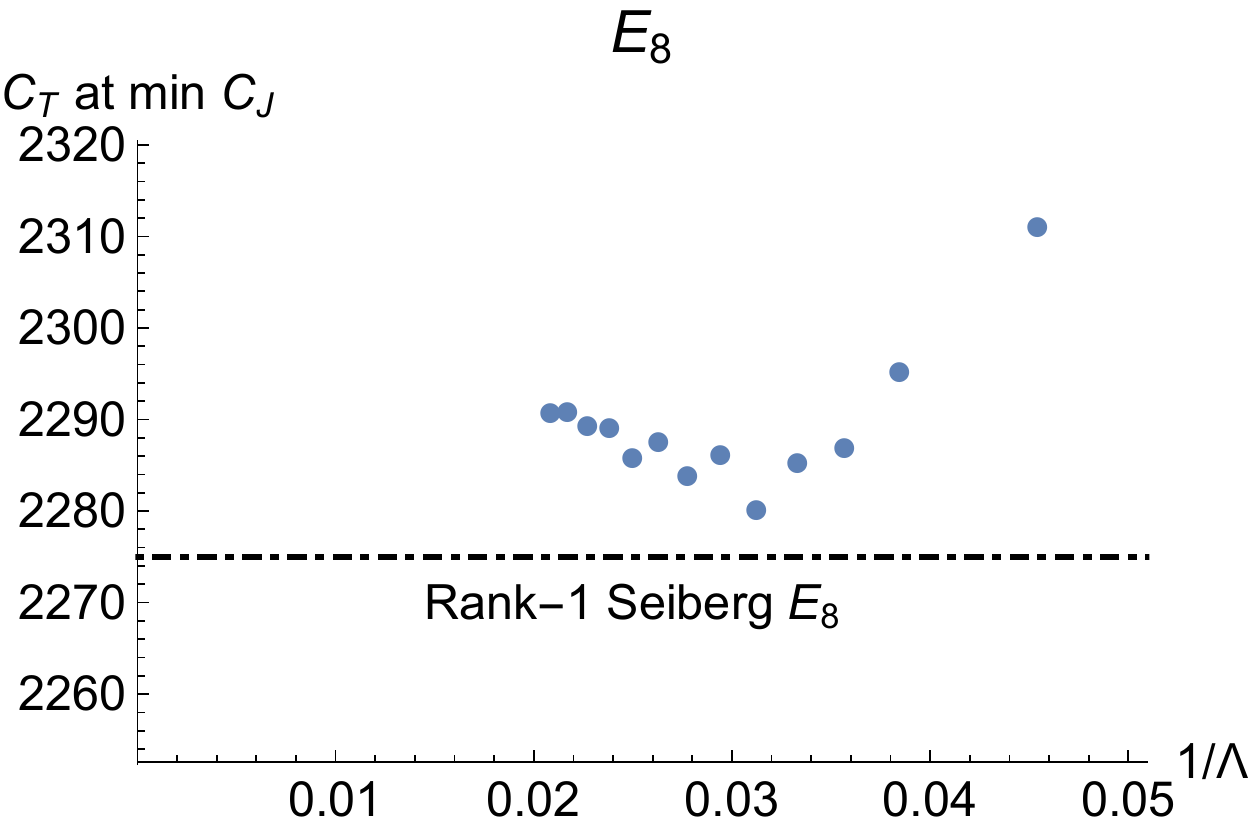}
}
\caption{{\bf Left:} The values of $C_J$ when the lower bounds on $C_T$ are saturated for interacting theories with $E_6$, $E_7$, and $E_8$ flavor groups, at various derivative orders $\Lambda$.  Also shown are the values of $C_T$ and $C_J$ in the rank-one Seiberg exceptional theories (dotdashed). {\bf Right:} Zoomed-in plots.}
\label{Fig:MinEn}
\end{figure}

Finally, we bootstrap interacting unitary superconformal field theories with $E_6$, $E_7$, and $E_8$ flavor symmetries.  Figure~\ref{Fig:CTCJEn} maps out the allowed regions in the $C_T-C_J$ plane.
For each flavor group, the allowed region has two corners, corresponding to the minimal $C_J$ and the minimal $C_T$, and the four-instanton values of $(C_J, C_T)$ in the rank-one Seiberg exceptional theory sit near the corner with the minimal $C_J$.
To examine the corners further, Figure~\ref{Fig:CTandCJEn} shows the lower bounds on $C_T$ and on $C_J$ at various derivative orders, and extrapolations to infinite derivative order.  And Figure~\ref{Fig:MinEn} shows the values of $C_J$ and $C_T$ that minimize $C_T$ and $C_J$, respectively.  The results as summarized in Tables~\ref{Tab:EnBootstrapResults} and~\ref{Tab:EnBootstrapResults2} provide strong evidence for the rank-one Seiberg $E_6$, $E_7$, and $E_8$ theories having the minimal $C_J$ among all interacting theories with $E_6$, $E_7$, and $E_8$ flavor symmetry, respectively.  We are not aware of candidate theories that saturate the extrapolated bound on $C_T$.

\begin{table}[h]
\centering
\begin{tabular}{|c||c|c||c|c|}
\hline
$G$ & $\lim_{\Lambda \to \infty} {\rm min} \, C_T$ & $C_T^{\rm 4-inst}$ & $\lim_{\Lambda \to \infty} {\rm min} \, C_J$ & $C_J^{\rm 4-inst}$
\\\hline\hline
$E_6$ & $6.8(1) \times 10^2$ & $1.08 \times 10^3$ & $3.44(2) \times 10^2$ & $3.58 \times 10^2$
\\
$E_7$ & $9.7(2) \times 10^2$ & $1.46 \times 10^3$ & $4.42(3) \times 10^2$ & $4.47 \times 10^2$
\\
$E_8$ & $1.62(4) \times 10^3$ & $2.27 \times 10^3$ & $6.48(4) \times 10^2$ & $6.48 \times 10^2$
\\\hline
\end{tabular}
\caption{The extrapolated lower bounds on $C_T$ and $C_J$ for interacting theories with flavor groups $G = E_6,\,E_7,\,E_8$ at infinite derivative order, and compared to the four-instanton values in the rank-one Seiberg exceptional theories.
}
\label{Tab:EnBootstrapResults}
\end{table}

\begin{table}[h]
\centering
\begin{tabular}{|c||c|c|}
\hline
$G$ & $\lim_{\Lambda \to \infty} C_T$ at ${\rm min} \, C_J$ & $C_T^{\rm 4-inst}$
\\\hline
$E_6$ 
& $1.05(2) \times 10^3$ & $1.08 \times 10^3$ 
\\
$E_7$ & $1.47(1) \times 10^3$ & $1.46 \times 10^3$
\\
$E_8$ & $2.31(2) \times 10^3$ & $2.27 \times 10^3$ 
\\
\hline
\end{tabular}
\caption{The values of $C_T$ when the lower bound on $C_J$ is saturated for interacting theories with flavor groups $G = E_6,\,E_7,\,E_8$, and compared to the four-instanton values in the rank-one Seiberg exceptional theories.
}
\label{Tab:EnBootstrapResults2}
\end{table}

\newpage
\vspace*{\fill}

\vspace*{\fill}
\newpage
\vspace*{\fill}

\vspace*{\fill}
\newpage
\vspace*{\fill}

\vspace*{\fill}
\newpage

\begin{table}[t]
\centering
\begin{tabular}{|c|c|c|}
\hline
$G$ & $\Delta_{\rm gap}^{{\cal L}[0]}$ & Representation
\\\hline\hline
$E_6$ & 4.91(1) & {\bf 650}
\\
$E_7$ & 4.95(2) & {\bf 1539}
\\
$E_8$ & 4.98(2) & {\bf 3875}
\\\hline
\end{tabular}
\caption{Predicted gaps (the lowest scaling dimension) in the long multiplets in the Seiberg exceptional theories, and the flavor group representations in which they transform.}
\label{Tab:Gaps}
\end{table}

We suspect that the slight discrepancies between the extrapolated bootstrap bounds on $C_J$ and the four-instanton values in the rank-one Seiberg exceptional theories, as well as the discrepancies in the values of $C_T$ at min $C_J$, disappear when even higher derivative orders are included.\footnote{Our estimates for the errors in the bootstrap data due to spin truncation are around 1\% \cite{Chang:2017xmr}. Therefore, the only meaningful discrepancies occur in the $E_6$ case.
}
In the zoomed-in plots of Figure~\ref{Fig:MinEn}, the values of $C_T$ at min $C_J$ exhibit upward trends at high derivative orders that potentially diminish the discrepancies.\footnote{In light of the results for $E_1$ in Section~\ref{Sec:E1}, there is the possibility that the rank-one Seiberg theories generally do not saturate the absolute lower bound on $C_J$ (it is conceivable that only the $E_8$ case does), but lie on the lower boundaries of the allowed regions close to the kinks.  In this scenario, the extremal functional method still solves these theories.  Determining which scenario is correct requires considerably more computational power, and is left for future work.
}

\begin{figure}[H]
\centering
\subfloat{
\includegraphics[width=.45\textwidth]{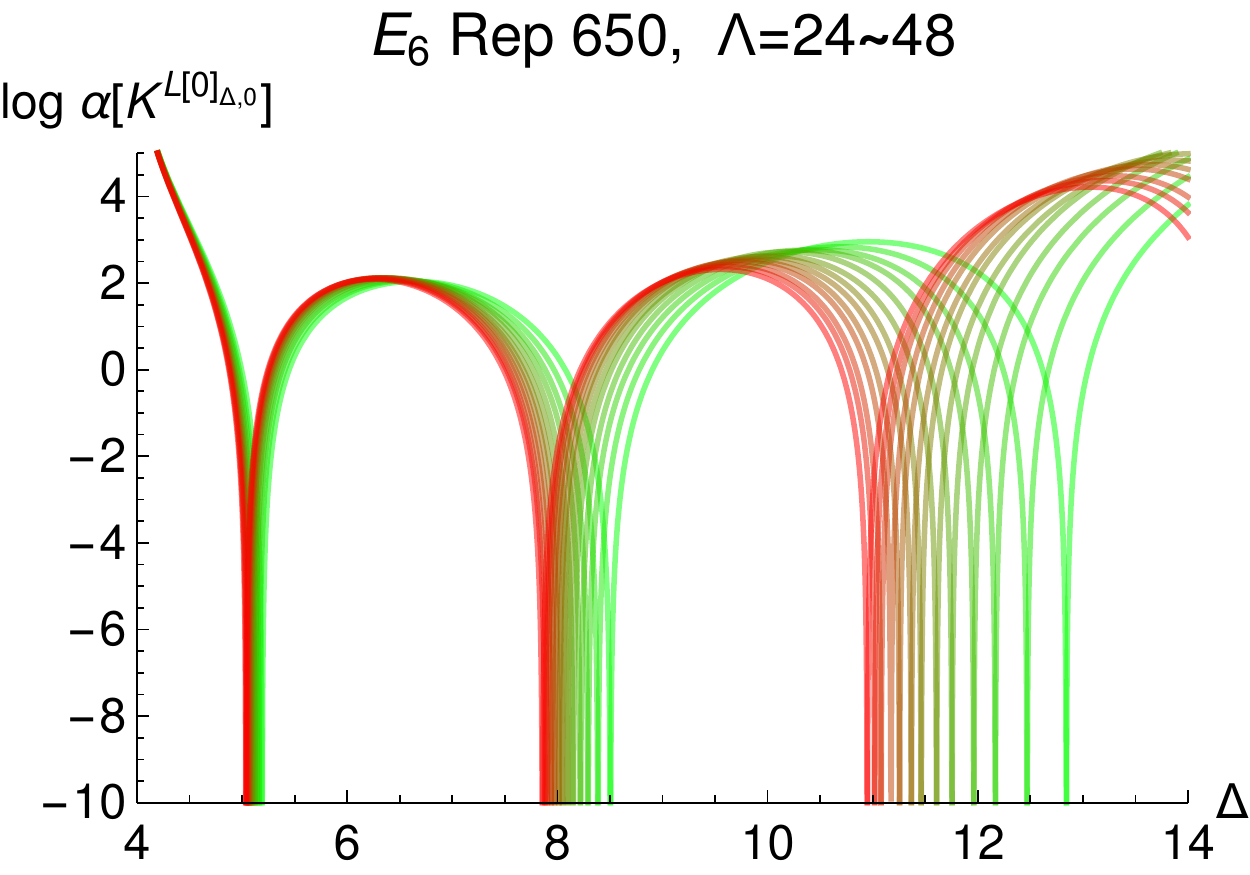}
}
\quad
\subfloat{
\includegraphics[width=.45\textwidth]{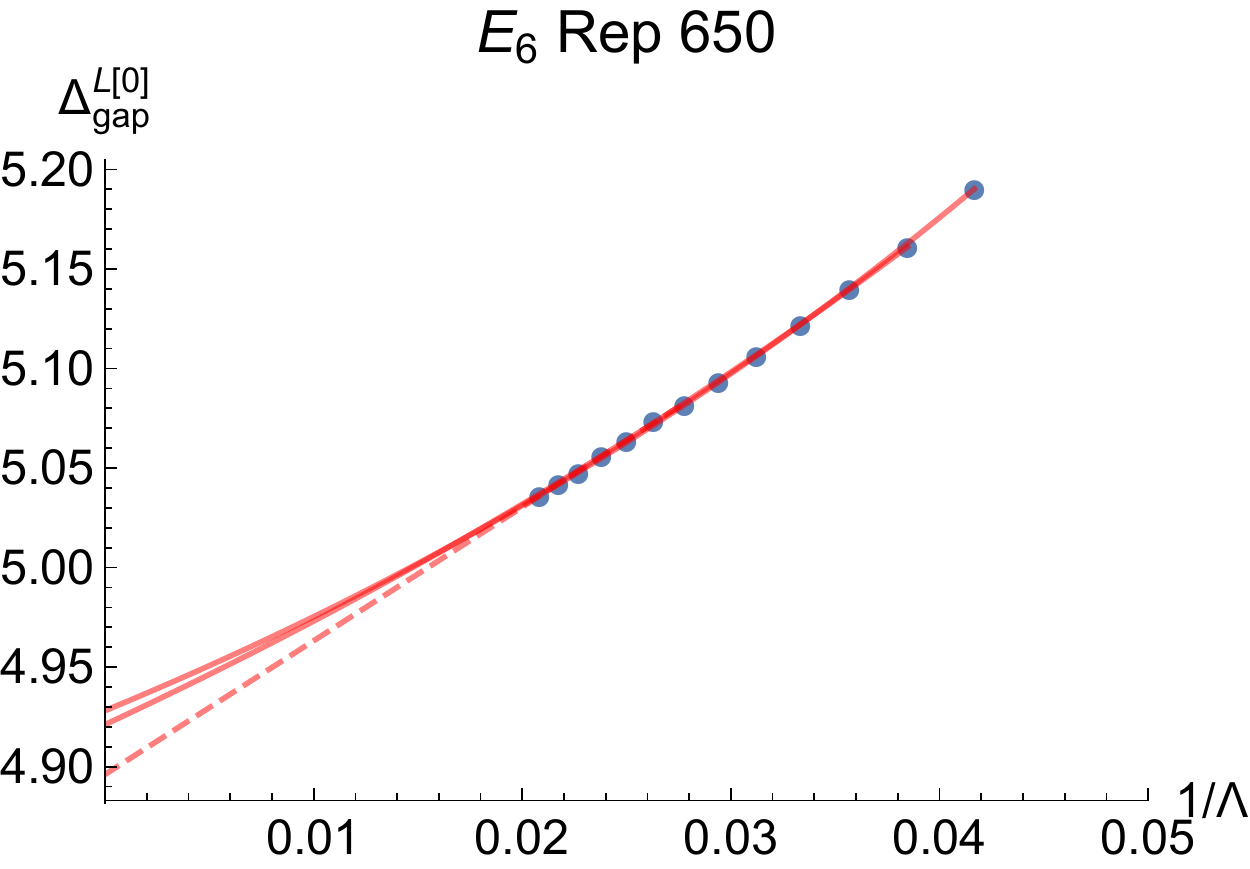}
}
\\
\subfloat{
\includegraphics[width=.45\textwidth]{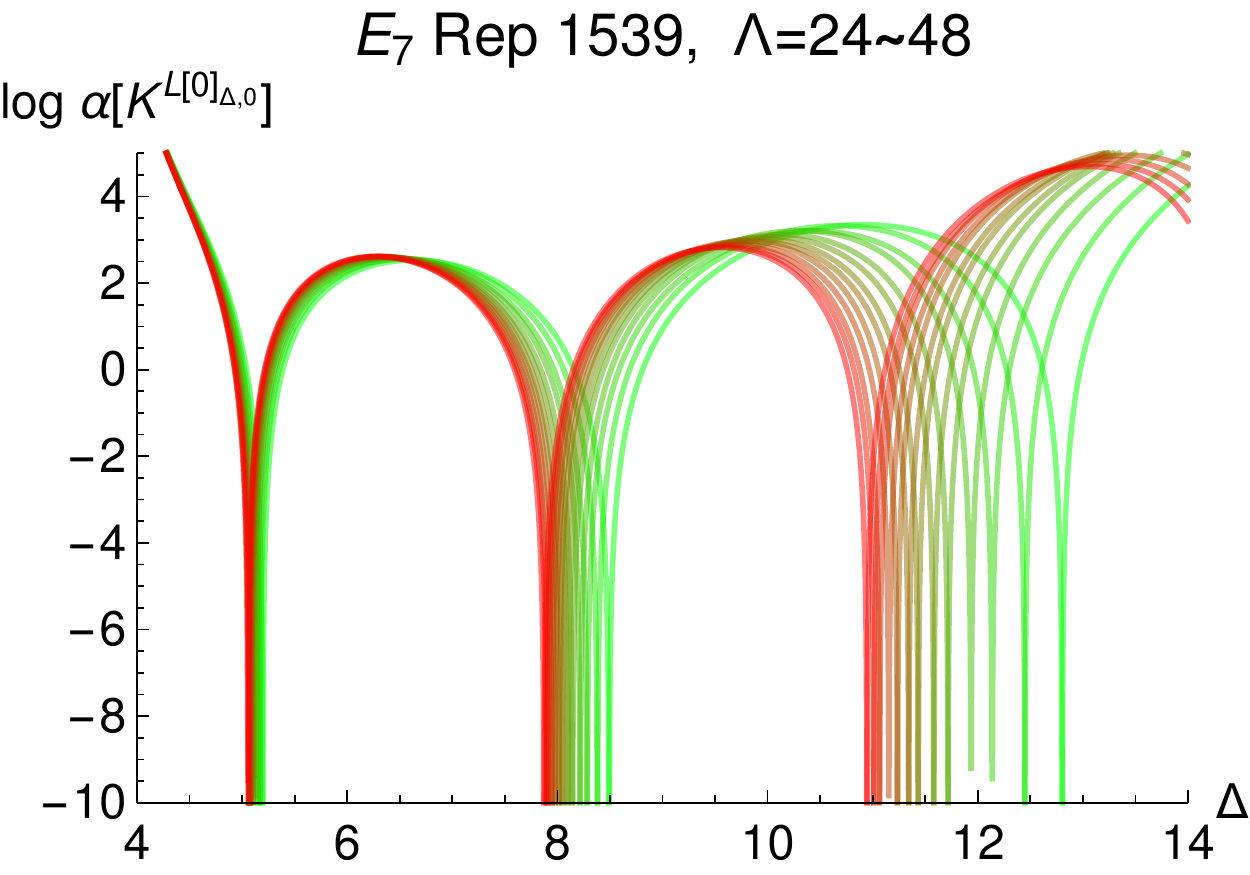}
}
\quad
\subfloat{
\includegraphics[width=.45\textwidth]{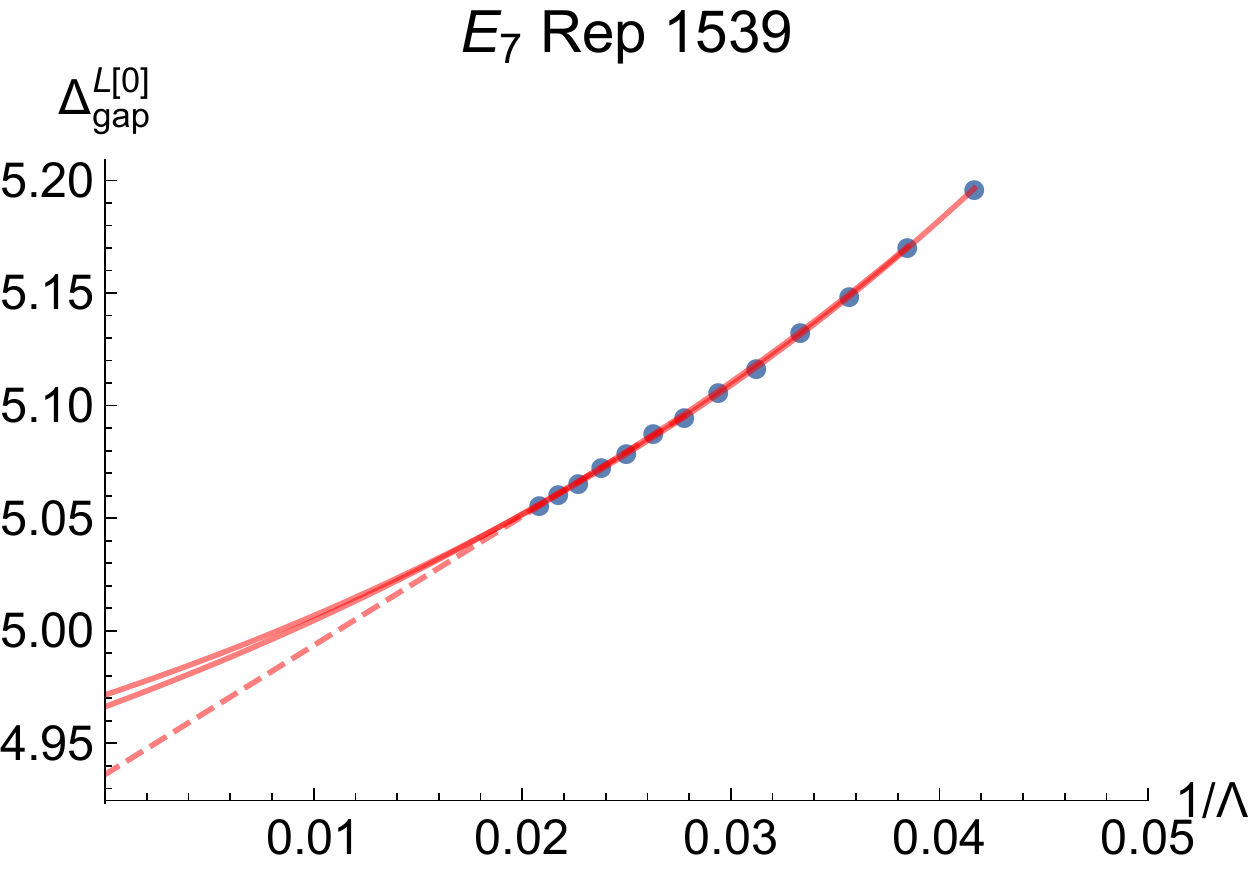}
}
\\
\subfloat{
\includegraphics[width=.45\textwidth]{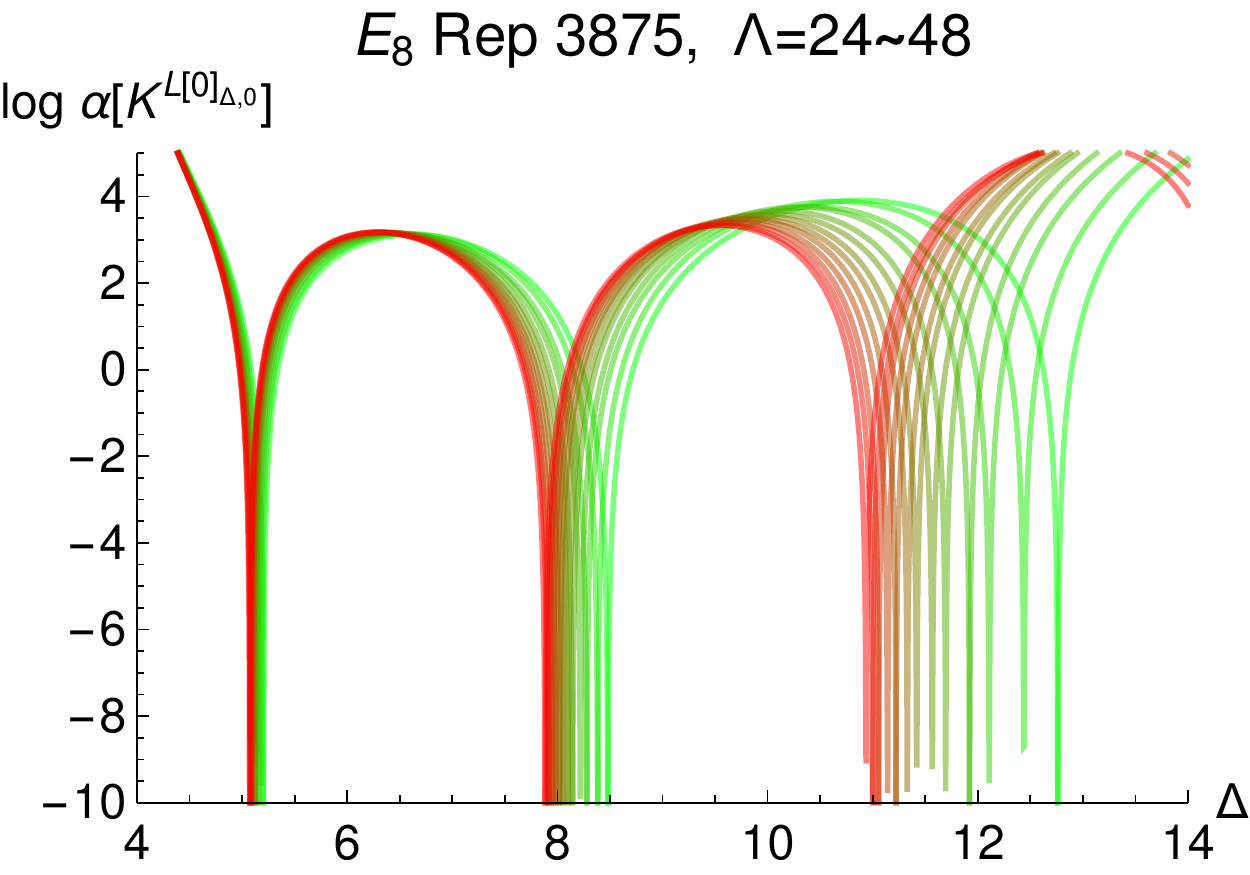}
}
\quad
\subfloat{
\includegraphics[width=.45\textwidth]{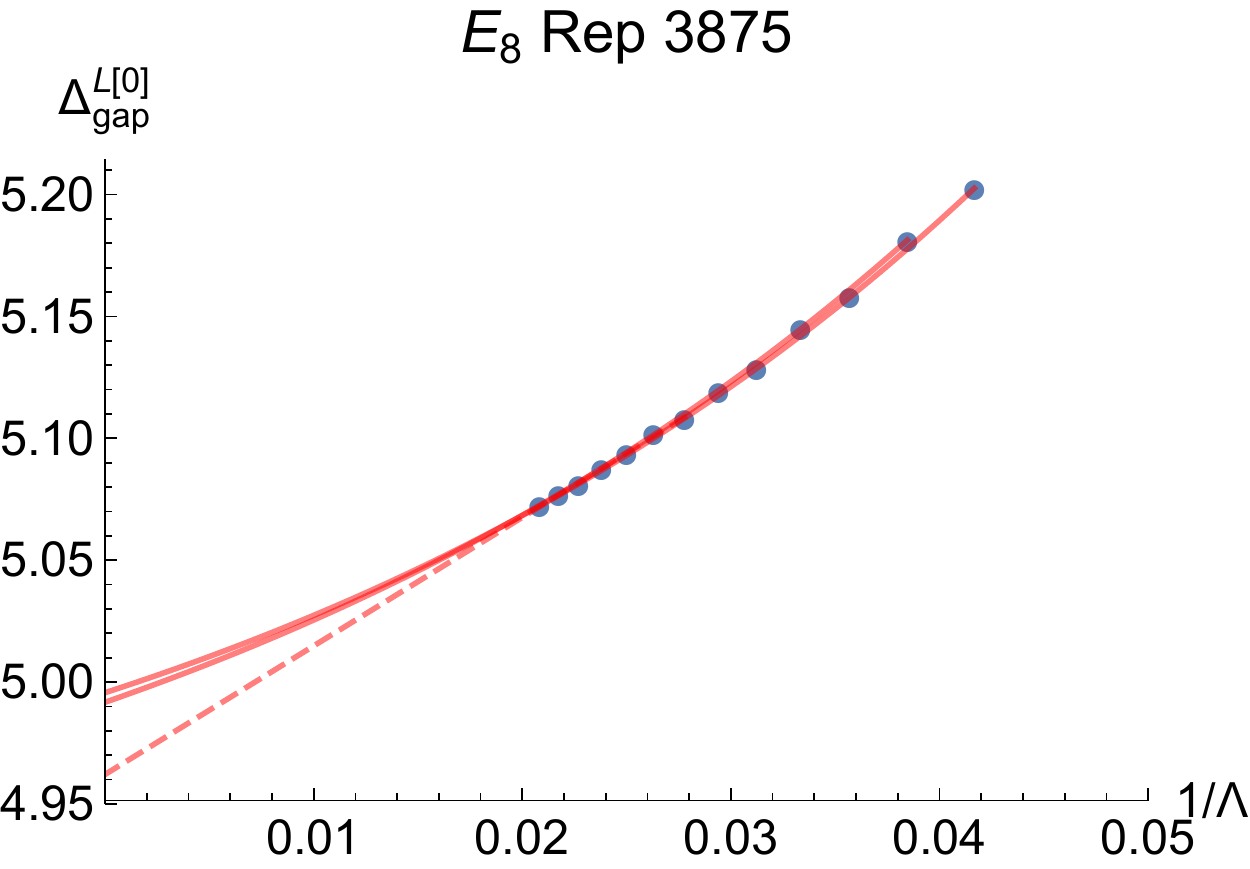}
}
\caption{{\bf Left:} The extremal functional optimizing the lower bound on $C_J$, acted on the contribution of the spin-zero long multiplet to the crossing equation, $\A_{{\cal D}[2],{\bf adj}}[{\cal K}^{{\cal L}[0]_{\Delta,0}}]$, in the {\bf 650} of $E_6$, the {\bf 1539} of $E_7$, and the {\bf 3875} of $E_8$, plotted in logarithmic scale.  These representations are chosen because they have the smallest gap (the lowest scaling dimension).  Increasing derivative orders $\Lambda = 24, 26, \dotsc, 48$ are shown from green to red.  {\bf Right:} The gap at different $\Lambda$, and extrapolations to $\Lambda \to \infty$ using the exponential ansatz \eqref{GapAnsatz} for $\Lambda \in 4\bZ$ and $\Lambda \in 4\bZ+2$, separately (solid), and using the linear ansatz \eqref{GapAnsatz} (dashed).}
\label{Fig:EnFunctional}
\end{figure}

Assuming that the rank-one Seiberg $E_6$, $E_7$, and $E_8$ theories saturate the lower bounds on $C_J$, we employ the extremal functional method to determine for each theory the spectrum appearing in the ${\cal D}[2] \times {\cal D}[2]$ OPE.  We find that the long multiplets that achieve the lowest scaling dimension have zero spin and appear in the representations ${\bf 650}$ of $E_6$, ${\bf 1539}$ of $E_7$, and ${\bf 3875}$ of $E_8$.  For these channels, Figure~\ref{Fig:EnFunctional} shows the gaps (the lowest scaling dimension) at various derivative orders, and Table~\ref{Tab:Gaps} summarizes the extrapolated gaps at infinite derivative order.

In  the rank-one Seiberg $E_{N_{\bf f}+1}$ theory, the structure of the Higgs branch moduli space $\cM_H$ is particularly simple: it is given by the one-instanton moduli space of $E_{N_{\bf f}+1}$~\cite{Seiberg:1996bd}, which can be described by a complex algebraic variety with holomorphic coordinates $L^a \equiv L_{11}^a$, subject to the Joseph ideal relations~\cite{Joseph1976,Cremonesi:2015lsa}
\ie
\left.L \otimes L \right|_{\mf J}=0,\quad {\rm Sym}^2 ({\bf adj})= (2\,{\bf adj}) \oplus {\mf J}.
\fe
Consequently, in the ${\cal D}[2]\times{\cal D}[2]$ OPE, the ${\cal D}[4]$ can only appear in the {\bf 5} of {\rm SU}(2), the {\bf 2430} of $E_6$, the {\bf 7371} of $E_7$, and the {\bf 27000} of $E_8$. We confirmed this expectation by numerically observing that $\A_{{\cal D}[2],{\bf adj}}[{\cal K}^{{\cal D}[4],\bf r}]$ are parametrically much smaller for $\bf r$ in the above representations than for $\bf r$ in the other representations.

\vspace*{\fill}
\newpage

\section{Discussion and outlook}

In this paper, we carefully analyzed the coupling of five-dimensional ${\cal N} = 1$ superconformal field theories with mass deformations to squashed-sphere backgrounds, and presented a precise triple factorization formula for computing the free energy in rank-one theories that incorporates instanton contributions.  Along the way, we classified the admissible supersymmetric counter-terms, discovered a new superconformal anomaly, and derived relations to the conformal and flavor central charges.  The knowledge of the counter-terms allowed us to elucidate the invariance of the five-sphere free energy under the enhanced Weyl group actions, both formally and supplemented with strong numerical evidence.  Using the triple factorization formula, we numerically computed the central charges in the rank-one Seiberg exceptional and Morrison-Seiberg $\widetilde E_1$ theories.  Finally, we made connections between the central charges and the OPE data, and studied the numerical bootstrap of the four-point function of moment map operators.  We found strong evidence for the saturation of the bootstrap bounds by the rank-one Seiberg theories, and extracted the spectra of long multiplets in these theories.

The Seiberg theories are holographically dual to type-I' string theory on ${\rm AdS}_6 \times {\rm HS}^4$, which in a certain low energy limit is captured by Romans $F(4)$ supergravity. In upcoming work~\cite{Chang:2017mxc}, we examine the higher-rank Seiberg theories and their holographic duals. In particular, we compute the central charges $C_T$ and $C_J$ at large-rank using matrix model techniques, and compare with certain couplings in Romans $F(4)$ supergravity.  In another paper~\cite{Chang1}, we investigate the new five-dimensional superconformal anomaly explained in Section~\ref{sec:ctt}, as well as its implications for dualities.  Other arenas for further exploration include a proof of the $F$- or a $C$-theorem for $\cN=1$ superconformal field theories in five dimensions, of which the numerical results of Section~\ref{Sec:InstantonNumerics} are suggestive, and a better (physical) understanding of our contour prescription for the gauge theory sphere partition function. A final observation is that the undeformed five-sphere free energy $F_0$ sits at a local maximum with respect to infinitesimal supersymmetric squashing and mass deformations, as the signs in \eqref{CTRelation}-\eqref{CJRelation2} show. This property is reminiscent of the $F$-maximization in three-dimensional $\cN=2$ superconformal field theories \cite{Jafferis:2010un}, except that here the maximization is {\it automatic} since there is no mixing between the non-Abelian ${\rm SU}(2)_{\rm R}$ R-symmetry and flavor symmetries.

\section*{Acknowledgments}

We are grateful to Daniel L. Jafferis, Hee-Cheol Kim, Igor R. Klebanov, Zohar Komargodski, Bruno Le Floch, Silviu S. Pufu, Nathan Seiberg, Shu-Heng Shao, David Simmons-Duffin, and Xi Yin for helpful discussions, and to Ori Ganor, Daniel L. Jafferis, Hee-Cheol Kim, and Igor R. Klebanov for comments on the first draft.  CC, YL, and YW thank the Aspen Center for Physics, MF thanks the Simons Summer Workshop, and YL thanks National Taiwan University for hospitality during the course of this work.  The numerical computations were performed on the Harvard Odyssey cluster
and the Caltech high energy theory group cluster.  We thank Tony Bartolotta for technical support with the Caltech cluster.  CC is supported in part by the U.S. Department of Energy grant DE-SC0009999. MF is supported by the David and Ellen Lee Postdoctoral Scholarship, YL is supported by the Sherman Fairchild Foundation, and both MF and YL by the U.S. Department of Energy, Office of Science, Office of High Energy Physics, under Award Number DE-SC0011632.  YW is supported in part by the US NSF under Grant No.~PHY-1620059 and by the Simons Foundation Grant No.~488653.  This work was partially performed at the Aspen Center for Physics, which is supported by National Science Foundation grant PHY-1607611. YW is also grateful to the Simons Collaboration on the Non-perturbative Bootstrap for generous support.

\appendix

\section{Conventions and normalization}
\label{sec:Convention}

This appendix summarizes the various conventions and normalization adopted in this paper.

\subsection{Normalization of the central charges $C_T$ and $C_J$}
\label{sec:NorCTCJ}

The conformally covariant structures ${\cal I}_{\m\n,\sigma\rho}(x)$ and $I_{\mu\nu}(x)$ that appear in the stress tensor two-point function \eqref{eqn:TT} and the flavor current two-point function \eqref{eqn:JJ}, are defined by
\ie\label{eqn:structureI}
&{\cal I}_{\m\n,\sigma\rho}(x)={1\over 2}\left[I_{\mu\sigma}(x)I_{\nu\rho}(x)+I_{\mu\rho}(x)I_{\nu\sigma}(x)\right]-{1\over d}\delta_{\m\n}\delta_{\sigma\rho},
\\
&I_{\mu\nu}(x)=\delta_{\mu\nu}-2{x_\mu x_\nu\over x^2}.
\fe
The stress tensor has a canonical normalization coming from the normalization of the dilatation operator $D$, which in radial quantization is defined by the following integral of the stress tensor,
\ie
D = - \int_{{\rm S}^{d-1}} {x_\m x_\n\over |x|} T^{\m\n} \diff S.
\fe
The normalization of $D$ (and hence $T^{\m\n}$) is such that the state $\ket{\cal O}$ corresponding to an operator ${\cal O}$ with scaling dimension $\Delta$ has eigenvalue $\Delta$ under $D$,
\ie
D\ket{\cal O}=\Delta\ket{\cal O}.
\fe
With this normalization, the coefficient $C_T$ in \eqref{eqn:TT} is physical and called the conformal central charge.

Similarly, the flavor currents have a canonical normalization coming from the normalization of the flavor charges $Q^a$, which in radial quantization are defined by the following integral of the flavor currents,
\ie\label{eqn:QJS}
Q^a=\int_{{\rm S}^{d-1}} J^a_{\m}(x) {x^\m\over |x|} \diff S.
\fe
The canonical normalization requires that the flavor charges acting on the states $\ket{J_\m^a}$ corresponding to the flavor currents under the state-operator correspondence give\footnote{In our convention, $Q^a$ is an anti-hermitian operator.  In particular, $f^{abc}$ is purely imaginary for compact flavor groups.
}
\ie\label{eqn:QfJ}
Q^a\ket{J_\m^b}=\ii f^{abc}\ket{J_\m^c},
\fe
where $f^{abc}$ is the structure constant of the flavor group $G$ normalized by
\ie\label{eqn:ff}
{ 1\over 2 h^\vee} f^{aed} f^{bde} =\D^{ab},
\fe
where $h^\vee$ is the dual Coxeter number.  With this normalization, the coefficient $C_J$ in \eqref{eqn:JJ} is physical and called the flavor central charge. When the flavor symmetry is U(1), we normalize the current by
\ie\label{eqn:U(1)Jnor}
Q\ket \phi=\ii\ket\phi,
\fe
where $\ket\phi$ is a state carrying the elementary U(1) charge in the theory.

Let $\mathfrak{g}$ be the Lie algebra of the flavor group $G$. For $x,y\in \mathfrak{g}$, we define the Killing form
\ie\label{eqn:Killing}
(x,y)={\rm Tr}(xy)\equiv {1\over 2h^\vee}{\rm tr}_{\bf adj}(xy),
\fe
and ${\rm tr}_{\bf adj}(\cdot)$ is the trace in the adjoint representation. We can pick a basis $\{T^a\}$ for the Lie algebra $\mathfrak{g}$, such that
\ie
{[}T^a,T^b]=f^{abc}T^c.
\fe
Then the normalization \eqref{eqn:ff} implies
\ie\label{eqn:GTnor}
(T^a,T^b)={\rm Tr}(T^a T^b)=\delta^{ab}.
\fe
The traces in other representations are linearly related to the Killing form. For example, we have ${\rm tr}_{\bf fund}(\cdot)={\rm Tr}(\cdot)$ for the fundamental representation of ${\rm SU}(N)$ or ${\rm USp}(2N)$, and ${\rm tr}_{\bf vec}(\cdot)=2{\rm Tr}(\cdot)$ for the vector representation of ${\rm SO}(N)$.

\subsection{Spinor conventions}\label{App:5dspinor}

The five-dimensional gamma matrices $\C_\m$ ($\m=1,\ldots,5$) satisfy the Clifford algebra 
\ie\label{eqn:Clifford}
\{\C_\m,\C_\n\}=2\D_{\m\n}.
\fe
The transpose $\C_\m^t$ of gamma matrices also satisfy the same Clifford algebra, and are related to the gamma matrices by
\ie
\C_\m^t=C\C_\m C^{-1},
\fe
where $C$ is the charge conjugation matrix, which is real, antisymmetric, and satisfies $C^2=-1$.  With the spinor indices $\A=1,\ldots,4$, the gamma matrices and the charge conjugation matrix are
\ie
(\C_\m)^\A{}_\B,\quad C_{\A\B},\quad
C^{\A\B}\equiv C_{\A\B}.
\fe
The higher rank gamma matrices are
\ie
(C\C_{\m})_{\A\B},~~(C\C_{\m\n\rho\sigma})_{\A\B},~~(C\C_{\m\n\rho\sigma\lambda})_{\A\B},~~
(C\C_{\m\n})_{\A\B},~~(C\C_{\m\n\rho})_{\A\B},
\fe
where $\gamma_{\m_1\cdots\m_n}\equiv\gamma_{[\m_1}\cdots\gamma_{\m_n]}$, and the first three are antisymmetric under exchanging $\A$ and $\B$, whereas the last two are symmetric. The symplectic-Majorana spinor $\psi_i^\A$ also carries an {\rm SU}(2) fundamental index $i=1,2$. It satisfies the symplectic-Majorana condition
\ie
(\psi_i^\A)^*=\epsilon^{ij}\psi_j^\B C_{\B\A}.
\fe

The spinor index $\A$ and the {\rm SU}(2) fundamental index $i$ are raised and lowered as follows
\ie
\psi^\A&=C^{\A\B}\psi_\B, &\psi_\A&=\psi^\B C_{\B\A},
\\
\psi^i&=\epsilon^{ij}\psi_j, &\psi_i&=\psi^j \epsilon_{ji},
\fe
with $\epsilon_{12}=\epsilon^{12}=1$.
The spinor index contraction is by default southwest to northeast, whereas the {\rm SU}(2) fundamental index contraction is by default northwest to southeast,
\ie
\chi \psi\equiv \chi^i_\A \psi_i^\A, \quad \chi M \psi\equiv \chi_\A^i M^\A{}_{\B} \psi_i^{\B}.
\fe

\subsection{Five-dimensional supersymmetry conventions}
\label{App:5dSUSY}

In flat space, the five-dimensional supersymmetry algebra is
\ie
& \{Q^i_\A,Q^j_\B\}=2\epsilon^{ij} (C\C^\m )_{\A\B}i\pa_\m , 
\\
& [U_i{}^j,Q^k]=\D_i^k Q^j-{1\over 2} \D_i^j Q^k,\quad [U_i{}^j,U_k{}^\ell]=\D_i^\ell U_k{}^j-\D_k^j U_i{}^\ell ,
\fe
where $U_i{}^j$ are the ${\rm SU}(2)_{\rm R}$ R-symmetry generators.

\section{Supersymmetric backgrounds for squashed five-spheres}
\label{App:sqbg}

In this Appendix, we compute the five-dimensional backgrounds preserving rigid (global) supersymmetry for a generic squashed five-sphere with the minimal ${\rm U}(1)\times {\rm U}(1)\times {\rm U}(1)$ isometry. The goal is to write down the background fields in the standard (irreducible) Weyl multiplet (see~\cite{Fujita:2001kv} and also~\cite{Hanaki:2006pj}). As a matter of fact, one can explicitly show that the standard Weyl multiplet is precisely equivalent to the conformal boundary of Romans $F(4)$ gauged supergravity. We start by providing the appropriate field redefinitions which relate the study of supersymmetric backgrounds performed in reference~\cite{Alday:2015lta}, arising at the conformal boundary of Romans $F(4)$ supergravity, to the standard Weyl multiplet.\footnote{See also~\cite{Pini:2015xha}.}

\subsection{Standard Weyl multiplet and its relation to Romans $F(4)$ Supergravity}
 
The irreducible standard Weyl multiplet consists of $32+32$ bosonic and fermionic degrees of freedom given in terms of the vielbein $e_{\mu}{}^{a}$, two ${\mf{ su}}(2)$ Majorana fermions $\psi_{\mu}^{i}$ and $\chi_{i}$, two real bosons $b_{\mu}$ and $D$, an ${\mf{ su}}(2)$ gauge field $V_{\mu}^{ij}$, and a real antisymmetric tensor $v_{ab}$. The supersymmetry conditions for the standard Weyl multiplet read
\ie
\label{SUSYFO1}
\D\psi_\m^i &= D_\m \ve^i +{1\over 2} v^{ab}\C_{\m ab} \ve^i-\C_\m \eta^i=0,
\\
\D\chi^i &= \ve^i D-2\C^c \C^{ab}\ve^i D_a v_{bc} + \C^{ab} F_{ab}{}^i{}_j(V) \ve^j-2\C^a \ve^i \epsilon_{abcde}v^{bc}v^{de}+4\C^{ab} v_{ab} \eta^i=0,
\fe
where $\ve^{i}$ and $\eta^{i}$ are the Killing spinors and conformal Killing spinors, respectively, $K_a$ the generators of special conformal transformation, and
\ie
 \delta \ = \ \bar \ve^{i} Q_{i} + \bar \eta^{i} S_{i} + \xi^{a}_{K} K_{a}
\fe
with $Q_{i}$ and $S_i$ the supercharges and their conformal cousins. The covariant derivatives are defined as follows,
\ie
D_{\mu} \ve^{i} \ = \ & \partial_{\mu} \ve^{i} + \frac{1}{2} b_{\mu} \ve^{i} + \frac{1}{4} \omega_{\mu}{}^{ab}\gamma_{ab} \ve^{i} - V_{\mu}{}^{ij} \ve_{j},\\
D_{\mu} \eta^{i} \ = \ & \partial_{\mu} \eta^{i} - \frac{1}{2} b_{\mu} \eta^{i} + \frac{1}{4} \omega_{\mu}{}^{ab}\gamma_{ab} \eta^{i} - V_{\mu}{}^{ij} \ve_{j} ,
\fe
where $\omega_\mu{}^{ab}$ is the spin connection. Finally, $V_{\mu\nu}{}^{ij}$ is the field strength of $V_{\mu}{}^{ij}$ and for our purposes,
\ie
D_{a} v_{bc} \ = \ \nabla_{a} v_{bc} .
\fe

We may translate the conditions \eqref{SUSYFO1} into \emph{equivalent} supersymmetry conditions arising in the analysis of~\cite{Alday:2015lta}.  We explicitly write down the translation for the fields between \cite{Fujita:2001kv} and \cite{Alday:2015lta} in Table~\ref{Tbl:SWvsRomans}.  Notice that in the last line, we denote by $\mathscr{C}_{\mu}$ the generator of Weyl transformations; a field of given Weyl weight $\mathrm{w}$ is acted on by the covariant derivative as $D_\mu\equiv \partial_\mu + \mathrm{w} \, \mathscr{C}_\mu$. In the following, we shall set $\mathscr{C} = b = 0$. To avoid confusion, we also denote by $\tilde\ve$ and $\tilde\eta$ the Killing spinors in the analysis of \cite{Alday:2015lta}.

\begin{table}[h]
\begin{center}
{\renewcommand{\arraystretch}{1.5}
\begin{tabular}{ | c | c | }
\hline
\ Standard Weyl multiplet \ & \ Conformal boundary of Romans $F(4)$ \ \\
\hline
\hline
$V_{\mu \, ij}$ & $-\frac{\ii}{2} a \left( \sigma^{3} \right)_{ij} $\\
$v_{ab}$ & $\frac{\ii }{2\sqrt{2}} b_{ab}$\\
$\left( \ve_{i}, \eta_{i}\right)$ & $\left( \tilde\ve_{i} ,  -\frac{\ii\sqrt{2}}{3} \tilde\eta_{i}+\frac{\ii}{6 \sqrt{2}} b_{ab}\gamma^{ab} \tilde\ve_{i} \right)$\\
$D$ & $ \frac{8}{3} X_{2}-\frac{1}{3} b_{ab}b^{ab}$\\
$b_{\mu} $ & $\mathscr{C}_{\mu}$
\\
\hline
\end{tabular}}
\end{center}
\caption{Dictionary between the fields in the standard Weyl multiplet in the notation of~\cite{Fujita:2001kv}, and the fields arising from expanding Romans $F(4)$ gauged supergravity at the conformal boundary~\cite{Alday:2015lta}.}\label{Tbl:SWvsRomans}
\end{table}

\subsection{Generic squashed five-dimensional backgrounds}
\label{sec:SquashedBackground}

In this section, we shall present the supersymmetric backgrounds for generic five-dimensional squashed spheres with (at least) ${\rm U}(1)\times {\rm U}(1)\times {\rm U}(1)$ isometry.

We pick the following five-dimensional metric,
\ie
\diff s^{2} &= \sum_{i=1}^3 \left(\diff y_{i}^{2} + y_{i}^{2} \diff \phi_{i}^{2} \right)+ \tilde{\kappa}^{2} \left( \sum_{j=1}^{3} a_{j} y_{j}^{2} \diff \phi_{j} \right)^{2} ,
\quad
\tilde{\kappa}^{2} = \frac{1}{1- \sum_{j=1}^{3} y_j^{2} a_{j}^{2}} ,
\label{5dmetric}
\fe
where $\phi_i$ are periodic coordinates, $\phi_i \sim \phi_{i} + 2 \pi$, $y_i$ are constrained coordinates such that $y_1^{2}+y_2^{2}+y_3^{2}=1$, and 
\ie
\omega_i \ = \ 1+a_i , \quad \text{for} \quad i = 1,2,3 ,
\fe
are the squashing parameters, that govern the deformation away from the round sphere.  Now, let us introduce the following frame,
\ie
e^{1}  \ = \ & \frac{1}{y_3 \sqrt{1-y_2^2} }\left[ y_1 y_2 \,  \diff y_2 + (1-y_2^{2}) \diff y_1 \right],\\
e^{2}    \ = \  &
\frac{y_1y_3}{ \sqrt{1-y_2^2}} 
\left[\left( \diff\phi_1-\diff \phi_3 \right)+\frac{a_3-a_1}{\tilde\beta}\mathcal{X}\right],\\
e^{3}    \ = \  & \frac{1}{\sqrt{1-y_2^2}}\, \diff y_2,\\
e^{4}    \ = \  & 
\frac{ y_2}{ \sqrt{1-y_2^2}} \left[-  \diff \phi_2+\frac{1+a_2 }{\tilde{\beta}}\mathcal{X}  \right],\\
e^{5}    \ = \  & \frac{1}{\tilde\kappa \tilde \beta} \mathcal{X} + \frac{1}{\tilde\kappa} \mathcal{Y},
\fe
where for ease of notation we introduced the definitions
\ie
\mathcal{X} \ = \ \sum_{i=1}^{3} y_i^2\diff \phi_i , \qquad  
\mathcal{Y} \ = \ \tilde{\kappa}^{2} \sum_{i=1}^{3} a_i y_{i}^{2} \diff \phi_{i} , \qquad  
\tilde{\beta} \ = \ 1 + \sum_{i} a_i y_i^{2}.
\fe

We now present the ${\rm U}(2)$ structure, defined by a function $S$, a one-form $K_1$ and real and complex two-forms $J$ and $\Omega$, respectively, which are explicitly given by
\ie
  S \  = \ &   \tilde\beta \tilde\kappa  , &\qquad K_1 \ &=  \ e^{5} , \\
 J  \ =  \  & e^{1}\wedge e^{2}+ e^{3}\wedge e^{4}, & \qquad \Omega \ &= \ \left( e^{1}+\ii e^{2}\right) \wedge \left( e^{3} + \ii e^{4} \right) .
\fe
Furthermore, we introduce the Killing vector $\xi$ via
\ie
 g\left( \xi, \cdot \right) \ = \ S K_1,
\fe
which can be explicitly written as
\ie
 \xi \ = \  \omega_{1} \partial_{\phi_1}+\omega_{2} \partial_{\phi_2}+\omega_{3} \partial_{\phi_3} .
\fe

Using the ${\rm U}(2)$ structure equations introduced in~\cite{Alday:2015lta},\footnote{Notice that the field $D$ -- or equivalently $X_2$ -- does not appear in these equations. However, it is completely fixed in terms of the other fields and the ${\rm U}(2)$ structure by (for instance) solving the second equation in \eqref{SUSYFO1}. Given this and by taking a ${\rm U}(1) \subset {\rm SU}(2)$ truncation of the background gauge field, equation~\eqref{SUSYFO1} is \emph{equivalent} to \eqref{Eqn:BilinearEqs} as proved in~\cite{Alday:2015lta}.}
\ie\label{Eqn:BilinearEqs}
 \diff S \ = \ & -\frac{\sqrt{2}}{3} \left( SK_2 + \ii i_{\xi} b \right) ,\\
 S \alpha \ = \ & \frac{1}{2 \sqrt{2}} i_{\xi} a ,\\
 \diff \left(  SK_1 \right) \ = \ & \frac{2 \sqrt{2}}{3} \left( 2 \alpha S J + S K_1 \wedge K_2 + \ii S b - \frac{\ii}{2} i_\xi (*b) \right) ,\\
 \diff \left( SK_2 \right) \ = \ & \ii i_{\xi} \diff b - \ii \cL_{\xi} \left( \log S \right)b ,\\
 \diff \left( SJ \right) \ = \ & - \sqrt{2} K_2 \wedge (SJ),\\
 \diff \left( S\Omega \right) \ = \ & - \ii \left( a - 2 \sqrt{2} \alpha K_1 - \ii \sqrt{2} K_2 \right)\wedge \left( S\Omega\right) ,
\fe
we can solve for the background fields -- the function $\alpha$, the one-form $K_2$, the ${\rm U}(1) \subset {\rm SU}(2)$ gauge field $a$ and the two-form $b$. We find the general solution\footnote{Here we pick a particular gauge for the ${\rm U}(1) \subset {\rm SU}(2)$ gauge field $a$, such that $i_{\xi} a = 0$.
}
\ie
K_2 \ = \ & - \frac{1}{\sqrt{2}}\diff \log\left(  \tilde{\beta}^2 \tilde\kappa \right),\\
a \ = \ & (1- a_\mathrm{tot})\mathcal{Y}+\left( (a_{\mathrm{tot}}-1)\tilde\kappa+\frac{4}{\tilde\kappa \tilde \beta}+2 \sqrt{2} \alpha \right)K_1-\diff (\phi_1+\phi_2+\phi_3) ,\\
b \ = \ &-\frac{\ii}{\sqrt{2} \tilde\kappa} \diff \mathcal{Y}-\sqrt{2} \ii \left( (a_{\mathrm{tot}}-1) \tilde\kappa  +\frac{4}{\tilde\kappa \tilde\beta} +2\sqrt{2} \alpha \right) J,
\fe
which is parameterized by the arbitrary (real) function $\alpha$.  Furthermore, we used the notation (recall that $\omega_i = 1+a_i$)
\ie
a_\mathrm{tot} \ = \ \sum_{i=1}^{3}a_i ,\quad
\omega_\mathrm{tot} \ = \ \sum_{i=1}^{3}\omega_i.
\fe
We can compute $X_2$ or (equivalently) the scalar field $D$ in the standard Weyl multiplet
\ie
D \ = \ & \frac{8}{3} X_{2}-\frac{1}{3} b_{ab}b^{ab} \\
\ = \ &
2 \left[ 3-2 a_\mathrm{tot} (2-a_\mathrm{tot}) +\sum_i a_i^2\right] \tilde\kappa^2+\frac{42 (a_\mathrm{tot}-1)}{\tilde\beta}+\frac{72}{(\tilde\kappa\tilde\beta)^2} \\
&+12\sqrt{2} \left( (a_\mathrm{tot}-1)\tilde\kappa+\frac{3 }{\tilde\kappa \tilde\beta}\right) \alpha.
\fe
A convenient choice for $\alpha$ is 
\ie\label{alphasimp}
 \alpha \ = \ \frac{1}{2\sqrt{2}}\left( (1-a_{\mathrm{tot}}) \tilde\kappa - \frac{4}{\tilde\kappa\tilde\beta} \right) .
\fe
The background fields then read
\ie
a  \ = \ &  (1-a_\mathrm{tot}) \mathcal{Y} - \diff (\phi_1+\phi_2+\phi_3) ,\\
b \ = \  & - \frac{\ii}{\sqrt{2} \tilde\kappa} \diff \mathcal{Y},
\\
 D  \ = \ & 4  (a_\mathrm{tot}- a_1a_2-a_1a_3 -a_2 a_3)\tilde\kappa^2 . 
\fe

We can now translate those background fields $\left( \alpha, a_{\mu} , b_{\mu\nu}, X_2 \right)$ into the ones appearing in the standard Weyl multiplet, $\left(  V_{\mu \, ij} , v_{\mu\nu}, D \right)$, by Table~\ref{Tbl:SWvsRomans}. In particular, one can show that the Killing spinor
\bea
\ve  \ = \
\frac{\sqrt{\rm S}}{2 \sqrt{2}}\left(
\begin{array}{c}
 -\ii  \\
\ii \\
1\\
 -1\\
\end{array}
\right) ,
\eea
and the conformal Killing spinor
\bea
\eta \ = \ \left[ \frac{\sqrt{2} \ii}{3}\alpha -\frac{1}{6 \tilde{\beta}^{2} \tilde\kappa}  \partial_a \big( \tilde{\beta}^{2} \tilde\kappa \big) \gamma^{a} + \frac{\ii}{6\sqrt{2}} b_{ab}\gamma^{ab} \right] \ve ,
\eea
(as well as their complex conjugates) solve the Killing spinor equations for the standard Weyl multiplet given in~\eqref{SUSYFO1}, with the explicit five-dimensional gamma matrices
\ie
\gamma_1 & = 
\sigma_3 \otimes \mathbf{1}_{2}
,\quad 
&\gamma_2 & =
\sigma_1 \otimes \mathbf{1}_{2}
,\quad 
&\gamma_3 & =
- \sigma_2 \otimes \sigma_3,
\\ 
\gamma_4 &= -\sigma_2 \otimes \sigma_2
 ,\quad 
&\gamma_5 &= - \sigma_2 \otimes \sigma_1.
&&
\fe

\subsection{Expansions in the standard Weyl multiplet}

Finally, let us write down the leading order deformations of background fields away from the round sphere. To do so, we pick the choice of $\alpha$ in~\eqref{alphasimp}, and thus only require the expansions of the following forms,
\ie
\mathcal{Y} \ = \ & \sum_{i} a_i y_i^{2} \diff \phi_i + \cO\left(a_i^{2}\right) ,\\
\diff \mathcal{Y} \ = \ & 2 \sum_{i} a_i y_i \diff y_i \wedge \diff \phi_i + \cO\left(a_i^{2}\right) ,\\
\tilde\kappa^{-1} \ = \ & 1 + \cO\left(a_i^{2}\right) .
\fe
We find the following explicit expansions for the background fields in the standard Weyl multiplet\footnote{The ${\rm SU}(2)$ gauge field $V_{\mu}{}^{ij}$ is pure gauge at leading order. We shall pick a gauge here in which it is vanishing (it will leave the remaining background fields invariant).
}
\ie
 V_{\mu}{}^{ij} \diff y^{\mu} \ = \ & \left[ (1-a_{\mathrm{tot}}) \sum_{i} a_i y_i^{2} \diff \phi_i + \cO\left(a_i^{2}\right) \right]   \left( \sigma_{3} \right)^{ij} ,\\
 \frac{1}{2} v_{\mu\nu} \diff y^{\mu} \wedge \diff y^{\nu} \ = \ & \frac{1}{2} \sum_{i} a_i y_i \diff y_i \wedge \diff \phi_i + \cO\left(a_i^{2}\right) ,\\
 D \ = \ & 4  (a_\mathrm{tot}- a_1a_2-a_1a_3 -a_2 a_3) + \cO\left(a_i^{2}\right) .
\fe

\subsection{Stereographic coordinates}
\label{app:stereo}

For the evaluation of integrated two point functions on the round five-sphere in Section~\ref{sec:CTderiv}, it is more convenient to use the stereographic coordinates $x_{1,2,3,4,5}$. The relation to the $\{y_i,\phi_i\}$ coordinates is
\ie
x_1=&{y_1 \cos \phi_1 \over 1+ y_3 \sin \phi_3},\quad
x_2={y_1 \sin \phi_1 \over 1+ y_3 \sin \phi_3},\quad 
x_3=&{y_2 \cos \phi_2 \over 1+ y_3 \sin \phi_3},
\\
x_4=&{y_2 \sin \phi_2 \over 1+ y_3 \sin \phi_3},
\quad
x_5= {y_3 \cos \phi_3 \over 1+ y_3 \sin \phi_3}.
\fe
We also have
\ie
\diff \phi_1=&{x_1 \diff x_2 -x_2 \diff x_1\over x_1^2+x_2^2},
\quad 
\diff \phi_2=  {x_3 \diff x_4 -x_4 \diff x_3\over x_1^3+x_2^4},
\\
\diff \phi_3=& 2{-2x_5 x_i \diff x_i +  (x^2-2 x_5^2-1) \diff x_5 \over (1-x^2)^2+4x_5^2}.
\fe

\section{Embedding Index}
\label{Sec:Embedding}

Let $\mathfrak{g}$ be a Lie algebra. The Killing form $(x,y)$ for $x,y\in\mathfrak{g}$ is defined in \eqref{eqn:Killing}. We denote the Cartan subalgebra of ${\mathfrak g}$ by ${\mathfrak h}$, and the dual vector space by ${\mathfrak h}^*$ (the space of linear functions from ${\mathfrak h}$ to $\bR$). The Cartan element $H_\A\in {\mathfrak h}$ associated to a vector $\A\in{\mathfrak h}^*$ is defined by $(H_\A, H)=\A(H)$ for any $H\in {\mathfrak h}$. The Killing form on the vector space ${\mathfrak h}^*$ is defined by $\la\A,\B\ra=(H_\A,H_\B)$ for $\A,\B\in{\mathfrak h}^*$.

Consider a subgroup $G'$ of the flavor group $G$, and let $\mathfrak{g}'$ be the Lie algebra of $G'$, with $\iota:\mathfrak{g}'\hookrightarrow \mathfrak{g}$ the embedding map.  The Killing forms of $\mathfrak{g}'$ and $\mathfrak{g}$ are linearly related by
\ie\label{eqn:EmbInd}
I_{\mathfrak{g}'\hookrightarrow \mathfrak{g}} \times (x,y)= (\iota x,\iota y)\quad\text{for any}\quad x,y\in \mathfrak{g}',
\fe
where $I_{\mathfrak{g}'\hookrightarrow \mathfrak{g}}$ is called the embedding index. Since the flavor current two-point function \eqref{eqn:JJ} is proportional to the Killing form, the flavor central charge $C_J^{G'}$ associated to the subgroup $G'$ is related to $C_J^{G}$ by
\ie
C_J^{G'} = I_{\mathfrak{g}'\hookrightarrow \mathfrak{g}} C_J^{G}.
\fe
In this appendix, we compute the embedding indices $I_{ \mathfrak{so}(2N_{\bf f}) \hookrightarrow \mathfrak{e}_{N_{\bf f}+1}} $ and $I_{ \mathfrak{u}(1)_{\rm I} \hookrightarrow \mathfrak{e}_{N_{\bf f}+1}}  $.

\begin{figure}[h]
\begin{picture}(400,240)
\put(100,20){\begin{picture}(0,0)
	\multiput(0,0)(30,0){7}{\circle{4}}
	\multiput(2,0)(30,0){6}{\line(1,0){26}}
	\put(-4,-10){$\alpha_1$}
	\put(26,-10){$\alpha_3$}
	\put(56,-10){$\alpha_4$}
	\put(86,-10){$\alpha_5$}
	\put(116,-10){$\alpha_6$}
	\put(146,-10){$\alpha_7$}
	\put(176,-10){$\alpha_8$}
	\put(60,30){\circle{4}}
	\put(60,2){\line(0,1){26}}
	\put(64,28){$\alpha_2$}
		
	\multiput(15,-15)(15.5,0){12}{\line(1,0){5}}
	\multiput(15,40)(15.5,0){12}{\line(1,0){5}}
	\multiput(15,-15)(0,10){6}{\line(0,1){5}}
	\multiput(190,-15)(0,10){6}{\line(0,1){5}}
	\end{picture}}

\put(100,100){\begin{picture}(0,0)
	\multiput(0,0)(30,0){6}{\circle{4}}
	\multiput(2,0)(30,0){5}{\line(1,0){26}}
	\put(-4,-10){$\alpha_1$}
	\put(26,-10){$\alpha_3$}
	\put(56,-10){$\alpha_4$}
	\put(86,-10){$\alpha_5$}
	\put(116,-10){$\alpha_6$}
	\put(146,-10){$\alpha_7$}
	\put(60,30){\circle{4}}
	\put(60,2){\line(0,1){26}}
	\put(64,28){$\alpha_2$}
	
	\multiput(15,-15)(15.5,0){10}{\line(1,0){5}}
	\multiput(15,40)(15.5,0){10}{\line(1,0){5}}
	\multiput(15,-15)(0,10){6}{\line(0,1){5}}
	\multiput(160,-15)(0,10){6}{\line(0,1){5}}	
	\end{picture}}
	
\put(100,180){\begin{picture}(0,0)
	\multiput(0,0)(30,0){5}{\circle{4}}
	\multiput(2,0)(30,0){4}{\line(1,0){26}}
	\put(-4,-10){$\alpha_1$}
	\put(26,-10){$\alpha_3$}
	\put(56,-10){$\alpha_4$}
	\put(86,-10){$\alpha_5$}
	\put(116,-10){$\alpha_6$}
	\put(60,30){\circle{4}}
	\put(60,2){\line(0,1){26}}
	\put(64,28){$\alpha_2$}
	
	\multiput(15,-15)(15.5,0){8}{\line(1,0){5}}
	\multiput(15,40)(15.5,0){8}{\line(1,0){5}}
	\multiput(15,-15)(0,10){6}{\line(0,1){5}}
	\multiput(130,-15)(0,10){6}{\line(0,1){5}}	
	\end{picture}}

\end{picture}
\caption{
The Dynkin diagrams for $\mathfrak{e}_6$, $\mathfrak{e}_7$, $\mathfrak{e}_8$, and their subdiagrams corresponding to the subalgebra $\mathfrak{so}(2N_{\bf f})\subset \mathfrak{e}_{N_{\bf f}+1}$.
}
\label{fig:DynkinE8}
\end{figure}
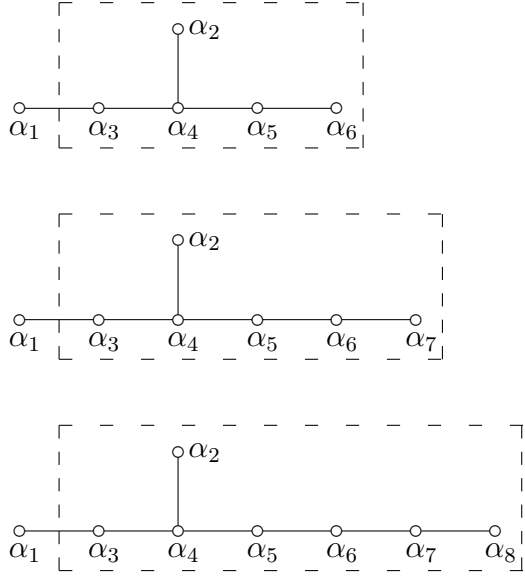

Consider a sublattice inside the root lattice of $\mathfrak{e}_{N_{\bf f}+1}$ generated by the simple roots $\left\{ \A_2,\A_3,\ldots,\A_{N_{\bf f}+1} \right\}$, the labeling of which is specified in Figure~\ref{fig:DynkinE8}. The roots $\left\{ \A_2,\A_3,\ldots,\A_{N_{\bf f}+1} \right\}$ form a subdiagram of Dynkin type $D_{N_{\bf f}}$, and the roots of $\mathfrak{e}_{N_{\bf f}+1}$ that are inside this sublattice generate an $\mathfrak{so}(2N_{\bf f})$ subalgebra.  The embedding $\mathfrak{so}(2N_{\bf f})\hookrightarrow \mathfrak{e}_{N_{\bf f}+1}$ of the Cartan elements is given by
\ie
\iota: H_{\A^{\mathfrak{so}}_i} &\mapsto H_{\A_{i+1}},
\fe 
where $\A^{\mathfrak{so}}_i$, $i=1,\ldots N_{\bf f}$ denote the simple roots of $\mathfrak{so}(2N_{\bf f})$. And the embedding index \eqref{eqn:EmbInd} of $\mathfrak{so}(2N_{\bf f}) \hookrightarrow \mathfrak{e}_{N_{\bf f}+1}$ can be computed by
\ie
I_{ \mathfrak{so}(2N_{\bf f}) \hookrightarrow \mathfrak{e}_{N_{\bf f}+1}}  = {(H_{\A_{2}},H_{\A_{2}}) \over (H_{\A^{\mathfrak{so}}_1},H_{\A^{\mathfrak{so}}_1})}={A_{22}\over A^{\mathfrak{so}}_{11}}=1,
\fe
where $A_{ij}=\la\A_i,\A_j\ra$ and $A^{\mathfrak{so}}_{ij}=\la\A^{\mathfrak{so}}_i,\A^{\mathfrak{so}}_j\ra$ are the Cartan matrices of $\mathfrak{e}_{N_{\bf f}+1}$ and $\mathfrak{so}(2N_{\bf f})$, respectively.

The instanton number ${\rm U}(1)_{\rm I}$ is defined as the commutant of this subgroup. Note that with this definition, the simple root of $\mathfrak{u}(1)_{\rm I}$ is not $\A_1$, which has nontrivial intersection with $\A_3$. Instead, we identify
\bea
\A_{\rm I} \equiv \rho_1,
\eea
where $\rho_i$ for $i=1,\ldots,N_{\bf f}+1$ are the fundamental weights that satisfy $\la\rho_i,\A_j\ra=\delta_{ij}$. Under this identification, $\A_{\rm I}$ is orthogonal to the roots $\A_2,\A_3,\ldots,\A_{N_{\bf f}+1}$, and with a normalization fixed by the condition \eqref{eqn:U(1)Jnor}. The embedding $\mathfrak{u}(1)_{\rm I}\hookrightarrow \mathfrak{e}_{N_{\bf f}+1}$ of the Cartan elements is given by
\ie
\iota: H_{\mathfrak{u}(1)_{\rm I}} &\mapsto H_{\A_{\rm I}}=H_{\rho_1}.
\fe 
The embedding index of $\mathfrak{u}(1)_{\rm I} \hookrightarrow \mathfrak{e}_{N_{\bf f}+1}$ can be computed by
\ie
I_{ \mathfrak{u}(1)_{\rm I} \hookrightarrow \mathfrak{e}_{N_{\bf f}+1}} ={ (H_{\rho_1},H_{\rho_1})\over (H_{\mathfrak{u}(1)_{\rm I}},H_{\mathfrak{u}(1)_{\rm I}})} =  A^{-1}_{11} = {4\over 8-N_{\bf f}},
\fe
where the inner product of the $\mathfrak{u}(1)_{\rm I}$ Lie algebra is normalized as $(H_{\mathfrak{u}(1)_{\rm I}},H_{\mathfrak{u}(1)_{\rm I}})=1$.

\section{Review of superconformal representation theory}
\label{App:SuperRep}

The five-dimensional superconformal algebra is $F(4)$, which contains the bosonic subalgebra $\mathfrak{so}(2,5)\times \mathfrak{su}(2)_{\rm R}$. There are sixteen fermonic generators: eight supercharges $Q^A_\A$ and eight superconformal supercharges $S^\A_A$, where $\A=1,\dotsc,4$ and $A=1,2$ are the $\mathfrak{so}(5)$ and $\mathfrak{su}(2)_{\rm R}$ spinor indices, respectively.  Superconformal primaries are operators that are annihilated by all the superconformal charges $S^\A_A$. A highest weight state of $F(4)$ is a superconformal primary that is also a highest weight state of the maximal compact subalgebra $\mathfrak{so}(2)\times\mathfrak{so}(5)\times \mathfrak{su}(2)_{\rm R}$. Representations of the superconformal algebra are generated by successively acting with the supercharges $Q^A_\A$ and the lowering generators of $\mathfrak{so}(5)\times \mathfrak{su}(2)_{\rm R}$ on the highest weight states. While some descendants of a highest weight state can appear to have zero norm, in unitary theories, they must be decoupled, and the shortened multiplets are referred to as short multiplets.

Each superconformal multiplet can be labeled by the charges $\Delta, J_\pm, J_{\rm R}$ of its highest weight state under the Cartan of $\mathfrak{so}(2) \times \mathfrak{so}(5)\times \mathfrak{su}(2)_{\rm R}$, where $J_\pm$ are the Cartan generators of the $\mathfrak{su}(2)_+\times \mathfrak{su}(2) _-\subset  \mathfrak{so}(4) \subset \mathfrak{so}(5)$. All the charges are real for unitary representations of the Lorentzian conformal algebra $\mathfrak{so}(2,5) \times \mathfrak{su}(2)_{\rm R}$. The short representations are classified into ${\cal A}, {\cal B}, {\cal D}$ types, satisfying the following conditions\cite{Minwalla:1997ka,Bhattacharya:2008zy,Buican:2016hpb,Cordova:2016emh}
\ie\label{ShortDelta}
{\cal A}&:&&\Delta=2J_++3J_{\rm R}+4,~~{\rm for}~~J_+,J_-,J_{\rm R}\ge0,
\\
{\cal B}&:&&\Delta=2J_++3J_{\rm R}+3,~~{\rm for}~~J_+=J_-~{\rm and}~J_+,J_{\rm R}\ge0,
\\
{\cal D}&:&&\Delta=3J_{\rm R},~~{\rm for}~~J_+=J_-=0~{\rm and}~J_{\rm R}\ge0.
\fe
The ${\cal D}$-type highest weight states are also annihilated by the four supercharges with positive R-charge, and are therefore ${1\over 2}$-BPS. The ${\cal A}$- and ${\cal B}$-type multiplets always contain BPS operators, although the highest weight states of them are not BPS. And the long representations satisfy the inequality
\ie
{\cal L}:\Delta>2J_++3J_{\rm R}+4.
\fe

Let us denote the multiplets by\footnote{We use $2J_{\rm R}$ since it is the Dynkin label of $\mathfrak{su}(2)_{\rm R}$.
}
\ie
{\cal X}[\Delta;d_1,d_2;2J_{\rm R}],~~~{\cal X}={\cal L},{\cal A},{\cal B},{\cal D},
\fe
where $d_1=2J_-$ and $d_2=2J_+ - 2J_-$ are the $\mathfrak{so}(5)$ Dynkin labels. Due to OPE selection rules, in the bootstrap analysis, we only have to consider multiplets whose superconformal primaries are in the symmetric rank-$\ell$ representation of $\mathfrak{so}(5)$. We denote such representations by
\ie
\label{ell5}
{\cal X}[2J_{\rm R}]_{\Delta,\ell}={\cal X}[\Delta;\ell,0;2J_{\rm R}].
\fe
The $\Delta,\ell$ subscripts for ${\cal D}$-type multiplets and the $\Delta$ subscript for ${\cal B}$-type will be omitted since their values are fixed by \eqref{ShortDelta} and \eqref{ell5}.

\section{Instanton particle mass term}
\label{App:mI}

As discussed in Section~\ref{Sec:En}, the Seiberg exceptional theories can be mass-deformed by a weight-four scalar in the flavor current multiplet. Consequently, they flow to {\rm USp}($2N$) gauge theories in the infrared. The general form of the mass deformation is given in \eqref{eqn:SUSYMassDeformation}. In this appendix, we derive the relation between the mass parameter and the Yang-Mills coupling of the infrared gauge theory.

Let us consider a {\rm USp}($2N$) vector multiplet, which contains a gauge field $A_\m^a$, a symplectic-Majorana spinor $\lambda^{ai}_\A$, a scalar $\phi^a$, and scalars $D^a_{ij}$ all in the adjoint representation of {\rm SU}(2)$_{\rm R}$.  The index $a \in \{ 1,\ldots, (2N+1)N\}$ labels the adjoint representation of the gauge group {\rm USp}($2N$). The leading terms in the supersymmetry variation of the component fields are
\ie
{[}Q^i_\A,A^a_\m]&\ = \ 2\ii\lambda^{ai}_\B(\gamma_\m)^\B{}_\A + \cdots,
\\
\{Q^i_\A,\lambda^{aj}_\B\}&\ = \ {1\over 4}\epsilon^{ij}(C\gamma^{\m\n})_{\A\B}F^a_{\m\n} + {\ii\over 2}\epsilon^{ij}(C\gamma^\m)_{\A\B}\partial_\m\phi^a + C_{\A\B}D^{aij} + \cdots,
\\
{[}Q^i_\A,\phi^a]&\ = \ 2\lambda^{ai}_\A + \cdots,
\\
{[}Q^i_\A,D^{ajk}]&\ = \ -2\ii\epsilon^{i(\underline{j}} (\gamma^\m)^\B{}_\A\partial_\m \lambda^{a\underline{k})}_\B + \cdots,
\\
{[}Q^i_\A,F^a_{\m\n}]&\ = \ 4\ii\partial_{[\underline{\m}}\lambda^{ai}_\B(\gamma_{\underline{\n}]})^\B{}_\A + \cdots.
\fe
The ${\rm U}(1)_{\rm I}$ instanton current multiplet can be constructed from bilinears of the vector multiplet together with higher order corrections. The superconformal primary $L^{ij}$ of the instanton current multiplet takes the form
\ie
\label{Lij}
&L^{ij}={1\over 4\pi^2}\left(C^{\A\B}\lambda^{ai}_\A\lambda^{aj}_\B + D^{aij} \phi^a\right) + \cdots.
\fe
The other two primaries $N$ and $J_\lambda$ are related to $L^{ij}$ via \eqref{Qdflavor}, and therefore given by
\ie
\label{NJ}
 N&={1\over 4\pi^2}\bigg[-2(\gamma^\m C)^{\A\B} \lambda_\A^{aj}\partial_{\m}\lambda^a_{j\B}-{\ii\over 4} F^a_{\m\n} F^{a\m\n}+{\ii\over 2}\partial_\m\phi^a \partial^\m\phi^a
 \\
& \hspace{1in} 
+\ii D^{aij}D^a_{ij}+\ii \phi^a\Box\phi^a\bigg]+\cdots,
\\
 J_\lambda&={1\over 4\pi^2}\left[{\ii\over 8} \epsilon^{\lambda\m\n\rho\sigma} F^a_{\m\n} F^a_{\rho\sigma} - \partial^\m (\phi^a F^a_{\m\lambda} )+{1\over 2}(\gamma_{\lambda\m} C)^{\A\B}\partial_{\m}(\lambda_\A^{aj}\lambda^a_{j\B})\right]+\cdots.
\fe
Note that the operators in \eqref{Lij} and \eqref{NJ} are all properly normalized, in the sense that the instanton number current $J_\lambda$ has the canonical normalization \eqref{eqn:U(1)Jnor}.\footnote{Consider a local operator of nonzero ${\rm U}(1)_{\rm I}$ charge $n$ inserted at the origin.  The boundary condition of the gauge field $A_\m^a$ near the origin is modified to
\ie
{1\over 8\pi^2}\int_{{\rm S}^4} {\rm Tr}(F\wedge F)=n,
\fe
where S$^4$ is a small four-sphere centered at the origin \cite{Lambert:2014jna}.  The definition of the charge \eqref{eqn:QJS} gives
\ie
Q=\int_{{\rm S}^4}  {x^\lambda\over |x|} J_\lambda \diff S={\ii\over 32\pi^2} \int_{{\rm S}^4}  {x^\lambda\over |x|}\epsilon^{\lambda\m\n\rho\sigma} F^a_{\m\n} F^a_{\rho\sigma} \diff {\rm S} =  \ii  n,
\fe
which agrees with \eqref{eqn:U(1)Jnor}.
}
Substituting the above into the action \eqref{eqn:SUSYMassDeformation}, and comparing with the standard Yang-Mills kinetic term, we find
\ie
M=\ii m_{\rm I} \equiv {4\pi^2\ii\over g_{\text{\tiny YM}}^2}.
\fe

\section{Triple sine function}
\label{App:triplesine}

In this appendix, we define the ``Barnes triple sine" function, and list a plethora of properties that are useful in the main text.

Let us start by defining the Barnes multiple zeta function, 
\bea
\zeta_N \left( s,w \mid {\vec\omega} \right)  & = & \sum_{m_1, \ldots, m_N = 0}^{\infty} \left( w+m_1 \omega_1 + \cdots + m_N \omega_N \right)^{-s}  ,
\eea
where $\Real w>0$, $\Real s >N$ and $\omega_1, \ldots, \omega_N>0$. This function is meromorphic in $s$, with simple poles at $s=1, \ldots, N$. One can then define the Barnes multiple gamma function 
\bea
\Gamma_{N}(w \mid {\vec\omega}) & =  & \exp \left[  \Psi_N\left( w \mid {\vec\omega} \right) \right],
\eea
 where
\bea
\Psi_N\left( w \mid {\vec\omega} \right)  & = &\left. \frac{\diff}{\diff s}\right|_{s=0} \zeta_N \left( s,w \mid {\vec\omega} \right) \ .
\eea
Finally, the multiple sine function is defined in terms of the Barnes gamma function as
\bea
 S_N (w \mid {\vec\omega}) & = & \Gamma_{N}(w \mid {\vec\omega})^{-1} \ \Gamma_{N} (a_{\mathrm{tot}} - w \mid {\vec\omega})^{(-1)^{N}} ,
\eea
where $\omega_{\mathrm{tot}} = \sum_{i=1}^{N} \omega_i$. For the purpose of this paper, there is a more convenient (equivalent) way of writing the triple sine function. One can prove that the above definition is equivalent to~\cite{Faddeev:1995nb,Narukawa:aa,Faddeev:2012zu}
\bea
 S_N (w \mid {\vec\omega}) & = & \exp \bigg[ \left( -1 \right)^{N} \frac{\pi \ii}{N!} B_{N,N}\left( z \mid \vec{\omega}\right)+ \left( -1 \right)^{N} \cI_{N}\left( z \mid \vec{\omega} \right) \bigg] , \label{SNint}
\eea
where we have introduced the integral
\bea
\cI_{N}\left( z \mid \vec{\omega} \right) & = & \int_{\mathbb{R}+\ii 0^+} \frac{\diff x}{x} \frac{e^{z x}}{ \prod_{k=1}^{N} \left( e^{\omega_k x}-1 \right)} .
\eea
The integral is over the real axis with the exclusion of the (essential) singularity at $x=0$ by a small half-circle reaching into the positive half-plane. Furthermore, we denote by $B_{N,N}\left( z \mid \vec{\omega}\right)$ the generalized/multiple Bernoulli polynomials, which can be explicitly computed by expanding and solving
\bea
 \frac{t^{N} e^{z t}}{\prod^{N}_{j=1} \left( e^{\omega_j t} -1 \right)} & = &  \sum_{n=0}^{\infty}\frac{t^{n}}{n!} B_{N,n}\left( z \mid \vec\omega \right)
\eea
order-by-order. In our case of interest, $N=3$, we have
\bea
B_{3,3}\left( z \mid \vec\omega \right) & = & \frac{z^3}{\omega_1 \omega_2 \omega_3}-\frac{3 \omega_{\mathrm{tot}}}{2 \omega_1 \omega_2 \omega_3}z^2+\frac{\omega_{\mathrm{tot}}^2+\left(\omega_1 \omega_2+\omega_1 \omega_3 +\omega_2 \omega_3\right)}{2 \omega_1 \omega_2 \omega_3} z\nn\\
&&-\frac{\omega_{\mathrm{tot}}\left(\omega_1 \omega_2+\omega_1 \omega_3+\omega_2 \omega_3\right)}{4 \omega_1 \omega_2 \omega_3} .
\eea

There is yet another definition of the triple sine function in terms of the generalized $q$-Pochhammer symbols, which are defined as\footnote{Notice as compared with \cite{Narukawa:aa}, we write $(p;q_1,q_2) \equiv (p;q_1,q_2)^{(3)}_\infty$.
}
\ie
(p;q_1,q_2) =\begin{cases} 
\prod\limits_{j,k=0}^\infty(1-p q_1^j q_2^k )&{\rm for}\quad |q_1|,|q_2|<1,
\\
\prod\limits_{j,k=0}^\infty(1-p q_1^{-j-1} q_2^k )^{-1}&{\rm for}\quad |q_2|<1<|q_1|,
\\
\prod\limits_{j,k=0}^\infty(1-p q_1^{j} q_2^{-k-1} )^{-1}&{\rm for}\quad |q_1|<1<|q_2|,
\\
\prod\limits_{j,k=0}^\infty(1-p q_1^{-j-1} q_2^{-k-1} )&{\rm for}\quad 1<|q_1|,|q_2|,
\end{cases}
\fe
which is a meromorphic function of $z$ ($p=e^{2\pi i z}$) \cite{Narukawa:aa}.  The triple sine function can be written as~\cite{Narukawa:aa}
\bea
\hspace{-1in}S_3(z\mid \vec{\omega}) &= &e^{-{\pi i\over 6}B_{3,3}(z\mid \vec{\omega})}\left[(e^{2\pi i{z\over \omega_1}}; e^{2\pi i{\omega_2\over \omega_1}},e^{2\pi i{\omega_3\over \omega_1}})\times \text{(2 cyclic perms on $\omega_i$)}\right]
\label{eqn:S3inShiftedFactorial1}
\\ 
&=& e^{{\pi i\over 6}B_{3,3}(z\mid \vec{\omega})}\left[(e^{-2\pi i{z\over \omega_1}}; e^{-2\pi i{\omega_2\over \omega_1}},e^{-2\pi i{\omega_3\over \omega_1}})\times \text{(2 cyclic perms on $\omega_i$)}\right].\label{eqn:S3inShiftedFactorial2}
\eea
Similarly, we can rewrite it as
\ie\label{eqn:S3inProd}
  S_3(z\mid \vec{\omega})=&(e^{2\pi i{z\over \omega_1}}; e^{2\pi i{\omega_2\over \omega_1}},e^{2\pi i{\omega_3\over \omega_1}})^{1\over 2}(e^{-2\pi i{z\over \omega_1}}; e^{-2\pi i{\omega_2\over \omega_1}},e^{-2\pi i{\omega_3\over \omega_1}})^{1\over 2}
\\
&\times \text{(2 cyclic perms on $\omega_i$)}.
\fe
Without loss of generality, let us assume ${\rm Im}({\omega_1\over\omega_2}),{\rm Im}({\omega_2\over\omega_3}),{\rm Im}({\omega_1\over\omega_3})>0$. Then we have the following formula for the derivative $S^{\prime}_3 \left(0 \mid \vec{\omega} \right)$,
\ie\label{eqn:S3PrimeinProd}
S_3'(0\mid \vec{\omega}) &= {2\pi\over \sqrt{\omega_1\omega_3}} \Big[ (1;e^{2\pi i\frac{\omega_1}{ \omega_2}},e^{2\pi i\frac{\omega_3}{  \omega_2}})^{{1\over 2}}(1;e^{-2\pi i\frac{\omega_1}{ \omega_2}},e^{-2\pi i\frac{\omega_3}{  \omega_2}})^{{1\over 2}}
\\
& \hspace{1in} \times \text{(2 cyclic perms on $\omega_i$)} \Big].
\fe

\section{Sphere partition function for five-dimensional gauge theories}
\label{sec:PartitionFunction}

\subsection{Perturbative partition function}
\label{App:Factorization}

The perturbative part of the Nekrasov partition function on ${\rm S}^1 \times \mathbb{R}^4$ was computed in~\cite{Kim:2012gu} using the Atiyah-Singer equivariant index theorem. It can be expressed in terms of the plethystic exponential 
\ie\label{PertIndex}
{\cal Z}^{\rm pert}_{{\rm S}^1 \times \mathbb{R}^4}(\epsilon_1,\epsilon_2,\A,{\mathfrak m}_f)&={\cal Z}_{\rm vec}(\epsilon_1,\epsilon_2,\A){\cal Z}_{\rm hyper}(\epsilon_1,\epsilon_2,\A,{\mathfrak m}_f),
\\
{\cal Z}_{\rm vec}(\epsilon_1,\epsilon_2,\A)&=\exp\Big[\sum_{n=1}^\infty{1\over n}f_{\rm vec}(n\epsilon_1,n\epsilon_2,n\A)\Big],
\\
{\cal Z}_{\rm hyper}(\epsilon_1,\epsilon_2,\A,{\mathfrak m}_f)&=\exp\Big[\sum_{n=1}^\infty{1\over n}f_{\rm hyper}(n\epsilon_1,n\epsilon_2,n\A,n{\mathfrak m}_f)\Big],
\fe
where $f_{\rm vec}$ and $f_{\rm hyper}$ are the ``single particle" indices,
\ie\label{eqn:SingleLetterPF}
f_{\rm vec}(\epsilon_1,\epsilon_2,\A)
&=-{1\over 2}{(e^{-\epsilon_1}+e^{-\epsilon_2})(1 +2e^{-2\A})\over (1-e^{-\epsilon_1})(1-e^{-\epsilon_2})}-e^ {-2\A},
\\
f_{\rm hyper}(\epsilon_1,\epsilon_2,\A,{\mathfrak m}_f)
&={2e^{-{1\over 2}(\epsilon_1+\epsilon_2)}e^{-\A}\over (1-e^{-\epsilon_1})(1-e^{-\epsilon_2})}   \sum_{f=1}^{N_{\bf f}}\cosh ( {\mathfrak m}_f).
\fe
Here, ${\cal Z}_{\rm vec}$ and ${\cal Z}_{\rm hyper}$ are the one-loop determinants of the vector and hypermultiplets on the zero-instanton background. Similarly, they can also be expressed in terms of the $q$-shifted factorials as
\ie
{\cal Z}_{\rm vec}(\epsilon_1,\epsilon_2,\A)&=(e^{- 2\A};e^{-\epsilon_1},e^{-\epsilon_2})(e^{- 2 \A};e^{\epsilon_1},e^{\epsilon_2}){\cal Z}_{\rm Cartan}(\epsilon_1,\epsilon_2),
\\
{\cal Z}_{\rm hyper}(\epsilon_1,\epsilon_2,\A,{\mathfrak m}_f)&=(e^{-\A \pm  {\mathfrak m}_f + {1\over 2}(\epsilon_1+\epsilon_2)};e^{\epsilon_1},e^{\epsilon_2}),
\fe
where ${\cal Z}_{\rm Cartan}$ is the contribution from the Cartan gluons of the ${\rm USp}(2)$ gauge group, explicitly 
\ie
{\cal Z}_{\rm Cartan}(\epsilon_1,\epsilon_2)=(1;e^{-\epsilon_1},e^{\epsilon_2})^{-{1\over 2}}(1;e^{\epsilon_1},e^{-\epsilon_2})^{-{1\over 2}}.
\fe

Following~\cite{Nieri:2013vba,Pasquetti:2016dyl}, we show that the full partition function \eqref{eqn:S5partition} in the weak coupling $m_{\rm I}\to\infty$ limit agrees with the perturbative partition function \eqref{eqn:PertPartition} expressed in terms of triple sine functions. We first notice that when ${\rm Im}({\omega_1\over\omega_2}),{\rm Im}({\omega_2\over\omega_3}),{\rm Im}({\omega_1\over\omega_3})>0$, the Cartan partition function ${\cal Z}_{\rm Cartan}({2\pi \ii\omega_1\over \omega_2},{2\pi \ii\omega_3\over \omega_2})$ diverges.\footnote{We have $(1;q_1,q_2)=0$ for $|q_1|,|q_2|<0$.
}
Formally, it can be written as
\ie
{\cal Z}_{\rm Cartan}\left({2\pi \ii\omega_1\over \omega_2},{2\pi \ii\omega_3\over \omega_2}\right)=\exp\Bigg[{1\over 2}\sum_{n=1}^\infty{1\over n}\bigg({e^{2\pi \ii{\omega_1\over \omega_2}}+e^{-2\pi \ii{\omega_3\over \omega_2}}\over \big(1-e^{2\pi \ii{\omega_1\over \omega_2}}\big)\big(1-e^{-2\pi \ii{\omega_3\over \omega_2}}\big)}+1\bigg)\Bigg].
\fe
We then define the regularized partition function of the Cartan gluons as
\ie
{\cal Z}'_{\rm Cartan}\left({2\pi \ii\omega_1\over \omega_2},{2\pi \ii\omega_3\over \omega_2}\right)&={2\pi\over \sqrt{\omega_1\omega_3}}\exp\Bigg[{1\over 2}\sum_{n=1}^\infty{1\over n}{e^{2\pi \ii{\omega_1\over \omega_2}}+e^{-2\pi \ii{\omega_3\over \omega_2}}\over \big(1-e^{2\pi \ii{\omega_1\over \omega_2}}\big)\big(1-e^{-2\pi \ii{\omega_3\over \omega_2}}\big)}\Bigg]
\\
&={2\pi\over \sqrt{\omega_1\omega_3}}(1;e^{2\pi \ii\frac{\omega_1}{ \omega_2}},e^{2\pi \ii\frac{\omega_3}{  \omega_2}})^{{1\over 2}}(1;e^{-2\pi \ii\frac{\omega_1}{ \omega_2}},e^{-2\pi \ii\frac{\omega_3}{  \omega_2}})^{{1\over 2}}.
\fe
Using the formulae \eqref{eqn:S3inShiftedFactorial1}, \eqref{eqn:S3inShiftedFactorial2} and \eqref{eqn:S3PrimeinProd}, we can write the second line of \eqref{eqn:PertPartition} as\footnote{Note that we used \eqref{eqn:S3inShiftedFactorial1} for $S_3(2\ii\lambda\mid \vec{\omega})$ and $S_3( \ii \lambda +  \ii m_f +{\omega_1+\omega_2+\omega_3\over 2}\mid \vec{\omega} )$, and \eqref{eqn:S3inShiftedFactorial2} for $S_3(-2\ii\lambda\mid \vec{\omega})$ and $S_3( -\ii \lambda +  \ii m_f +{\omega_1+\omega_2+\omega_3\over 2}\mid \vec{\omega} )$.}
\ie\label{eqn:S5perIntFac}
&\hspace{-.15in} {S_3(\pm 2\ii \lambda\mid \vec{\omega})
\over
 \prod_f S_3( \pm \ii \lambda +  \ii m_f +{\omega_1+\omega_2+\omega_3\over 2}\mid \vec{\omega} )}
\\
&\hspace{-.15in} = \exp\Bigg\{ -{\pi \ii\over 6} \bigg[ B_{3,3}(2\ii\lambda\mid \vec{\omega})-B_{3,3}(-2\ii\lambda\mid \vec{\omega})
\\
& -\sum_{f=1}^{N_{\bf f}} B_{3,3}(\ii\lambda+  \ii m_f +{\textstyle {\omega_1+\omega_2+\omega_3\over 2}}\mid \vec{\omega}) + \sum_{f=1}^{N_{\bf f}} B_{3,3}(-\ii\lambda+  \ii m_f + {\textstyle{\omega_1+\omega_2+\omega_3\over 2}}\mid \vec{\omega}) \bigg] \Bigg\}
\\
&\quad\times \bigg[ (e^{- 4\pi {\lambda\over \omega_1}}; e^{2\pi \ii{\omega_2\over \omega_1}},e^{2\pi \ii{\omega_3\over \omega_1}}) (e^{- 4\pi {\lambda\over \omega_1}}; e^{-2\pi \ii{\omega_2\over \omega_1}},e^{-2\pi \ii{\omega_3\over \omega_1}})
\\
&\hspace{.5in} \times \prod_{f=1}^{N_{\bf f}}(-e^{2\pi {- \lambda\pm m_f\over \omega_1}+\pi \ii{\omega_2+\omega_3\over \omega_1}}; e^{2\pi \ii{\omega_2\over \omega_1}},e^{2\pi \ii{\omega_3\over \omega_1}}) \times \text{(2 cyclic perms on $\omega_i$)} \bigg]
\\
& \hspace{-.15in} =\exp\Bigg\{-\frac{(8-N_{\bf f})\pi  \lambda ^3 }{3 \omega _1 \omega _2 \omega _3}+\frac{  \left[12 m^2 N_{\bf f}+(N_{\bf f}+4) \sum_{i=1}^3 \omega_i^2 +12 \sum_{i<j} \omega_i \omega_j \right] \pi  \lambda}{12 \omega _1 \omega _2 \omega
   _3}\Bigg\}
\\
&\hspace{.5in} \times{1\over S_3'(0\mid \vec{\omega})} \times {\cal Z}^{\rm pert}_{{\rm S}^1 \times \mathbb{R}^4}\left(\frac{2\pi \ii\omega_2}{\omega_1},\frac{2\pi \ii\omega_3}{\omega_1},\frac{2\pi\lambda}{\omega_1},{2\pi m_f\over \omega_1}-\pi \ii\right)
\\
&\hspace{1in} \times{\cal Z}'^{\rm pert}_{{\rm S}^1 \times \mathbb{R}^4}\left(\frac{2\pi \ii\omega_1}{\omega_2},\frac{2\pi \ii\omega_3}{\omega_2},\frac{2\pi\lambda}{\omega_2},{2\pi m_f\over \omega_2}-\pi \ii\right)
\\
&\hspace{1in} \times{\cal Z}^{\rm pert}_{{\rm S}^1 \times \mathbb{R}^4}\left(\frac{2\pi \ii\omega_1}{\omega_3},\frac{2\pi \ii\omega_2}{\omega_3},\frac{2\pi\lambda}{\omega_3},{2\pi m_f\over \omega_3}-\pi \ii\right),
\fe
where ${\cal Z}'^{\rm pert}_{{\rm S}^1 \times \mathbb{R}^4}$ is the perturbative partition function with ${\cal Z}_{\rm Cartan}$ replaced by ${\cal Z}'_{\rm Cartan}$. The perturbative five-sphere partition function is now written as
\ie
\mathcal{Z}^{\text{pert}}_{{\rm S}^5} 
= &  \int_{\cal C} {\diff\lambda \over 4\pi} \, e^{-{\cal F}^\vee_{\rm eff}} {\cal Z}^{\rm pert}_{{\rm S}^1 \times \mathbb{R}^4}\left(\frac{2\pi \ii\omega_2}{\omega_1},\frac{2\pi \ii\omega_3}{\omega_1},\frac{2\pi\lambda}{\omega_1},{2\pi m_f\over \omega_1}-\pi \ii\right)
\\
&\quad\times{\cal Z}'^{\rm pert}_{{\rm S}^1 \times \mathbb{R}^4}\left(\frac{2\pi \ii\omega_1}{\omega_2},\frac{2\pi \ii\omega_3}{\omega_2},\frac{2\pi\lambda}{\omega_2},{2\pi m_f\over \omega_2}-\pi \ii\right)
\\
& \quad \times{\cal Z}^{\rm pert}_{{\rm S}^1 \times \mathbb{R}^4}\left(\frac{2\pi \ii\omega_1}{\omega_3},\frac{2\pi \ii\omega_2}{\omega_3},\frac{2\pi\lambda}{\omega_3},{2\pi m_f\over \omega_3}-\pi \ii\right),
\fe
where
\ie
{\cal F}^\vee_{\rm eff} \ = \ & \frac{(8-N_{\bf f})\pi  \lambda ^3 }{3 \omega _1 \omega _2 \omega _3}+\frac{ 2\pi m_{\rm I}\lambda^2}{\omega_1\omega_2\omega_3}
\\
& -\frac{ \left[ 12 \sum_{f=1}^{N_{\bf f}}m^2_f +(N_{\bf f}+4) \sum_{i=1}^{3}\omega _i^2+12\sum_{i<j} \omega_{i}\omega_j \right] \pi  \lambda}{12 \omega _1 \omega _2 \omega
   _3}.
\fe
Following~\cite{Kim:2012ava,Lockhart:2012vp,Kim:2012qf}, we conjecture that the full five-sphere partition function $\mathcal{Z}_{{\rm S}^5}$ has a similar triply-factorized form as does the perturbative five-sphere partition function.  For ease of notation, we keep the prime on one of the ${\cal Z}_{{\rm S}^1 \times \mathbb{R}^4}$-factors implicit in \eqref{eqn:S5partition}.

\subsection{ADHM Quantum Mechanics}
\label{sec:ADHM}

In the string theory realization of five-dimensional Seiberg exceptional theories, the instantons are described by D0-branes moving in a D4-D8/O8 background~\cite{Hwang:2014uwa}. The directions of the various branes are summarized in Table~\ref{table:BraneDirections}.  The low energy theory on $k$ D0-branes is an ${\cal N}=4$ O($k$) gauged quantum mechanics with ${\rm SU}(2)^{\rm R}_+\times{\rm SU}(2)^{\rm R}_-$ R-symmetry and ${\rm SU}(2)_+\times{\rm SU}(2)_-$ flavor symmetry corresponding to the rotations on the four-planes $\bR^{1234}$ and $\bR^{5678}$, respectively. The field content of the supersymmetric quantum mechanics is summarized in Table~\ref{table:D0-D4-D8/O8}. The vector and Fermi multiplets, which arise from the D0-D0 strings are in the adjoint representation of the gauge group ${\rm O}(k)$, and the hyper- and twisted hypermultiplets are in the symmetric representation.  The D0-D4 and D0-D8 strings are in the bi-fundamental representation of the gauge group ${O}(k)$ and their flavor groups, ${\rm USp}(2N)$ and ${\rm SO}(2N_{\bf f})$, respectively.

\begin{table}[h]
\centering
\begin{tabular}{|c|c|c|c|}
\hline
strings &${\cal N}=4$ multiplets & fields & {\footnotesize {\rm SU}(2)$_-\times${\rm SU}(2)$_+\times${\rm SU}(2)$^{\rm R}_-\times${\rm SU}(2)$^{\rm R}_+$}
\\ \hline
\multirow{8}{*}{D0-D0 strings}& \multirow{3}{*}{vector} & gauge field &$({\bf 1},{\bf 1},{\bf 1},{\bf 1})$
\\ \cline{3-4}
& & scalar & $({\bf 1},{\bf 1},{\bf 1},{\bf 1})$ 
\\ \cline{3-4}
&  & fermions &$({\bf 1},{\bf 2},{\bf 1},{\bf 2})$ 
\\ \cline{2-4}
&Fermi & fermions & $({\bf 2},{\bf 1},{\bf 2},{\bf 1})$
\\ \cline{2-4}
&\multirow{2}{*}{twisted hyper}  & scalars & $({\bf 1},{\bf 1},{\bf 2},{\bf 2})$
\\ \cline{3-4}
&& fermions & $({\bf 1},{\bf 2},{\bf 2},{\bf 1})$ 
\\ \cline{2-4}
&\multirow{2}{*}{hyper} & scalars & $({\bf 2},{\bf 2},{\bf 1},{\bf 1})$ 
\\ \cline{3-4}
&& fermions & $({\bf 2},{\bf 1},{\bf 1},{\bf 2})$ 
\\ \hline
\multirow{3}{*}{D0-D4 strings}& \multirow{2}{*}{hyper} & scalars & $({\bf 1},{\bf 2},{\bf 1},{\bf 1})$
\\ \cline{3-4}
& & fermions & $({\bf 1},{\bf 1},{\bf 1},{\bf 2})$ 
\\ \cline{2-4}
& Fermi & fermions & $({\bf 1},{\bf 1},{\bf 2},{\bf 1})$ 
\\ \hline
D0-D8 strings & Fermi & fermions & $({\bf 1},{\bf 1},{\bf 1},{\bf 1})$
\\ \hline
\end{tabular}
\caption{The field content of the D0-D4-D8/O8 quantum mechanics.}
\label{table:D0-D4-D8/O8}
\end{table}

Consider an ${\cal N}=2$ subalgebra inside the ${\cal N}=4$ supersymmetry algebra. The Witten index is defined as
\ie\label{eqn:WittenIndex}
&{\cal Z}^{k}_{\text{\tiny D0-D4-D8/O8}}(\epsilon_1,\epsilon_2,\epsilon^{\rm R}_-,\A_i,{\mathfrak m}_f)
\\
&={\rm Tr}_{{\cal H}_{\text{QM}}} \left[(-1)^F e^{-\beta\{Q^\dagger,Q\}-2\epsilon_+(J_++J^{\rm R}_+)-2 \epsilon_-J_- - 2\epsilon^{\rm R}_- J^{\rm R}_- -\sum_f F_f {\mathfrak m}_f- \sum_i\A_i H_i}\right],
\fe
where $\epsilon_\pm={\epsilon_1\pm\epsilon_2\over 2}$, $Q$ and $Q^\dagger$ are the supercharges in the ${\cal N}=2$ subalgebra, $J^{\rm R}_+$, $J^{\rm R}_-$ are the Cartan generators of the {\rm SU}(2)$^{\rm R}_+$ and {\rm SU}(2)$^{\rm R}_-$ R-symmetry groups, and $J_\pm$, $F_f$ and $H_i$ of the {\rm SU}(2)$_\pm$, {\rm SO}($2N_{\bf f}$), and {\rm USp}(2$N$) flavor groups respectively.  The eigenstates with the same nonzero eigenvalues of the Hamiltonian $H=\{Q^\dagger,Q\}$ are paired up and exchanged by the action of the supercharges $Q$ and $Q^\dagger$. Their contributions to the trace \eqref{eqn:WittenIndex} cancel, and hence the Witten index is independent of the inverse temperature $\beta$.  

The Witten index was computed using supersymmetric localization in~\cite{Hwang:2014uwa,Cordova:2014oxa}. The result can be  expressed as a multi-dimensional contour integral over a variable $\phi$ valued in the maximal torus of the complexified gauge group. The real part of $\phi$ is the scalar zero mode inside the vector multiplet, and the imaginary part is the holonomy of the gauge field along the time circle.  The integrand is then given by the one-loop determinants of the field content listed in Table~\ref{table:D0-D4-D8/O8}. It was shown in~\cite{Hwang:2014uwa,Cordova:2014oxa,Benini:2013xpa} that the precise contour prescription is given by the Jeffrey-Kirwan residue.  The Witten indices for the quantum mechanics of different $k$ can be combined into a generating function
\be
{\cal Z}_{\text{\tiny D0-D4-D8/O8}}(\epsilon_1,\epsilon_2,\epsilon^{\rm R}_-,\A_i,{\mathfrak m}_f,{\mathfrak m}_{\rm I})=1+\sum_{k=1}^\infty e^{-k {\mathfrak m}_{\rm I}}{\cal Z}^{k}_{\text{\tiny D0-D4-D8/O8}}(\epsilon_1,\epsilon_2,\epsilon^{\rm R}_-,\A_i,{\mathfrak m}_f),
\ee
where we introduced the chemical potential ${\mathfrak m}_{\rm I}$ for the ${\rm U}(1)_{\rm I}$ instanton symmetry.  In the following, we shall restrict to rank-one.

The instantons in the five-dimensional gauge theory have the interpretation as D0-branes bound with D4-branes.  However, the Witten index receives contributions from both the bound and unbound D0-branes. The instanton part of the Nekrasov partition function is a ratio of two generating functions of Witten indices~\cite{Hwang:2014uwa},
\ie\label{InstantonFromQM}
&{\cal Z}^{\rm inst}_{{\rm S}^1 \times \mathbb{R}^4}(\epsilon_1,\epsilon_2,\A,{\mathfrak m}_f,{\mathfrak m}_{\rm I})={{\cal Z}_{\text{\tiny D0-D4-D8/O8}}(\epsilon_1,\epsilon_2,\epsilon^{\rm R}_-,\A,{\mathfrak m}_f,{\mathfrak m}_{\rm I})\over {\cal Z}_{\text{\tiny D0-D8/O8}}(\epsilon_1,\epsilon_2,\epsilon^{\rm R}_-,{\mathfrak m}_f,{\mathfrak m}_{\rm I})},
\fe
which effectively removes the contribution of the unbound D0-branes. The denominator ${\cal Z}_{\text{\tiny D0-D8/O8}}$ is the generating function of the Witten indices of the system \emph{without} D4-branes. Explicitly it is given by the $\A \to \infty$ limit, \Ie,
\ie
{\cal Z}_{\text{\tiny D0-D8/O8}}(\epsilon_1,\epsilon_2,\epsilon^{\rm R}_-,{\mathfrak m}_f,{\mathfrak m}_{\rm I})=\lim_{\A\to \infty}{\cal Z}_{\text{\tiny D0-D4-D8/O8}}(\epsilon_1,\epsilon_2,\epsilon^{\rm R}_-,\A,{\mathfrak m}_f,{\mathfrak m}_{\rm I}).
\fe
Notice that ${\cal Z}_{\rm inst}$ is independent of the chemical $\epsilon^{\rm R}_-$ associated with the Cartan of {\rm SU}(2)$^{\rm R}_-$, which is expected because {\rm SU}(2)$^{\rm R}_-$ is not part of the flavor symmetry of the rank-one $E_n$ theories. The instanton partition function can be re-expanded as
\ie\label{InstantonFromQM}
&{\cal Z}^{\rm inst}_{{\rm S}^1 \times \mathbb{R}^4}(\epsilon_1,\epsilon_2,\A,{\mathfrak m}_f,{\mathfrak m}_{\rm I})=1+\sum_{k=1}^\infty e^{- k {\mathfrak m}_{\rm I}}{\cal Z}^{{\rm inst},k}_{{\rm S}^1 \times \mathbb{R}^4}(\epsilon_1,\epsilon_2,\A,{\mathfrak m}_f).
\fe
We reproduce the resulting single-instanton ($k=1$) partition function
\ie
& {\cal Z}^{{\rm inst},1}_{{\rm S}^1 \times \mathbb{R}^4}(\epsilon_1,\epsilon_2,\A,{\mathfrak m}_f)
= -{1\over 16 \sinh^2{\frac{ (\epsilon_1+\epsilon_2)}{4}}\sinh{{\epsilon_1\over 2}} \sinh{{\epsilon_2\over 2}} } 
\\
&\hspace{.5in} \times\Bigg[\left(\frac{\cosh^2{ \A\over 2}}{\cosh\frac{(2\A+\epsilon_1+\epsilon_2)}{4}\cosh\frac{(2\A-\epsilon_1-\epsilon_2)}{4}}-1\right)\prod_{f=1}^{N_{\bf f}}\cosh{{\mathfrak m}_f\over 2} 
\\
&\hspace{1in} +\left(\frac{\sinh^2{ \A\over 2}}{\sinh\frac{(2\A+\epsilon_1+\epsilon_2)}{4}\sinh\frac{(2\A-\epsilon_1-\epsilon_2)}{4}}-1\right)\prod_{f=1}^{N_{\bf f}}\sinh { {\mathfrak m}_f\over 2} \Bigg].
\fe

\section{Weyl group action on the mass parameters}
\label{sec:Basis}

In this appendix, we specify our choice of basis for the root systems of $E_{N_{\bf f}+1}$. For $N_{\bf f}=1,\ldots,7$, the simple roots are given by the rows in the following matrices,
\ie\label{eqn:SimpleRoots}
&\begin{pmatrix}
 2 & 0 \\
 -\frac{1}{2} & -\frac{\sqrt{7}}{2} \\
\end{pmatrix},
~
\begin{pmatrix}
 1 & -1 & 0 \\
 -\frac{1}{2} & -\frac{1}{2} & {\sqrt{6}\over 2} \\
 1 & 1 & 0 \\
\end{pmatrix},
~
\begin{pmatrix}
 1 & -1 & 0 & 0 \\
 0 & 1 & -1 & 0 \\
 -\frac{1}{2} & -\frac{1}{2} & -\frac{1}{2} & -\frac{\sqrt{5}}{2} \\
 0 & 1 & 1 & 0 \\
\end{pmatrix},
~
\begin{pmatrix}
 1 & -1 & 0 & 0 & 0 \\
 0 & 1 & -1 & 0 & 0 \\
 0 & 0 & 1 & -1 & 0 \\
 -\frac{1}{2} & -\frac{1}{2} & -\frac{1}{2} & -\frac{1}{2} & 1 \\
 0 & 0 & 1 & 1 & 0 \\
\end{pmatrix},
\\
&
\begin{pmatrix}
 1 & -1 & 0 & 0 & 0 & 0 \\
 0 & 1 & -1 & 0 & 0 & 0 \\
 0 & 0 & 1 & -1 & 0 & 0 \\
 0 & 0 & 0 & 1 & -1 & 0 \\
 -\frac{1}{2} & -\frac{1}{2} & -\frac{1}{2} & -\frac{1}{2} & -\frac{1}{2} &
   -\frac{\sqrt{3}}{2} \\
 0 & 0 & 0 & 1 & 1 & 0 \\
\end{pmatrix},
\quad
\begin{pmatrix}
 1 & -1 & 0 & 0 & 0 & 0 & 0 \\
 0 & 1 & -1 & 0 & 0 & 0 & 0 \\
 0 & 0 & 1 & -1 & 0 & 0 & 0 \\
 0 & 0 & 0 & 1 & -1 & 0 & 0 \\
 0 & 0 & 0 & 0 & 1 & -1 & 0 \\
 -\frac{1}{2} & -\frac{1}{2} & -\frac{1}{2} & -\frac{1}{2} & -\frac{1}{2} &
   -\frac{1}{2} & \frac{\sqrt{2}}{2} \\
 0 & 0 & 0 & 0 & 1 & 1 & 0 \\
\end{pmatrix},
\\
& \begin{pmatrix}
 1 & -1 & 0 & 0 & 0 & 0 & 0 & 0 \\
 0 & 1 & -1 & 0 & 0 & 0 & 0 & 0 \\
 0 & 0 & 1 & -1 & 0 & 0 & 0 & 0 \\
 0 & 0 & 0 & 1 & -1 & 0 & 0 & 0 \\
 0 & 0 & 0 & 0 & 1 & -1 & 0 & 0 \\
 0 & 0 & 0 & 0 & 0 & 1 & -1 & 0 \\
 -\frac{1}{2} & -\frac{1}{2} & -\frac{1}{2} & -\frac{1}{2} & -\frac{1}{2} &
   -\frac{1}{2} & -\frac{1}{2} & -\frac{1}{2} \\
 0 & 0 & 0 & 0 & 0 & 1 & 1 & 0 \\
\end{pmatrix}.
\fe
The simple roots span the root lattice $\Lambda^{{\mathfrak e}_{N_{\bf f}+1}}_{\rm root}$, whose dual is the weight lattice $\Lambda^{{\mathfrak e}_{N_{\bf f}+1}}_{\rm weight}=(\Lambda^{{\mathfrak e}_{N_{\bf f}+1}}_{\rm root})^*$.

\bibliography{refs} 

\providecommand{\href}[2]{#2}\begingroup\raggedright\begin{thebibliography}{100}

\bibitem{Wilson:1971dc}
K.~G. Wilson and M.~E. Fisher, \emph{{Critical exponents in 3.99 dimensions}},
  \href{http://dx.doi.org/10.1103/PhysRevLett.28.240}{\emph{Phys. Rev. Lett.}
  {\bf 28} (1972) 240--243}.

\bibitem{Caswell:1974gg}
W.~E. Caswell, \emph{{Asymptotic Behavior of Nonabelian Gauge Theories to Two
  Loop Order}}, \href{http://dx.doi.org/10.1103/PhysRevLett.33.244}{\emph{Phys.
  Rev. Lett.} {\bf 33} (1974) 244}.

\bibitem{Banks:1981nn}
T.~Banks and A.~Zaks, \emph{{On the Phase Structure of Vector-Like Gauge
  Theories with Massless Fermions}},
  \href{http://dx.doi.org/10.1016/0550-3213(82)90035-9}{\emph{Nucl. Phys.} {\bf
  B196} (1982) 189--204}.

\bibitem{Fei:2014yja}
L.~Fei, S.~Giombi and I.~R. Klebanov, \emph{{Critical $O(N)$ models in
  $6-\epsilon$ dimensions}},
  \href{http://dx.doi.org/10.1103/PhysRevD.90.025018}{\emph{Phys. Rev.} {\bf
  D90} (2014) 025018}, [\href{http://arxiv.org/abs/1404.1094}{{\tt
  1404.1094}}].

\bibitem{Mati:2014xma}
P.~Mati, \emph{{Vanishing beta function curves from the functional
  renormalization group}},
  \href{http://dx.doi.org/10.1103/PhysRevD.91.125038}{\emph{Phys. Rev.} {\bf
  D91} (2015) 125038}, [\href{http://arxiv.org/abs/1501.00211}{{\tt
  1501.00211}}].

\bibitem{Mati:2016wjn}
P.~Mati, \emph{{Critical scaling in the large-$N$ $O(N)$ model in higher
  dimensions and its possible connection to quantum gravity}},
  \href{http://dx.doi.org/10.1103/PhysRevD.94.065025}{\emph{Phys. Rev.} {\bf
  D94} (2016) 065025}, [\href{http://arxiv.org/abs/1601.00450}{{\tt
  1601.00450}}].

\bibitem{Seiberg:1996bd}
N.~Seiberg, \emph{{Five-dimensional SUSY field theories, nontrivial fixed
  points and string dynamics}},
  \href{http://dx.doi.org/10.1016/S0370-2693(96)01215-4}{\emph{Phys. Lett.}
  {\bf B388} (1996) 753--760}, [\href{http://arxiv.org/abs/hep-th/9608111}{{\tt
  hep-th/9608111}}].

\bibitem{Morrison:1996xf}
D.~R. Morrison and N.~Seiberg, \emph{{Extremal transitions and five-dimensional
  supersymmetric field theories}},
  \href{http://dx.doi.org/10.1016/S0550-3213(96)00592-5}{\emph{Nucl. Phys.}
  {\bf B483} (1997) 229--247}, [\href{http://arxiv.org/abs/hep-th/9609070}{{\tt
  hep-th/9609070}}].

\bibitem{Intriligator:1997pq}
K.~A. Intriligator, D.~R. Morrison and N.~Seiberg, \emph{{Five-dimensional
  supersymmetric gauge theories and degenerations of Calabi-Yau spaces}},
  \href{http://dx.doi.org/10.1016/S0550-3213(97)00279-4}{\emph{Nucl. Phys.}
  {\bf B497} (1997) 56--100}, [\href{http://arxiv.org/abs/hep-th/9702198}{{\tt
  hep-th/9702198}}].

\bibitem{Jefferson:2017ahm}
P.~Jefferson, H.-C. Kim, C.~Vafa and G.~Zafrir, \emph{{Towards Classification
  of 5d SCFTs: Single Gauge Node}},
  \href{http://arxiv.org/abs/1705.05836}{{\tt 1705.05836}}.

\bibitem{Witten:1996qb}
E.~Witten, \emph{{Phase transitions in M theory and F theory}},
  \href{http://dx.doi.org/10.1016/0550-3213(96)00212-X}{\emph{Nucl. Phys.} {\bf
  B471} (1996) 195--216}, [\href{http://arxiv.org/abs/hep-th/9603150}{{\tt
  hep-th/9603150}}].

\bibitem{Douglas:1996xp}
M.~R. Douglas, S.~H. Katz and C.~Vafa, \emph{{Small instantons, Del Pezzo
  surfaces and type I-prime theory}},
  \href{http://dx.doi.org/10.1016/S0550-3213(97)00281-2}{\emph{Nucl. Phys.}
  {\bf B497} (1997) 155--172}, [\href{http://arxiv.org/abs/hep-th/9609071}{{\tt
  hep-th/9609071}}].

\bibitem{Katz:1996fh}
S.~H. Katz, A.~Klemm and C.~Vafa, \emph{{Geometric engineering of quantum field
  theories}},
  \href{http://dx.doi.org/10.1016/S0550-3213(97)00282-4}{\emph{Nucl. Phys.}
  {\bf B497} (1997) 173--195}, [\href{http://arxiv.org/abs/hep-th/9609239}{{\tt
  hep-th/9609239}}].

\bibitem{Aharony:1997ju}
O.~Aharony and A.~Hanany, \emph{{Branes, superpotentials and superconformal
  fixed points}},
  \href{http://dx.doi.org/10.1016/S0550-3213(97)00472-0}{\emph{Nucl. Phys.}
  {\bf B504} (1997) 239--271}, [\href{http://arxiv.org/abs/hep-th/9704170}{{\tt
  hep-th/9704170}}].

\bibitem{Aharony:1997bh}
O.~Aharony, A.~Hanany and B.~Kol, \emph{{Webs of (p,q) five-branes,
  five-dimensional field theories and grid diagrams}},
  \href{http://dx.doi.org/10.1088/1126-6708/1998/01/002}{\emph{JHEP} {\bf 01}
  (1998) 002}, [\href{http://arxiv.org/abs/hep-th/9710116}{{\tt
  hep-th/9710116}}].

\bibitem{Leung:1997tw}
N.~C. Leung and C.~Vafa, \emph{{Branes and toric geometry}}, {\emph{Adv. Theor.
  Math. Phys.} {\bf 2} (1998) 91--118},
  [\href{http://arxiv.org/abs/hep-th/9711013}{{\tt hep-th/9711013}}].

\bibitem{Xie:2017pfl}
D.~Xie and S.-T. Yau, \emph{{Three dimensional canonical singularity and five
  dimensional $ \mathcal{N} $ = 1 SCFT}},
  \href{http://dx.doi.org/10.1007/JHEP06(2017)134}{\emph{JHEP} {\bf 06} (2017)
  134}, [\href{http://arxiv.org/abs/1704.00799}{{\tt 1704.00799}}].

\bibitem{Iqbal:2002we}
A.~Iqbal, \emph{{All genus topological string amplitudes and five-brane webs as
  Feynman diagrams}},  \href{http://arxiv.org/abs/hep-th/0207114}{{\tt
  hep-th/0207114}}.

\bibitem{Aganagic:2003db}
M.~Aganagic, A.~Klemm, M.~Marino and C.~Vafa, \emph{{The Topological vertex}},
  \href{http://dx.doi.org/10.1007/s00220-004-1162-z}{\emph{Commun. Math. Phys.}
  {\bf 254} (2005) 425--478}, [\href{http://arxiv.org/abs/hep-th/0305132}{{\tt
  hep-th/0305132}}].

\bibitem{Awata:2005fa}
H.~Awata and H.~Kanno, \emph{{Instanton counting, Macdonald functions and the
  moduli space of D-branes}},
  \href{http://dx.doi.org/10.1088/1126-6708/2005/05/039}{\emph{JHEP} {\bf 05}
  (2005) 039}, [\href{http://arxiv.org/abs/hep-th/0502061}{{\tt
  hep-th/0502061}}].

\bibitem{Iqbal:2007ii}
A.~Iqbal, C.~Kozcaz and C.~Vafa, \emph{{The Refined topological vertex}},
  \href{http://dx.doi.org/10.1088/1126-6708/2009/10/069}{\emph{JHEP} {\bf 10}
  (2009) 069}, [\href{http://arxiv.org/abs/hep-th/0701156}{{\tt
  hep-th/0701156}}].

\bibitem{Taki:2007dh}
M.~Taki, \emph{{Refined Topological Vertex and Instanton Counting}},
  \href{http://dx.doi.org/10.1088/1126-6708/2008/03/048}{\emph{JHEP} {\bf 03}
  (2008) 048}, [\href{http://arxiv.org/abs/0710.1776}{{\tt 0710.1776}}].

\bibitem{Awata:2008ed}
H.~Awata and H.~Kanno, \emph{{Refined BPS state counting from Nekrasov's
  formula and Macdonald functions}},
  \href{http://dx.doi.org/10.1142/S0217751X09043006}{\emph{Int. J. Mod. Phys.}
  {\bf A24} (2009) 2253--2306}, [\href{http://arxiv.org/abs/0805.0191}{{\tt
  0805.0191}}].

\bibitem{Awata:2011ce}
H.~Awata, B.~Feigin and J.~Shiraishi, \emph{{Quantum Algebraic Approach to
  Refined Topological Vertex}},
  \href{http://dx.doi.org/10.1007/JHEP03(2012)041}{\emph{JHEP} {\bf 03} (2012)
  041}, [\href{http://arxiv.org/abs/1112.6074}{{\tt 1112.6074}}].

\bibitem{Nekrasov:2002qd}
N.~A. Nekrasov, \emph{{Seiberg-Witten prepotential from instanton counting}},
  {\emph{Adv.Theor.Math.Phys.} {\bf 7} (2004) 831--864},
  [\href{http://arxiv.org/abs/hep-th/0206161}{{\tt hep-th/0206161}}].

\bibitem{Nekrasov:2003rj}
N.~Nekrasov and A.~Okounkov, \emph{{Seiberg-Witten theory and random
  partitions}},  \href{http://arxiv.org/abs/hep-th/0306238}{{\tt
  hep-th/0306238}}.

\bibitem{Pestun:2007rz}
V.~Pestun, \emph{{Localization of gauge theory on a four-sphere and
  supersymmetric Wilson loops}},
  \href{http://dx.doi.org/10.1007/s00220-012-1485-0}{\emph{Commun. Math. Phys.}
  {\bf 313} (2012) 71--129}, [\href{http://arxiv.org/abs/0712.2824}{{\tt
  0712.2824}}].

\bibitem{Kapustin:2009kz}
A.~Kapustin, B.~Willett and I.~Yaakov, \emph{{Exact Results for Wilson Loops in
  Superconformal Chern-Simons Theories with Matter}},
  \href{http://dx.doi.org/10.1007/JHEP03(2010)089}{\emph{JHEP} {\bf 03} (2010)
  089}, [\href{http://arxiv.org/abs/0909.4559}{{\tt 0909.4559}}].

\bibitem{Kallen:2012cs}
J.~K{\"a}ll{\'e}n and M.~Zabzine, \emph{{Twisted supersymmetric 5D Yang-Mills
  theory and contact geometry}},
  \href{http://dx.doi.org/10.1007/JHEP05(2012)125}{\emph{JHEP} {\bf 05} (2012)
  125}, [\href{http://arxiv.org/abs/1202.1956}{{\tt 1202.1956}}].

\bibitem{Hosomichi:2012ek}
K.~Hosomichi, R.-K. Seong and S.~Terashima, \emph{{Supersymmetric Gauge
  Theories on the Five-Sphere}},
  \href{http://dx.doi.org/10.1016/j.nuclphysb.2012.08.007}{\emph{Nucl. Phys.}
  {\bf B865} (2012) 376--396}, [\href{http://arxiv.org/abs/1203.0371}{{\tt
  1203.0371}}].

\bibitem{Kallen:2012va}
J.~K{\"a}ll{\'e}n, J.~Qiu and M.~Zabzine, \emph{{The perturbative partition
  function of supersymmetric 5D Yang-Mills theory with matter on the
  five-sphere}}, \href{http://dx.doi.org/10.1007/JHEP08(2012)157}{\emph{JHEP}
  {\bf 08} (2012) 157}, [\href{http://arxiv.org/abs/1206.6008}{{\tt
  1206.6008}}].

\bibitem{Kim:2012ava}
H.-C. Kim and S.~Kim, \emph{{M5-branes from gauge theories on the 5-sphere}},
  \href{http://dx.doi.org/10.1007/JHEP05(2013)144}{\emph{JHEP} {\bf 05} (2013)
  144}, [\href{http://arxiv.org/abs/1206.6339}{{\tt 1206.6339}}].

\bibitem{Kim:2012gu}
H.-C. Kim, S.-S. Kim and K.~Lee, \emph{{5-dim Superconformal Index with
  Enhanced En Global Symmetry}},
  \href{http://dx.doi.org/10.1007/JHEP10(2012)142}{\emph{JHEP} {\bf 10} (2012)
  142}, [\href{http://arxiv.org/abs/1206.6781}{{\tt 1206.6781}}].

\bibitem{Terashima:2012ra}
S.~Terashima, \emph{{Supersymmetric gauge theories on $S^4$ x $S^1$}},
  \href{http://dx.doi.org/10.1103/PhysRevD.89.125001}{\emph{Phys. Rev.} {\bf
  D89} (2014) 125001}, [\href{http://arxiv.org/abs/1207.2163}{{\tt
  1207.2163}}].

\bibitem{Jafferis:2012iv}
D.~L. Jafferis and S.~S. Pufu, \emph{{Exact results for five-dimensional
  superconformal field theories with gravity duals}},
  \href{http://dx.doi.org/10.1007/JHEP05(2014)032}{\emph{JHEP} {\bf 05} (2014)
  032}, [\href{http://arxiv.org/abs/1207.4359}{{\tt 1207.4359}}].

\bibitem{Imamura:2012xg}
Y.~Imamura, \emph{{Supersymmetric theories on squashed five-sphere}},
  \href{http://dx.doi.org/10.1093/ptep/pts052}{\emph{PTEP} {\bf 2013} (2013)
  013B04}, [\href{http://arxiv.org/abs/1209.0561}{{\tt 1209.0561}}].

\bibitem{Iqbal:2012xm}
A.~Iqbal and C.~Vafa, \emph{{BPS Degeneracies and Superconformal Index in
  Diverse Dimensions}},
  \href{http://dx.doi.org/10.1103/PhysRevD.90.105031}{\emph{Phys. Rev.} {\bf
  D90} (2014) 105031}, [\href{http://arxiv.org/abs/1210.3605}{{\tt
  1210.3605}}].

\bibitem{Lockhart:2012vp}
G.~Lockhart and C.~Vafa, \emph{{Superconformal Partition Functions and
  Non-perturbative Topological Strings}},
  \href{http://arxiv.org/abs/1210.5909}{{\tt 1210.5909}}.

\bibitem{Imamura:2012bm}
Y.~Imamura, \emph{{Perturbative partition function for squashed $S^5$}},
  \href{http://dx.doi.org/10.1093/ptep/ptt044}{\emph{PTEP} {\bf 2013} (2013)
  073B01}, [\href{http://arxiv.org/abs/1210.6308}{{\tt 1210.6308}}].

\bibitem{Kim:2012qf}
H.-C. Kim, J.~Kim and S.~Kim, \emph{{Instantons on the 5-sphere and
  M5-branes}},  \href{http://arxiv.org/abs/1211.0144}{{\tt 1211.0144}}.

\bibitem{Assel:2012nf}
B.~Assel, J.~Estes and M.~Yamazaki, \emph{{Wilson Loops in 5d N=1 SCFTs and
  AdS/CFT}}, \href{http://dx.doi.org/10.1007/s00023-013-0249-5}{\emph{Annales
  Henri Poincare} {\bf 15} (2014) 589--632},
  [\href{http://arxiv.org/abs/1212.1202}{{\tt 1212.1202}}].

\bibitem{Bergman:2013ala}
O.~Bergman, D.~Rodr{\'\i}guez-G{\'o}mez and G.~Zafrir, \emph{{Discrete $\theta$
  and the 5d superconformal index}},
  \href{http://dx.doi.org/10.1007/JHEP01(2014)079}{\emph{JHEP} {\bf 01} (2014)
  079}, [\href{http://arxiv.org/abs/1310.2150}{{\tt 1310.2150}}].

\bibitem{Bao:2013pwa}
L.~Bao, V.~Mitev, E.~Pomoni, M.~Taki and F.~Yagi, \emph{{Non-Lagrangian
  Theories from Brane Junctions}},
  \href{http://dx.doi.org/10.1007/JHEP01(2014)175}{\emph{JHEP} {\bf 01} (2014)
  175}, [\href{http://arxiv.org/abs/1310.3841}{{\tt 1310.3841}}].

\bibitem{Hayashi:2013qwa}
H.~Hayashi, H.-C. Kim and T.~Nishinaka, \emph{{Topological strings and 5d $T_N$
  partition functions}},
  \href{http://dx.doi.org/10.1007/JHEP06(2014)014}{\emph{JHEP} {\bf 06} (2014)
  014}, [\href{http://arxiv.org/abs/1310.3854}{{\tt 1310.3854}}].

\bibitem{Taki:2014pba}
M.~Taki, \emph{{Seiberg Duality, 5d SCFTs and Nekrasov Partition Functions}},
  \href{http://arxiv.org/abs/1401.7200}{{\tt 1401.7200}}.

\bibitem{Alday:2014rxa}
L.~F. Alday, M.~Fluder, P.~Richmond and J.~Sparks, \emph{{Gravity Dual of
  Supersymmetric Gauge Theories on a Squashed Five-Sphere}},
  \href{http://dx.doi.org/10.1103/PhysRevLett.113.141601}{\emph{Phys. Rev.
  Lett.} {\bf 113} (2014) 141601}, [\href{http://arxiv.org/abs/1404.1925}{{\tt
  1404.1925}}].

\bibitem{Alday:2014bta}
L.~F. Alday, M.~Fluder, C.~M. Gregory, P.~Richmond and J.~Sparks,
  \emph{{Supersymmetric gauge theories on squashed five-spheres and their
  gravity duals}}, \href{http://dx.doi.org/10.1007/JHEP09(2014)067}{\emph{JHEP}
  {\bf 09} (2014) 067}, [\href{http://arxiv.org/abs/1405.7194}{{\tt
  1405.7194}}].

\bibitem{Hwang:2014uwa}
C.~Hwang, J.~Kim, S.~Kim and J.~Park, \emph{{General instanton counting and 5d
  SCFT}}, \href{http://dx.doi.org/10.1007/JHEP07(2015)063,
  10.1007/JHEP04(2016)094}{\emph{JHEP} {\bf 07} (2015) 063},
  [\href{http://arxiv.org/abs/1406.6793}{{\tt 1406.6793}}].

\bibitem{Mitev:2014jza}
V.~Mitev, E.~Pomoni, M.~Taki and F.~Yagi, \emph{{Fiber-Base Duality and Global
  Symmetry Enhancement}},
  \href{http://dx.doi.org/10.1007/JHEP04(2015)052}{\emph{JHEP} {\bf 04} (2015)
  052}, [\href{http://arxiv.org/abs/1411.2450}{{\tt 1411.2450}}].

\bibitem{Hayashi:2015xla}
H.~Hayashi and G.~Zoccarato, \emph{{Topological vertex for Higgsed 5d T$_{N}$
  theories}}, \href{http://dx.doi.org/10.1007/JHEP09(2015)023}{\emph{JHEP} {\bf
  09} (2015) 023}, [\href{http://arxiv.org/abs/1505.00260}{{\tt 1505.00260}}].

\bibitem{Bergman:2015dpa}
O.~Bergman and G.~Zafrir, \emph{{5d fixed points from brane webs and
  O7-planes}}, \href{http://dx.doi.org/10.1007/JHEP12(2015)163}{\emph{JHEP}
  {\bf 12} (2015) 163}, [\href{http://arxiv.org/abs/1507.03860}{{\tt
  1507.03860}}].

\bibitem{Alday:2015lta}
L.~F. Alday, P.~Benetti~Genolini, M.~Fluder, P.~Richmond and J.~Sparks,
  \emph{{Supersymmetric gauge theories on five-manifolds}},
  \href{http://dx.doi.org/10.1007/JHEP08(2015)007}{\emph{JHEP} {\bf 08} (2015)
  007}, [\href{http://arxiv.org/abs/1503.09090}{{\tt 1503.09090}}].

\bibitem{Alday:2015jsa}
L.~F. Alday, M.~Fluder, C.~M. Gregory, P.~Richmond and J.~Sparks,
  \emph{{Supersymmetric solutions to Euclidean Romans supergravity}},
  \href{http://dx.doi.org/10.1007/JHEP02(2016)100}{\emph{JHEP} {\bf 02} (2016)
  100}, [\href{http://arxiv.org/abs/1505.04641}{{\tt 1505.04641}}].

\bibitem{Gaiotto:2015una}
D.~Gaiotto and H.-C. Kim, \emph{{Duality walls and defects in 5d $
  \mathcal{N}=1 $ theories}},
  \href{http://dx.doi.org/10.1007/JHEP01(2017)019}{\emph{JHEP} {\bf 01} (2017)
  019}, [\href{http://arxiv.org/abs/1506.03871}{{\tt 1506.03871}}].

\bibitem{Zafrir:2015ftn}
G.~Zafrir, \emph{{Brane webs and $O5$-planes}},
  \href{http://dx.doi.org/10.1007/JHEP03(2016)109}{\emph{JHEP} {\bf 03} (2016)
  109}, [\href{http://arxiv.org/abs/1512.08114}{{\tt 1512.08114}}].

\bibitem{Chang:2016iji}
C.-M. Chang, O.~Ganor and J.~Oh, \emph{{An index for ray operators in 5d
  $E_{n}$ SCFTs}}, \href{http://dx.doi.org/10.1007/JHEP02(2017)018}{\emph{JHEP}
  {\bf 02} (2017) 018}, [\href{http://arxiv.org/abs/1608.06284}{{\tt
  1608.06284}}].

\bibitem{DHoker:2016ysh}
E.~D'Hoker, M.~Gutperle and C.~F. Uhlemann, \emph{{Holographic duals for
  five-dimensional superconformal quantum field theories}},
  \href{http://dx.doi.org/10.1103/PhysRevLett.118.101601}{\emph{Phys. Rev.
  Lett.} {\bf 118} (2017) 101601}, [\href{http://arxiv.org/abs/1611.09411}{{\tt
  1611.09411}}].

\bibitem{DHoker:2017mds}
E.~D'Hoker, M.~Gutperle and C.~F. Uhlemann, \emph{{Warped $AdS_6\times S^2$ in
  Type IIB supergravity II: Global solutions and five-brane webs}},
  \href{http://dx.doi.org/10.1007/JHEP05(2017)131}{\emph{JHEP} {\bf 05} (2017)
  131}, [\href{http://arxiv.org/abs/1703.08186}{{\tt 1703.08186}}].

\bibitem{Hayashi:2017jze}
H.~Hayashi and K.~Ohmori, \emph{{5d/6d DE instantons from trivalent gluing of
  web diagrams}}, \href{http://dx.doi.org/10.1007/JHEP06(2017)078}{\emph{JHEP}
  {\bf 06} (2017) 078}, [\href{http://arxiv.org/abs/1702.07263}{{\tt
  1702.07263}}].

\bibitem{Lambert:2014jna}
N.~Lambert, C.~Papageorgakis and M.~Schmidt-Sommerfeld, \emph{{Instanton
  Operators in Five-Dimensional Gauge Theories}},
  \href{http://dx.doi.org/10.1007/JHEP03(2015)019}{\emph{JHEP} {\bf 03} (2015)
  019}, [\href{http://arxiv.org/abs/1412.2789}{{\tt 1412.2789}}].

\bibitem{Tachikawa:2015mha}
Y.~Tachikawa, \emph{{Instanton operators and symmetry enhancement in 5d
  supersymmetric gauge theories}},
  \href{http://dx.doi.org/10.1093/ptep/ptv040}{\emph{PTEP} {\bf 2015} (2015)
  043B06}, [\href{http://arxiv.org/abs/1501.01031}{{\tt 1501.01031}}].

\bibitem{Zafrir:2015uaa}
G.~Zafrir, \emph{{Instanton operators and symmetry enhancement in 5d
  supersymmetric USp, SO and exceptional gauge theories}},
  \href{http://dx.doi.org/10.1007/JHEP07(2015)087}{\emph{JHEP} {\bf 07} (2015)
  087}, [\href{http://arxiv.org/abs/1503.08136}{{\tt 1503.08136}}].

\bibitem{Yonekura:2015ksa}
K.~Yonekura, \emph{{Instanton operators and symmetry enhancement in 5d
  supersymmetric quiver gauge theories}},
  \href{http://dx.doi.org/10.1007/JHEP07(2015)167}{\emph{JHEP} {\bf 07} (2015)
  167}, [\href{http://arxiv.org/abs/1505.04743}{{\tt 1505.04743}}].

\bibitem{Cremonesi:2015lsa}
S.~Cremonesi, G.~Ferlito, A.~Hanany and N.~Mekareeya, \emph{{Instanton
  Operators and the Higgs Branch at Infinite Coupling}},
  \href{http://dx.doi.org/10.1007/JHEP04(2017)042}{\emph{JHEP} {\bf 04} (2017)
  042}, [\href{http://arxiv.org/abs/1505.06302}{{\tt 1505.06302}}].

\bibitem{Bergman:2016avc}
O.~Bergman and D.~Rodriguez-Gomez, \emph{{A Note on Instanton Operators,
  Instanton Particles, and Supersymmetry}},
  \href{http://dx.doi.org/10.1007/JHEP05(2016)068}{\emph{JHEP} {\bf 05} (2016)
  068}, [\href{http://arxiv.org/abs/1601.00752}{{\tt 1601.00752}}].

\bibitem{Cardy:1988cwa}
J.~L. Cardy, \emph{{Is There a c Theorem in Four-Dimensions?}},
  \href{http://dx.doi.org/10.1016/0370-2693(88)90054-8}{\emph{Phys. Lett.} {\bf
  B215} (1988) 749--752}.

\bibitem{Capper:1974ic}
D.~M. Capper and M.~J. Duff, \emph{{Trace anomalies in dimensional
  regularization}}, \href{http://dx.doi.org/10.1007/BF02748300}{\emph{Nuovo
  Cim.} {\bf A23} (1974) 173--183}.

\bibitem{Zamolodchikov:1986gt}
A.~B. Zamolodchikov, \emph{{Irreversibility of the Flux of the Renormalization
  Group in a 2D Field Theory}}, {\emph{JETP Lett.} {\bf 43} (1986) 730--732}.

\bibitem{Komargodski:2011vj}
Z.~Komargodski and A.~Schwimmer, \emph{{On Renormalization Group Flows in Four
  Dimensions}}, \href{http://dx.doi.org/10.1007/JHEP12(2011)099}{\emph{JHEP}
  {\bf 12} (2011) 099}, [\href{http://arxiv.org/abs/1107.3987}{{\tt
  1107.3987}}].

\bibitem{Myers:2010tj}
R.~C. Myers and A.~Sinha, \emph{{Holographic c-theorems in arbitrary
  dimensions}}, \href{http://dx.doi.org/10.1007/JHEP01(2011)125}{\emph{JHEP}
  {\bf 01} (2011) 125}, [\href{http://arxiv.org/abs/1011.5819}{{\tt
  1011.5819}}].

\bibitem{Jafferis:2010un}
D.~L. Jafferis, \emph{{The Exact Superconformal R-Symmetry Extremizes Z}},
  \href{http://dx.doi.org/10.1007/JHEP05(2012)159}{\emph{JHEP} {\bf 05} (2012)
  159}, [\href{http://arxiv.org/abs/1012.3210}{{\tt 1012.3210}}].

\bibitem{Casini:2011kv}
H.~Casini, M.~Huerta and R.~C. Myers, \emph{{Towards a derivation of
  holographic entanglement entropy}},
  \href{http://dx.doi.org/10.1007/JHEP05(2011)036}{\emph{JHEP} {\bf 05} (2011)
  036}, [\href{http://arxiv.org/abs/1102.0440}{{\tt 1102.0440}}].

\bibitem{Jafferis:2011zi}
D.~L. Jafferis, I.~R. Klebanov, S.~S. Pufu and B.~R. Safdi, \emph{{Towards the
  F-Theorem: N=2 Field Theories on the Three-Sphere}},
  \href{http://dx.doi.org/10.1007/JHEP06(2011)102}{\emph{JHEP} {\bf 06} (2011)
  102}, [\href{http://arxiv.org/abs/1103.1181}{{\tt 1103.1181}}].

\bibitem{Klebanov:2011gs}
I.~R. Klebanov, S.~S. Pufu and B.~R. Safdi, \emph{{F-Theorem without
  Supersymmetry}}, \href{http://dx.doi.org/10.1007/JHEP10(2011)038}{\emph{JHEP}
  {\bf 10} (2011) 038}, [\href{http://arxiv.org/abs/1105.4598}{{\tt
  1105.4598}}].

\bibitem{Elvang:2012st}
H.~Elvang, D.~Z. Freedman, L.-Y. Hung, M.~Kiermaier, R.~C. Myers and
  S.~Theisen, \emph{{On renormalization group flows and the a-theorem in 6d}},
  \href{http://dx.doi.org/10.1007/JHEP10(2012)011}{\emph{JHEP} {\bf 10} (2012)
  011}, [\href{http://arxiv.org/abs/1205.3994}{{\tt 1205.3994}}].

\bibitem{Giombi:2014xxa}
S.~Giombi and I.~R. Klebanov, \emph{{Interpolating between $a$ and $F$}},
  \href{http://dx.doi.org/10.1007/JHEP03(2015)117}{\emph{JHEP} {\bf 03} (2015)
  117}, [\href{http://arxiv.org/abs/1409.1937}{{\tt 1409.1937}}].

\bibitem{Cordova:2015vwa}
C.~Cordova, T.~T. Dumitrescu and X.~Yin, \emph{{Higher Derivative Terms,
  Toroidal Compactification, and Weyl Anomalies in Six-Dimensional (2,0)
  Theories}},  \href{http://arxiv.org/abs/1505.03850}{{\tt 1505.03850}}.

\bibitem{Cordova:2015fha}
C.~Cordova, T.~T. Dumitrescu and K.~Intriligator, \emph{{Anomalies,
  renormalization group flows, and the a-theorem in six-dimensional (1, 0)
  theories}}, \href{http://dx.doi.org/10.1007/JHEP10(2016)080}{\emph{JHEP} {\bf
  10} (2016) 080}, [\href{http://arxiv.org/abs/1506.03807}{{\tt 1506.03807}}].

\bibitem{Pufu:2016zxm}
S.~S. Pufu, \emph{{The F-Theorem and F-Maximization}},
  \href{http://dx.doi.org/10.1088/1751-8121/aa6765}{\emph{J. Phys.} {\bf A50}
  (2017) 443008}, [\href{http://arxiv.org/abs/1608.02960}{{\tt 1608.02960}}].

\bibitem{Rattazzi:2008pe}
R.~Rattazzi, V.~S. Rychkov, E.~Tonni and A.~Vichi, \emph{{Bounding scalar
  operator dimensions in 4D CFT}},
  \href{http://dx.doi.org/10.1088/1126-6708/2008/12/031}{\emph{JHEP} {\bf 12}
  (2008) 031}, [\href{http://arxiv.org/abs/0807.0004}{{\tt 0807.0004}}].

\bibitem{Rychkov:2009ij}
V.~S. Rychkov and A.~Vichi, \emph{{Universal Constraints on Conformal Operator
  Dimensions}}, \href{http://dx.doi.org/10.1103/PhysRevD.80.045006}{\emph{Phys.
  Rev.} {\bf D80} (2009) 045006}, [\href{http://arxiv.org/abs/0905.2211}{{\tt
  0905.2211}}].

\bibitem{Poland:2010wg}
D.~Poland and D.~Simmons-Duffin, \emph{{Bounds on 4D Conformal and
  Superconformal Field Theories}},
  \href{http://dx.doi.org/10.1007/JHEP05(2011)017}{\emph{JHEP} {\bf 05} (2011)
  017}, [\href{http://arxiv.org/abs/1009.2087}{{\tt 1009.2087}}].

\bibitem{Poland:2011ey}
D.~Poland, D.~Simmons-Duffin and A.~Vichi, \emph{{Carving Out the Space of 4D
  CFTs}}, \href{http://dx.doi.org/10.1007/JHEP05(2012)110}{\emph{JHEP} {\bf 05}
  (2012) 110}, [\href{http://arxiv.org/abs/1109.5176}{{\tt 1109.5176}}].

\bibitem{ElShowk:2012ht}
S.~El-Showk, M.~F. Paulos, D.~Poland, S.~Rychkov, D.~Simmons-Duffin and
  A.~Vichi, \emph{{Solving the 3D Ising Model with the Conformal Bootstrap}},
  \href{http://dx.doi.org/10.1103/PhysRevD.86.025022}{\emph{Phys. Rev.} {\bf
  D86} (2012) 025022}, [\href{http://arxiv.org/abs/1203.6064}{{\tt
  1203.6064}}].

\bibitem{ElShowk:2012hu}
S.~El-Showk and M.~F. Paulos, \emph{{Bootstrapping Conformal Field Theories
  with the Extremal Functional Method}},
  \href{http://dx.doi.org/10.1103/PhysRevLett.111.241601}{\emph{Phys. Rev.
  Lett.} {\bf 111} (2013) 241601}, [\href{http://arxiv.org/abs/1211.2810}{{\tt
  1211.2810}}].

\bibitem{Beem:2013qxa}
C.~Beem, L.~Rastelli and B.~C. van Rees, \emph{{The $\mathcal N=4$
  Superconformal Bootstrap}},
  \href{http://dx.doi.org/10.1103/PhysRevLett.111.071601}{\emph{Phys. Rev.
  Lett.} {\bf 111} (2013) 071601}, [\href{http://arxiv.org/abs/1304.1803}{{\tt
  1304.1803}}].

\bibitem{Kos:2013tga}
F.~Kos, D.~Poland and D.~Simmons-Duffin, \emph{{Bootstrapping the $O(N)$ vector
  models}}, \href{http://dx.doi.org/10.1007/JHEP06(2014)091}{\emph{JHEP} {\bf
  06} (2014) 091}, [\href{http://arxiv.org/abs/1307.6856}{{\tt 1307.6856}}].

\bibitem{El-Showk:2014dwa}
S.~El-Showk, M.~F. Paulos, D.~Poland, S.~Rychkov, D.~Simmons-Duffin and
  A.~Vichi, \emph{{Solving the 3d Ising Model with the Conformal Bootstrap II.
  c-Minimization and Precise Critical Exponents}},
  \href{http://dx.doi.org/10.1007/s10955-014-1042-7}{\emph{J. Stat. Phys.} {\bf
  157} (2014) 869}, [\href{http://arxiv.org/abs/1403.4545}{{\tt 1403.4545}}].

\bibitem{Chester:2014fya}
S.~M. Chester, J.~Lee, S.~S. Pufu and R.~Yacoby, \emph{{The $ \mathcal{N}=8 $
  superconformal bootstrap in three dimensions}},
  \href{http://dx.doi.org/10.1007/JHEP09(2014)143}{\emph{JHEP} {\bf 09} (2014)
  143}, [\href{http://arxiv.org/abs/1406.4814}{{\tt 1406.4814}}].

\bibitem{Kos:2014bka}
F.~Kos, D.~Poland and D.~Simmons-Duffin, \emph{{Bootstrapping Mixed Correlators
  in the 3D Ising Model}},
  \href{http://dx.doi.org/10.1007/JHEP11(2014)109}{\emph{JHEP} {\bf 11} (2014)
  109}, [\href{http://arxiv.org/abs/1406.4858}{{\tt 1406.4858}}].

\bibitem{Caracciolo:2014cxa}
F.~Caracciolo, A.~Castedo~Echeverri, B.~von Harling and M.~Serone,
  \emph{{Bounds on OPE Coefficients in 4D Conformal Field Theories}},
  \href{http://dx.doi.org/10.1007/JHEP10(2014)020}{\emph{JHEP} {\bf 10} (2014)
  020}, [\href{http://arxiv.org/abs/1406.7845}{{\tt 1406.7845}}].

\bibitem{Chester:2014mea}
S.~M. Chester, J.~Lee, S.~S. Pufu and R.~Yacoby, \emph{{Exact Correlators of
  BPS Operators from the 3d Superconformal Bootstrap}},
  \href{http://dx.doi.org/10.1007/JHEP03(2015)130}{\emph{JHEP} {\bf 03} (2015)
  130}, [\href{http://arxiv.org/abs/1412.0334}{{\tt 1412.0334}}].

\bibitem{Beem:2014zpa}
C.~Beem, M.~Lemos, P.~Liendo, L.~Rastelli and B.~C. van Rees, \emph{{The $
  \mathcal{N}=2 $ superconformal bootstrap}},
  \href{http://dx.doi.org/10.1007/JHEP03(2016)183}{\emph{JHEP} {\bf 03} (2016)
  183}, [\href{http://arxiv.org/abs/1412.7541}{{\tt 1412.7541}}].

\bibitem{Simmons-Duffin:2015qma}
D.~Simmons-Duffin, \emph{{A Semidefinite Program Solver for the Conformal
  Bootstrap}}, \href{http://dx.doi.org/10.1007/JHEP06(2015)174}{\emph{JHEP}
  {\bf 06} (2015) 174}, [\href{http://arxiv.org/abs/1502.02033}{{\tt
  1502.02033}}].

\bibitem{Kos:2015mba}
F.~Kos, D.~Poland, D.~Simmons-Duffin and A.~Vichi, \emph{{Bootstrapping the
  O(N) Archipelago}},
  \href{http://dx.doi.org/10.1007/JHEP11(2015)106}{\emph{JHEP} {\bf 11} (2015)
  106}, [\href{http://arxiv.org/abs/1504.07997}{{\tt 1504.07997}}].

\bibitem{Chester:2015qca}
S.~M. Chester, S.~Giombi, L.~V. Iliesiu, I.~R. Klebanov, S.~S. Pufu and
  R.~Yacoby, \emph{{Accidental Symmetries and the Conformal Bootstrap}},
  \href{http://dx.doi.org/10.1007/JHEP01(2016)110}{\emph{JHEP} {\bf 01} (2016)
  110}, [\href{http://arxiv.org/abs/1507.04424}{{\tt 1507.04424}}].

\bibitem{Iliesiu:2015qra}
L.~Iliesiu, F.~Kos, D.~Poland, S.~S. Pufu, D.~Simmons-Duffin and R.~Yacoby,
  \emph{{Bootstrapping 3D Fermions}},
  \href{http://dx.doi.org/10.1007/JHEP03(2016)120}{\emph{JHEP} {\bf 03} (2016)
  120}, [\href{http://arxiv.org/abs/1508.00012}{{\tt 1508.00012}}].

\bibitem{Chang:2015qfa}
C.-M. Chang and Y.-H. Lin, \emph{{Bootstrapping 2D CFTs in the Semiclassical
  Limit}}, \href{http://dx.doi.org/10.1007/JHEP08(2016)056}{\emph{JHEP} {\bf
  08} (2016) 056}, [\href{http://arxiv.org/abs/1510.02464}{{\tt 1510.02464}}].

\bibitem{Lemos:2015awa}
M.~Lemos and P.~Liendo, \emph{{Bootstrapping $ \mathcal{N}=2 $ chiral
  correlators}}, \href{http://dx.doi.org/10.1007/JHEP01(2016)025}{\emph{JHEP}
  {\bf 01} (2016) 025}, [\href{http://arxiv.org/abs/1510.03866}{{\tt
  1510.03866}}].

\bibitem{Kim:2015oca}
H.~Kim, P.~Kravchuk and H.~Ooguri, \emph{{Reflections on Conformal Spectra}},
  \href{http://dx.doi.org/10.1007/JHEP04(2016)184}{\emph{JHEP} {\bf 04} (2016)
  184}, [\href{http://arxiv.org/abs/1510.08772}{{\tt 1510.08772}}].

\bibitem{Lin:2015wcg}
Y.-H. Lin, S.-H. Shao, D.~Simmons-Duffin, Y.~Wang and X.~Yin, \emph{{N=4
  Superconformal Bootstrap of the K3 CFT}},
  \href{http://arxiv.org/abs/1511.04065}{{\tt 1511.04065}}.

\bibitem{Kos:2016ysd}
F.~Kos, D.~Poland, D.~Simmons-Duffin and A.~Vichi, \emph{{Precision islands in
  the Ising and O(N ) models}},
  \href{http://dx.doi.org/10.1007/JHEP08(2016)036}{\emph{JHEP} {\bf 08} (2016)
  036}, [\href{http://arxiv.org/abs/1603.04436}{{\tt 1603.04436}}].

\bibitem{Chang:2016ftb}
C.-M. Chang and Y.-H. Lin, \emph{{Bootstrap, universality and horizons}},
  \href{http://dx.doi.org/10.1007/JHEP10(2016)068}{\emph{JHEP} {\bf 10} (2016)
  068}, [\href{http://arxiv.org/abs/1604.01774}{{\tt 1604.01774}}].

\bibitem{Collier:2016cls}
S.~Collier, Y.-H. Lin and X.~Yin, \emph{{Modular Bootstrap Revisited}},
  \href{http://arxiv.org/abs/1608.06241}{{\tt 1608.06241}}.

\bibitem{Lin:2016gcl}
Y.-H. Lin, S.-H. Shao, Y.~Wang and X.~Yin, \emph{{(2,2) Superconformal
  Bootstrap in Two Dimensions}},  \href{http://arxiv.org/abs/1610.05371}{{\tt
  1610.05371}}.

\bibitem{Lemos:2016xke}
M.~Lemos, P.~Liendo, C.~Meneghelli and V.~Mitev, \emph{{Bootstrapping
  $\mathcal{N}=3$ superconformal theories}},
  \href{http://dx.doi.org/10.1007/JHEP04(2017)032}{\emph{JHEP} {\bf 04} (2017)
  032}, [\href{http://arxiv.org/abs/1612.01536}{{\tt 1612.01536}}].

\bibitem{Kravchuk:2016qvl}
P.~Kravchuk and D.~Simmons-Duffin, \emph{{Counting Conformal Correlators}},
  \href{http://arxiv.org/abs/1612.08987}{{\tt 1612.08987}}.

\bibitem{Li:2017ddj}
D.~Li, D.~Meltzer and A.~Stergiou, \emph{{Bootstrapping Mixed Correlators in 4D
  $\mathcal{N}=1$ SCFTs}},  \href{http://arxiv.org/abs/1702.00404}{{\tt
  1702.00404}}.

\bibitem{Collier:2017shs}
S.~Collier, P.~Kravchuk, Y.-H. Lin and X.~Yin, \emph{{Bootstrapping the
  Spectral Function: On the Uniqueness of Liouville and the Universality of
  BTZ}},  \href{http://arxiv.org/abs/1702.00423}{{\tt 1702.00423}}.

\bibitem{Li:2017agi}
W.~Li, \emph{{Inverse Bootstrapping Conformal Field Theories}},
  \href{http://arxiv.org/abs/1706.04054}{{\tt 1706.04054}}.

\bibitem{Cuomo:2017wme}
G.~F. Cuomo, D.~Karateev and P.~Kravchuk, \emph{{General Bootstrap Equations in
  4D CFTs}},  \href{http://arxiv.org/abs/1705.05401}{{\tt 1705.05401}}.

\bibitem{Karateev:2017jgd}
D.~Karateev, P.~Kravchuk and D.~Simmons-Duffin, \emph{{Weight Shifting
  Operators and Conformal Blocks}},
  \href{http://arxiv.org/abs/1706.07813}{{\tt 1706.07813}}.

\bibitem{Dymarsky:2017yzx}
A.~Dymarsky, F.~Kos, P.~Kravchuk, D.~Poland and D.~Simmons-Duffin, \emph{{The
  3d Stress-Tensor Bootstrap}},  \href{http://arxiv.org/abs/1708.05718}{{\tt
  1708.05718}}.

\bibitem{Kravchuk:2017dzd}
P.~Kravchuk, \emph{{Casimir recursion relations for general conformal blocks}},
   \href{http://arxiv.org/abs/1709.05347}{{\tt 1709.05347}}.

\bibitem{Nakayama:2014yia}
Y.~Nakayama and T.~Ohtsuki, \emph{{Five dimensional $O(N)$-symmetric CFTs from
  conformal bootstrap}},
  \href{http://dx.doi.org/10.1016/j.physletb.2014.05.058}{\emph{Phys. Lett.}
  {\bf B734} (2014) 193--197}, [\href{http://arxiv.org/abs/1404.5201}{{\tt
  1404.5201}}].

\bibitem{Bae:2014hia}
J.-B. Bae and S.-J. Rey, \emph{{Conformal Bootstrap Approach to O(N) Fixed
  Points in Five Dimensions}},  \href{http://arxiv.org/abs/1412.6549}{{\tt
  1412.6549}}.

\bibitem{Chester:2014gqa}
S.~M. Chester, S.~S. Pufu and R.~Yacoby, \emph{{Bootstrapping $O(N)$ vector
  models in 4 $< d <$ 6}},
  \href{http://dx.doi.org/10.1103/PhysRevD.91.086014}{\emph{Phys. Rev.} {\bf
  D91} (2015) 086014}, [\href{http://arxiv.org/abs/1412.7746}{{\tt
  1412.7746}}].

\bibitem{Li:2016wdp}
Z.~Li and N.~Su, \emph{{Bootstrapping Mixed Correlators in the Five Dimensional
  Critical O(N) Models}},  \href{http://arxiv.org/abs/1607.07077}{{\tt
  1607.07077}}.

\bibitem{Beem:2015aoa}
C.~Beem, M.~Lemos, L.~Rastelli and B.~C. van Rees, \emph{{The (2, 0)
  superconformal bootstrap}},
  \href{http://dx.doi.org/10.1103/PhysRevD.93.025016}{\emph{Phys. Rev.} {\bf
  D93} (2016) 025016}, [\href{http://arxiv.org/abs/1507.05637}{{\tt
  1507.05637}}].

\bibitem{Chang:2017xmr}
C.-M. Chang and Y.-H. Lin, \emph{{Carving Out the End of the World or
  (Superconformal Bootstrap in Six Dimensions)}},
  \href{http://dx.doi.org/10.1007/JHEP08(2017)128}{\emph{JHEP} {\bf 08} (2017)
  128}, [\href{http://arxiv.org/abs/1705.05392}{{\tt 1705.05392}}].

\bibitem{Closset:2012ru}
C.~Closset, T.~T. Dumitrescu, G.~Festuccia and Z.~Komargodski,
  \emph{{Supersymmetric Field Theories on Three-Manifolds}},
  \href{http://dx.doi.org/10.1007/JHEP05(2013)017}{\emph{JHEP} {\bf 05} (2013)
  017}, [\href{http://arxiv.org/abs/1212.3388}{{\tt 1212.3388}}].

\bibitem{Bobev:2017asb}
N.~Bobev, P.~Bueno and Y.~Vreys, \emph{{Comments on Squashed-sphere Partition
  Functions}}, \href{http://dx.doi.org/10.1007/JHEP07(2017)093}{\emph{JHEP}
  {\bf 07} (2017) 093}, [\href{http://arxiv.org/abs/1705.00292}{{\tt
  1705.00292}}].

\bibitem{Closset:2012vg}
C.~Closset, T.~T. Dumitrescu, G.~Festuccia, Z.~Komargodski and N.~Seiberg,
  \emph{{Contact Terms, Unitarity, and F-Maximization in Three-Dimensional
  Superconformal Theories}},
  \href{http://dx.doi.org/10.1007/JHEP10(2012)053}{\emph{JHEP} {\bf 10} (2012)
  053}, [\href{http://arxiv.org/abs/1205.4142}{{\tt 1205.4142}}].

\bibitem{Giombi:2015haa}
S.~Giombi, I.~R. Klebanov and G.~Tarnopolsky, \emph{{Conformal QED$_d$,
  $F$-Theorem and the $\epsilon$ Expansion}},
  \href{http://dx.doi.org/10.1088/1751-8113/49/13/135403}{\emph{J. Phys.} {\bf
  A49} (2016) 135403}, [\href{http://arxiv.org/abs/1508.06354}{{\tt
  1508.06354}}].

\bibitem{Diab:2016spb}
K.~Diab, L.~Fei, S.~Giombi, I.~R. Klebanov and G.~Tarnopolsky, \emph{{On
  ${C}_{J}$ and ${C}_{T}$ in the Gross?Neveu and O(N) models}},
  \href{http://dx.doi.org/10.1088/1751-8113/49/40/405402}{\emph{J. Phys.} {\bf
  A49} (2016) 405402}, [\href{http://arxiv.org/abs/1601.07198}{{\tt
  1601.07198}}].

\bibitem{Giombi:2016fct}
S.~Giombi, G.~Tarnopolsky and I.~R. Klebanov, \emph{{On $C_{J}$ and $C_{T}$ in
  Conformal QED}}, \href{http://dx.doi.org/10.1007/JHEP08(2016)156}{\emph{JHEP}
  {\bf 08} (2016) 156}, [\href{http://arxiv.org/abs/1602.01076}{{\tt
  1602.01076}}].

\bibitem{Festuccia:2011ws}
G.~Festuccia and N.~Seiberg, \emph{{Rigid Supersymmetric Theories in Curved
  Superspace}}, \href{http://dx.doi.org/10.1007/JHEP06(2011)114}{\emph{JHEP}
  {\bf 06} (2011) 114}, [\href{http://arxiv.org/abs/1105.0689}{{\tt
  1105.0689}}].

\bibitem{Closset:2012vp}
C.~Closset, T.~T. Dumitrescu, G.~Festuccia, Z.~Komargodski and N.~Seiberg,
  \emph{{Comments on Chern-Simons Contact Terms in Three Dimensions}},
  \href{http://dx.doi.org/10.1007/JHEP09(2012)091}{\emph{JHEP} {\bf 09} (2012)
  091}, [\href{http://arxiv.org/abs/1206.5218}{{\tt 1206.5218}}].

\bibitem{Gerchkovitz:2014gta}
E.~Gerchkovitz, J.~Gomis and Z.~Komargodski, \emph{{Sphere Partition Functions
  and the Zamolodchikov Metric}},
  \href{http://dx.doi.org/10.1007/JHEP11(2014)001}{\emph{JHEP} {\bf 11} (2014)
  001}, [\href{http://arxiv.org/abs/1405.7271}{{\tt 1405.7271}}].

\bibitem{Bergshoeff:2004kh}
E.~Bergshoeff, S.~Cucu, T.~de~Wit, J.~Gheerardyn, S.~Vandoren and
  A.~Van~Proeyen, \emph{{N = 2 supergravity in five-dimensions revisited}},
  \href{http://dx.doi.org/10.1088/0264-9381/23/23/C01,
  10.1088/0264-9381/21/12/013}{\emph{Class. Quant. Grav.} {\bf 21} (2004)
  3015--3042}, [\href{http://arxiv.org/abs/hep-th/0403045}{{\tt
  hep-th/0403045}}].

\bibitem{Ozkan:2013nwa}
M.~Ozkan and Y.~Pang, \emph{{All off-shell $R^{2}$ invariants in five
  dimensional $\mathcal{N} =$ 2 supergravity}},
  \href{http://dx.doi.org/10.1007/JHEP08(2013)042}{\emph{JHEP} {\bf 08} (2013)
  042}, [\href{http://arxiv.org/abs/1306.1540}{{\tt 1306.1540}}].

\bibitem{Zucker:1999ej}
M.~Zucker, \emph{{Minimal off-shell supergravity in five-dimensions}},
  \href{http://dx.doi.org/10.1016/S0550-3213(99)00750-6}{\emph{Nucl. Phys.}
  {\bf B570} (2000) 267--283}, [\href{http://arxiv.org/abs/hep-th/9907082}{{\tt
  hep-th/9907082}}].

\bibitem{Kugo:2000af}
T.~Kugo and K.~Ohashi, \emph{{Off-shell D = 5 supergravity coupled to matter
  Yang-Mills system}}, \href{http://dx.doi.org/10.1143/PTP.105.323}{\emph{Prog.
  Theor. Phys.} {\bf 105} (2001) 323--353},
  [\href{http://arxiv.org/abs/hep-ph/0010288}{{\tt hep-ph/0010288}}].

\bibitem{Kugo:2000hn}
T.~Kugo and K.~Ohashi, \emph{{Supergravity tensor calculus in 5-D from 6-D}},
  \href{http://dx.doi.org/10.1143/PTP.104.835}{\emph{Prog. Theor. Phys.} {\bf
  104} (2000) 835--865}, [\href{http://arxiv.org/abs/hep-ph/0006231}{{\tt
  hep-ph/0006231}}].

\bibitem{Bergshoeff:2011xn}
E.~A. Bergshoeff, J.~Rosseel and E.~Sezgin, \emph{{Off-shell D=5, N=2 Riemann
  Squared Supergravity}},
  \href{http://dx.doi.org/10.1088/0264-9381/28/22/225016}{\emph{Class. Quant.
  Grav.} {\bf 28} (2011) 225016}, [\href{http://arxiv.org/abs/1107.2825}{{\tt
  1107.2825}}].

\bibitem{Ozkan:2013uk}
M.~Ozkan and Y.~Pang, \emph{{Supersymmetric Completion of Gauss-Bonnet
  Combination in Five Dimensions}},
  \href{http://dx.doi.org/10.1007/JHEP03(2013)158,
  10.1007/JHEP07(2013)152}{\emph{JHEP} {\bf 03} (2013) 158},
  [\href{http://arxiv.org/abs/1301.6622}{{\tt 1301.6622}}].

\bibitem{Butter:2014xxa}
D.~Butter, S.~M. Kuzenko, J.~Novak and G.~Tartaglino-Mazzucchelli,
  \emph{{Conformal supergravity in five dimensions: New approach and
  applications}}, \href{http://dx.doi.org/10.1007/JHEP02(2015)111}{\emph{JHEP}
  {\bf 02} (2015) 111}, [\href{http://arxiv.org/abs/1410.8682}{{\tt
  1410.8682}}].

\bibitem{Hanaki:2006pj}
K.~Hanaki, K.~Ohashi and Y.~Tachikawa, \emph{{Supersymmetric Completion of an
  R**2 term in Five-dimensional Supergravity}},
  \href{http://dx.doi.org/10.1143/PTP.117.533}{\emph{Prog. Theor. Phys.} {\bf
  117} (2007) 533}, [\href{http://arxiv.org/abs/hep-th/0611329}{{\tt
  hep-th/0611329}}].

\bibitem{Fujita:2001kv}
T.~Fujita and K.~Ohashi, \emph{{Superconformal tensor calculus in
  five-dimensions}}, \href{http://dx.doi.org/10.1143/PTP.106.221}{\emph{Prog.
  Theor. Phys.} {\bf 106} (2001) 221--247},
  [\href{http://arxiv.org/abs/hep-th/0104130}{{\tt hep-th/0104130}}].

\bibitem{Gerchkovitz:2016gxx}
E.~Gerchkovitz, J.~Gomis, N.~Ishtiaque, A.~Karasik, Z.~Komargodski and S.~S.
  Pufu, \emph{{Correlation Functions of Coulomb Branch Operators}},
  \href{http://dx.doi.org/10.1007/JHEP01(2017)103}{\emph{JHEP} {\bf 01} (2017)
  103}, [\href{http://arxiv.org/abs/1602.05971}{{\tt 1602.05971}}].

\bibitem{Chang1}
C.-M. Chang, M.~Fluder, Y.-H. Lin and Y.~Wang, ``Work in progress.''.

\bibitem{Osborn:1993cr}
H.~Osborn and A.~C. Petkou, \emph{{Implications of conformal invariance in
  field theories for general dimensions}},
  \href{http://dx.doi.org/10.1006/aphy.1994.1045}{\emph{Annals Phys.} {\bf 231}
  (1994) 311--362}, [\href{http://arxiv.org/abs/hep-th/9307010}{{\tt
  hep-th/9307010}}].

\bibitem{Buican:2016hpb}
M.~Buican, J.~Hayling and C.~Papageorgakis, \emph{{Aspects of Superconformal
  Multiplets in $D>4$}},
  \href{http://dx.doi.org/10.1007/JHEP11(2016)091}{\emph{JHEP} {\bf 11} (2016)
  091}, [\href{http://arxiv.org/abs/1606.00810}{{\tt 1606.00810}}].

\bibitem{Cordova:2016emh}
C.~Cordova, T.~T. Dumitrescu and K.~Intriligator, \emph{{Multiplets of
  Superconformal Symmetry in Diverse Dimensions}},
  \href{http://arxiv.org/abs/1612.00809}{{\tt 1612.00809}}.

\bibitem{Chang:2017mxc}
C.-M. Chang, M.~Fluder, Y.-H. Lin and Y.~Wang, \emph{{Romans Supergravity from
  Five-Dimensional Holograms}},  \href{http://arxiv.org/abs/1712.10313}{{\tt
  1712.10313}}.

\bibitem{Movshev:2003ib}
M.~Movshev and A.~S. Schwarz, \emph{{On maximally supersymmetric Yang-Mills
  theories}},
  \href{http://dx.doi.org/10.1016/j.nuclphysb.2003.12.033}{\emph{Nucl. Phys.}
  {\bf B681} (2004) 324--350}, [\href{http://arxiv.org/abs/hep-th/0311132}{{\tt
  hep-th/0311132}}].

\bibitem{Movshev:2004aw}
M.~Movshev and A.~S. Schwarz, \emph{{Algebraic structure of Yang-Mills
  theory}}, \href{http://dx.doi.org/10.1007/0-8176-4467-9_14}{\emph{Prog.
  Math.} {\bf 244} (2006) 473--523},
  [\href{http://arxiv.org/abs/hep-th/0404183}{{\tt hep-th/0404183}}].

\bibitem{Movshev:2005ei}
M.~Movshev, \emph{{Deformation of maximally supersymmetric Yang-Mills theory in
  dimensions 10. An Algebraic approach}},
  \href{http://arxiv.org/abs/hep-th/0601010}{{\tt hep-th/0601010}}.

\bibitem{Movshev:2009ba}
M.~Movshev and A.~Schwarz, \emph{{Supersymmetric Deformations of Maximally
  Supersymmetric Gauge Theories}},
  \href{http://dx.doi.org/10.1007/JHEP09(2012)136}{\emph{JHEP} {\bf 09} (2012)
  136}, [\href{http://arxiv.org/abs/0910.0620}{{\tt 0910.0620}}].

\bibitem{Chang:2014kma}
C.-M. Chang, Y.-H. Lin, Y.~Wang and X.~Yin, \emph{{Deformations with Maximal
  Supersymmetries Part 1: On-shell Formulation}},
  \href{http://arxiv.org/abs/1403.0545}{{\tt 1403.0545}}.

\bibitem{Qiu:2016dyj}
J.~Qiu and M.~Zabzine, \emph{{Review of localization for 5d supersymmetric
  gauge theories}}, \href{http://dx.doi.org/10.1088/1751-8121/aa5ef0}{\emph{J.
  Phys.} {\bf A50} (2017) 443014}, [\href{http://arxiv.org/abs/1608.02966}{{\tt
  1608.02966}}].

\bibitem{Kim:2016usy}
S.~Kim and K.~Lee, \emph{{Indices for 6 dimensional superconformal field
  theories}}, \href{http://dx.doi.org/10.1088/1751-8121/aa5cbf}{\emph{J. Phys.}
  {\bf A50} (2017) 443017}, [\href{http://arxiv.org/abs/1608.02969}{{\tt
  1608.02969}}].

\bibitem{Nishioka:2013gza}
T.~Nishioka and K.~Yonekura, \emph{{On RG Flow of $tau_{RR}$ for Supersymmetric
  Field Theories in Three-Dimensions}},
  \href{http://dx.doi.org/10.1007/JHEP05(2013)165}{\emph{JHEP} {\bf 05} (2013)
  165}, [\href{http://arxiv.org/abs/1303.1522}{{\tt 1303.1522}}].

\bibitem{Bobev:2017jhk}
N.~Bobev, E.~Lauria and D.~Mazac, \emph{{Superconformal Blocks for SCFTs with
  Eight Supercharges}},  \href{http://arxiv.org/abs/1705.08594}{{\tt
  1705.08594}}.

\bibitem{cvitanovic2008group}
P.~Cvitanovi{\'c}, \emph{Group Theory: Birdtracks, Lies, and Exceptional
  Groups}.
\newblock Princeton University Press, 41 William St, Princeton, NJ 08540 USA,
  2008.

\bibitem{Joseph1976}
A.~Joseph, \emph{The minimal orbit in a simple lie algebra and its associated
  maximal ideal}, .

\bibitem{Pini:2015xha}
A.~Pini, D.~Rodriguez-Gomez and J.~Schmude, \emph{{Rigid Supersymmetry from
  Conformal Supergravity in Five Dimensions}},
  \href{http://dx.doi.org/10.1007/JHEP09(2015)118}{\emph{JHEP} {\bf 09} (2015)
  118}, [\href{http://arxiv.org/abs/1504.04340}{{\tt 1504.04340}}].

\bibitem{Minwalla:1997ka}
S.~Minwalla, \emph{{Restrictions imposed by superconformal invariance on
  quantum field theories}}, {\emph{Adv. Theor. Math. Phys.} {\bf 2} (1998)
  781--846}, [\href{http://arxiv.org/abs/hep-th/9712074}{{\tt
  hep-th/9712074}}].

\bibitem{Bhattacharya:2008zy}
J.~Bhattacharya, S.~Bhattacharyya, S.~Minwalla and S.~Raju, \emph{{Indices for
  Superconformal Field Theories in 3,5 and 6 Dimensions}},
  \href{http://dx.doi.org/10.1088/1126-6708/2008/02/064}{\emph{JHEP} {\bf 02}
  (2008) 064}, [\href{http://arxiv.org/abs/0801.1435}{{\tt 0801.1435}}].

\bibitem{Faddeev:1995nb}
L.~D. Faddeev, \emph{{Discrete Heisenberg-Weyl group and modular group}},
  \href{http://dx.doi.org/10.1007/BF01872779}{\emph{Lett. Math. Phys.} {\bf 34}
  (1995) 249--254}, [\href{http://arxiv.org/abs/hep-th/9504111}{{\tt
  hep-th/9504111}}].

\bibitem{Narukawa:aa}
A.~Narukawa, \emph{The modular properties and the integral representations of
  the multiple elliptic gamma functions},
  \href{http://arxiv.org/abs/math/0306164}{{\tt math/0306164}}.

\bibitem{Faddeev:2012zu}
L.~D. Faddeev, \emph{{Volkov's Pentagon for the Modular Quantum Dilogarithm}},
  \href{http://dx.doi.org/10.1007/s10688-011-0031-8}{\emph{Funct. Anal. Appl.}
  {\bf 45} (2011) 291}, [\href{http://arxiv.org/abs/1201.6464}{{\tt
  1201.6464}}].

\bibitem{Nieri:2013vba}
F.~Nieri, S.~Pasquetti, F.~Passerini and A.~Torrielli, \emph{{5D partition
  functions, q-Virasoro systems and integrable spin-chains}},
  \href{http://dx.doi.org/10.1007/JHEP12(2014)040}{\emph{JHEP} {\bf 12} (2014)
  040}, [\href{http://arxiv.org/abs/1312.1294}{{\tt 1312.1294}}].

\bibitem{Pasquetti:2016dyl}
S.~Pasquetti, \emph{{Holomorphic blocks and the 5d AGT correspondence}},
  \href{http://dx.doi.org/10.1088/1751-8121/aa60fe}{\emph{J. Phys.} {\bf A50}
  (2017) 443016}, [\href{http://arxiv.org/abs/1608.02968}{{\tt 1608.02968}}].

\bibitem{Cordova:2014oxa}
C.~Cordova and S.-H. Shao, \emph{{An Index Formula for Supersymmetric Quantum
  Mechanics}},  \href{http://arxiv.org/abs/1406.7853}{{\tt 1406.7853}}.

\bibitem{Benini:2013xpa}
F.~Benini, R.~Eager, K.~Hori and Y.~Tachikawa, \emph{{Elliptic Genera of 2d
  ${\mathcal{N}}$ = 2 Gauge Theories}},
  \href{http://dx.doi.org/10.1007/s00220-014-2210-y}{\emph{Commun. Math. Phys.}
  {\bf 333} (2015) 1241--1286}, [\href{http://arxiv.org/abs/1308.4896}{{\tt
  1308.4896}}].

\end{thebibliography}\endgroup
\bibliographystyle{JHEP}

\end{document}